\documentclass{elsart}
\usepackage{natbib}
\usepackage{latexsym}
\usepackage{graphicx}
\usepackage{amsmath}
\usepackage{amssymb}
\usepackage{subfigure}

\usepackage{epsf}

\def\<{\langle}
\def\>{\rangle}

\newcounter{step}

\def\centerps#1#2#3{\vskip#2\relax\centerline{\hbox to#1{\special
  {ps:#3 x=#1, y=#2}\hfil}}}

\def\q{{\bf q}}
\def\0{{\bf 0}}

\newtheorem{theorem}{Theorem}

\newtheorem{proposition}{Proposition}

\newcommand{\raisenumber}[3]{\hspace{#1}\raisebox{#2}{#3}}

\newcommand{\raisenumbertwo}[3]{\hspace{#1} \raisebox{#2}{#3}}

\newcommand{\boundnum}[1]{\hspace{-10pt}\rule[#1]{11pt}{0.3pt} \hspace{-5.3pt}
\rule[#1]{0.3pt}{9.5pt} }

\newcommand{\mA}{\mathcal{A}}
\newcommand{\mB}{\mathcal{B}}
\newcommand{\Id}{\textrm{Id}}
\newcommand{\Lower}[1]{\smash{\lower 1.5ex \hbox{#1}}}
\newcommand{\MLower}[1]{\smash{\lower 4.5ex \hbox{#1}}}
\newcommand{\TLower}[1]{\smash{\lower 2.5ex \hbox{#1}}}

\begin{document}

\begin{frontmatter}

\date{26 December 2006}
\title{Signature Sequence of\\
Intersection Curve of Two Quadrics for\\
Exact Morphological Classification}

\author[SD]{Changhe Tu}
%\ead{chtu@sdu.edu.cn}
\author[HK]{Wenping Wang\corauthref{cor}}
\corauth[cor]{Corresponding author, Department of Computer Science,
The University of Hong Kong, Pokfulam Road, Hong Kong, China.} \ead
{wenping@cs.hku.hk}
\author[INRIA]{Bernard Mourrain}
\author[SD]{Jiaye Wang}
\address[SD]{Shandong University}
\address[HK]{University of Hong Kong}
\address[INRIA]{INRIA}
%\thanks{The work is ...}

\begin{abstract}
%Computing the intersection curve of two quadric
%surfaces is an important problem in geometric computation, ranging
%from shape modeling in computer graphics and CAD/CAM, collision
%detection in robotics and computational physics, to arrangement
%computation in computational geometry.
%
We present an efficient method for classifying the morphology of the
intersection curve of two quadrics (QSIC) in $\mathbb{PR}^3$, 3D
real projective space; here, the term {\em morphology} is used in a
broad sense to mean the shape, topological, and algebraic properties
of a QSIC, including singularity, reducibility, the number of
connected components, and the degree of each irreducible component,
etc.
There are in total 35 different QSIC morphologies with
non-degenerate quadric pencils.
For each of these 35 QSIC morphologies, through a detailed study of
the eigenvalue curve and the index function jump we establish a
characterizing algebraic condition expressed in terms of the Segre
characteristics and the signature sequence of a quadric pencil.
We show how to compute a signature sequence with rational arithmetic
so as to determine the morphology of the intersection curve of any
two given quadrics.
Two immediate applications of our results are the robust topological
classification of QSIC in computing B-rep surface representation in
solid modeling and the derivation of algebraic conditions for
collision detection of quadric primitives.
\end{abstract}

%\category{I.3.5}{Computer Graphics}{Computational Geometry and
%Object Modeling}[Curve, surface, solid, and object representations]

%\category{J.6}{Computer-Aided Engineering}{Computer-aided design
%(CAD)}

%\terms{surface-surface intersection, robust computation}
\end{frontmatter}

\maketitle

\section{Introduction}

Quadric surface, being the simplest curved surfaces, are widely used
in computational science for shape representation.
It is therefore often necessary to compute the intersection or
detect the interference of two quadrics.
In computer graphics and CAD/CAM, the intersection curve of two
quadrics needs to be found for computing a boundary representation
of a 3D shape defined by quadrics.
In robotics~\citep{Rimon_Boyd_1997} and computational
physics~\citep{Lin,Perram_96} one often needs to perform
interference analysis between ellipsoids modeling the shape of
various objects.
There have recently been rising interests in computing the
arrangements of quadric surfaces in computational
geometry~\citep{Mourrain2005,Berberich2005}, a field traditionally
focused on linear primitives.

The intersection curve of two quadric surfaces will be abbreviated
as {\em QSIC}.
Exact determination of the morphology of a QSIC is critical to the
robust computation of its parametric description.
We study the problem of classifying the morphology of a QSIC in
$\mathbb{PR}^3$ (3D real projective space);  here, we use the term
{\em morphology} in a broad sense to mean the shape, topological,
and algebraic properties of a QSIC, including singularity, the
number of irreducible or connected components, and the degree of
each irreducible component, etc.
There are many types of QSIC in
$\mathbb{PR}^3$~\citep{Sommerville1947}.
A nonsingular QSIC can have zero, one, or two components.
When a QSIC is singular, it can be either irreducible or reducible.
A singular but irreducible QSIC may have three different types of
singular points, i.e., acnode, cusp, and crunode, while a reducible
QSIC may be planar or nonplanar.
A planar QSIC consists of only lines or conics, which are planar
curves, while a reducible but non-planar QSIC always consists of a
real line and a real space cubic curve.
Among planar QSICs, further distinction can be made according to how
many of the linear or conic components are imaginary, i.e., not
present in the real projective space.

There are mainly three basic problems in studying the morphology of
a QSIC: 1)~{\it Enumeration}: listing all possible morphologically
different types of QSICs; 2)~{\it Classification}: determining the
morphology of the QSIC of two given quadrics; 3)~{\it
Representation}: determining the transformation which brings a given
problem QSIC into a canonical representative of its class. We
emphasize on the second problem of classification, which is an
algorithmic issue, while also having the first problem solved as a
by-product of our results. Specifically, we enumerate all 35
different morphologies of QSIC, and characterize each of these
morphologies using a signature sequence that can exactly be computed
using rational arithmetic for the purpose of classification. The
third problem, not handled here, leads to a lengthy case by case
study which depends a lot on the application behind.

Consider the intersection curve of two quadrics given by ${\mathcal
A}$: $X^TAX=0$ and ${\mathcal B}$: $X^TBX=0$, where $X=(x,y,z,w)^T
\in \mathbb{PR}^3$ and $A, B$ are $4\times 4$ real symmetric
matrices.
The {\em characteristic polynomial} of ${\mathcal A}$ and ${\mathcal
B}$ is defined as
\begin{equation}
f(\lambda)=\det(\lambda A-B),
\end{equation}
and $f(\lambda) =0$ is called the {\em characteristic equation} of
${\mathcal A}$ and ${\mathcal B}$.

The characteristic polynomial $f(\lambda)$ is defined with a
projective variable $\lambda \in \mathbb{PR}$; thus it is either a
quartic polynomial or vanishes identically.
The latter case of $f(\lambda)$ vanishing identically occurs if and
only if ${\mathcal A}$ and ${\mathcal B}$ are two singular quadrics
sharing a singular point; thus, all the quadrics in the pencil
formed by ${\mathcal A}$ and ${\mathcal B}$ are singular. In this
case, the pencil of ${\mathcal A}$ and ${\mathcal B}$ is said to be
{\em degenerate}; otherwise, the pencil is {\em non-degenerate}.
For example, if ${\mathcal A}$ and ${\mathcal B}$ are two cones with
their vertices at the same point, then they form a degenerate
pencil.
When two quadrics form a degenerate pencil, by projecting the two
quadrics from one of their common singular points to a plane
${\mathbb P}$ not passing through the center of projection, we
reduce the problem of computing the QSIC to one of computing the
intersection of two conics in the plane ${\mathbb P}$, which is a
separate and relatively simple problem.
For this reason and the sake of space, we will not cover this case
in the present paper.
Hence, we assume throughout that $f(\lambda)$ does not vanish
identically.

Our contributions are as follows. We consider a new characterization
of the QSIC of a pencil, namely the signature sequence, and show how
it can be computed effectively and efficiently, using only rational
arithmetic operations. We establish a complete correspondence among
the QSIC morphologies, the Segre characterization over the real
numbers, the Quadric Pair Canonical Form~\citep{Muth1905,Will1935,Uhlig1976}
and the signature sequence, which allows us to derive a direct algorithm
based on exact arithmetic for the classification of QSIC. Based on this
correspondence, a simplified analysis of the morphology of different
QSIC's is described. We obtain a complete table of all the possible
morphologies of QSIC, with their Segre characterizations, signature
sequences and Quadric Pair Canonical Forms. These results apply to any
quadric pencil whose characteristic polynomial $f(\lambda)$ does not
vanish identically. The case of $f(\lambda)\equiv 0$ leads to the
classification of conics in $\mathbb {PR}^{2}$, which is not treated
here.
Tables 1, 2 and 3 give the complete list of all 35 different types
of QSICs in $\mathbb{PR}^3$ with non-degenerate quadric pencils.
A detailed explanation of these tables is given in
Section~\ref{sec:list of morphologies}.

We stress that this paper is {\em not} about affine classification
of QSICs, although the results of this paper can be used for an
implementation of affine classification by further considering the
intersection of a QSIC with the plane at infinity.

A few words are in order about our approach. Since any pair of
quadrics can be put in the Quadric Pair Canonical Form, we obtain all possible
QSIC morphologies by an exhaustive enumeration of all Quadric Pair Canonical Forms,
with distinct Jordan chains and sign combinations. 
For each pair of the Quadric Pair Canonical Forms, on one hand, we obtain its
index sequence, and on the other hand, we determine its
corresponding morphology.
The derivation of the index sequence necessitates the study on
eigenvalue curves and index jumps at real roots of a characteristics
equation, while the determination of the QSIC morphology is largely
based on case-by-case geometric analysis of two quadrics in their
Quadric Pair Canonical Forms. Finally, we convert all index sequences to their
corresponding signature sequences for efficient and exact
computation. In this way we establish a complete correspondence
among the QSIC morphologies, Quadric Pair Canonical Forms and signature
sequences. Overall, the paper is mainly about an algorithm for
determining the type of an input QSIC. The algorithm itself is very
simple, but it is based on a new framework of using the signature
sequences of different QSICs. Therefore, the large portion of the
paper is devoted to identifying the signature sequence of each of
the 35 QSICs, rather than to describing the flow of the simple
algorithm.

The remainder of the paper is organized as follows.
We discuss related work in the rest of this section.
Uhlig's method and other preliminaries, including a careful study of
the eigenvalue curves of a quadric pencil, are introduced in
Section~\ref{sec:Preliminaries}.
For an organized presentation, characterizing conditions for
different QSIC morphologies are grouped into three sections:
nonsingular QSIC (Section~\ref{sec:nonsingular}), singular but
non-planar QSIC (Section~\ref{sec:singular}), and planar QSIC
(Section~\ref{sec:planar}).
In Section~\ref{sec:complete classification} we discuss how to use
the obtained results for complete classification of QSIC
morphologies.
We conclude the paper in Section~\ref{sec:conclusion}.

For a better flow of discussion, in the main body of the paper we
will include only the proofs of theorems for the first few cases of
QSICs, so as to give the gist of the techniques employed. The proofs
for the rest cases will be given in the appendix. 

\subsection{Related work}\label{sec:Related work}

Literature on quadrics abounds, including both classical results
from algebraic geometry and modern ones from computer graphics,
computer-aided geometric design (CAGD) and computational geometry.
Classifying the QSIC is a classical problem in algebraic geometry,
but the solutions found therein are given in $\mathbb{PC}^3$ (3D
complex projective space), and therefore provide only a partial
solution to our classification problem posed in $\mathbb{PR}^3$.
Some methods for computing the QSIC in the computer graphics and
CAGD literature do not classify the QSIC morphology completely,
while others use a procedural approach to computing the QSIC
morphology.
The procedural approach is usually lengthy, therefore prone to
erroneous classification if floating point arithmetic is used or
leading to exceedingly large integer values or complicated algebraic
numbers if exact arithmetic is used.

When the input quadrics are assumed to be the so-called {\it natural
quadrics}, i.e., special quadrics including spheres, circular right
cones and cylinders, there are several methods that exploit
geometric observations to yield robust methods for computing the
QSIC~\citep{Miller1987,Miller_Goldman1995,Shene_Johnstone1994}.
However, we shall consider only methods for computing the QSIC of
two {\em arbitrary} quadrics, and focus on how these methods
classify the QSIC morphology.

In algebraic geometry the QSIC morphology is classified in
$\mathbb{PC}^3$, the complex projective space using the Segre
characteristic~\citep{Bromwich1906}.
The Segre characteristic is defined by the multiplicities of the
roots of $f(\lambda) =0$ with respect to $f(\lambda)$ as well as the
sub-determinants of the matrix $\lambda A-B$.
The Segre characteristic assumes the complex field, i.e., assuming
that the input quadrics are defined with complex coefficients, and
therefore it does not distinguish whether a root of $f(\lambda)=0$
is real or imaginary.
When applying the Segre characteristic in $\mathbb{PR}^3$, several
different types of QSICs in $\mathbb{PR}^3$ may correspond to the
same Segre characteristic, thus cannot be distinguished.
An example is the case where four morphologically different types of
nonsingular QSICs correspond to the same Segre characteristic
$[1111]$, meaning that $f(\lambda) =0$ has four distinct roots; (see
cases 1 through 4 in Table 1).

QSICs in $\mathbb{PR}^3$, real projective space, are studied
comprehensively in \citep{Killing1872,Staude1914}, but the
algorithmic aspect of classification is not considered.
In this paper we obtain a complete classification by signature
sequences of quadric pencils and apply this result to efficient
classification of QSICs in $\mathbb{PR}^3$.

A well-known method for computing QSIC in 3D real space is proposed
by Levin~\citep{Levin1976,Levin1979}, based on the observation that
there exists a ruled surface in the pencil of any two distinct
quadrics in $\mathbb{PR}^3$.
Levin's method substitutes a parameterization of this ruled quadric
to the equation of one of the two input quadrics to obtain a
parameterization of the QSIC.
However, this method does not classify the morphology of the QSIC;
consequently, it does not produce a rational parameterization for a
degenerate QSIC, which is known to be a rational curve or consist of
lower-degree rational components.

There have been proposed several methods that improve upon Levin's
method.
Sarraga~\citep{Sarraga1983} refines Levin's method in several
aspects but does not attempt to completely classify the QSIC.
Wilf and Manor~\citep{Wilf1993} combine Levin's method with the
Segre characteristic to devise a hybrid method, which, however, is
still not capable of completely classifying the QSIC in
$\mathbb{PR}^3$; for example, the four different types of
nonsingular QSICs are not classified in $\mathbb{PR}^3$.
Wang, Goldman and Tu~\citep{Wenping2003} show how to classify the
QSICs within the framework of Levin's method.
DuPont et al ~\citep{Dupont2003} proposed a variant of Levin's
method in exact arithmetic by selecting a special ruled quadric in
the pencil of two quadrics, in order to minimize the number of
radicals used in representing the QSIC; an implementation of this
method is described in~\citep{Lazard2004}.
The methods in~\citep{Wenping2003} and~\citep{Dupont2003} both adopt
a lengthy procedural approach, with no systematic approach for a
complete classification.

A different idea of computing the QSIC, again using a procedural
approach, is to project a QSIC into a planar algebraic curve and
analyze this projection curve to deduce the properties of the QSIC,
including its morphology and parameterization.
Farouki, Neff and O'Connor~\citep{Farouki1989} project a QSIC to a
planar quartic curve and factorize this quartic curve to determine
the morphology of the QSIC. (Note that only degenerate QSICs are
considered in ~\citep{Farouki1989}.)
Wang, Joe and Goldman~\citep{Wenping2002} project a QSIC to a planar
cubic curve using a point of the QSIC as the center of projection;
this cubic curve is then analyzed to compute the morphology and
parameterization of the QSIC. However, exact computation is
difficult with this method, since the center of projection is
computed with Levin's method.

The work of Ocken et al~\citep{Ocken1987}, Dupont et al
\citep{Dupont2004, Dupont2005}, Tu et al~\citep{Tu2002} and
\citep{Tu2005} all use simultaneous matrix diagonalization for
computing or classifying the QSIC. The diagonalization procedure
used in~\citep{Ocken1987} is not based on any established canonical
form, such as the Uhlig form~\citep{Muth1905,Will1935,Uhlig1976}, and the analysis
in~\citep{Ocken1987} is incomplete -- it leaves some cases of QSIC
morphology missing and some other cases classified incorrectly; for
example, the case of a QSIC consisting of a line and a space cubic
curve is missing and the cases where $f(\lambda) =0$ has exactly two
real roots or four real roots are not distinguished. The
classification by Dupont\citep{Dupont2004, Dupont2005} is based on
the Quadric Pair Canonical Form and involves criteria such as signature and sign of
deflated polynomials at specific roots of the characteristic
polynomial, leading to a complete procedure to determine the type of
a QSIC, covering also the case where the characteristic polynomial
vanishes identically.

%The computation of this procedure is not completely, though largely,
%based on checking algebraic conditions, since some cases need to be
%resolved resorting to geometric computation.

%%<BM> modification
In the above methods some cases of different QSIC morphologies need
to be distinguished using procedures involving geometric
computation,  such as extracting singular points or
%transforming the input quadrics into canonical forms or 
intersecting a line with a quadric. 
Application of such procedural methods is not uniform and follows a case by
case study, which is very specific to the tridimensional problem.
%Application of such procedural methods is prone to relatively large accumulated errors
%with floating point implementation or tends to generate complicated
%algebraic numbers when using exact arithmetic. 

It is therefore natural to ask if it is possible to determine the morphology
of a QSIC by checking some simple algebraic conditions, rather than
invoking a long computational procedure.

Several arguments are in favor of more algebra. First, a description of the
configurations of QSIC by algebraic conditions allows us to introduce easily
new parameters in our problem. For instance, introducing the time, it has
direct application in collision detection problems. Secondly, it provides a
computational framework to analyse the space of configurations of QSIC and
the stratification induced by this classification, that is how the different
families are related and what happen when we move on the ``border'' of these
families.  Moreover, the correlation between the canonical form of pencils
and the algebraic characterisation can be extended in higher dimension.
%%</BM> modification

Algebraic conditions have recently been established for QSIC
morphology or configuration formed by two quadrics in some special
cases. The goal here is to characterize each possible morphology or
configuration using a simple algebraic condition, which can be
tested or evaluated easily and exactly to determine the type of an
input morphology or configuration. In related topics, a simple
condition in terms of the number of positive real roots of the
characteristic equation $f(\lambda)$ is given by Wang et al
in~\citep{Wenping2001} for the separation of two ellipsoids in 3D
affine space. Similar algebraic conditions are obtained by Wang and
Krasauskas in~\citep{Wenping2004} for characterizing non-degenerate
configurations formed by two ellipses in 2D affine plane or
ellipsoids in 3D affine space.

As for QSICs, the Quadric Pair Canonical Form form is used in~\citep{Tu2002} to derive
simple characterizing algebraic conditions for the four types of
non-singular QSICs in terms of the number of real roots of the
characteristic polynomial; however, two of the four types are not
distinguished, i.e., they are covered by the same condition. This
pursuit of algebraic conditions is extended to cover all 35 QSICs of
non-degenerate pencils in the report~\citep{Tu2005}, which again
uses the Quadric Pair Canonical Form to derive characterizing conditions in
terms of signature sequences. The present paper is based on~\citep{Tu2005}.

Finally, we  mention that Chionh, Goldman and
Miller~\citep{Chionh1991} uses multivariate resultants to compute
the intersection of three quadrics.

%\small
%\begin{minipage}{0.070\textwidth}
%\vspace*{0.005\textheight}\centerline{${\mathcal A}:\; x^2 + y^2 + z^2 - w^2 =0$}
%\centerline{${\mathcal B}:\; 2x^2 + 4y^2 - w^2 =0$}\end{minipage}
%
%\begin{minipage}{0.070\textwidth}
%\vspace*{0.005\textheight}\centerline{(0,((0,2)),2,(2,1),3,(3,0),4)}
%\centerline{(1,((1,1)),3,(3,0),4,(3,0),3)}\end{minipage}
%
%
%%%%%%%%%%%%%%%%%%%%%%%%%%%%%%%%%%%%%%%%%%%%%%%%%%%%%%%%
%%%%%Table I %%%%%%%%%%%%%%%

\begin{table}[htbp]\caption{Classification of nonplanar QSIC in $\mathbb{PR}^3$}
\begin{center}
\begin{scriptsize}
\begin{tabular}{l|l|c|l|l}
\hline
$[{\bf Segre}]_r$ &&&&\\
$r$ = the \# &{\bf Index}&{\bf Signature Sequence}&{\bf Illus-}&{\bf Representative}\\
of real roots &{\bf Sequence}&&{\bf tration}& {\bf Quadric Pair}\\
\hline
%[1111]
\Lower{$[1111]_4$}&\raisenumber{-2.8pt}{3.4 ex}{1 }\boundnum{2.6 ex}
$\langle1|2|1|2|3\rangle$ &(1,(1,2),2,(1,2),1,(1,2),2,(2,1),3)&
\begin{minipage}{0.075\textwidth}
\vspace{-0.2cm}\centerline{\epsfxsize=1.0\textwidth
\epsfbox{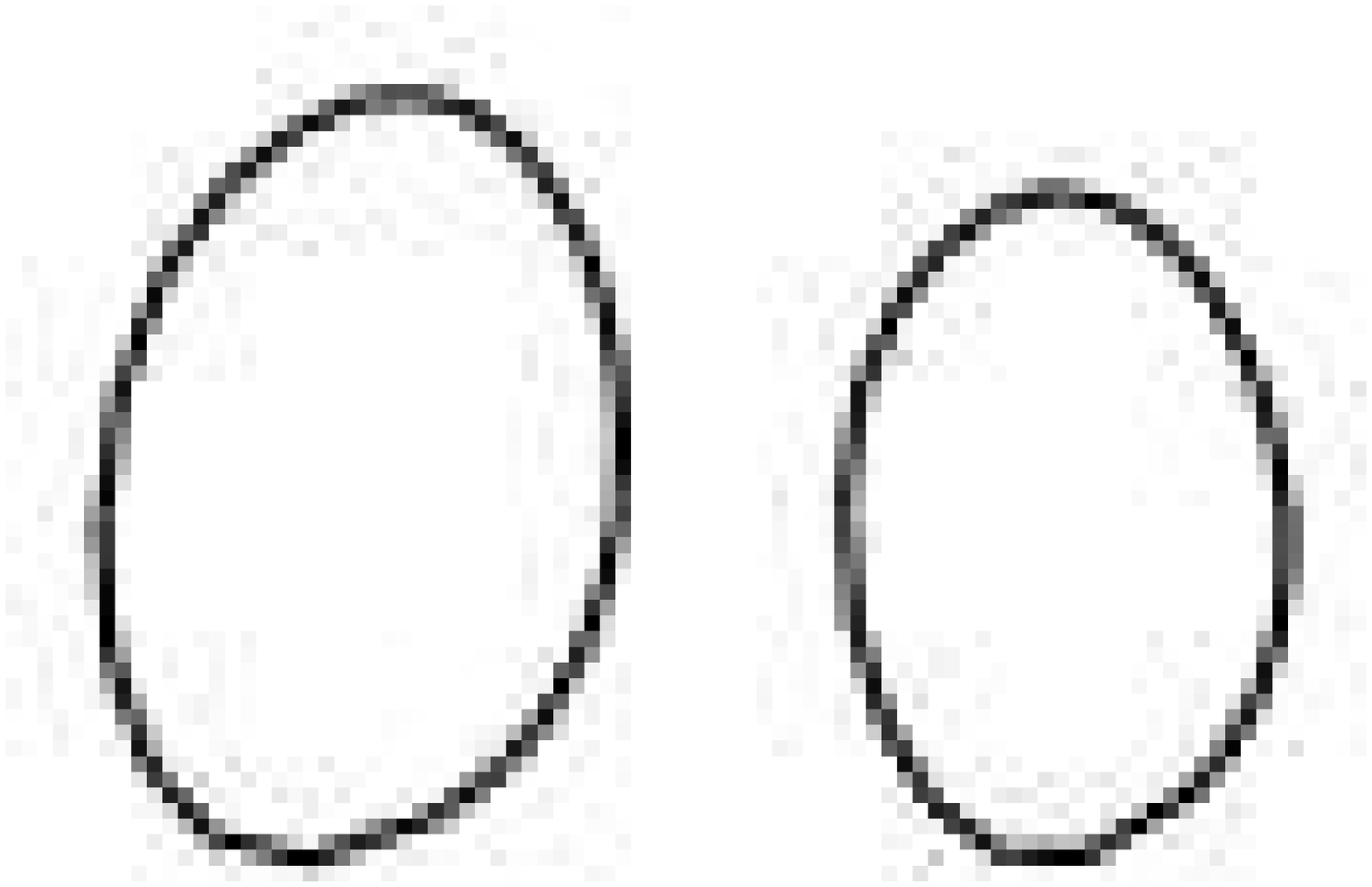}}
\end{minipage}&
%canonical forms
%\hspace{0.6in}
\begin{minipage}{0.28\textwidth}
\vspace*{0.005\textheight}\leftline{${\mathcal A}:\; x^2 + y^2 +z^2
- w^2 =0$} \leftline{${\mathcal B}:\; 2x^2 + 4y^2 - w^2
=0$}\end{minipage}
\\ \cline{2-5}
& \raisenumber{-2.8pt}{3.0 ex}{2 }\boundnum{2.2 ex} $\langle
0|1|2|3|4\rangle$ & (0,(0,3),1,(1,2),2,(2,1),3,(3,0),4)
&\begin{minipage}{0.075\textwidth}
\vspace{-0.2cm}\centerline{\epsfxsize=0.3\textwidth
\epsfbox{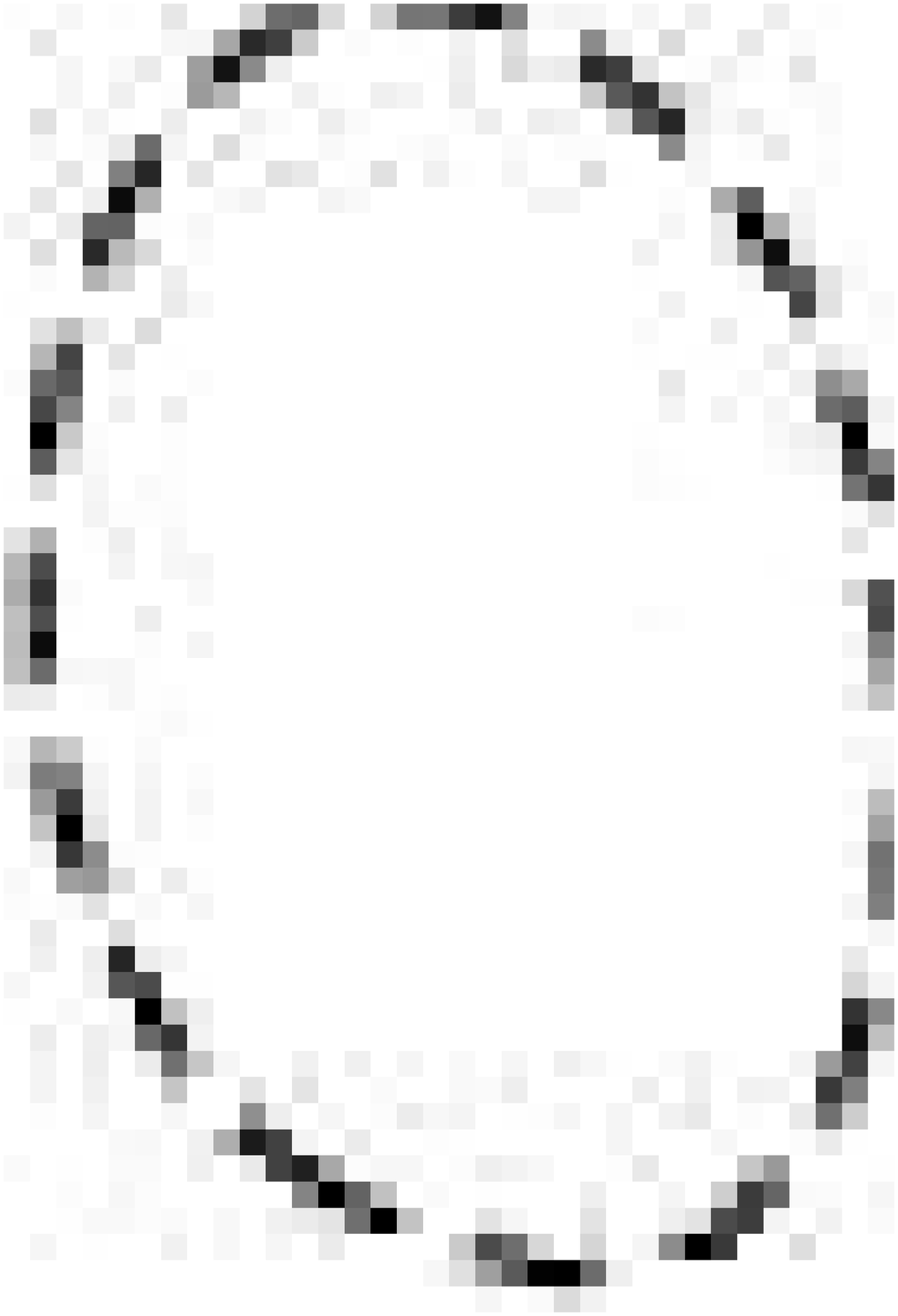}}
\end{minipage}
&
%canonical forms
%\hspace{0.7in}
\begin{minipage}{0.18\textwidth}
\vspace*{0.005\textheight}\leftline{${\mathcal A}:\; x^2 + y^2 + z^2
- w^2 =0$}\leftline{${\mathcal B}:\; 2x^2 + 4y^2 + 3z^2- w^2
=0$}\end{minipage}
\\
\cline{1-5}
$[1111]_2$& \raisenumber{-2.8pt}{3.4 ex}{3 }\boundnum{2.6 ex}
$\langle 1|2|3\rangle$ & (1,(1,2),2,(2,1),3)
&\begin{minipage}{0.075\textwidth}
\vspace{-0.2cm}\centerline{\epsfxsize=0.52\textwidth
\epsfbox{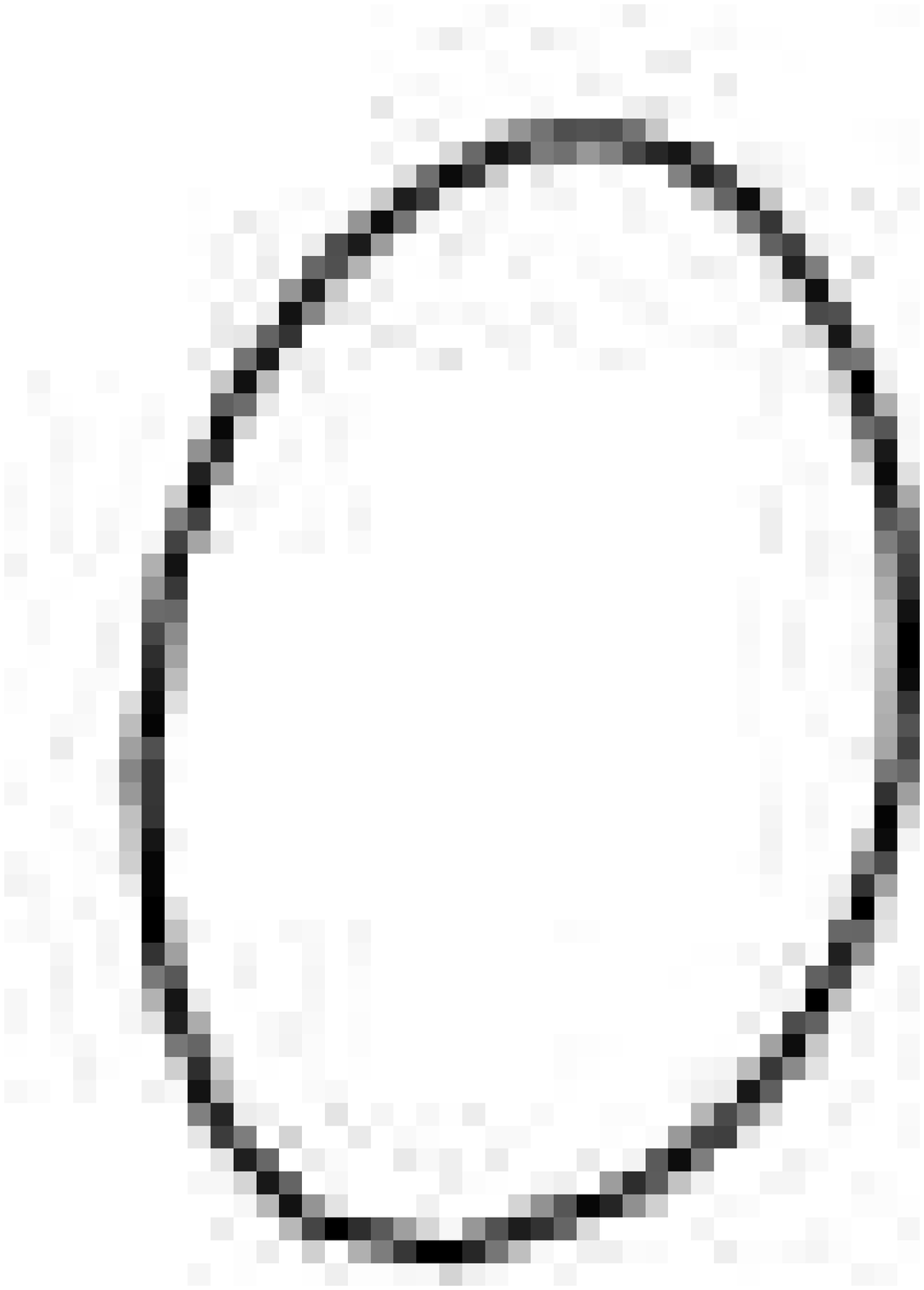}}
\end{minipage}&
%canonical forms
%\hspace{0.7in}
\begin{minipage}{0.18\textwidth}
\vspace*{0.005\textheight}\leftline{${\mathcal A}:\; 2xy + z^2 + w^2
=0$}  \leftline{${\mathcal B}:\; -x^2 + y^2 + z^2 + 2 w^2
=0$}\end{minipage}
\\ \cline{1-5}
$[1111]_0$& \raisenumber{-2.8pt}{3.4 ex}{4 }\boundnum{2.6 ex}
$\langle 2\rangle$ & (2) &\begin{minipage}{0.075\textwidth}
\vspace{-0.2cm}\centerline{\epsfxsize=0.7\textwidth
\epsfbox{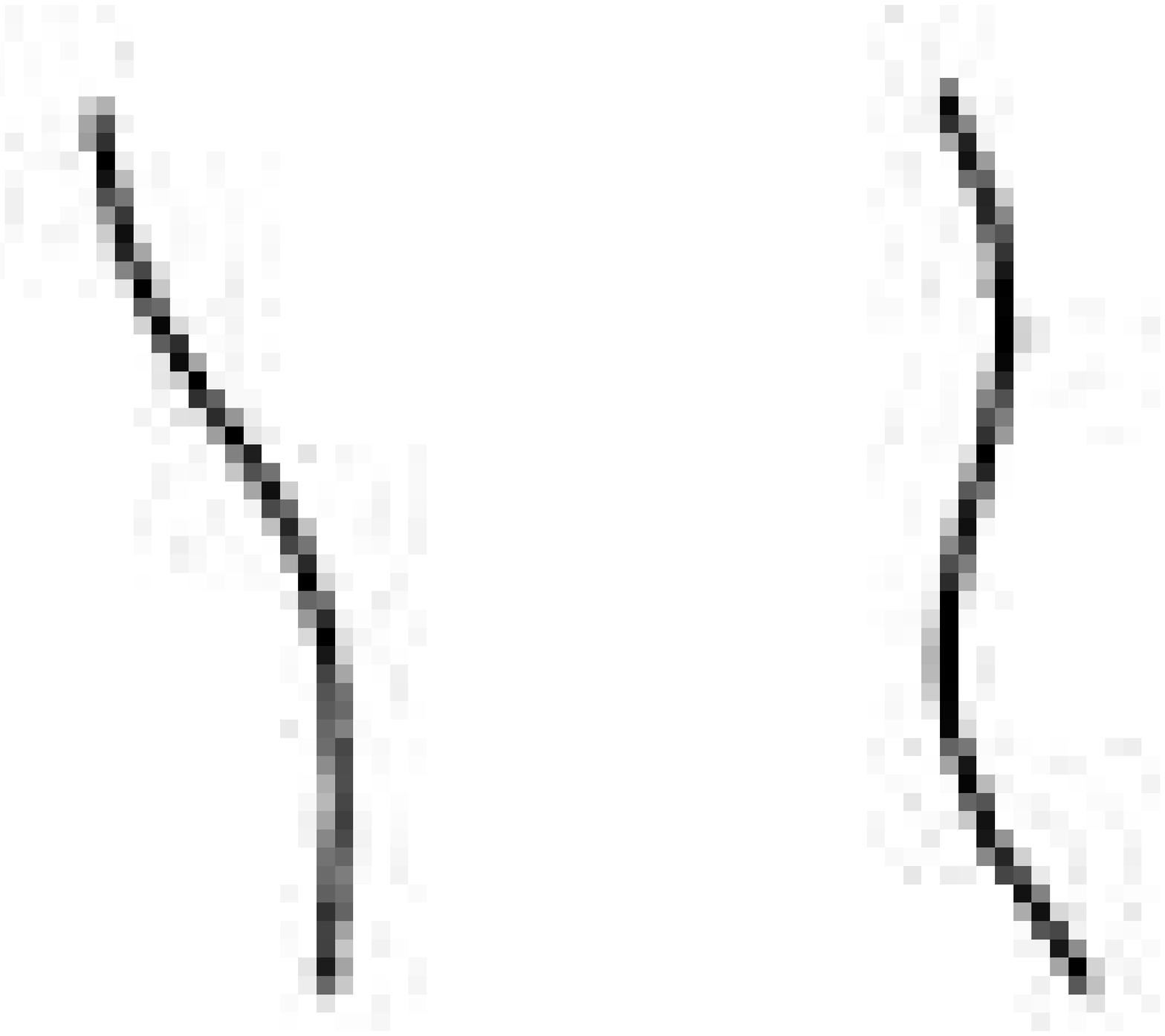}}
\end{minipage}&
%canonical forms
%\hspace{0.7in}
\begin{minipage}{0.18\textwidth}
\vspace*{0.005\textheight}\leftline{${\mathcal A}:\; xy + zw =0$}
\leftline{${\mathcal B}:\; -x^2 + y^2 - 2z^2 + zw + $}
 \leftline {$\;\;\;\;\;\; 2w^2 = 0$}\end{minipage}
\\ \hline \hline
%
%
%[211]
& \raisenumber{-2.8pt}{3 ex}{5 }\boundnum{2.2 ex}
\begin{minipage}{0.070\textwidth}
\vspace*{0.005\textheight}\centerline{ $\langle
2{\wr\wr}_{-}2|3|2\rangle$}\centerline{ $\langle
2{\wr\wr}_{+}2|3|2\rangle$}\end{minipage} &
\begin{minipage}{0.070\textwidth}
\vspace*{0.005\textheight}\centerline{(2,((2,1)),2,(2,1),3,(2,1),2)}
\centerline{(2,((1,2)),2,(2,1),3,(2,1),2)}\end{minipage}
&\begin{minipage}{0.075\textwidth}
\vspace{-0.2cm}\centerline{\epsfxsize=0.7\textwidth
\epsfbox{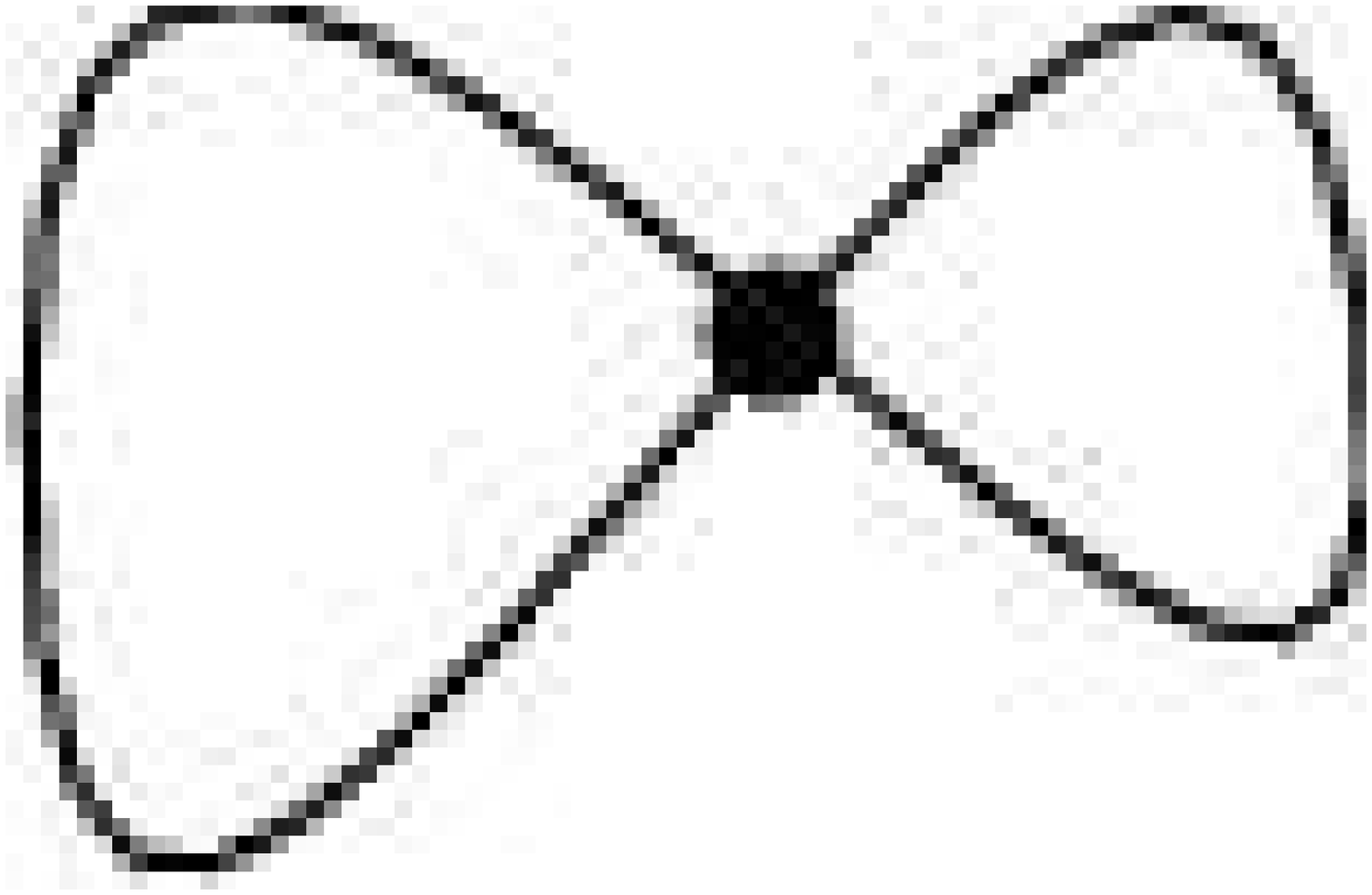}}
\end{minipage}&
%canonical forms
%\hspace{0.7in}
\begin{minipage}{0.18\textwidth}
\vspace*{0.005\textheight}\leftline{${\mathcal A}:\; x^2 - y^2 + z^2
+ 4yw =0$} \leftline{${\mathcal B}:\; -3x^2 + y^2 + z^2
=0$}\end{minipage}
\\ \cline{2-5}
\Lower{$[211]_3$}&\raisenumber{-2.8pt}{3.0 ex}{6 }\boundnum{2.2 ex}
{$\langle 1 {\wr\wr}_{-} 1 | 2 |3 \rangle$} &
(1,((1,2)),1,(1,2),2,(2,1),3) &\begin{minipage}{0.075\textwidth}
\vspace{-0.2cm}\centerline{\epsfxsize=0.7\textwidth
\epsfbox{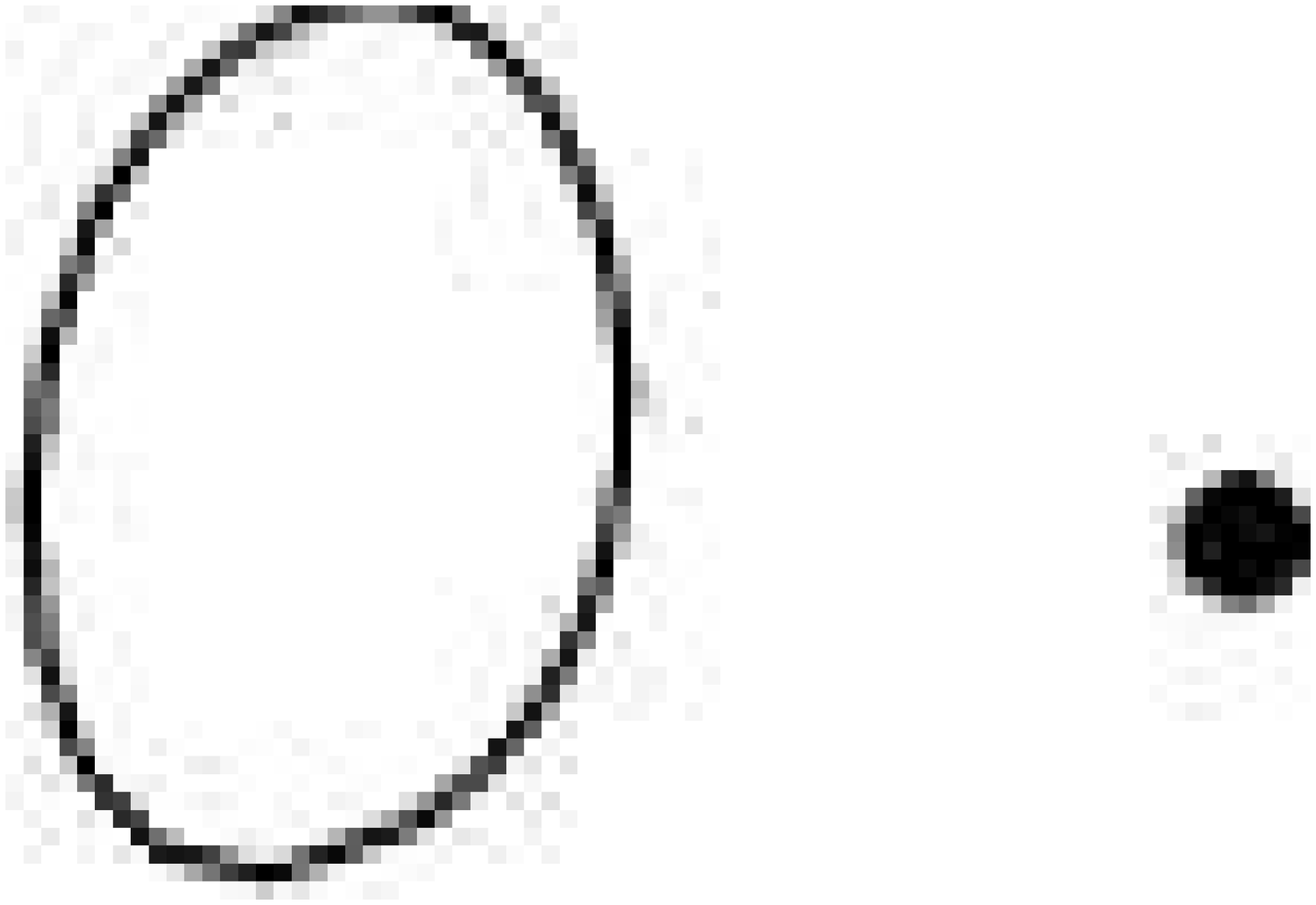}}
\end{minipage}&
%canonical forms
%\hspace{0.7in}
\begin{minipage}{0.18\textwidth}
\vspace*{0.005\textheight}\leftline{${\mathcal A}:\; -x^2 - z^2 +
2yw =0$} \leftline{${\mathcal B}:\; -3x^2 + y^2 - z^2
=0$}\end{minipage}
\\ \cline{2-5}
& \raisenumber{-2.8pt}{3.0 ex}{7 }\boundnum{2.2 ex} {$\langle 1
{\wr\wr}_{+} 1 | 2 |3 \rangle$} & (1,((0,3)),1,(1,2),2,(2,1),3)
&\begin{minipage}{0.075\textwidth}
\vspace{-0.2cm}\centerline{\epsfxsize=0.7\textwidth
\epsfbox{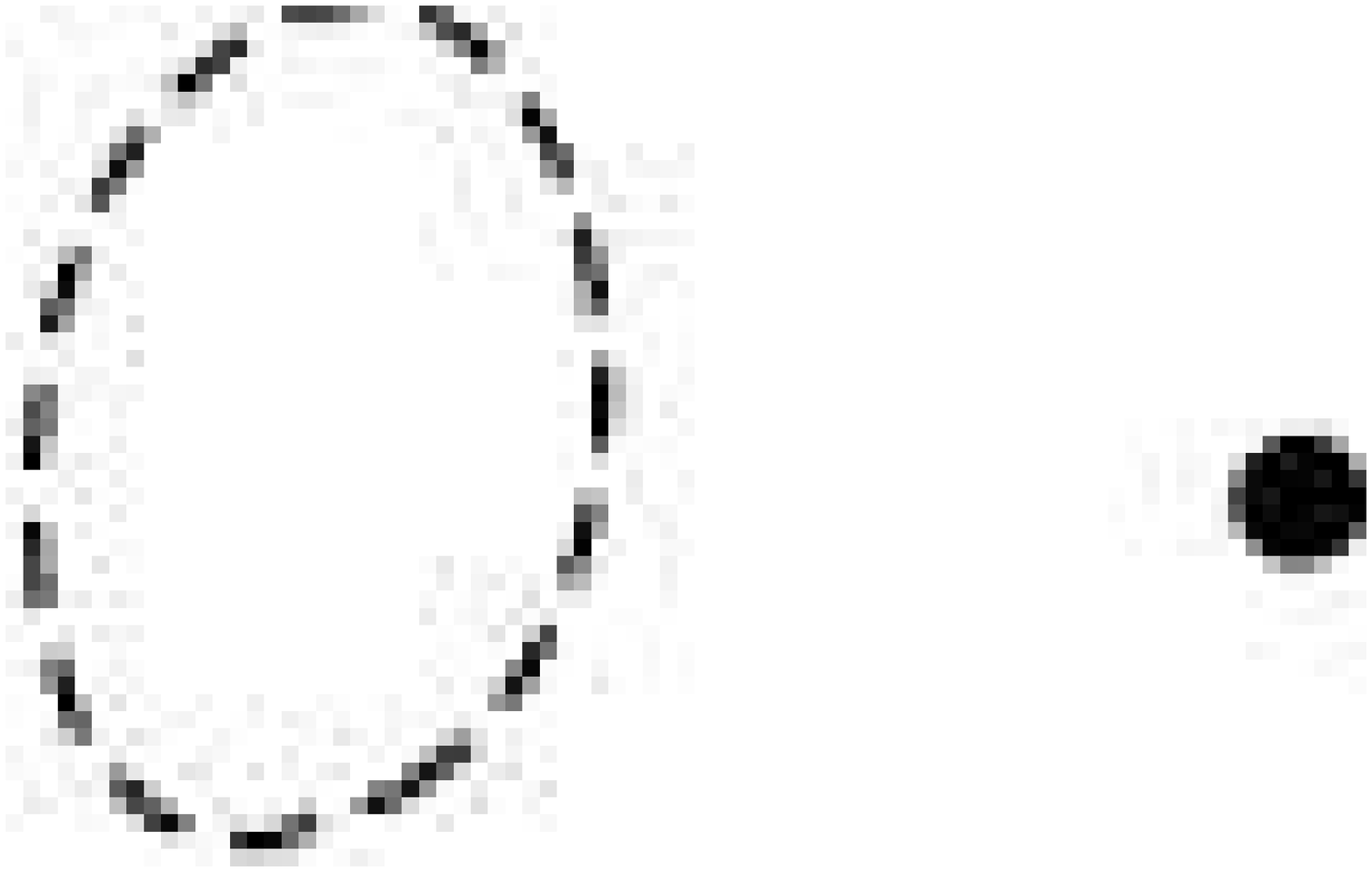}}
\end{minipage}&
%canonical forms
%\hspace{0.7in}
\begin{minipage}{0.18\textwidth}
\vspace*{0.005\textheight}\leftline{${\mathcal A}:\; x^2 + z^2 + 2yw
=0$} \leftline{${\mathcal B}:\; 3x^2 + y^2 + z^2 =0$}\end{minipage}
\\ \cline{1-5}
$[211]_1$& \raisenumber{-2.8pt}{3.0 ex}{8 }\boundnum{2.2 ex}
{$\langle 2 {\wr\wr}_{-} 2 \rangle$} & (2,((2,1)),2)
&\begin{minipage}{0.075\textwidth}
\vspace{-0.2cm}\centerline{\epsfxsize=0.8\textwidth
\epsfbox{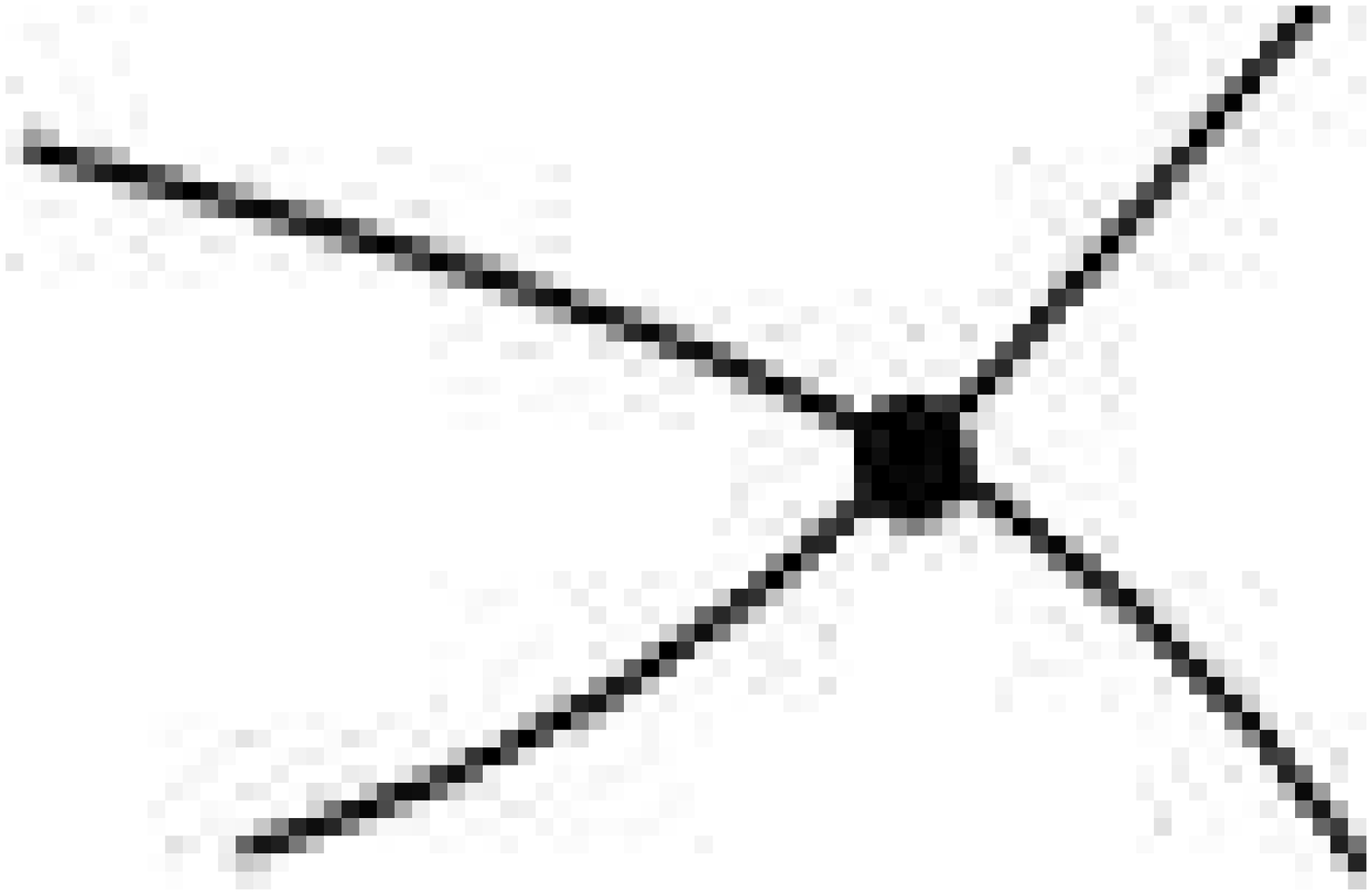}}
\end{minipage}&
%canonical forms
%\hspace{0.7in}
\begin{minipage}{0.18\textwidth}
\vspace*{0.005\textheight}\leftline{${\mathcal A}:\; xy + zw=0$}
\leftline{${\mathcal B}:\; 2xy + y^2 - z^2 + w^2=0$}\end{minipage}
\\ \hline\hline
%
%
%[22]
$[22]_2$& \raisenumber{-2.8pt}{3.0 ex}{9 }\boundnum{2.2 ex}
\begin{minipage}{0.070\textwidth}
\vspace*{0.005\textheight}\centerline{ $\langle 2 {\wr\wr}_{-} 2
{\wr\wr}_{-} 2 \rangle$}\centerline{ $\langle 2 {\wr\wr}_{-} 2
{\wr\wr}_{+} 2 \rangle$}\end{minipage} &
\begin{minipage}{0.070\textwidth}
\vspace*{0.005\textheight}\centerline{(2,((2,1)),2,((2,1)),2)}
\centerline{(2,((2,1)),2,((1,2)),2)}\end{minipage}
&\begin{minipage}{0.075\textwidth}
\vspace{-0.2cm}\centerline{\epsfxsize=0.7\textwidth
\epsfbox{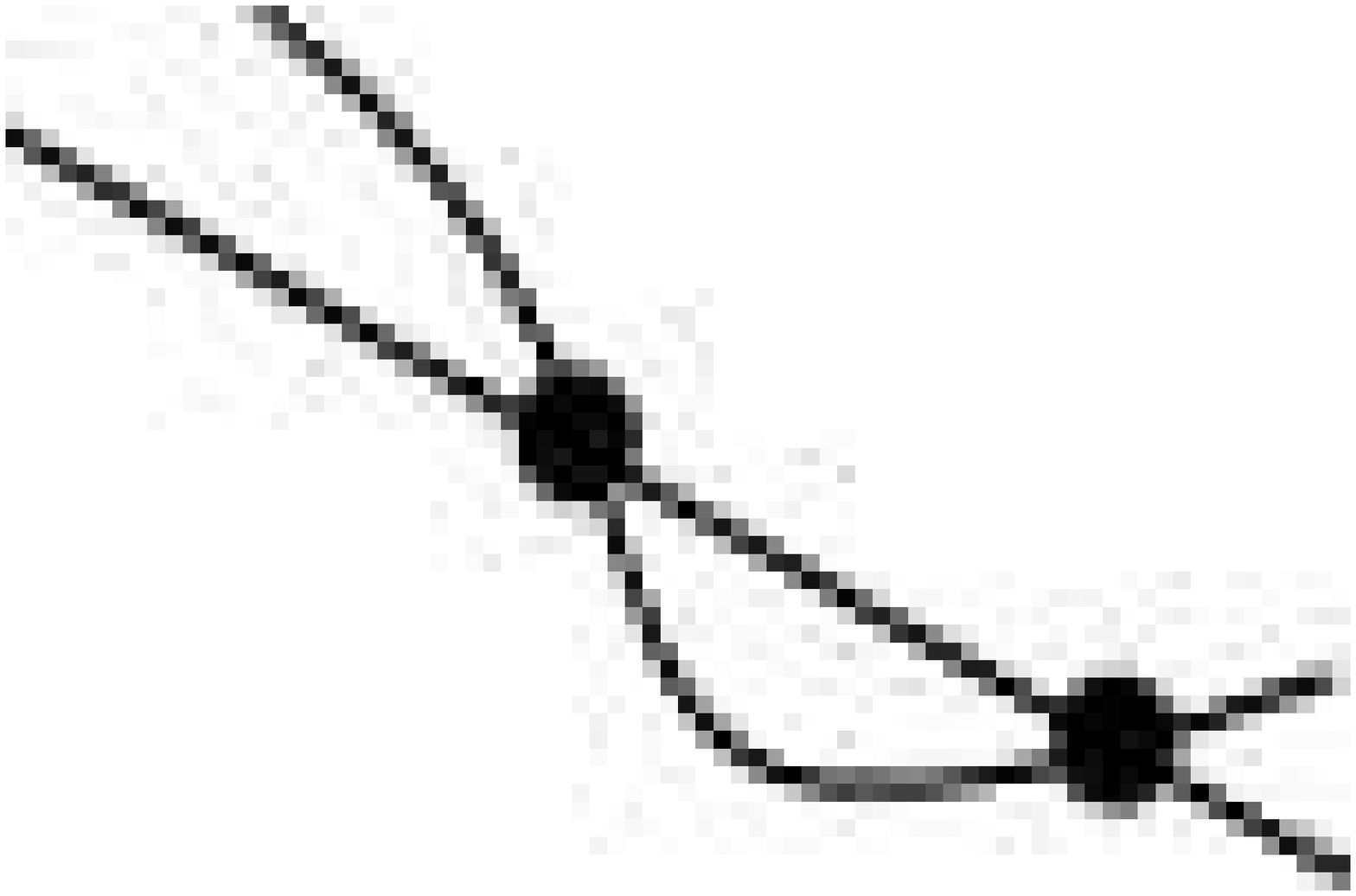}}
\end{minipage}&
%canonical forms
%\hspace{0.7in}
\begin{minipage}{0.18\textwidth}
\vspace*{0.005\textheight}\leftline{${\mathcal A}:\; xy + zw=0$}
\leftline{${\mathcal B}:\; y^2 + 2zw + w^2=0$}\end{minipage}
\\ \cline{1-5}
$[22]_0$& \raisenumbertwo{-6.8pt}{3.0 ex}{10}\boundnum{2.2 ex}
{$\langle 2 \rangle$} & (2) &\begin{minipage}{0.075\textwidth}
\vspace{-0.2cm}\centerline{\epsfxsize=.8\textwidth
\epsfbox{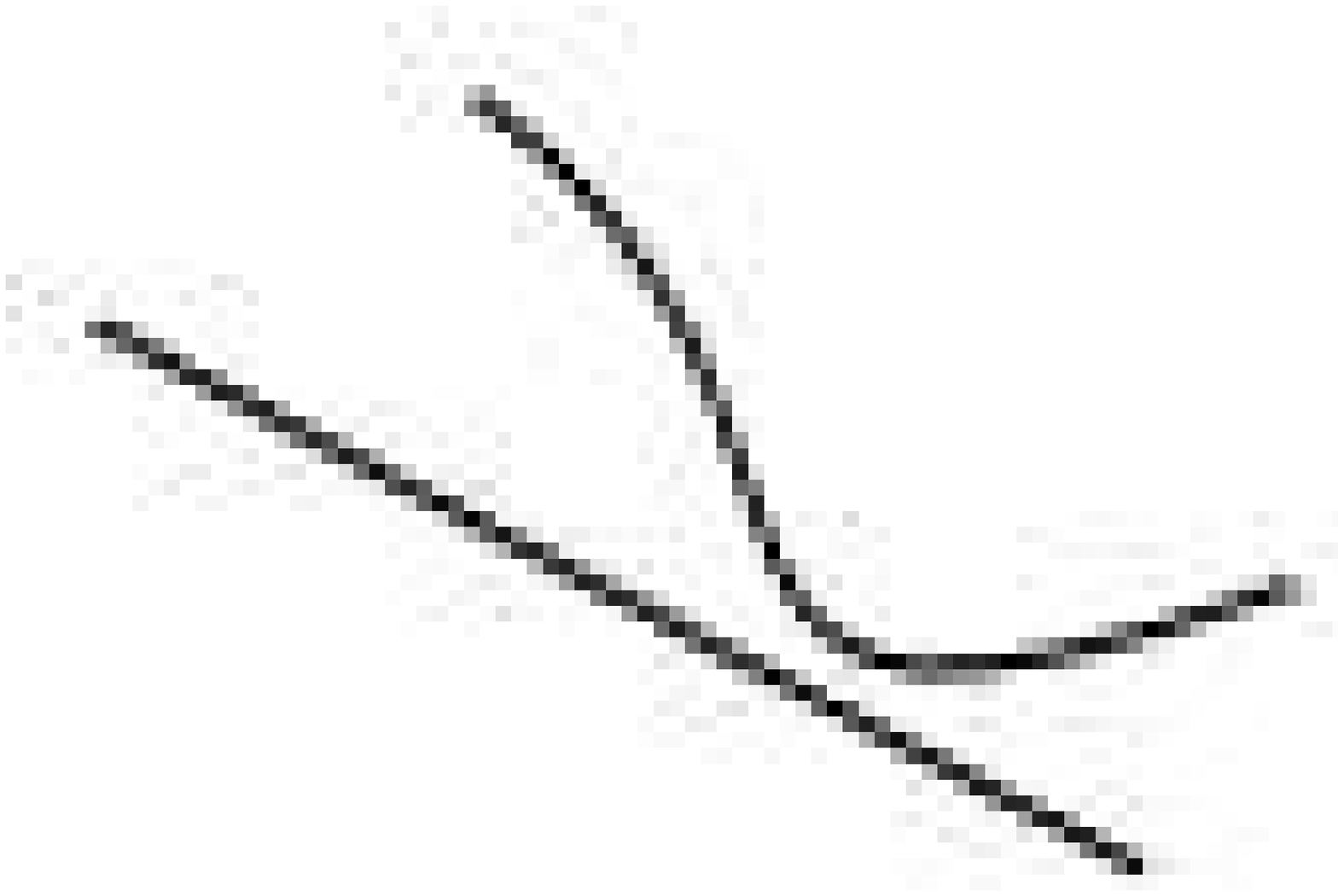}}
\end{minipage}&
%canonical forms
%\hspace{0.7in}
\begin{minipage}{0.18\textwidth}
\vspace*{0.005\textheight}\leftline{${\mathcal A}:\; xw + yz=0$}
\leftline{${\mathcal B}:\; xz - yw =0$}\end{minipage}
\\ \hline \hline
%
%
%[31]
$[31]_2$ &\raisenumbertwo{-6.8pt}{3.0 ex}{11}\boundnum{2.2 ex}
 {$\langle
1 {\wr\wr}{\wr}_{+} 2 | 3 \rangle$} & (1,(((1,2))),2,(2,1),3)
 &\begin{minipage}{0.075\textwidth}
\vspace{-0.3cm}\centerline{\epsfxsize=0.6\textwidth
\epsfbox{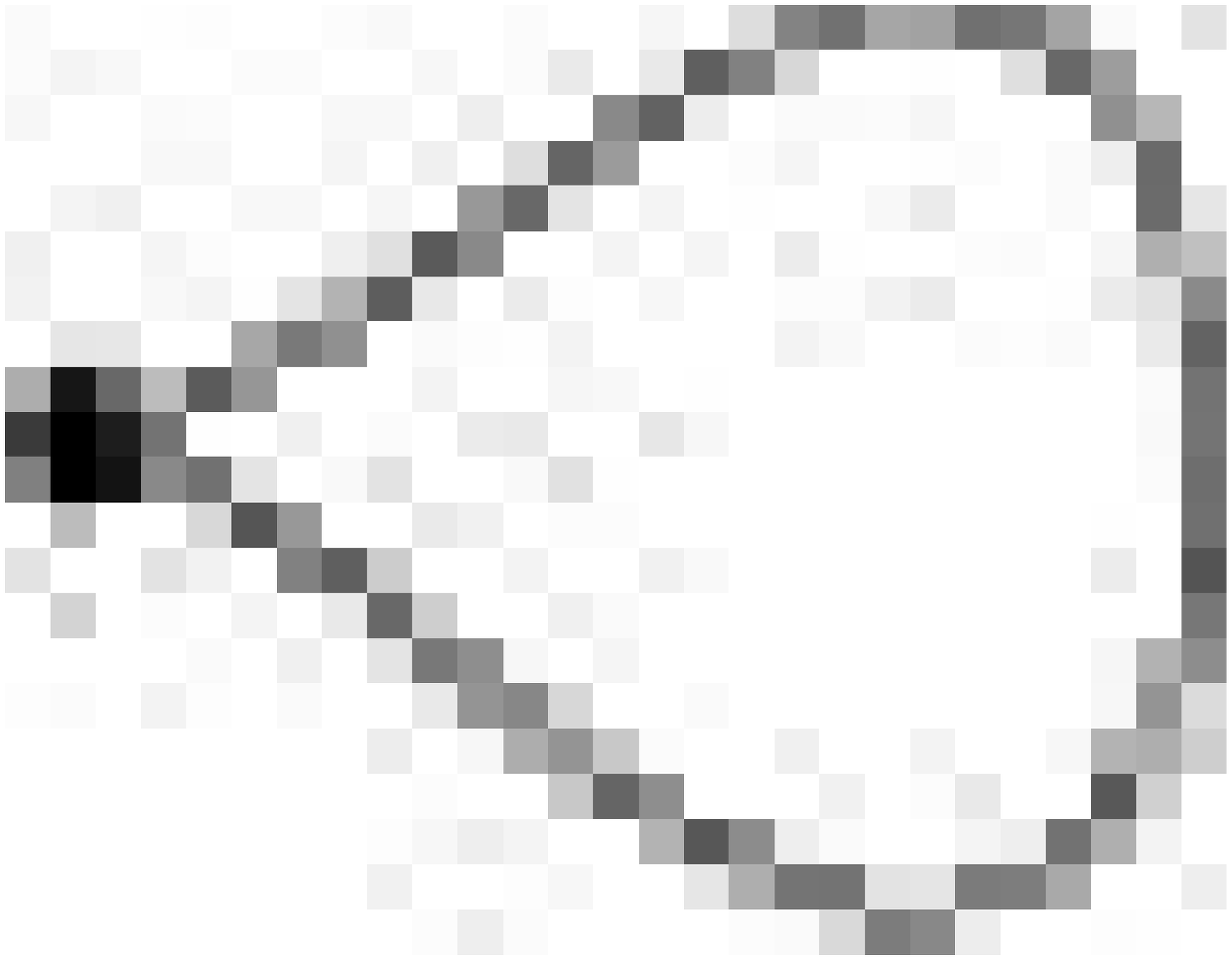}}
\end{minipage}&
%canonical forms
%\hspace{0.7in}
\begin{minipage}{0.18\textwidth}
\vspace{-0.2cm}\leftline{${\mathcal A}:\; y^2 + 2xz + w^2=0$}
\leftline{${\mathcal B}:\; 2yz + w^2=0$}\end{minipage}
\\ \hline \hline
%
%
%[4]
$[4]_1$&\raisenumbertwo{-6.8pt}{3.0 ex}{12}\boundnum{2.2 ex}
{$\langle 2 {\wr\wr}{\wr\wr}_{-} 2 \rangle$} & (2,((((2,1)))),2)
 &\begin{minipage}{0.075\textwidth}
\vspace{-0.4cm}\centerline{\epsfxsize=0.6\textwidth
\epsfbox{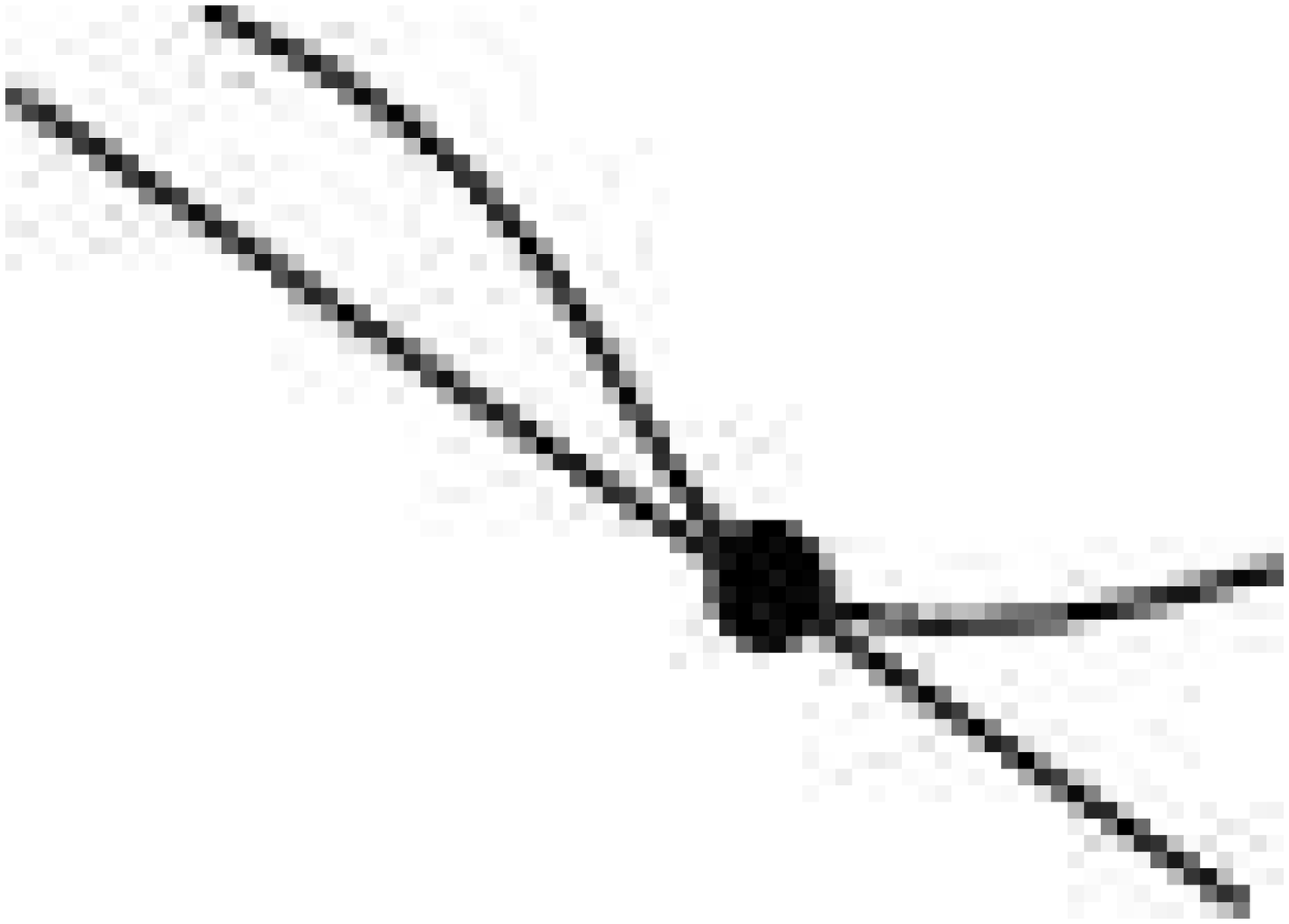}}
\end{minipage}&
%canonical forms
%\hspace{0.7in}
\begin{minipage}{0.18\textwidth}
\vspace{-0.2cm}\leftline{${\mathcal A}:\; xw + yz =0$}
\leftline{${\mathcal B}:\; z^2 + 2yw=0$}\end{minipage}
\\ \hline\hline
\end{tabular}
\end{scriptsize}
\end{center}
\end{table}
%
%
%
%
%%%%%%%%%%%%%%%%%%%%%%%%%%%%%%%%%%%%%%%%%%%%%%%%%%%%%%%%
%%%%%Table II%%%%%%%%%%%%%%%
\begin{table}[htbp]\caption{Classification of planar QSIC in $\mathbb{PR}^3$ - Part I}
\begin{center}
\begin{scriptsize}
\begin{tabular}{l|l|c|l|l}
\hline
$[{\bf Segre}]_r$ &&&&\\
$r$ = the \# &{\bf Index}&{\bf Signature Sequence}&{\bf Illus-}&{\bf Representative}\\
of real roots &{\bf Sequence}&&{\bf tration}& {\bf Quadric Pair}\\
\hline
%[(11)11]
&\raisenumbertwo{-6.8pt}{5.0 ex}{13}\boundnum{4.2 ex} {$\langle 2 ||
2 | 1 | 2 \rangle$} &(2,((1,1)),2,(1,2),1,(1,2),2)
&\begin{minipage}{0.075\textwidth} \vspace*{0.005\textheight}
\centerline{\epsfysize=0.8\textwidth \epsfbox{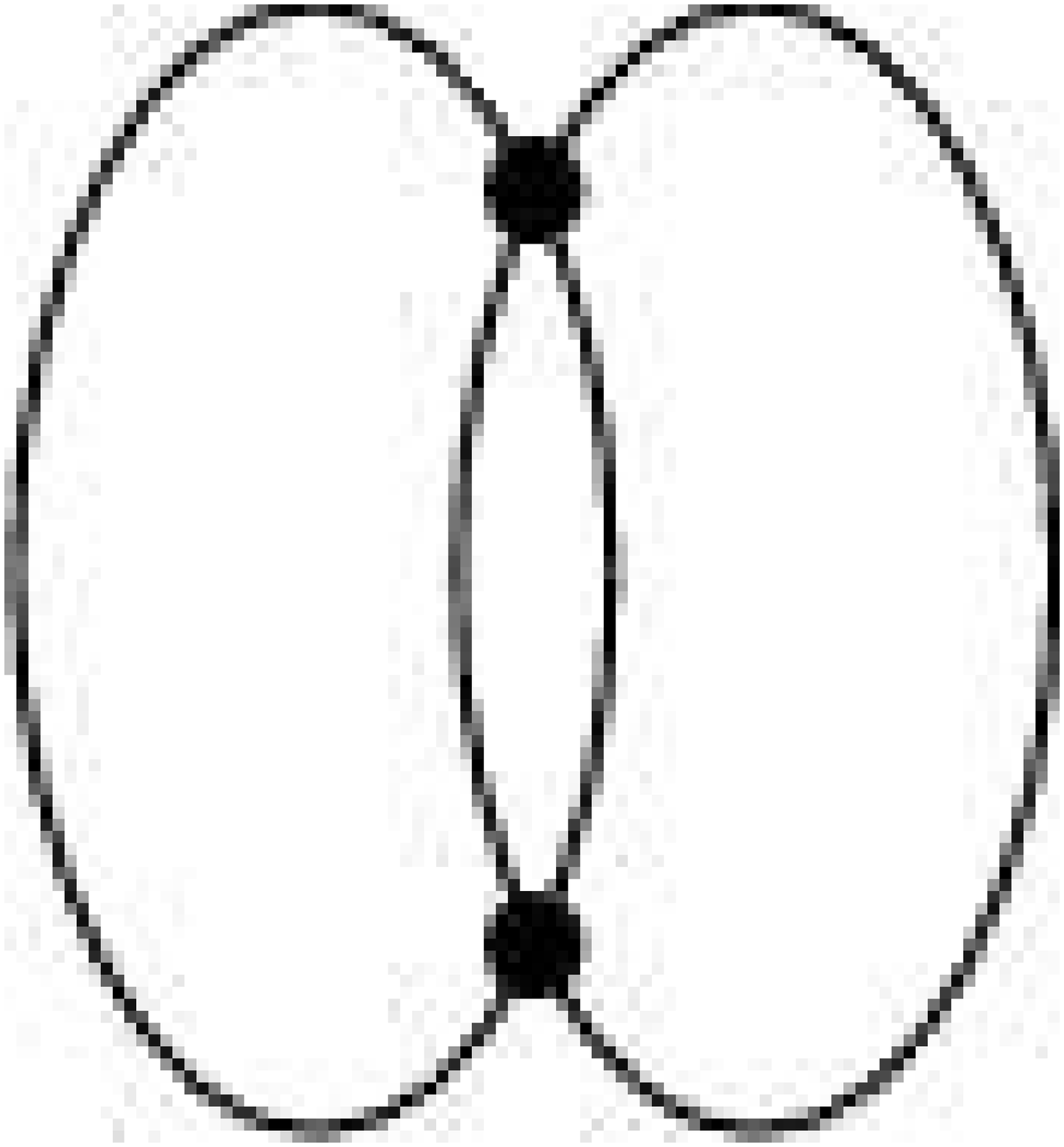}}
\vspace*{0.005\textheight}
\end{minipage}&
%canonical forms
%\hspace{0.7in}
\begin{minipage}{0.26\textwidth}
\vspace*{0.005\textheight}\leftline{${\mathcal A}:\; x^2 - y^2 +z^2
- w^2=0$} \leftline{${\mathcal B}:\; x^2 - 2y^2=0$}\end{minipage}
\\ \cline{2-5}

\MLower{$[(11)11]_3$}&\raisenumbertwo{-6.8pt}{5.0
ex}{14}\boundnum{4.2 ex} {$\langle 1 || 3 | 2 | 3 \rangle$}
&(1,((1,1)),3,(2,1),2,(2,1),3) &\begin{minipage}{0.075\textwidth}
\vspace*{0.005\textheight} \centerline{\epsfysize=0.8\textwidth
\epsfbox{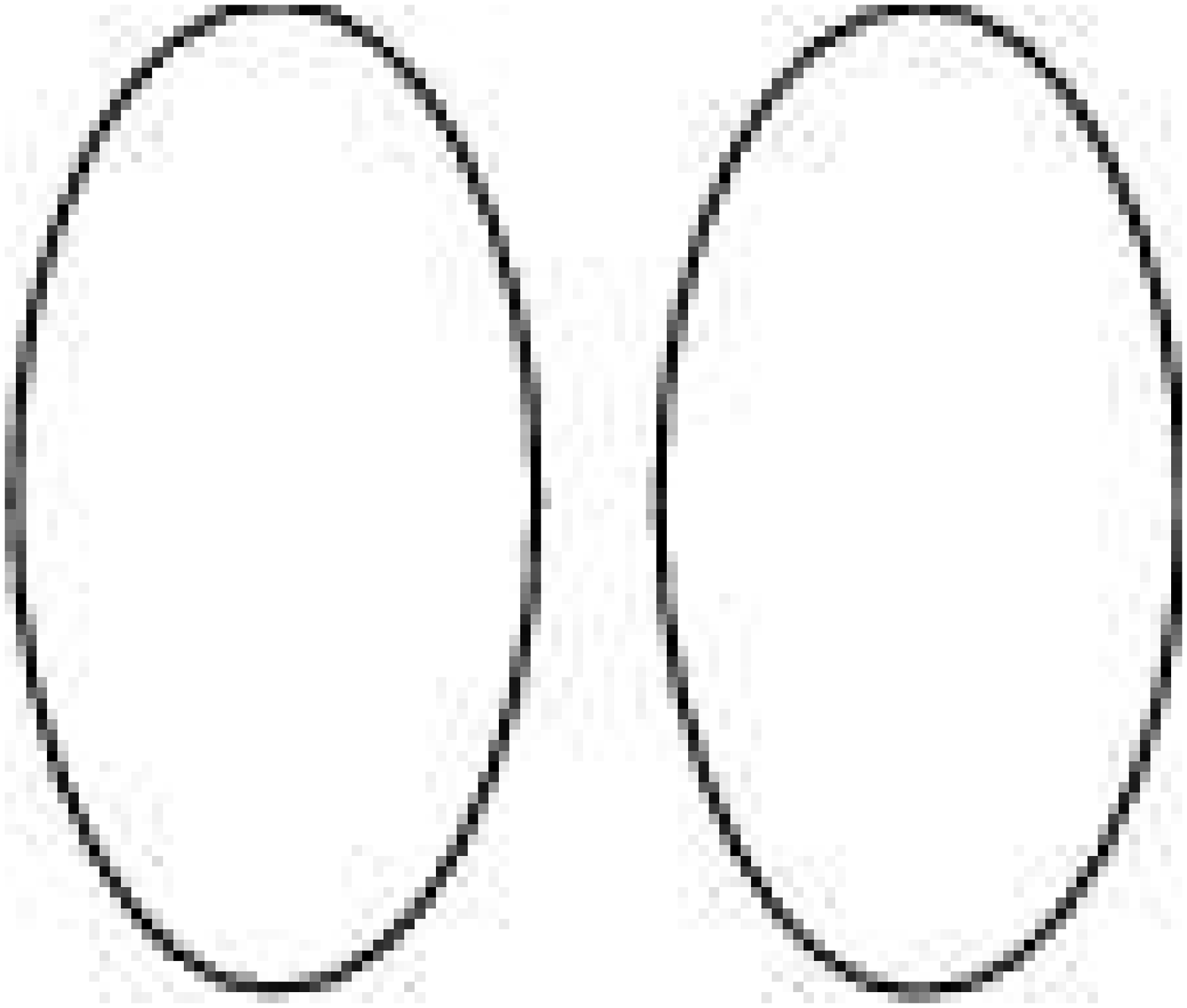}} \vspace*{0.005\textheight}
\end{minipage}&
%canonical forms
%\hspace{0.7in}
\begin{minipage}{0.26\textwidth}
\vspace*{0.005\textheight}\leftline{${\mathcal A}:\; -x^2 + y^2 +z^2
+ w^2=0$} \leftline{${\mathcal B}:\; -x^2 + 2y^2=0$}\end{minipage}
\\ \cline{2-5}
&\raisenumbertwo{-6.8pt}{5.0 ex}{15}\boundnum{4.2 ex} {$\langle 1 ||
1 | 2 | 3 \rangle$} &(1,((0,2)),1,(1,2),2,(2,1),3)
&\begin{minipage}{0.075\textwidth} \vspace*{0.005\textheight}
\centerline{\epsfysize=0.8\textwidth \epsfbox{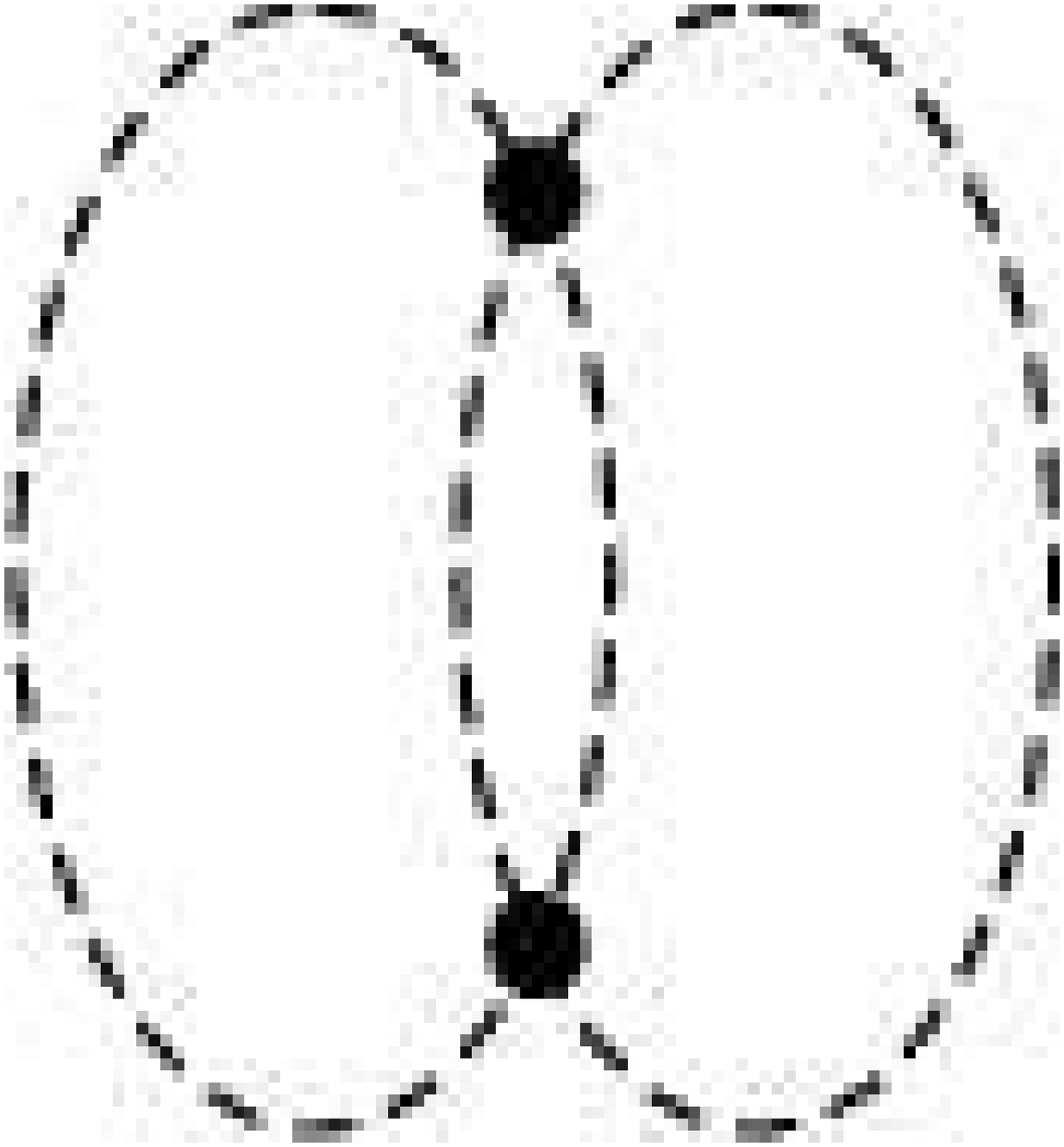}}
\vspace*{0.005\textheight}
\end{minipage}&
%canonical forms
%\hspace{0.7in}
\begin{minipage}{0.26\textwidth}
\vspace*{0.005\textheight}\leftline{${\mathcal A}:\; x^2 + y^2 +z^2
- w^2=0$} \leftline{${\mathcal B}:\; x^2 + 2y^2=0$}\end{minipage}
\\ \cline{2-5}

&\raisenumbertwo{-6.8pt}{5.0 ex}{16}\boundnum{4.2 ex}
\begin{minipage}{0.070\textwidth}
\vspace*{0.005\textheight}\centerline{$\langle 0 || 2  | 3 | 4
\rangle$}\centerline{$\langle 1 || 3  | 4 | 3
\rangle$}\end{minipage} &\begin{minipage}{0.070\textwidth}
\vspace*{0.005\textheight}\centerline{(0,((0,2)),2,(2,1),3,(3,0),4)}
\centerline{(1,((1,1)),3,(3,0),4,(3,0),3)}\end{minipage}
 &\begin{minipage}{0.075\textwidth}
\vspace*{0.005\textheight} \centerline{\epsfysize=0.8\textwidth
\epsfbox{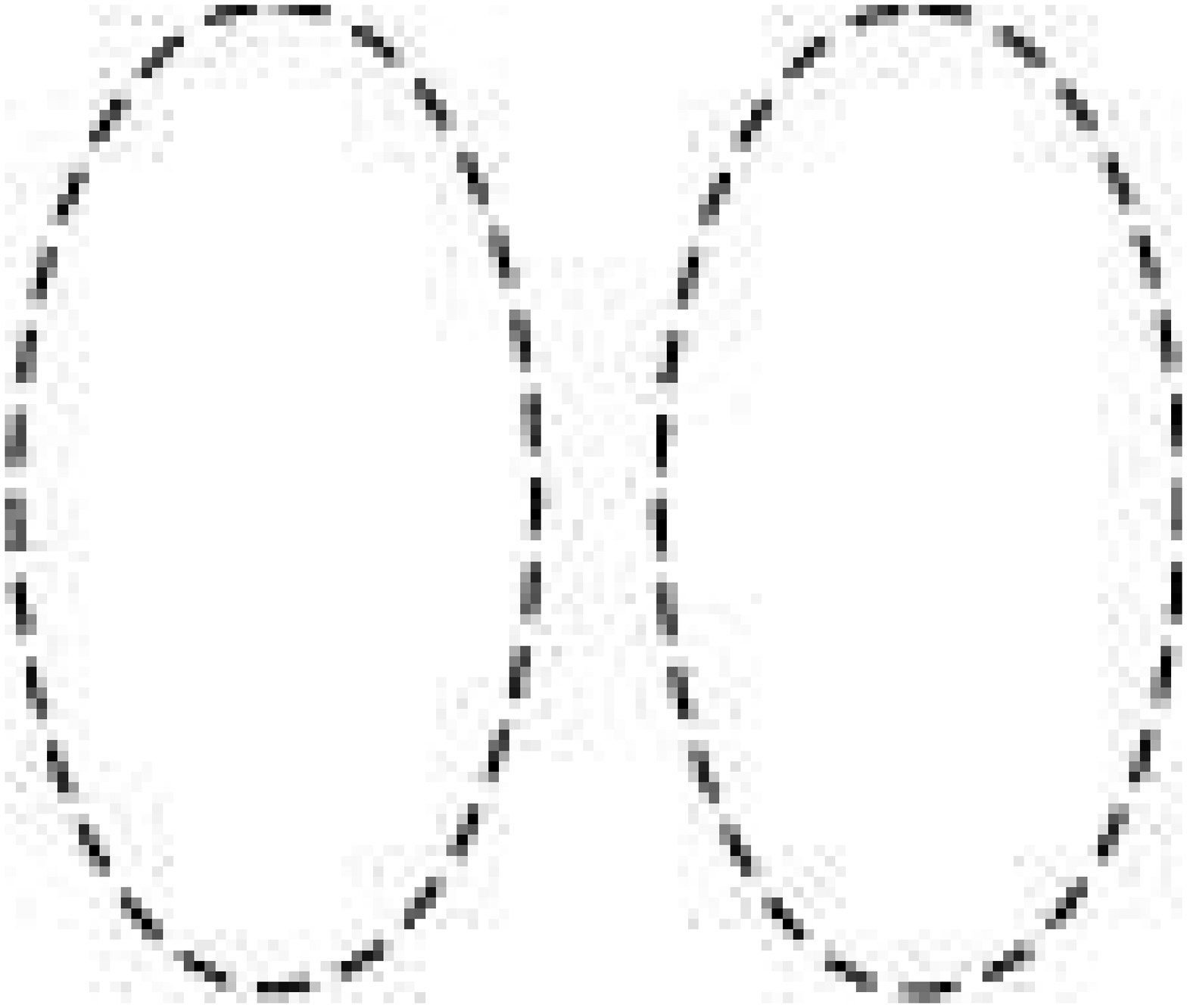}} \vspace*{0.005\textheight}
\end{minipage}&
%canonical forms
%\hspace{0.7in}
\begin{minipage}{0.26\textwidth}
\vspace*{0.005\textheight}\leftline{${\mathcal A}:\; x^2 + y^2 - z^2
- w^2=0$} \leftline{${\mathcal B}:\; x^2 + 2y^2=0$}\end{minipage}
\\ \cline{1-5}

\MLower{$[(11)11]_1$}&\raisenumbertwo{-6.8pt}{5.0
ex}{17}\boundnum{4.2 ex} {$\langle 1 || 3 \rangle$} & (1,((1,1)),3)
&\begin{minipage}{0.075\textwidth} \vspace*{0.005\textheight}
\centerline{\epsfxsize=1.0\textwidth \epsfbox{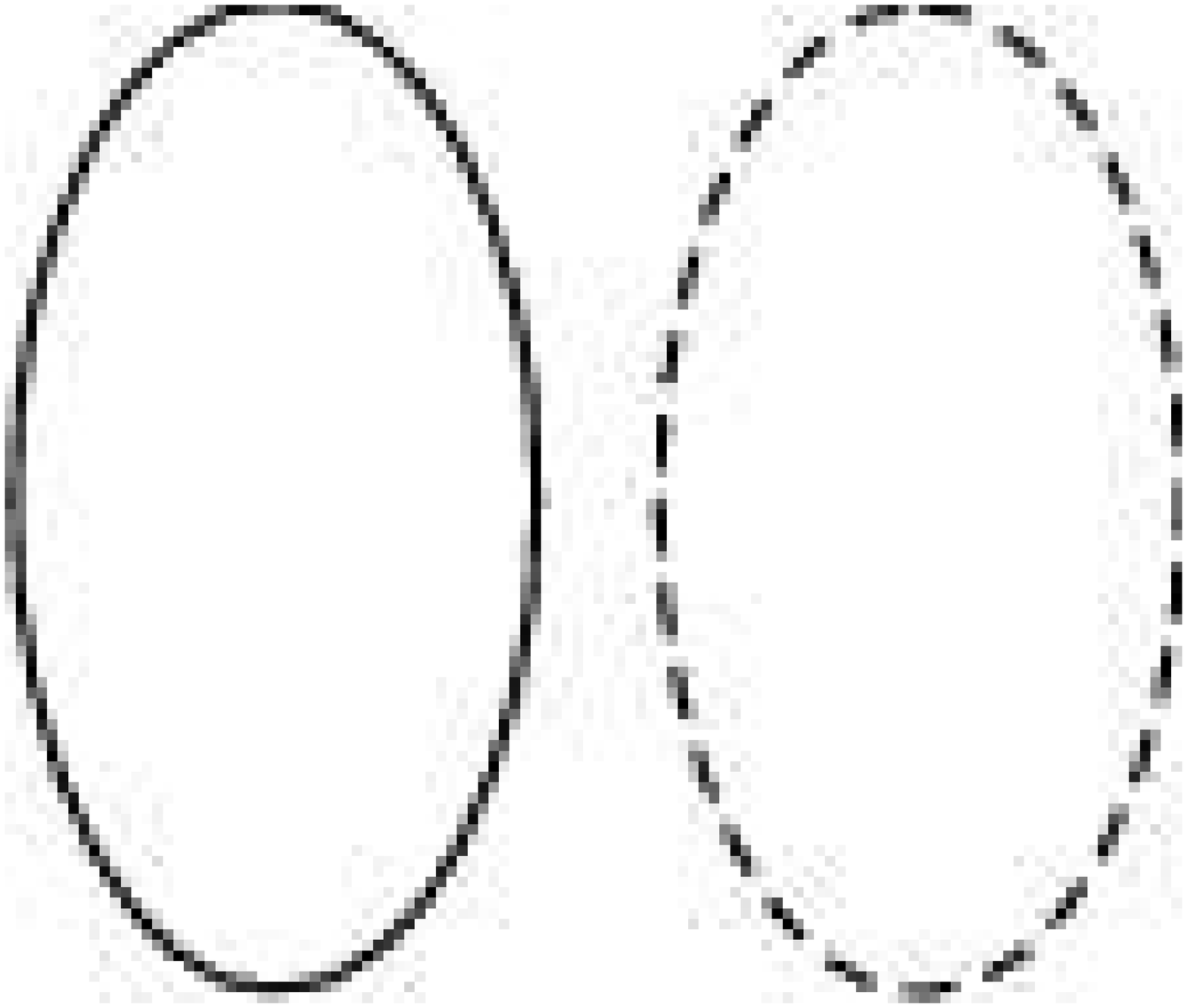}}
\vspace*{0.005\textheight}
\end{minipage}&
%canonical forms
%\hspace{0.7in}
\begin{minipage}{0.26\textwidth}
\vspace*{0.005\textheight}\leftline{${\mathcal A}:\; x^2 + y^2
+2zw=0$} \leftline{${\mathcal B}:\; -z^2 + w^2 +
2zw=0$}\end{minipage}
\\ \cline{2-5}

&\raisenumbertwo{-6.8pt}{5.0 ex}{18}\boundnum{4.2 ex} {$\langle 2 ||
2 \rangle$} & (2,((1,1)),2) &\begin{minipage}{0.075\textwidth}
\vspace*{0.005\textheight} \leftline{\epsfysize=0.55\textwidth
\epsfbox{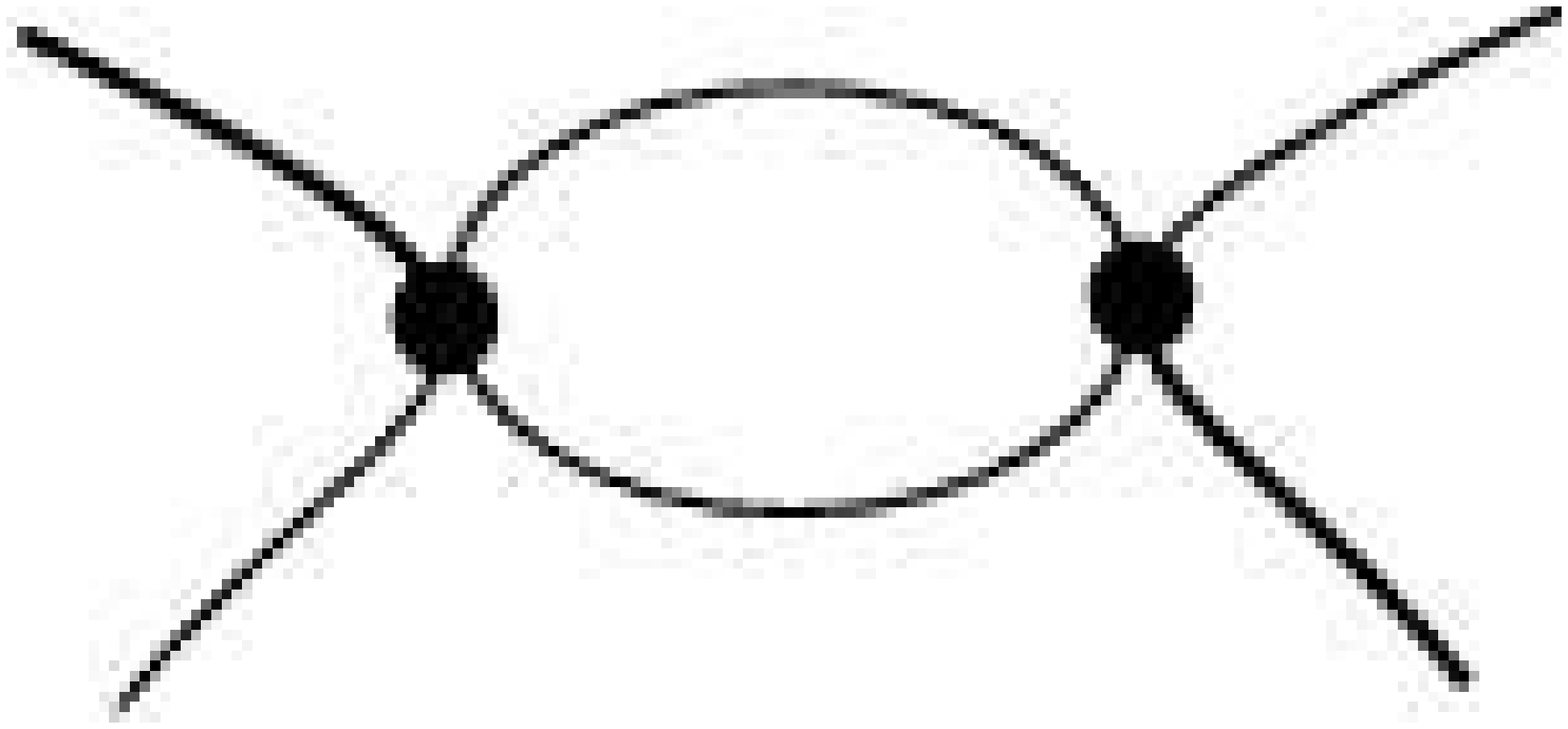}} \vspace*{0.005\textheight}
\end{minipage}&
%canonical forms
%\hspace{0.7in}
\begin{minipage}{0.26\textwidth}
\vspace*{0.005\textheight}\leftline{${\mathcal A}:\; x^2 - y^2
-2zw=0$} \leftline{${\mathcal B}:\; -z^2 + w^2 +
2zw=0$}\end{minipage}
\\ \hline\hline
%
%
%[(111)1]
\MLower{$[(111)1]_2$}&\raisenumbertwo{-6.8pt}{5.0
ex}{19}\boundnum{4.2 ex} {$\langle 1 ||| 2 | 3 \rangle$} &
(1,(((0,1))),2,(2,1),3) &\begin{minipage}{0.075\textwidth}
\vspace*{0.005\textheight} \centerline{\epsfysize=0.8\textwidth
\epsfbox{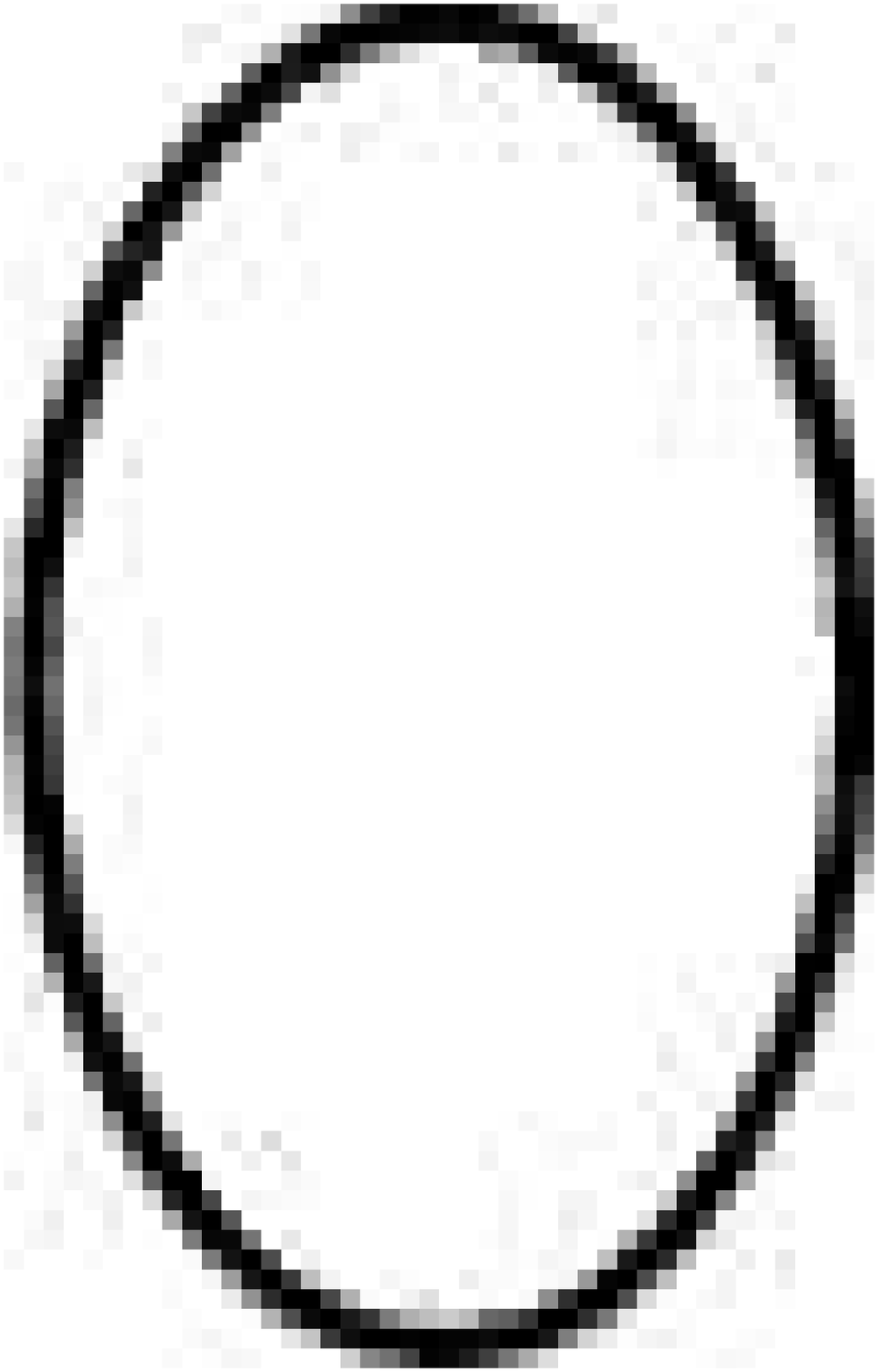} \vspace*{0.005\textheight} }
\end{minipage}&
%canonical forms
%\hspace{0.7in}
\begin{minipage}{0.26\textwidth}
\vspace*{0.005\textheight}\leftline{${\mathcal A}:\; y^2 + z^2 -w^2
=0$} \leftline{${\mathcal B}:\; x^2 =0$}\end{minipage}
\\  \cline{2-5}
&\raisenumbertwo{-6.8pt}{5.0 ex}{20}\boundnum{4.2 ex} {$\langle 0
||| 3 | 4 \rangle$} & (0,(((0,1))),3,(3,0),4)
&\begin{minipage}{0.075\textwidth} \vspace*{0.005\textheight}
\centerline{\epsfysize=0.8\textwidth \epsfbox{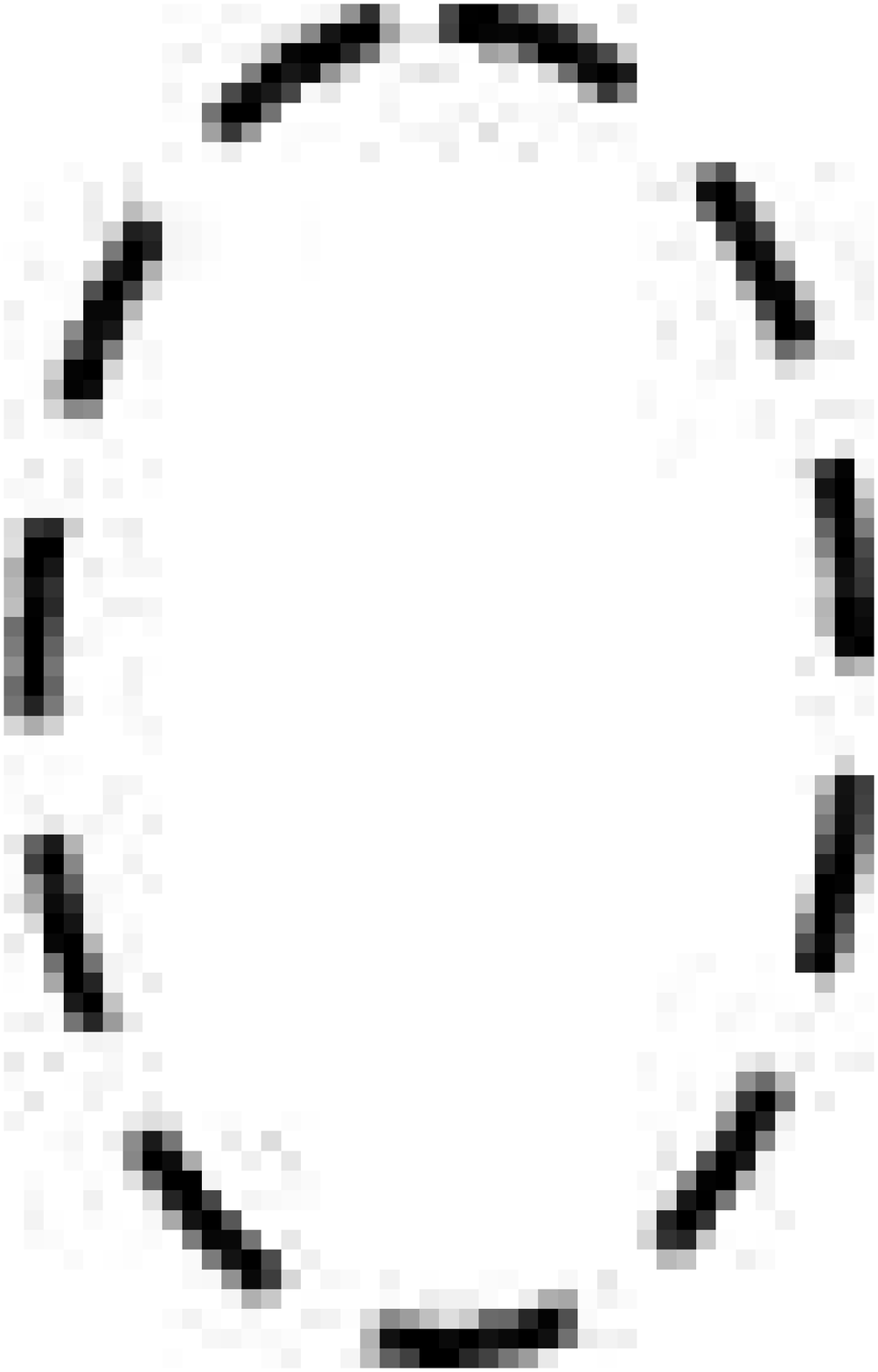}
\vspace*{0.005\textheight} }
\end{minipage}&
%canonical forms
%\hspace{0.7in}
\begin{minipage}{0.26\textwidth}
\vspace*{0.005\textheight}\leftline{${\mathcal A}:\; y^2 + z^2 +w^2
=0$} \leftline{${\mathcal B}:\; x^2 =0$}\end{minipage}
\\ \hline \hline
%
%
%
%[(21)1]
\MLower{$[(21)1]_2$}&\raisenumbertwo{-6.8pt}{5.0
ex}{21}\boundnum{4.2 ex}{$\langle 1 {\wr\wr}_{-}| 2 | 3 \rangle$}
&(1,(((1,1))),2,(2,1),3) &\begin{minipage}{0.075\textwidth}
\vspace*{0.005\textheight} \leftline{\epsfysize=.8\textwidth
\epsfbox{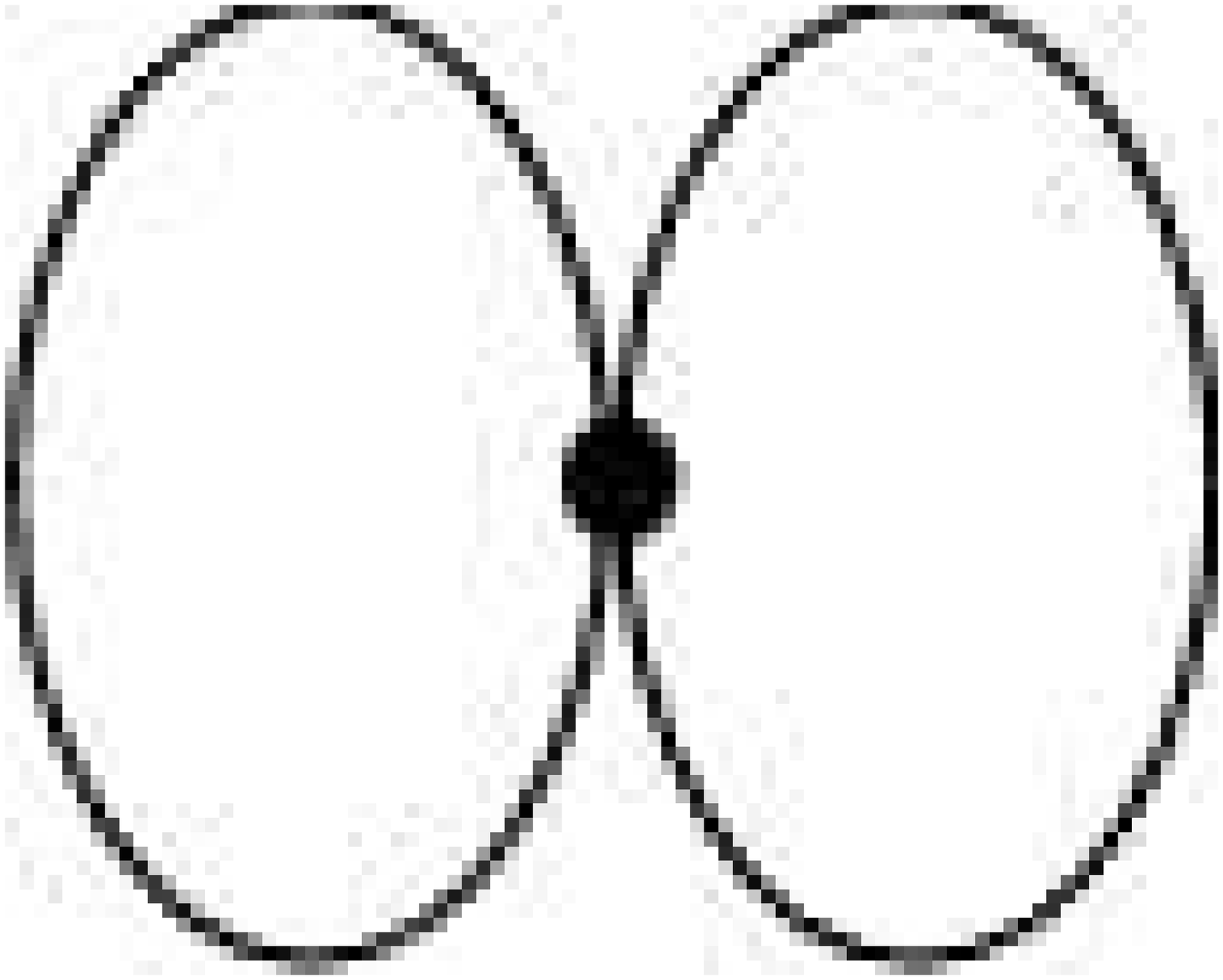} \vspace*{0.005\textheight} }
\end{minipage}&
%canonical forms
%\hspace{0.7in}
\begin{minipage}{0.26\textwidth}
\vspace*{0.005\textheight}\leftline{${\mathcal A}:\; y^2 - z^2 +2zw
=0$} \leftline{${\mathcal B}:\; -x^2 + z^2 =0$}\end{minipage}
\\  \cline{2-5}
&\raisenumbertwo{-6.8pt}{5.0 ex}{22}\boundnum{4.2 ex} {$\langle 1
{\wr\wr}_{+}| 2 | 3 \rangle$} & (1,(((0,2))),2,(2,1),3)
&\begin{minipage}{0.075\textwidth} \vspace*{0.005\textheight}
\leftline{\epsfysize=.8\textwidth \epsfbox{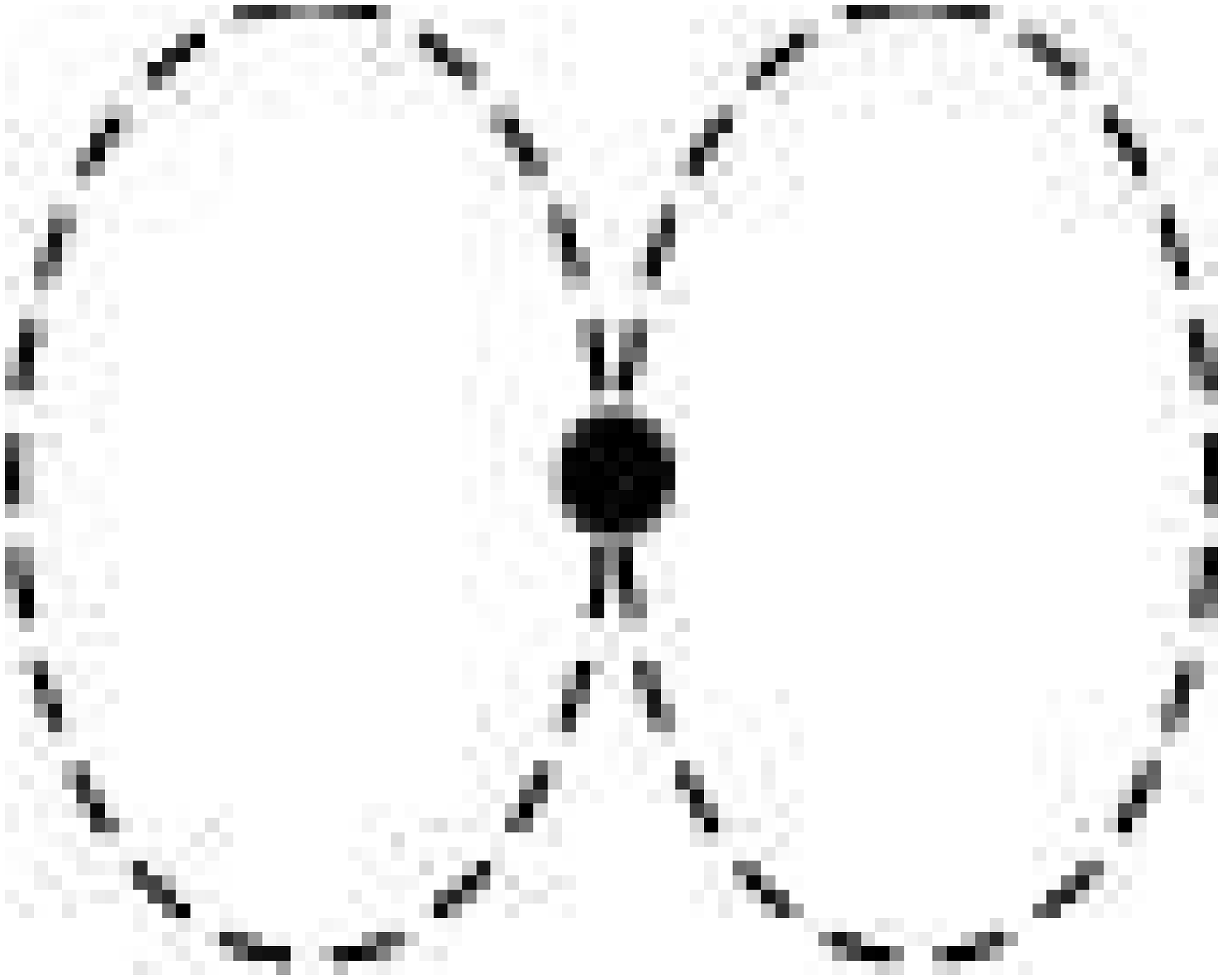}
\vspace*{0.005\textheight} }
\end{minipage}&
%canonical forms
%\hspace{0.7in}
\begin{minipage}{0.26\textwidth}
\vspace*{0.005\textheight}\leftline{${\mathcal A}:\; y^2 - z^2 +2zw
=0$} \leftline{${\mathcal B}:\; x^2 + z^2 =0$}\end{minipage}
\\ \hline \hline
\end{tabular}
\end{scriptsize}
\end{center}
\end{table}
%
%
%
%
%%%%%%%%%%%%%%%%%%%%%%%%%%%%%%%%%%%%%%%%%%%%%%%%%%%%%%%%
%%%%%Table III%%%%%%%%%%%%%%%
\begin{table}[htbp]\caption{Classification of planar QSIC in $\mathbb{PR}^3$ - Part II}
\begin{center}
\begin{scriptsize}
\begin{tabular}{l|l|c|l|l}
\hline
$[{\bf Segre}]_r$ &&&&\\
$r$ = the \# &{\bf Index}&{\bf Signature Sequence}&{\bf Illus-}&{\bf Representative}\\
of real roots &{\bf Sequence}&&{\bf tration}& {\bf Quadric Pair}\\
\hline
%
%
%[(11)2]
&\raisenumbertwo{-6.8pt}{6.0 ex}{23}\boundnum{5.2 ex}{$\langle 2
{\wr\wr}_{-} 2 || 2 \rangle$} &(2,((2,1)),2,((1,1)),2)
&\begin{minipage}{0.080\textwidth}
\vspace*{-0.01\textheight}\centerline{\epsfysize=.8\textwidth
\epsfbox{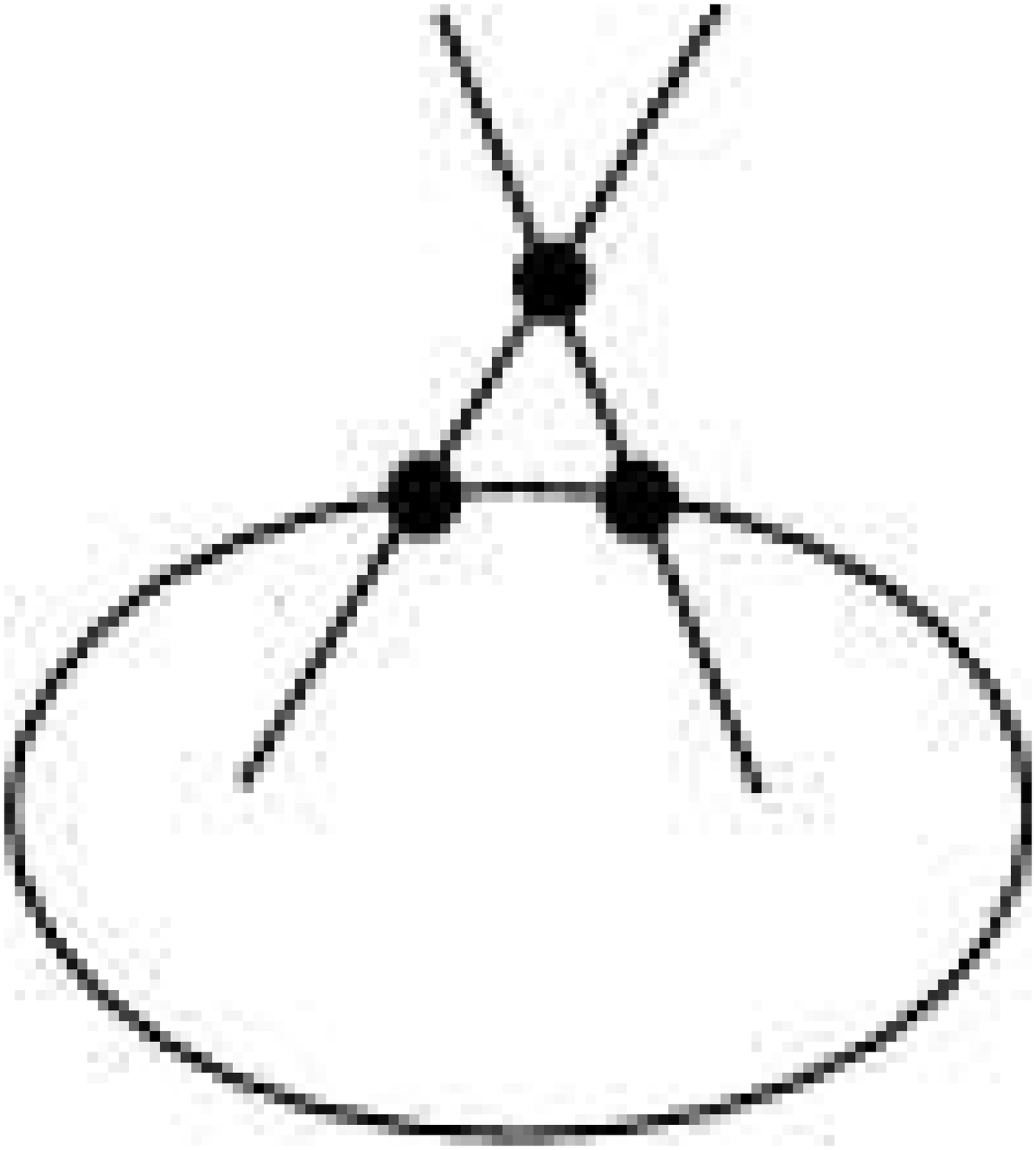} \vspace*{0.002\textheight}}
\end{minipage}&
%canonical forms
%\hspace{0.7in}
\begin{minipage}{0.28\textwidth}
\vspace*{-0.01\textheight}\leftline{${\mathcal A}:\; 2xy - y^2=0$}
\leftline{${\mathcal B}:\; y^2 + z^2 - w^2=0$}\end{minipage}
\\  \cline{2-5}
$[2(11)]_2$&\raisenumbertwo{-6.8pt}{6.0 ex}{24}\boundnum{5.2 ex}
{$\langle 1 {\wr\wr}_{-} 1 || 3 \rangle$} &
(1,((1,2)),1,((1,1)),3)&\begin{minipage}{0.080\textwidth}
\vspace*{-0.01\textheight} \centerline{\epsfysize=.8\textwidth
\epsfbox{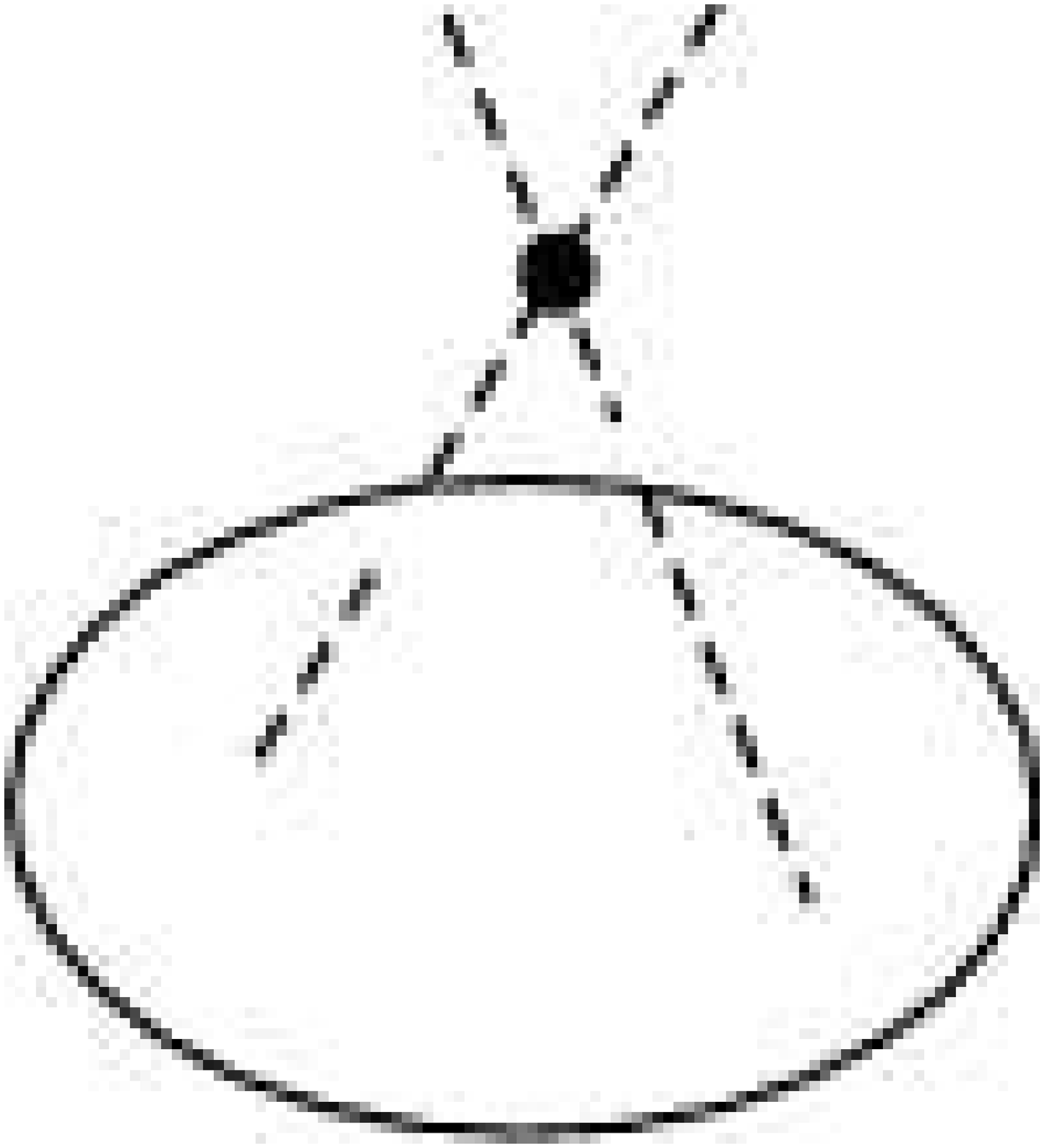} \vspace*{0.002\textheight} }
\end{minipage}&
%canonical forms
%\hspace{0.7in}
\begin{minipage}{0.28\textwidth}
\vspace*{-0.01\textheight}\leftline{${\mathcal A}:\; 2xy - y^2=0$}
\leftline{${\mathcal B}:\; y^2  - z^2 - w^2=0$}\end{minipage}
\\  \cline{2-5}
&\raisenumbertwo{-6.8pt}{6.0 ex}{25}\boundnum{5.2 ex} {$\langle 1
{\wr\wr}_{+} 1 || 3 \rangle$}& (1,((0,3)),1,((1,1)),3)
&\begin{minipage}{0.080\textwidth} \vspace*{-0.01\textheight}
\centerline{\epsfysize=.8\textwidth \epsfbox{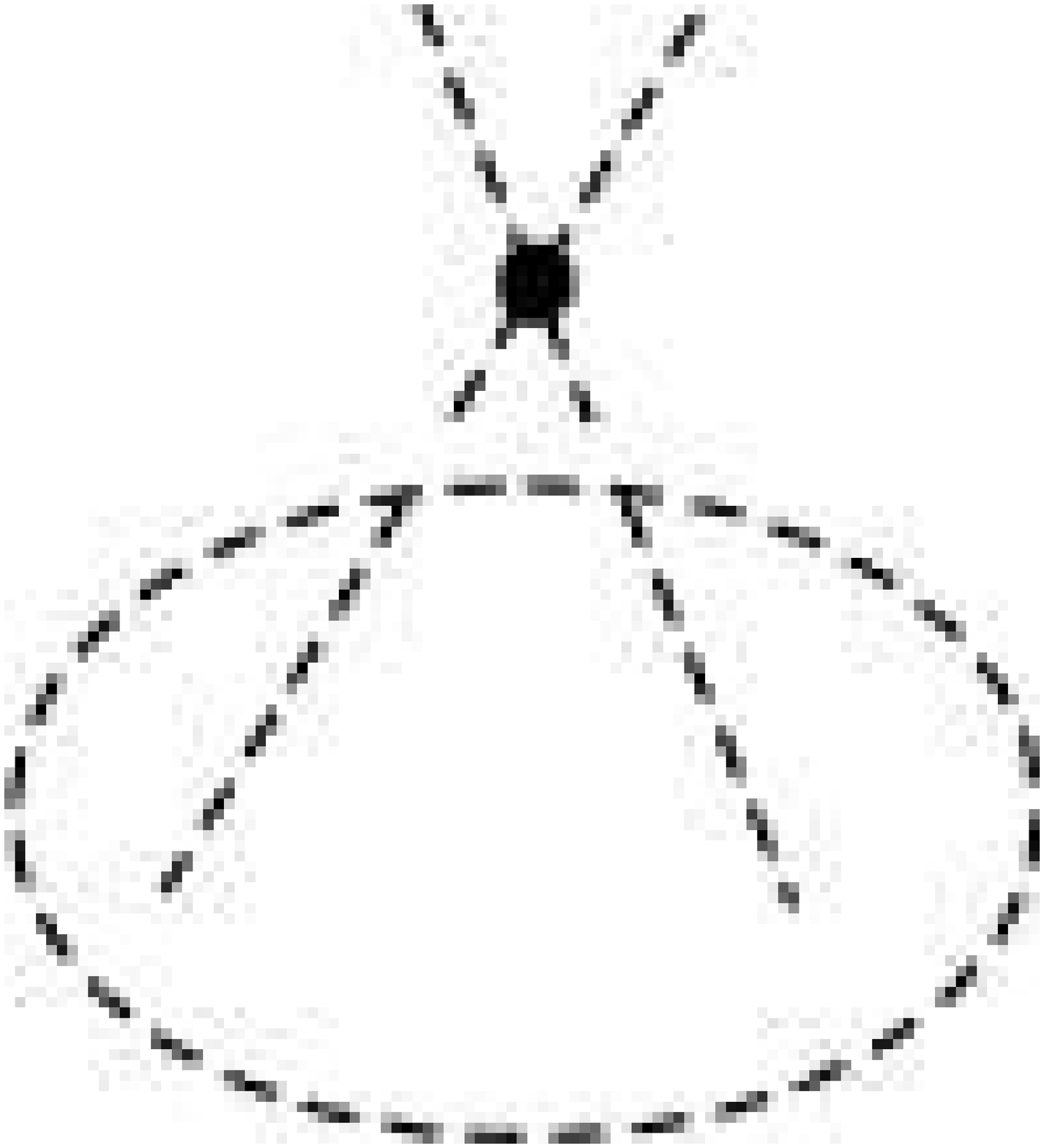}
\vspace*{0.002\textheight} }
\end{minipage}&
%canonical forms
%\hspace{0.7in}
\begin{minipage}{0.28\textwidth}
\vspace*{-0.01\textheight}\leftline{${\mathcal A}:\; 2xy - y^2=0$}
\leftline{${\mathcal B}:\; y^2 + z^2 + w^2=0$}\end{minipage}
\\
\hline \hline
%
%
%[(31)]
\MLower{$[(31)]_1$}&\raisenumbertwo{-6.8pt}{6.0 ex}{26}\boundnum{5.2
ex} {$\langle 2 {\wr\wr}{\wr}_{-} | 2 \rangle$}  & (2,((((1,1)))),2)
&\begin{minipage}{0.080\textwidth} \vspace*{-0.01\textheight}
\centerline{\epsfysize=.8\textwidth \epsfbox{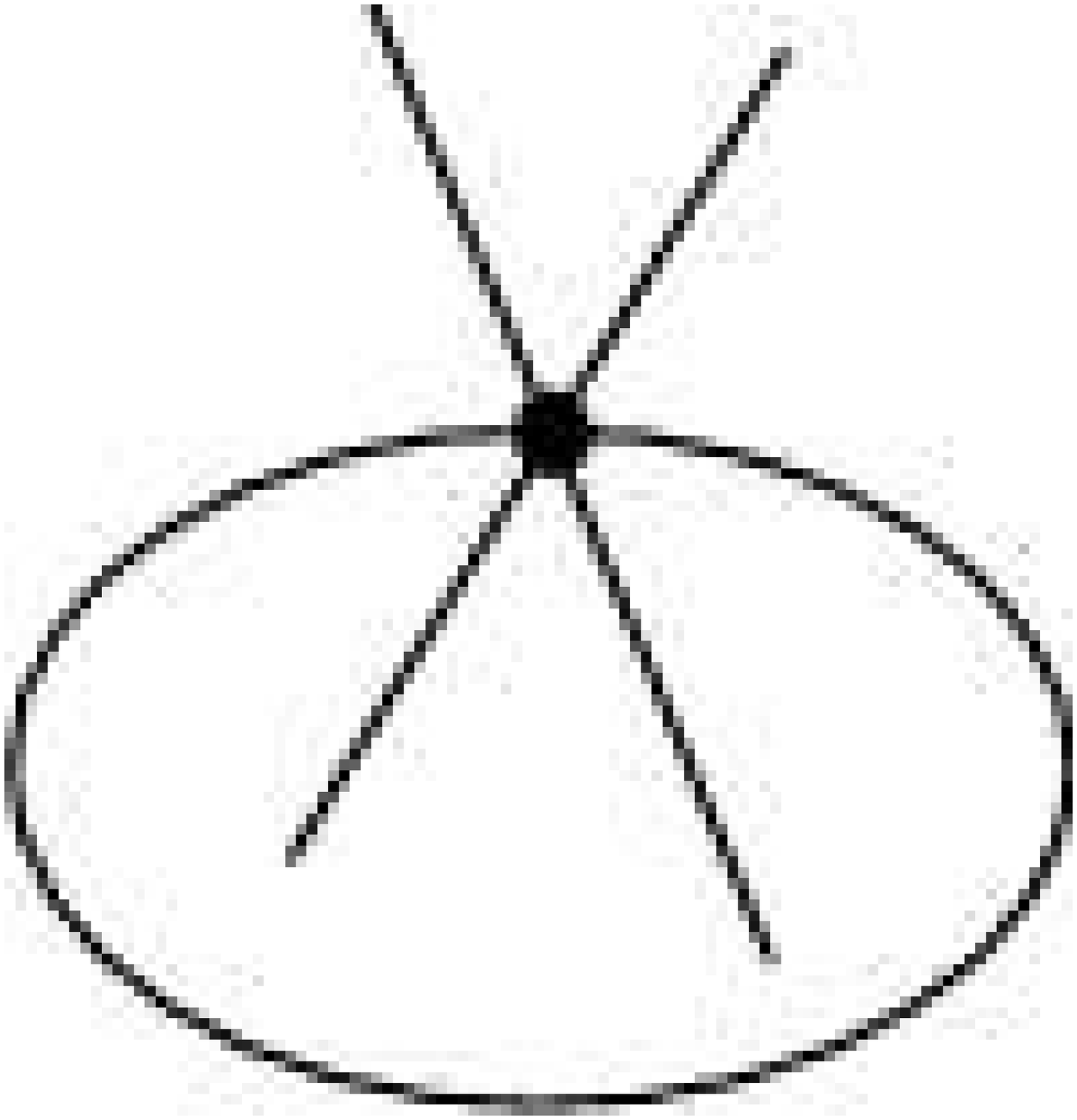}
\vspace*{0.002\textheight} }
\end{minipage}&
%canonical forms
%\hspace{0.7in}
\begin{minipage}{0.28\textwidth}
\vspace*{-0.01\textheight}\leftline{${\mathcal A}:\; y^2 + 2xz -
w^2=0$} \leftline{${\mathcal B}:\; yz=0$}\end{minipage}
\\  \cline{2-5}
&\raisenumbertwo{-6.8pt}{6.0 ex}{27}\boundnum{5.2 ex} {$\langle 1
{\wr\wr}{\wr}_{+} | 3 \rangle$} & (1,((((1,1)))),3)
&\begin{minipage}{0.080\textwidth} \vspace*{-0.01\textheight}
\leftline{\epsfysize=0.8\textwidth \epsfbox{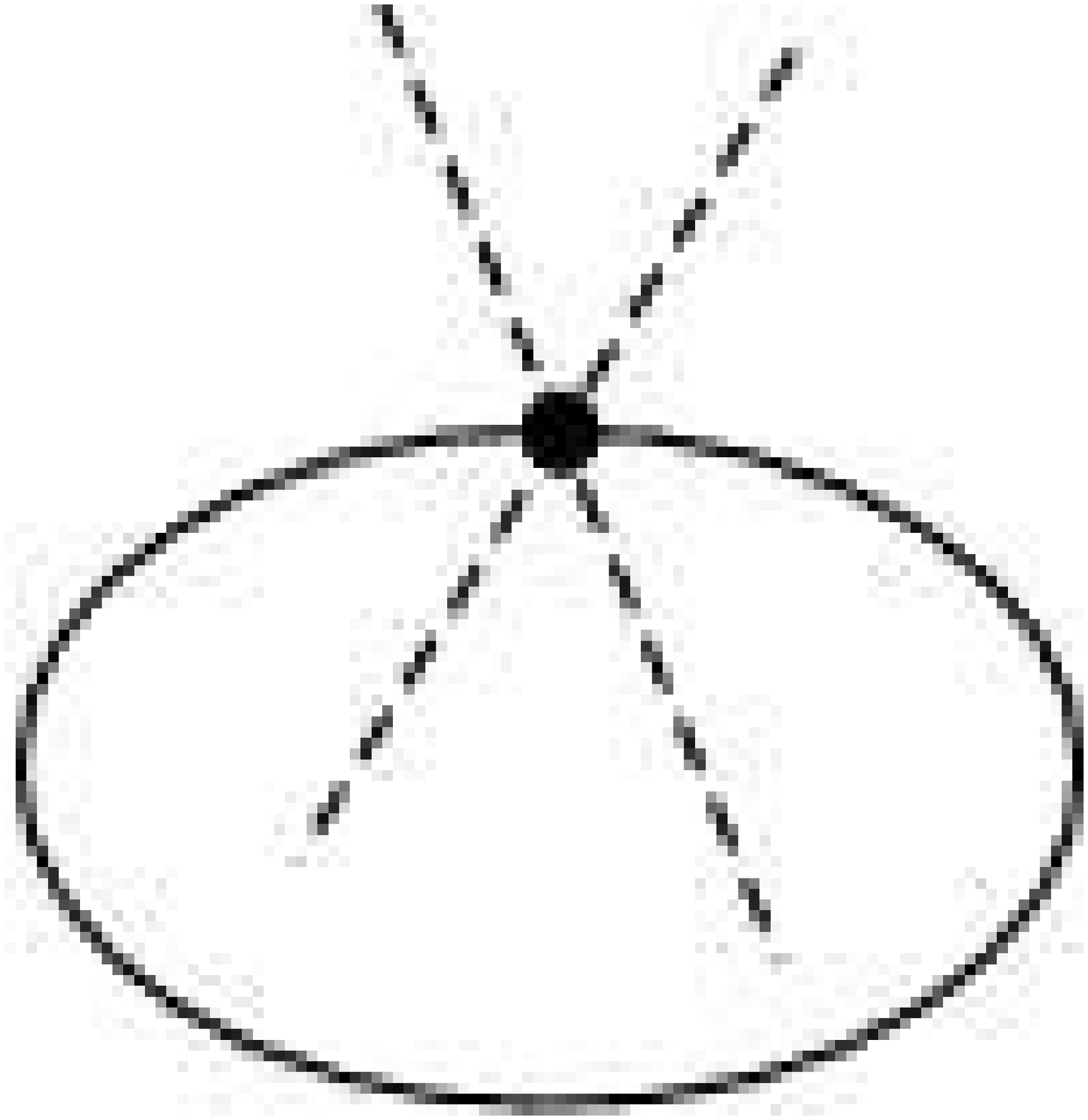}
\vspace*{0.002\textheight} }
\end{minipage}&
%canonical forms
%\hspace{0.7in}
\begin{minipage}{0.28\textwidth}
\vspace*{-0.01\textheight}\leftline{${\mathcal A}:\; y^2 + 2xz +
w^2=0$} \leftline{${\mathcal B}:\; yz=0$}\end{minipage}
\\ \hline\hline
%
%
%[(11)(11)]
&\raisenumbertwo{-6.8pt}{6.0 ex}{28}\boundnum{5.2 ex} {$\langle 2 ||
2 || 2 \rangle$}  & (2,((1,1)),2,((1,1)),2)
&\begin{minipage}{0.080\textwidth} \vspace*{-0.01\textheight}
\centerline{\epsfxsize=1.0\textwidth \epsfbox{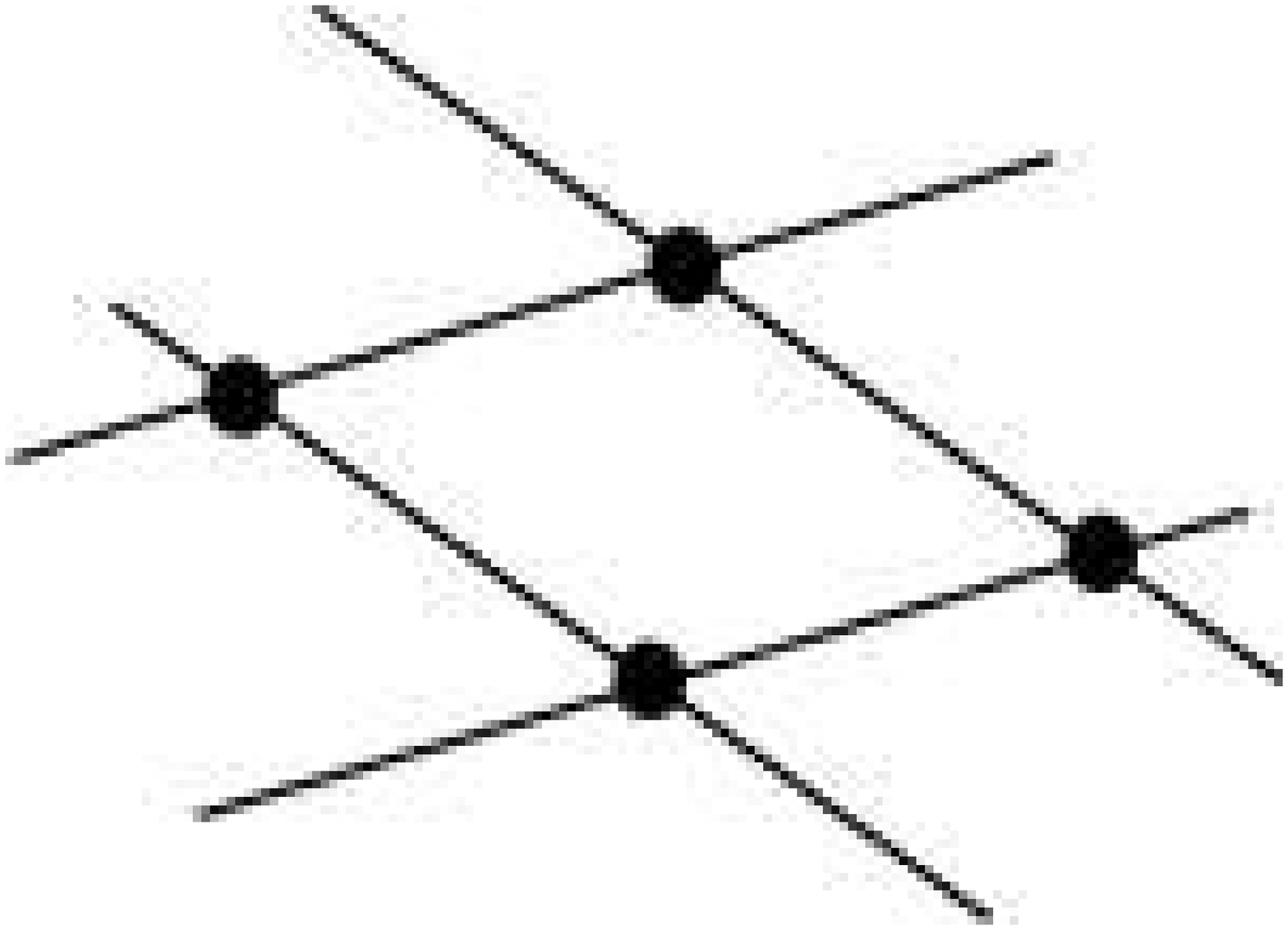}}
\vspace*{0.002\textheight}
\end{minipage}&
%canonical forms
%\hspace{0.7in}
\begin{minipage}{0.28\textwidth}
\vspace*{-0.01\textheight}\leftline{${\mathcal A}:\; x^2-y^2 =0$}
\leftline{${\mathcal B}:\; z^2-w^2=0$}\end{minipage}
\\ \cline{2-5}
{$[(11)(11)]_2$}&\raisenumbertwo{-6.8pt}{6.0 ex}{29}\boundnum{5.2
ex} {$\langle 0 || 2 || 4 \rangle$} & (0,((0,2)),2,((2,0)),4)
&\begin{minipage}{0.080\textwidth} \vspace*{-0.01\textheight}
\centerline{\epsfxsize=.85\textwidth \epsfbox{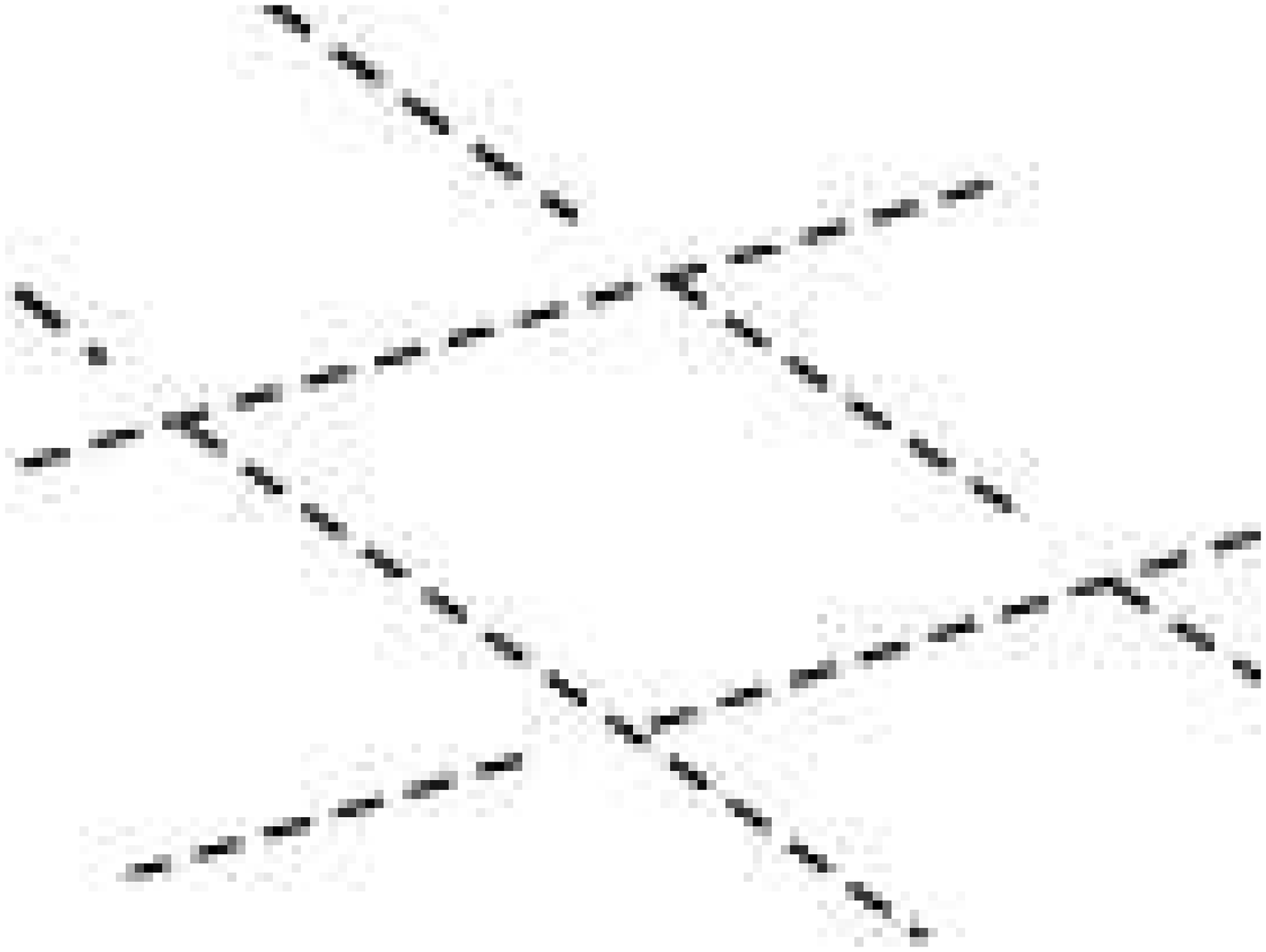}}
\vspace*{0.002\textheight}
\end{minipage}&
%canonical forms
%\hspace{0.7in}
\begin{minipage}{0.28\textwidth}
\vspace*{-0.01\textheight}\leftline{${\mathcal A}:\; x^2 + y^2 =0$}
\leftline{${\mathcal B}:\; z^2 + w^2=0$}\end{minipage}
\\ \cline{2-5}
 &\raisenumbertwo{-6.8pt}{6.0 ex}{30}\boundnum{5.2 ex} {$\langle 1 || 1 || 3
\rangle$} &(1,((0,2)),1,((1,1)),3)
&\begin{minipage}{0.080\textwidth} \vspace*{-0.01\textheight}
\centerline{\epsfxsize=.85\textwidth \epsfbox{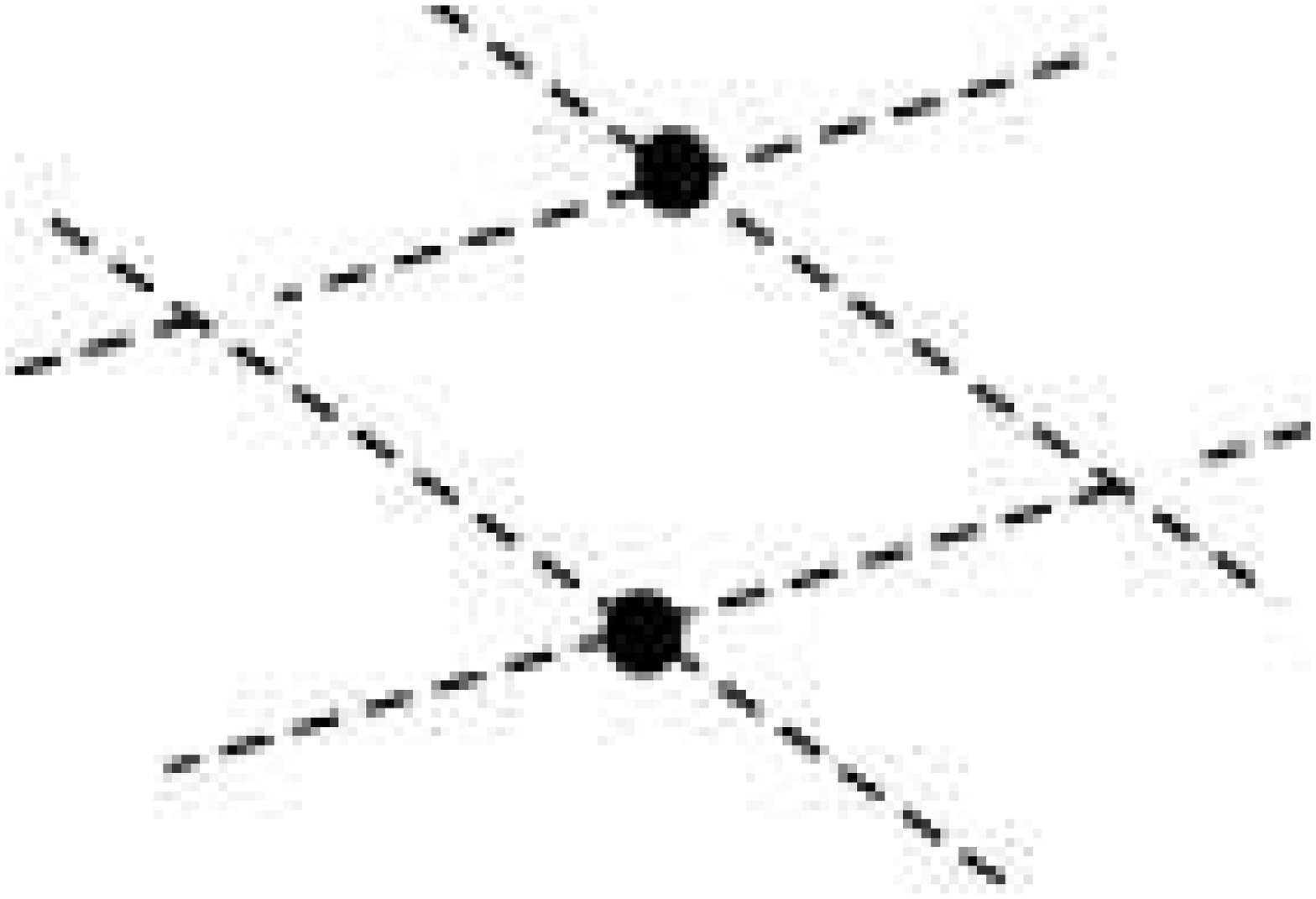}}
\vspace*{0.002\textheight}
\end{minipage}&
%canonical forms
%\hspace{0.7in}
\begin{minipage}{0.28\textwidth}
\vspace*{-0.01\textheight}\leftline{${\mathcal A}:\; x^2 + y^2 =0$}
\leftline{${\mathcal B}:\; z^2-w^2=0$}\end{minipage}
\\  \cline{1-5}
$[(11)(11)]_0$&\raisenumbertwo{-6.8pt}{6.0 ex}{31}\boundnum{5.2 ex}
{$\langle 2 \rangle$} &(2) &\begin{minipage}{0.080\textwidth}
\vspace*{-0.01\textheight} \centerline{\epsfysize=.85\textwidth
\epsfbox{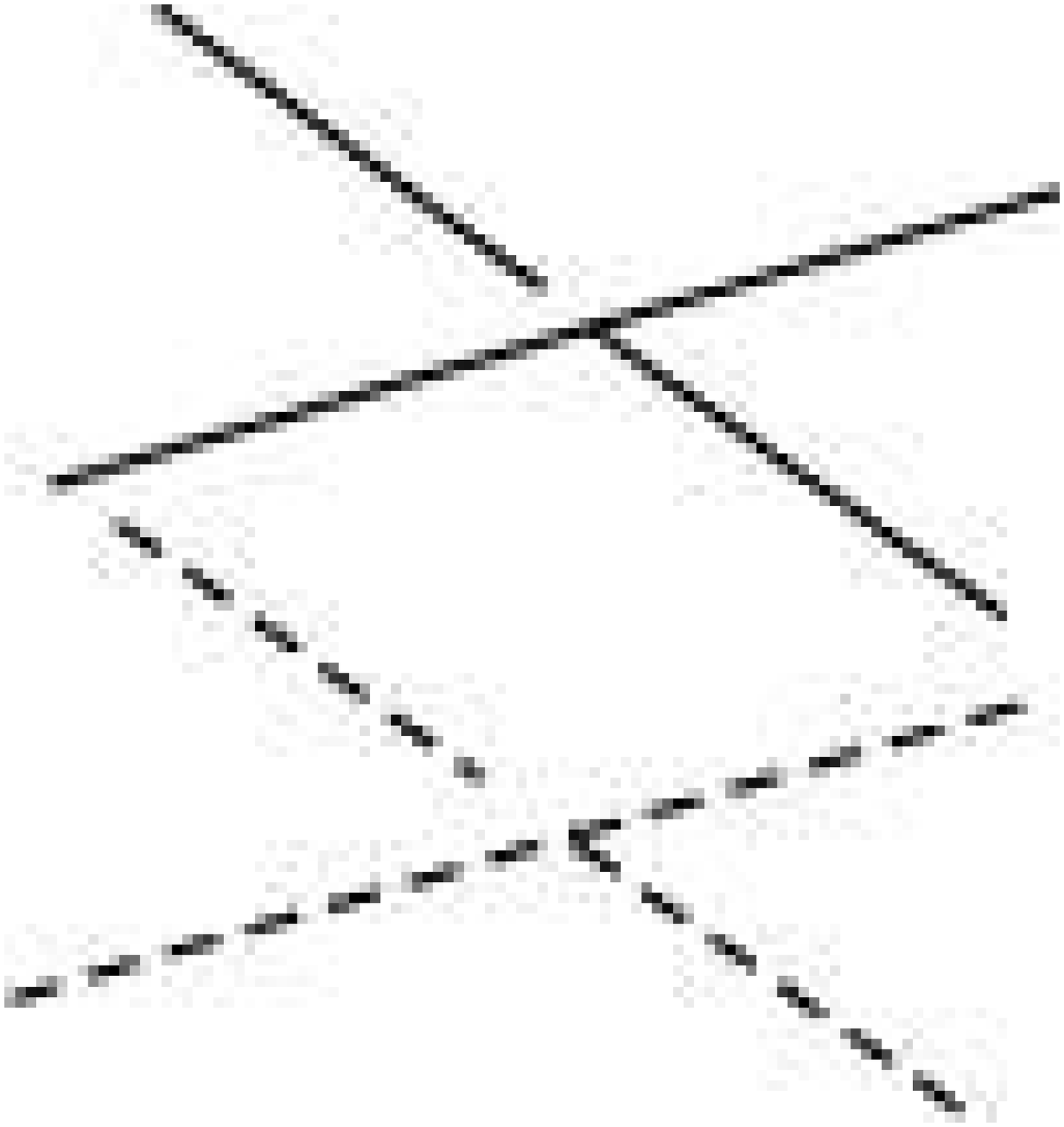} \vspace*{0.002\textheight} }
\end{minipage}&
%canonical forms
%\hspace{0.7in}
\begin{minipage}{0.28\textwidth}
\vspace*{-0.01\textheight}\leftline{${\mathcal A}:\; xy + zw =0$}
\leftline{${\mathcal B}:\; -x^2+y^2-z^2 + w^2=0$}\end{minipage}
\\ \hline\hline
%
%
%[(211)]
\MLower{$[(211)]_1$}&\raisenumbertwo{-6.8pt}{6.0
ex}{32}\boundnum{5.2 ex} {$\langle 2 {\wr\wr}_{-}|| 2 \rangle$}  &
(2,((((1,0)))),2)&\begin{minipage}{0.080\textwidth}
\vspace*{-0.01\textheight} \centerline{\epsfxsize=0.8\textwidth
\epsfbox{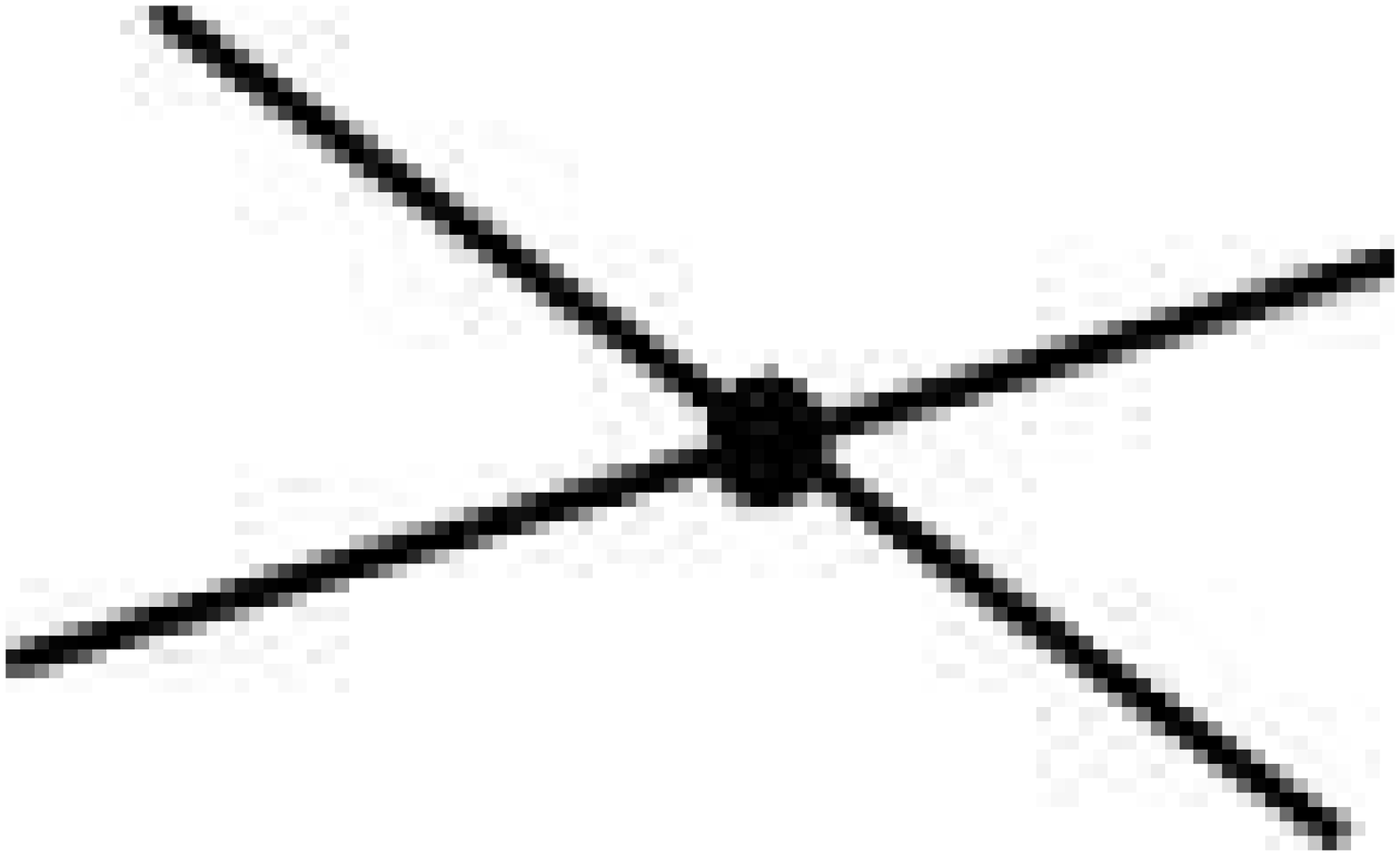}} \vspace*{0.002\textheight}
\end{minipage}&
%canonical forms
%\hspace{0.7in}
\begin{minipage}{0.28\textwidth}
\vspace*{-0.01\textheight}\leftline{${\mathcal A}:\; x^2-y^2+2zw
=0$} \leftline{${\mathcal B}:\; z^2=0$}\end{minipage}
\\ \cline{2-5}
&\raisenumbertwo{-6.8pt}{6.0 ex}{33}\boundnum{5.2 ex} {$\langle 1
{\wr\wr}_{-}|| 3 \rangle$}  & (1,((((1,0)))),3)
&\begin{minipage}{0.080\textwidth} \vspace*{-0.01\textheight}
\centerline{\epsfxsize=0.8\textwidth \epsfbox{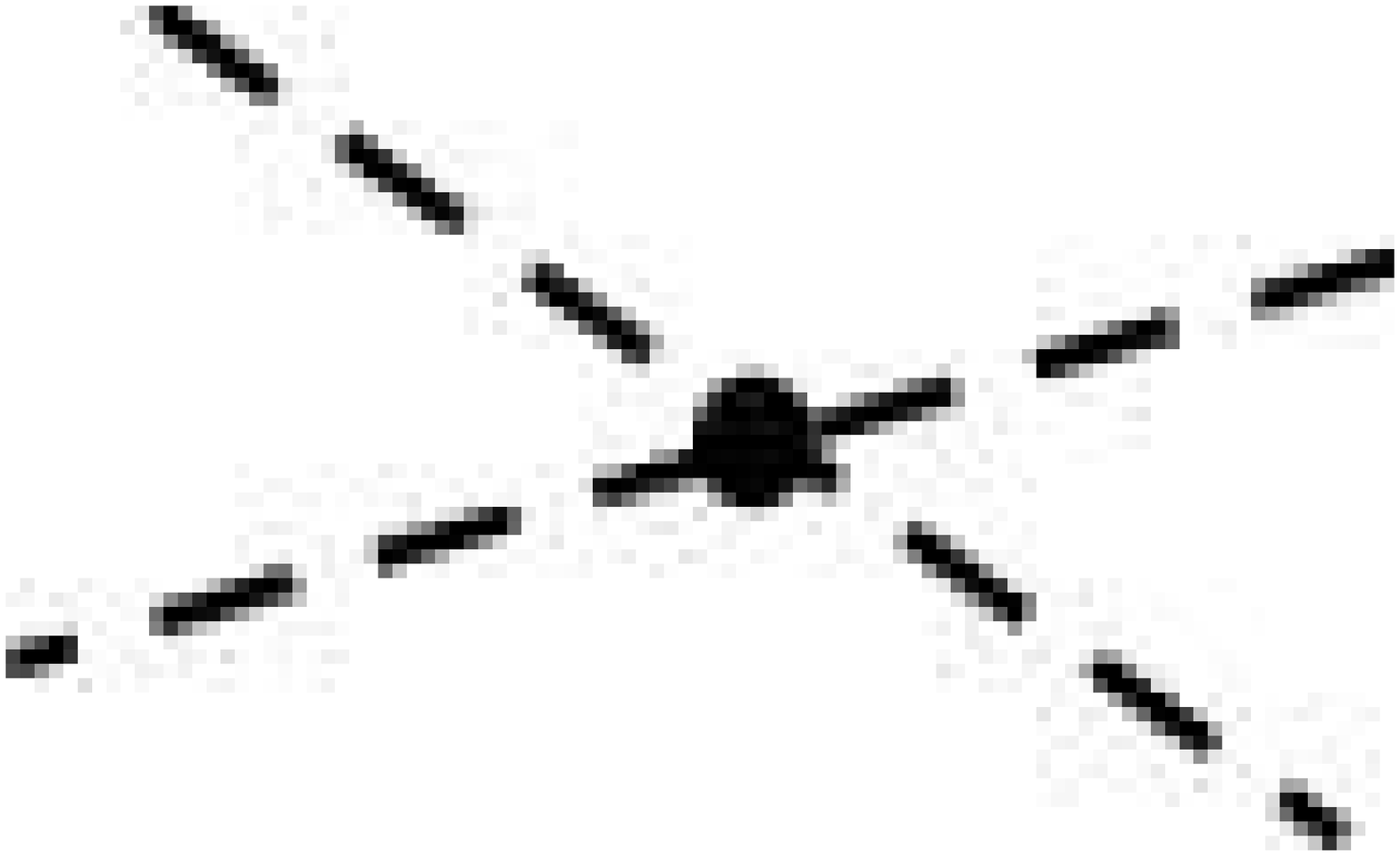}}
\vspace*{0.002\textheight}
\end{minipage}&
%canonical forms
%\hspace{0.7in}
\begin{minipage}{0.28\textwidth}
\vspace*{-0.01\textheight}\leftline{${\mathcal A}:\; x^2+y^2+2zw
=0$} \leftline{${\mathcal B}:\; z^2=0$}\end{minipage}
\\ \hline \hline
%
%
%[(22)]
\MLower{$[(22)]_1$}&\raisenumbertwo{-6.8pt}{6.0 ex}{34}\boundnum{5.2
ex} {$\langle 2 \widehat{\wr\wr}_{-}\widehat{\wr\wr}_{-} 2 \rangle$}
& (2,((((2,0)))),2) &\begin{minipage}{0.080\textwidth}
\vspace*{-0.01\textheight} \centerline{\epsfxsize=0.8\textwidth
\epsfbox{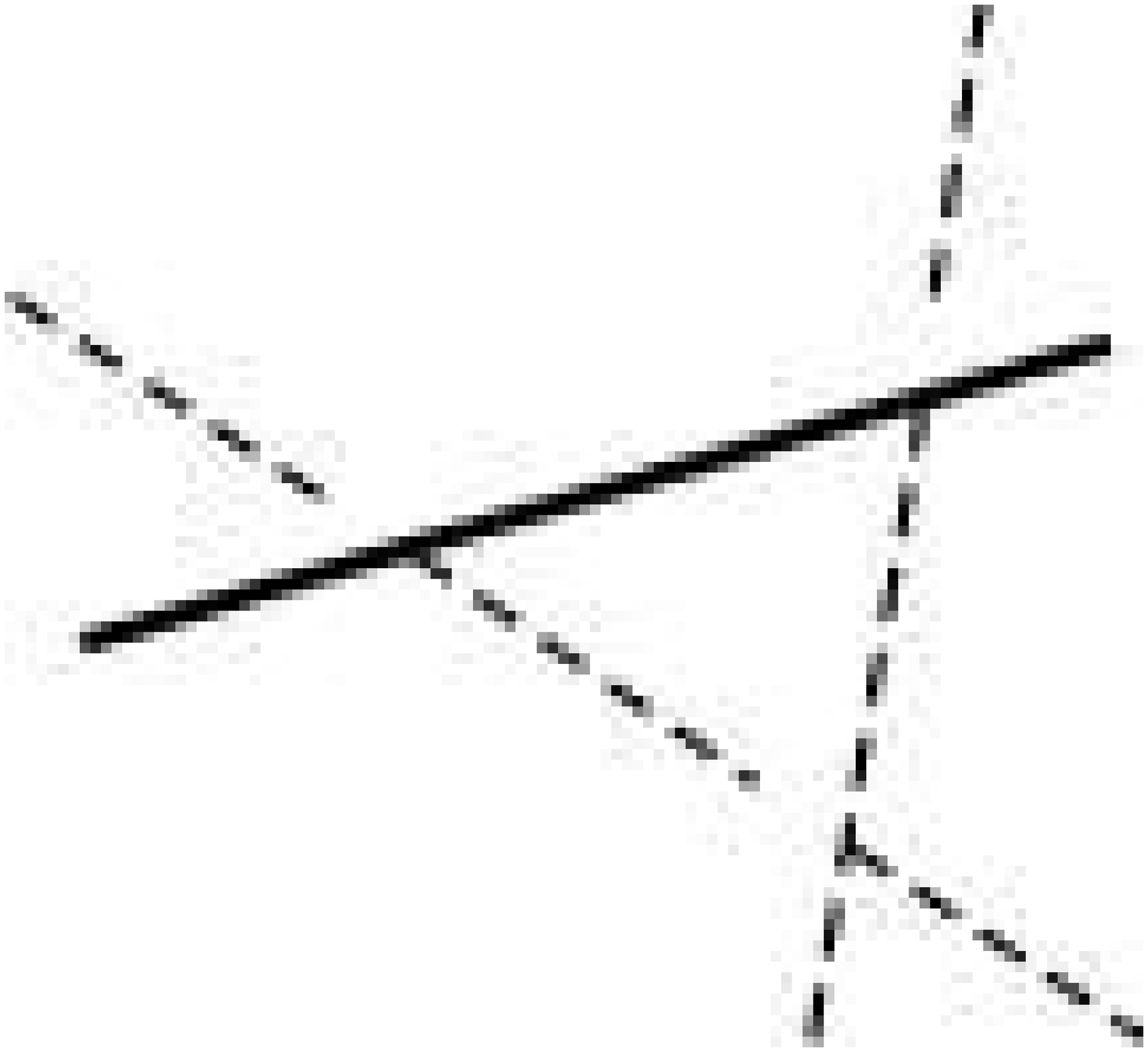}} \vspace*{0.002\textheight}
\end{minipage}&
%canonical forms
%\hspace{0.7in}
\begin{minipage}{0.28\textwidth}
\vspace*{-0.01\textheight}\leftline{${\mathcal A}:\; xy + zw =0$}
\leftline{${\mathcal B}:\; y^2 + w^2=0$}\end{minipage}
\\ \cline{2-5}
&\raisenumbertwo{-6.8pt}{6.0 ex}{35}\boundnum{5.2 ex} {$\langle 2
\widehat{\wr\wr}_{-}\widehat{\wr\wr}_{+} 2 \rangle$} &
(2,((((1,1)))),2) &\begin{minipage}{0.080\textwidth}
\vspace*{-0.01\textheight} \centerline{\epsfxsize=0.8\textwidth
\epsfbox{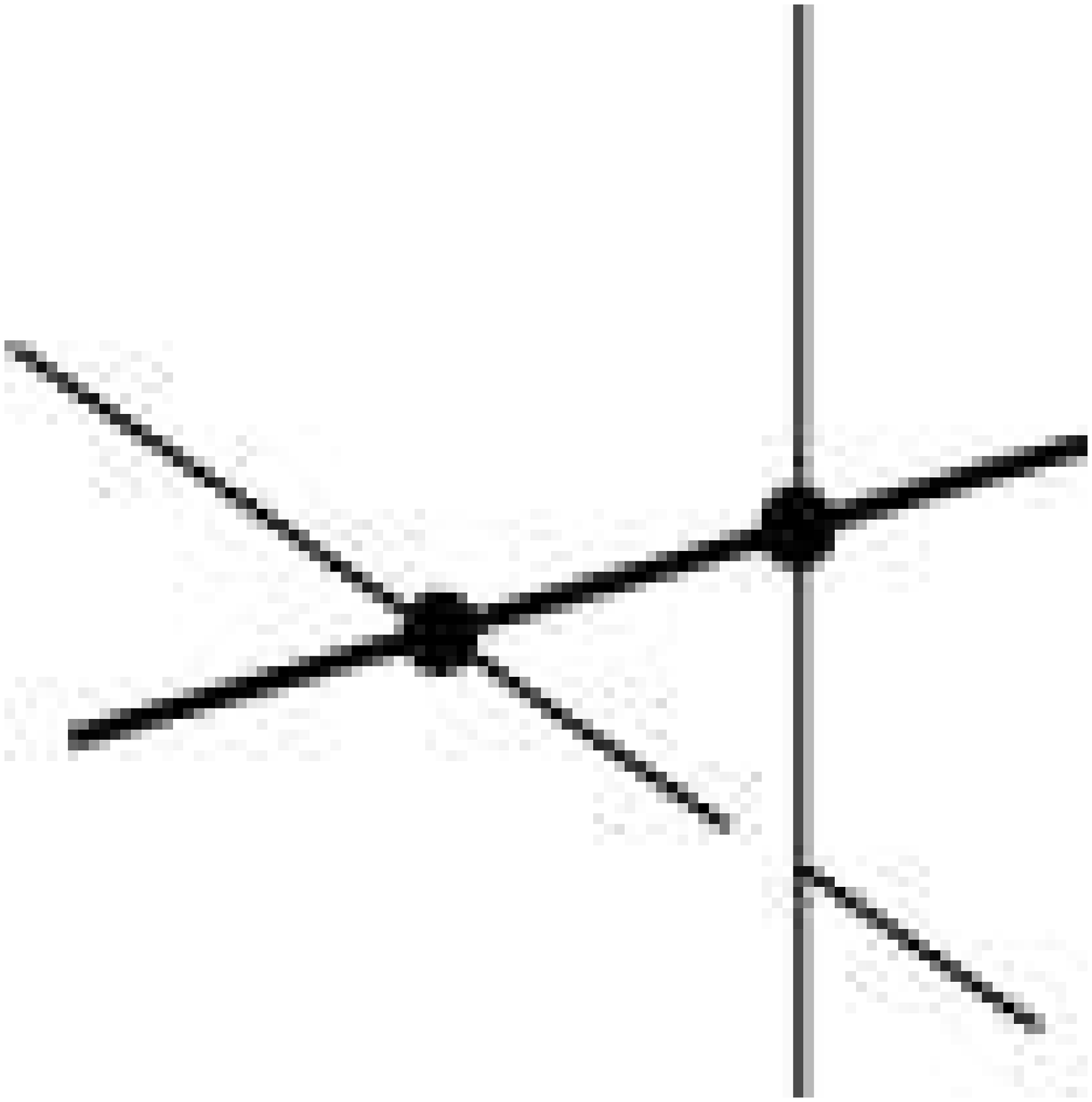}} \vspace*{0.002\textheight}
\end{minipage}&
%canonical forms
%\hspace{0.7in}
\begin{minipage}{0.28\textwidth}
\vspace*{-0.01\textheight}\leftline{${\mathcal A}:\; xy - zw =0$}
\leftline{${\mathcal B}:\; y^2 - w^2=0$}\end{minipage}

\\ \hline \hline
\end{tabular}
\end{scriptsize}
\end{center}
\end{table}

%\normalsize

\section{Preliminaries}\label{sec:Preliminaries}

\subsection{Simplification techniques}

There are two transformations that we will use frequently to
simplify the analysis of a QSIC. Based on Quadric Pair Canonical Form
results~\citep{Muth1905,Will1935,Uhlig1976} 
(see also Section~\ref{block diagonalization}), we
sometimes apply a projective transformation to both ${\mathcal A}$ and
${\mathcal B}$ to get a pair of quadrics ${\mathcal A'}: X^T(Q^TAQ)X=0$ and
${\mathcal B'}: X^T(Q^TBQ)X=0$ in simpler forms.
The transformed quadrics ${\mathcal A'}$ and ${\mathcal B'}$ are
projectively equivalent to ${\mathcal A}$ and ${\mathcal B}$,
therefore have the same QSIC morphology in ${\mathbb {PR}}^3$ and
the same characteristic equation as the pair ${\mathcal A}$ and
${\mathcal B}$.

We sometimes also consider two simpler quadrics in the pencil
spanned by ${\mathcal A}$ and ${\mathcal B}$.  Note that any two
distinct members of the pencil have the same QSIC as that of
${\mathcal A}$ and ${\mathcal B}$, and their characteristic
polynomial is only different from that of ${\mathcal A}$ and
${\mathcal B}$  by a projective (i.e., rational linear) variable
substitution.

\subsection{Open curve components}

If a connected or an irreducible component ${\mathcal C}$ of a QSIC
is intersected by {\em every} plane in $\mathbb{PR}^3$, then
${\mathcal C}$ is called an {\it open component}; otherwise,
${\mathcal C}$ is called a {\it closed component}.
A closed component curve ${\mathcal C}$ is compact in some affine
realization of $\mathbb{PR}^3$; such an affine realization is
obtained by designating the plane at infinity to be a plane in
$\mathbb{PR}^3$ that does not intersect ${\mathcal C}$.
For example, a real non-degenerate conic ${\mathcal C}$ is closed in
$\mathbb{PR}^3$.
In contrast, an open component curve is unbounded in any affine
realization of $\mathbb{PR}^3$; a real line, for example, is an open
curve in $\mathbb{PR}^3$.
Another familiar example in $\mathbb{PR}^2$, 2D projective plane, is
given by a nonsingular cubic curve with two connected components; it
is well known that one of the two components is open (i.e.,
intersected by every line in $\mathbb{PR}^2$) and the other one is
closed (i.e., not intersected by some line in $\mathbb{PR}^2$).
We will see that some higher order open curve components occur in
several QSIC morphologies.

We stress that whether a curve component is open or closed is a
projective property, i.e., this property is not changed by a
projective transformation to the curve. Therefore we need to
consider it for classification of QSICs in $\mathbb{PR}^3$. In fact,
the name ``{\em open component''} is used here due to the lack of a
more appropriate name, because any irreducible or connected
component of a QISC is always ``closed'' in the sense that it is
homomorphic to a circle. However, in this paper we consider the
equivalence of two curve components $\mathbb{PR}^3$ from the point
of view of isotopy, i.e., homotopy of homomorphisms, as used in for
knot theory~\citep{Sieradski1992}. In this sense, an open curve
component (i.e., intersected by every plane in $\mathbb{PR}^3$) and
a closed component (i.e., not intersected by some plane in
$\mathbb{PR}^3$) are not equivalent, because they cannot be mapped
into each other by an isotopy of $\mathbb{PR}^3$.

\subsection{Simultaneous block diagonalization}
\label{block diagonalization}

When given two arbitrary quadrics, we use a projective
transformation to simultaneously map the two quadrics to some
simpler quadrics having the same QSIC morphology and the same root
pattern of the characteristic equation.
Such a projective transformation is based on the standard results on
simultaneous block diagonalization of two real symmetric
matrices~\citep{Muth1905,Will1935,Uhlig1976}, which will be reviewed below.

{\bf Definition 1}: {\em Let $A$ and $B$ be two real symmetric
matrices with $A$ being nonsingular. Then $A$ and $B$ are called a
{\em nonsingular pair of real symmetric (r.s.) matrices}}.

{\bf Definition 2}: {\em A square matrix of the form
\[
M = \left(\begin{array}{cccc}\lambda \;& e\; & & \\
&.\;&.\;& \\
& &.\;& e \\
& & & \lambda\; \end{array} \right)_{k \times k}
\]
is called a {\em Jordan block of type I} if $\lambda \in R$ and
$e=1$ for $k \geq 2$ or $M = (\lambda)$  with $\lambda\in R$ for
$k=1$; $M$ is called a {\em Jordan block of type II} if
\[
\lambda = \left(\begin{array}{rr} a\;&-b\\
b\;&a\end{array} \right) \;\;\; a,b \in R, \; b \ne 0\;\; {\rm
and}\;\;
e = \left(\begin{array}{rr} 1\;&0 \\
0\;&1\end{array} \right),
\]
for $k\geq 4$ or
\[
M = \left(\begin{array}{rr} a&-b\\
b& a\end{array} \right)
\]
for $k=2$, with $a,b \in R$, $b\neq 0$.}

{\bf Definition 3}: {\em Let $J_1$,...,$J_k$ be all the Jordan
blocks (of type I or type II) associated with the same eigenvalue
$\lambda$ of a real matrix $A$. Then $$C=C(\lambda)={\rm
diag}(J_1,...,J_k),$$ where ${\rm dim} (J_i) \geq {\rm dim}
(J_{i+1})$, is called the {\em full chain of Jordan blocks} or {\em
full Jordan chain of length $k$} associated with $\lambda$.}

{\bf Definition 4}: {\em If $\lambda_1$,...,$\lambda_k$ are all
distinct eigenvalues of a real matrix $A$, with only one being
listed for each pair of complex conjugate eigenvalues,  then the
{\em real Jordan normal form} of $A$ is J={\rm
diag}(C($\lambda_1$),...,C($\lambda_k$))}.

Recall that two square matrices $C$ and $D$ are congruent if there
exists a nonsingular matrix $Q$ such that $C = Q^TDQ$; we also say
that $C$ and $D$ are related by a congruence transformation, which
amounts to a change of projective coordinates.

%%<BM>
\begin{theorem}(Quadric Pair Canonical Form)
\label{thm:canonical_form}

Let $A$ and $B$ be a nonsingular pair of real symmetric matrices of
size n. Suppose that $A^{-1}B$ has real Jordan normal form ${\rm
diag} (J_1,...J_r,J_{r+1},...J_m)$, where $J_1,...J_r$ are Jordan
blocks of type I corresponding to the real eigenvalues of $A^{-1}B$
and $J_{r+1},...J_m$ are Jordan blocks of type II corresponding to
the complex eigenvalues of $A^{-1}B$.
Then the following properties hold:
\begin{enumerate}
\item $A$ and $B$ are simultaneously congruent by a real congruence transformation to
$${\rm diag}(\varepsilon_1E_1,... \varepsilon_rE_r, E_{r+1},...E_m)$$ and
$${\rm diag}(\varepsilon_1E_1J_1,...\varepsilon_rE_rJ_r, E_{r+1}J_{r+1},...E_mJ_m),$$
respectively, where $\varepsilon_i=\pm 1$ and the $E_i$ are of the
form
\[
\left(\begin{array}{cccc} 0&.&0&1\\
.&.&.&.\\
1&0&.&0\end{array} \right)
\]
of the same size as $J_i$, $i=1,2,..,m$. The signs of
$\varepsilon_i$ are unique for each set of indices $i$ that are
associated with a set of identical Jordan blocks $J_i$ of type I.
\item The characteristic polynomial of $A^{-1}B$ and $det(\lambda
A-B)$ have the same roots $\lambda_j$ with the same multiplicities
$\gamma_i$.
\item The sum of the sizes of the Jordan blocks corresponding to a real root $\lambda_i$ is
the multiplicity $\gamma_i$ if $\lambda_i$ is real or twice this
multiplicity if $\lambda_i$ is complex. The number of the
corresponding blocks is $\rho_i = n-{\rm rank}(\lambda_i A - B)$,
and $1\leq\rho_i\leq\gamma_i$.
\end{enumerate}
\end{theorem}
As detailed in the review article \cite{LanRod2005}, this result has a long story. It was proved in \cite{Muth1905} for non-degenerate pencils,
and them further extended and rediscovered several times.
See~\citep{Trott1934,Will1935,Dieud1946,Uhlig1973a,Uhlig1976,Thomp1991,LanRod2005}. 
%%</BM>

In order to apply Theorem 1, we need to ensure that the matrix
${\mathcal A}$ is nonsingular. Since we assume that $f(\lambda) =
\det(\lambda A - B)$ does not vanish identically, $\lambda A - B$ is
nonsingular for infinitely many values of $\lambda$. Therefore,
given two quadrics ${\mathcal A}: X^TAX=0$ and ${\mathcal B}:
X^TBX=0$, we may assume that $A$ is nonsingular; for otherwise we
may replace $A$ by another nonsingular matrix $\tilde A$ such that $
\tilde {\mathcal A}: X^T\tilde AX=0$ and ${\mathcal B}$ have the
same QSIC as that of ${\mathcal A}$ and ${\mathcal B}$.

\subsection{Index sequences} \label{sec:index sequences}

\textbf{Signature and index}: Any $n \times n$ real symmetric matrix $D$ is
congruent to a unique diagonal form $D' = \textrm{diag} ( I_i, - I_j, 0_k )$.
The \textit{signature}, or \textit{inertia}, of $D$ is $( \sigma_+, \sigma_-,
\sigma_0 ) = ( i, j, k )$. The \textit{index} of $D$ is defined as
$\mathrm{index}( D ) = i$.

\textbf{Index function}: The index function of a quadric pencil $\lambda A -
B$ is defined as
\[ \Id ( \lambda ) = \mathrm{index} ( \lambda A - B ), \hspace{0.75em}
   \hspace{0.75em} \lambda \in \mathbb{PR} . \]
Since $A$ and $B$ are matrices of order 4 in our discussion, i.e.,
$n = 4$, we have $\Id ( \lambda ) \in \{ 0, 1, 2, 3, 4 \}$. Note
that $\Id ( \lambda )$ has a constant value in the interval between
any two consecutive real roots of $f ( \lambda ) = 0$. The index
function may have a jump across a real root of $f ( \lambda ) = 0$,
depending on the nature of the root. The index function is also
defined for $\lambda = \infty$ and $- \infty$. We have $\Id ( -
\infty ) + \Id ( + \infty ) = \textrm{rank} ( A )$.

\begin{figure}[htbp]
\begin{center}
\includegraphics[width=8cm]{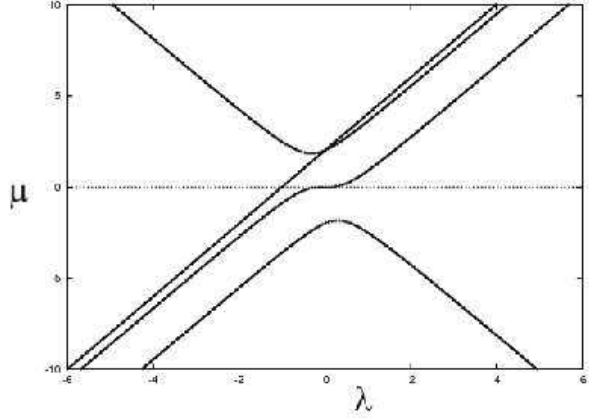}
\end{center}
\footnotesize \caption{The eigenvalue curve of the pencil of the
quadrics $(y^2+2\,x\,z+1,\ 2\, y\, z+ 1)$.} \label{eigencurve}
\end{figure}

\textbf{Eigenvalue Curve:} We consider the real eigenvalues of the pencil
$\lambda A - B$, defined by the equation
\[ C ( \lambda, u ) = \det ( \lambda A - B -u\, I) = 0. \]
We are going to see that the QSIC of a pencil $(A,B)$ can be
characterized by the geometry of the planar curve $\mathcal{C}$
defined by the equation $C ( \lambda, u ) = 0$. This curve
$\mathcal{C}$ is defined by a polynomial whose total and partial
degree in either $\lambda$ or $u$ is $4$. Since a $4 \times 4$
symmetric matrix has $4$ real eigenvalues, for any $\lambda \in
\mathbb{R}$, the number of real roots $C ( \lambda, u ) = 0$ in $u$
is $4$ (counted with multiplicities). Consequently, there are $4$
$\lambda$-monotone branches of $\mathcal{C}$. For any fixed $\lambda
\in \mathbb{R}$, the number of points of $\mathcal{C}$ not on the
$\lambda$-axis, i.e., with $u \neq 0$, is the rank of the quadratic
form $\lambda A - B$; the number of points of $\mathcal{C}$ above
the $\lambda$-axis and the number of points of $\mathcal{C}$ below
the $\lambda$-axis determine the signature of $( \lambda A - B )$.
Figure~\ref{eigencurve} shows the eigenvalue curve of the pencil of
quadrics $(y^2+2\,x\,z+1,\ 2\, y\, z+ 1)$.

\textbf{Index sequence}: Let $\lambda_j$, $j = 1, 2, \dots, r$, be
all distinct real roots of $f ( \lambda ) = 0$ in increasing order.
Let $\mu_k$, $k = 1, 2, \dots, r - 1$, be any real numbers
separating the $\lambda_j$, i.e.,
\[ - \infty < \lambda_1 < \mu_1 < \lambda_2 < \dots \mu_{r - 1} < \lambda_r <
   \infty . \]
Denote $s_j = \Id ( \mu_j )$, $j = 1, 2, \dots, r - 1$. Denote $s_0 = \Id ( -
\infty )$ and $s_r = \Id ( \infty )$. Then the \textit{index sequence} of
$\mA$ and $\mB$ is defined as
\[ \langle s_0 \uparrow s_1 \uparrow \dots \uparrow s_{r - 1} \uparrow s_r
   \rangle, \]
where $\uparrow$ stands for a real root, single or multiple, of $f ( \lambda )
= 0$.

To distinguish different types of multiplicity of a real root, we
use $|$ to denote a real root associated with a $1 \times 1$ Jordan
block, and use $\wr$ for $p$ consecutive times to denote a real root
associated with a $p \times p$ Jordan block. For example, a real
root with Segre characteristic $[ 11 ]$ will be denoted by $||$ in
place of an $\uparrow$ in the index sequence, and a real root with
the Segre characteristic $[ 21 ]$ will be denoted by ${\wr\wr}|$ in
place of an $\uparrow$. When the Segre characteristic is $( 22 )$,
we use $\widehat{\wr\wr} \widehat{\wr\wr}$ to distinguish it from
${\wr\wr}{\wr\wr}$, which has the Segre characteristic $[ 4 ]$.
%In particular,
Supposing that $\lambda_0$ is a real zero of $f ( \lambda )$ with a
Jordan block of size $k \times k$, we use $\wr \cdots \wr_+$ or
$\wr\cdots\wr_{-}$ to indicate that the corresponding sign
$\varepsilon_i$ of the block in the Quadric Pair Canonical Form is $+$ or
$-$.

Since $\lambda$ is a projective parameter, a projective
transformation $\lambda' = ( a \lambda + b ) / ( c \lambda + d )$
does not change the pencil but may change the index sequence of the
pencil. On the other hand, thinking of the projective real line of
$\lambda$ as a circle topologically, such a transformation induces
either a rotation or a reversal of order of the index sequence of
the pencil. Therefore we need to define an equivalence relation of
all index sequences of a quadric pencil under projective
transformations of $\lambda$. In addition, replacing $A$ and $B$ by
$- A$ and $- B$ changes each index $s_i$ to $\textrm{rank} ( \lambda
A - B ) - s_i$ but essentially does not change the pencil $\lambda A
- B$.
Note that the above replacement changes of the sign associated with
a Jordan block of a root; for instance, if the quadrics $A$ and $B$
have the index sequence $\langle 2{\wr\wr}_{-}2|3|2\rangle$, then
$-A$ and $-B$ have the index sequence $\langle
2{\wr\wr}_{+}2|1|2\rangle$.

We choose a representative in an equivalence class such that $A$ is
nonsingular; therefore, $\infty$ is not a root of $f(\lambda)=0$ and
$s_0 + s_r =4$.
Taking these observations and conventions into consideration and
denoting the equivalence relation by $\sim$, this equivalence of
index sequences is then defined by the following three rules:

{\noindent} \textbf{1) Rotation equivalence}:
\begin{eqnarray}
  \langle s_0 \uparrow s_1 \uparrow \dots \uparrow s_{r - 1} \uparrow s_r
  \rangle & \sim & \langle 4 - s_{r - 1} \uparrow s_0 \uparrow s_1 \uparrow
  \dots \uparrow s_{r - 1} \rangle,  \label{rule}\\
  \langle s_0 \uparrow s_1 \uparrow \dots \uparrow s_{r - 1} \uparrow s_r
  \rangle & \sim & \langle s_1 \uparrow s_2 \uparrow \dots \uparrow s_r
  \uparrow 4 - s_1 \rangle . \nonumber
\end{eqnarray}
\textbf{2) Reversal equivalence}:
\begin{equation}
  \label{rule1} \langle s_0 \uparrow s_1 \uparrow \dots \uparrow s_{r - 1}
  \uparrow s_r \rangle \sim \langle s_r \uparrow s_{r - 1} \uparrow \dots
  \uparrow s_1 \uparrow s_0 \rangle.
\end{equation}
\textbf{3) Complement equivalence}:
\begin{equation}
  \label{rule2} \langle s_0 \uparrow s_1 \uparrow \dots \uparrow s_{r - 1}
  \uparrow s_r \rangle \sim \langle 4 - s_0 \uparrow 4 - s_1 \uparrow \dots
  \uparrow 4 - s_{r - 1} \uparrow 4 - s_r \rangle.
\end{equation}

\subsection{Signature variation}\label{signvariation}

In this section we analyze the behavior of the eigenvalues of the
pencil $H ( \lambda ) = \lambda A - B$, near the roots of $f (
\lambda ) = \det ( H ( \lambda ) ) = 0$.
This analysis amounts to analyzing the eigenvalue curves at a real
root of $f(\lambda)$, and is needed for computing the jump of the
index function at the real root.

Consider a transformation $H' ( \lambda ) = P^T H ( \lambda ) P$ of
$H ( \lambda )$, where $P$ is an invertible matrix. First, we
compare the behavior of the eigenvalues of $H' ( \lambda )$ and $H (
\lambda )$. For any real symmetric matrix $Q$ of size $n$, we denote
by $\rho_k ( Q )$ the $k^{\text{\textrm{th}}}$ real eigenvalue of
$Q$, so that $\rho_1 ( Q ) \leqslant \rho_2 ( Q ) \leqslant \cdots
\leqslant \rho_n ( Q )$. Using the Courant-Fischer Maximin Theorem
(see {\citep{gl-mc-89}} p. 403), we have the following result:
\begin{equation}
  \rho_k ( Q ) \sigma_1 ( P )^2 \leqslant \rho_k ( P^T QP ) \leqslant \rho_k (
  Q ) \sigma_n ( P )^2, \label{CrFish}
\end{equation}
where $\sigma_1 ( P )$ (resp. $\sigma_n(P)$) is the smallest (resp. largest)
singular value of $P$.

\begin{proposition}
  \label{proppuiseux} Let $P$ be an invertible matrix and $H' ( \lambda ) =
  P^T H ( \lambda ) P$. If $\rho_k ( H ( \lambda ) ) = a \lambda^{\mu} ( 1 + o
  ( \lambda ) )$ with $a \neq 0$, then $\rho_k ( H' ( \lambda ) ) = a'
  \lambda^{\mu} ( 1 + o ( \lambda ) )$ with ${\mathrm{sign}} ( a ) =
  {\mathrm{sign}} ( a' )$.
\end{proposition}

%\begin{proof}
{\bf Proof} As the eigenvalue $\rho_k ( H' ( \lambda ) )$ has a
Puiseux expansion
  ~\citep{Abh90,w-ac-62} near $\lambda = 0$ of the form $\rho_k ( H'
  ( \lambda ) ) = \rho' + a' \lambda^{\mu'} ( 1 + o ( \lambda ) )$ with
  $\rho', a' \in \mathbb{R}$ and $\mu' \in \mathbb{Q}$, we deduce from the
  inequalities (\ref{CrFish}) that $\rho' = 0$, $\mu' = \mu$ and
  ${\mathrm{sign}} ( a' ) = {\mathrm{sign}} ( a )$.
%\end{proof}

Proposition 1 allows us to deduce the behavior of the eigenvalues of
the pencil $H ( \lambda )$, from its normal form. Indeed, by Theorem
1, $H ( \lambda )$ is equivalent to
\begin{equation}
  \label{normform} D ( \lambda ) = \textrm{diag} ( \varepsilon_1 E_1 ( \lambda
  I_1 - J_1 ( \lambda_1 ) ), \varepsilon_2 E_2 ( \lambda I_2 - J_2 ( \lambda_2 )
  ), \dots, \varepsilon_r E_r ( \lambda I_r - J_r ( \lambda_r ) ), D' ( \lambda )
  ),
\end{equation}
where $I_i$ is the identity matrix of the same size as that of the Jordan
block $J_i ( \lambda_i )$ of eigenvalue $\lambda_i$, and $\det ( D' ( \lambda
) )$ has no real roots. Let us denote by $N_{k_{}} ( \lambda, \rho,
\varepsilon ) = \varepsilon_{} E^k ( \lambda_{} I^k - J^k ( \rho^{} ) )$ a
block of the preceding form, where $k$ is the size of the corresponding
matrices. Then we have the following property:

\begin{proposition}
  The eigenvalue branch $\rho ( \lambda )$ corresponding to $N_{k_{}} ( \lambda, \rho_0,
  \varepsilon )$ which vanishes at $\lambda = \rho_0$ is of the form
  \[ \rho = \varepsilon \nu^k ( 1 + o ( \nu ) ) \]
  where $\lambda = \rho_0 + \nu$.
\end{proposition}

%\begin{proof}
{\bf Proof} By an explicit expansion of the determinant $\mathcal{N}
( \lambda, u ) =   \det ( N_{k_{}} ( \lambda, \rho_0, \varepsilon )
- uI_k )$ and denoting $\nu = \lambda - \rho_0$,  we obtain
  \[ \mathcal{N} ( \lambda, u ) = \widetilde{\mathcal{N}} ( \nu, u ) = ( - 1 )^k u^k + \cdots + ( -
     \varepsilon )^{k - 1} ( - 1 )^{\frac{( k - 1 ) ( k - 2 )}{2} + 1} u +
     \varepsilon^k ( - 1 )^{\frac{k ( k - 1 )}{2}} \nu^k. \]

The vertices of the lower envelop of the Newton polygon of
$\widetilde{\mathcal{N}} ( \nu, u )$ in the $( u, \nu )$-monomial
space are the points  $( k,  0 ), ( 1, 0 ), ( 0, k )$. By Newton's
theorem (see \citep{Abh90} p. 89), the Puiseux expansion of the root
branch which vanishes near $\rho_0$ is of the form
  \[ \rho = \varepsilon \nu^k ( 1 + o ( \nu ) ), \]
which completes the proof.
%\end{proof}

According to Proposition \ref{proppuiseux}, if the pencil $H (
\lambda )$ is equivalent to (\ref{normform}), then near each root
$\lambda_i$, the eigenvalue branches approaching $0$ are of the form
$\varepsilon_i ( \lambda - \lambda_i )^{k_i} ( 1 + o ( \lambda -
\lambda_i ) )$, where $k_{i}$ is the size of a block of the Quadric Pair Canonical 
Form (\ref{normform}) of the eigenvalue $\lambda_{i}$ and
$\varepsilon_{i}$ is the corresponding sign.

\textbf{Index Jump:} The preceding analysis explains how the index
function can change around the real roots of $f ( \lambda ) = 0$.
Let $\alpha$ be a real root of $f ( \lambda ) = 0$. Let $\alpha_-$
and $\alpha_+$ be values sufficiently close to $\alpha$, with
$\alpha_- < \alpha$ and $\alpha_+ > \alpha$. Then the index jumps of
$\Id ( \lambda )$ at $\alpha$ are denoted as
$$
\begin{array}{l}
\Delta^- ( \alpha ) = \Id ( \alpha_{} ) - \Id ( \alpha_- ),\;\;
\Delta^+ (
   \alpha ) = \Id ( \alpha_+ ) - \Id ( \alpha ),\\
\Delta ( \alpha ) = \Id (
   \alpha_+ ) - \Id ( \alpha_- ) = \Delta^- ( \alpha ) + \Delta^+ ( \alpha ) .
\end{array}
$$
We denote by $\Delta_i^{\pm} ( \alpha )$ the changes of signature functions of
the blocks $N_{k_i} ( \lambda, \lambda_i, \varepsilon_i )$ at $\alpha .$
Clearly, we have
\begin{equation}
  \label{index} \Delta^{\pm} ( \alpha ) = \sum_{i = 1}^k \Delta_i^{\pm} (
  \alpha ),\;\; \Delta ( \alpha ) = \sum_{i = 1}^k \Delta_i ( \alpha ) .
\end{equation}
Let us describe each $\Delta_i^{\pm} ( \alpha )$ separately. For any $a \in
\mathbb{R}$, we denote $a^+ = \max ( a, 0 )$ and $a^- = \min ( a, 0 )$.
Note that $a^+ + a^- = a$.

\textbf{(1) Jordan block of size $1 \times 1$}: In this case,
clearly, we have the following signature sequence $(
\varepsilon_i^-, ( 0, 0 ), \varepsilon_i^+ )$ and the jumps are
$\Delta^- ( \lambda_i ) = \varepsilon_i^-, \Delta^+ ( \lambda_i ) =
\varepsilon_{i^{}}^+$ and $\Delta ( \lambda_i ) = \varepsilon_i$.

\textbf{(2) Jordan block of size $2 \times 2$}:
\[ N_{k_i} ( \lambda, \lambda_i, \varepsilon_i ) = \varepsilon_i \left[
   \begin{array}{cc}
     0 & \lambda - \lambda_i\\
     \lambda - \lambda_i & - 1
   \end{array} \right] . \]
In this case the corresponding eigenvalue branch vanishing at
$\lambda_i$ is equivalent to $\varepsilon_i ( \lambda - \lambda_i
)^2$; therefore its sign is the same before and after $\lambda_i$.
There is one positive eigenvalue and one negative eigenvalue before
and after $\lambda_i$. If $\varepsilon_i
> 0$,  we have a positive eigenvalue branch which goes to $0$ at $\lambda_i$;
otherwise, we have a negative one. Thus, the signature sequence of
$N_{k_i} ( \lambda, \lambda_i, \varepsilon_i )$ is $( 1, ( 1 -
\varepsilon_i^+, 1 + \varepsilon_i^- ), 1^{} )$ and the jumps are
$\Delta^- ( \lambda_i ) = - \varepsilon_i^+, \Delta^+ ( \lambda_i )
= \varepsilon_i^+$ and $\Delta ( \lambda_i ) = 0$.

\textbf{(3) Jordan block of size $3 \times 3$}: Since
\[ N_{k_i} ( \lambda, \lambda_i, \varepsilon_i ) = \varepsilon_i \left[
   \begin{array}{ccc}
     0 & 0 & \lambda - \lambda_i\\
     0 & \lambda - \lambda_i & - 1\\
     \lambda - \lambda_i & - 1 & 0
   \end{array} \right] . \]
The corresponding eigenvalue branch is equivalent to $\varepsilon_i
( \lambda - \lambda_i )^3$, whose sign changes before and after
$\lambda_i$. If $\varepsilon_i > 0$, the signature of $N_{k_i} (
\lambda, \lambda_i, \varepsilon_i)$  is $( 1, 2 )$ before
$\lambda_i$ and $( 2, 1 )$ after. If $\varepsilon_i < 0$, we
exchange the order of the two signatures. Thus, we have the
signature sequence $( \varepsilon_i^+ - 2 \varepsilon_i^-, ( 1, 1 ),
2 \varepsilon_i^+ - \varepsilon_{i^{}}^- ) = ( 1 - \varepsilon_i^-,
( 1, 1 ), 1 + \varepsilon_i^+ )$ and $\Delta^- ( \lambda_i ) =
\varepsilon_i^-, \Delta^+ ( \lambda_i ) = \varepsilon_i^+$ and
$\Delta ( \lambda_i ) = \varepsilon_i$.

\textbf{(4) Jordan block of size $4 \times 4$}: Using a similar
argument, we can show that there are two positive eigenvalues and
two negative eigenvalues before and after $\lambda_i$ and the
eigenvalue curve approaching zero has the form $\varepsilon_i (
\lambda - \lambda_i )^4$. Thus, the signature sequence of $N_{k_i} (
\lambda, \lambda_i, \varepsilon_i )$ is $( 2, ( 2 - \varepsilon_i^+,
2 + \varepsilon_i^- ), 2 )$ and $\Delta^- ( \lambda_i ) = -
\varepsilon_i^+, \Delta^+ ( \lambda_i ) = \varepsilon_i^+, \Delta (
\lambda_i ) = 0$.

To summarize, taking into account the sign $\varepsilon_i = \pm 1$,
we have $\Delta_i ( \alpha ) = \varepsilon_i$ if $J_i$ has the size
$1 \times 1$ or $3 \times 3$, and $\Delta_i^{} ( \alpha ) = 0$ if
$J_i$ has the size $2 \times 2$ or $4 \times 4$. The rank of $H (
\lambda )$ drops by $1$ at $\lambda = \lambda_i $ for each block of
the form $N_{k_i} ( \lambda, \lambda_i, \varepsilon_i )$. Thus, the
signature of $H ( \lambda_i )$ can be deduced directly from its
index $\mathrm{Id} ( \lambda_i )$ and the number of Jordan blocks
with eigenvalue $\lambda_i$.

The above rules can be used to decide the permissible index jumps of
$\Id ( \lambda )$ at a real root of $f ( \lambda ) = 0$, through
Eqn. (\ref{index}) and the signature of $H (\lambda_{i})$. In
particular, in the case of a simple root $\lambda_{i}$ of
$f(\lambda)=0$, the sign $\varepsilon_{i}$ in the Quadric Pair Canonical Form
can be deduced directly from the index before and after the root.
For instance, an index sequence of the form $\langle 1|
2|1|2|3\rangle$ corresponds to a sequence of signs
$\varepsilon_{1}=+1,\varepsilon_{2}=-1,\varepsilon_{3}=+1,
\varepsilon_{4}=+1$, and the signatures at the roots are $(1,2),
(1,2), (1,2), (2,1)$, respectively.

\textbf{Signature sequence:} The previous analysis allows us to
completely determine  the signature sequence of the pencil $H (
\lambda ) = \lambda A - B$, from its Quadric Pair Canonical Form. For most of
the cases, this signature sequence is, as we will see, a
characterization of the QSIC.  A signature sequence is defined as
\[ \langle s_0, ( \cdots ( p_1, n_1 ) \cdots ), s_1, \cdots s_{r - 1}, ( \cdots ( p_r,
   n_r ) \cdots ), s_r \rangle, \]
where $s_i$ is the index of $H ( \lambda )$ between two consecutive
real roots of $f ( \lambda ) = 0$, $( p_i, n_i )$ is the signature
of $H ( \lambda_i )$ at a root $\lambda_i$ and  the number of
parentheses is the multiplicity of $\lambda_i$. Note that $ p_i +
n_i = {\rm rank}(\lambda_i A - B)$.

The advantage of using the signature sequence over using the index sequence is
that we just need to compute the multiplicity of a real root and determine the
signature of $\lambda A - B$ at the root; this is a far simpler computation
than computing the Jordan block size, which is the information required by the
index sequence.
Conversion from an index sequence to the corresponding signature
sequence is straightforward.
For a given pair of quadrics, the signature sequence can be computed
easily using only rational arithmetic as described in Section
\ref{EffectiveIssue}.
Similar equivalence rules to those for index sequences apply to
signature sequences as well.
The signature sequences of all 35 QSIC morphologies are listed in
the third column of Tables 1, 2 and 3.

\subsection{Effective issues}\label{EffectiveIssue}

Now we discuss how to use rational arithmetic to compute the signature
sequence for classifying the QISC morphology of a given pair of quadrics.
Consider the polynomial
\[ C ( \lambda, u ) = \det (\lambda A - B - uI ) = u^4 + c_3 ( \lambda ) u^3
   + c_2 ( \lambda ) u^2 + c_1 ( \lambda ) u + c_0 ( \lambda ) .
   \]
The values where the signature changes are defined by $C ( \lambda,
0 ) = c_0 ( \lambda ) = f(\lambda)= 0$.
For a fixed $\lambda$, the
rank of the corresponding quadratic form is the number of non-zero
roots of $C (\lambda, u ) = 0$.
For any fixed $\lambda$, the number of real roots in $u$, counted
with multiplicity, is $4$.
The signature of $\lambda A - B$ is determined by the rank of
$\lambda A - B$ and the number of positive roots of $ C ( \lambda,
u)=0$ in $u$.
In the case where the number of real roots equals the degree of the
polynomial, the Descartes rule gives an exact counting of the number
of positive roots~\citep{Basu2003}, and we have the following
property:
\begin{theorem}\label{thm:signature}
  For any $\lambda \in \mathbb{R}$,
  \begin{itemize}
    \item the number of positive eigenvalues of $\lambda A - B$ is the number
    of sign variations of $[ 1, c_3 ( \lambda ), c_2 ( \lambda ), c_1 (
    \lambda ), c_0 ( \lambda ) ]$.

    \item the number of negative eigenvalues of $\lambda A - B$ is the number
    of sign variations of $[ 1, - c_3 ( \lambda ), c_2 ( \lambda ), - c_1 (
    \lambda ), c_0 ( \lambda ) ]$.
  \end{itemize}
\end{theorem}

Computing the signature $\lambda A - B$ for $\lambda \in \mathbb{Q}$
is straightforward. Computing its signature at a root of $C (
\lambda, 0 ) = f( \lambda ) = 0$ can also be performed using only
rational arithmetic. According to the previous propositions, this
reduces to evaluating the sign of $c_i ( \lambda )$, $i = 1 \ldots
3$. This problem can be transformed into rational computation as
follows. First, we represent a root $\alpha$ of $f ( \lambda ) = 0$
by
\begin{itemize}
  \item the square-free part $p ( \lambda )$ of $f ( \lambda ) = 0$ and

  \item an isolating interval $[ a, b ]$ with $a, b \in \mathbb{Q}$ such that
  $\alpha$ is the only root of $p ( \lambda )$ in $[ a, b ]$.
\end{itemize}
Isolating intervals can be obtained efficiently in several ways
(see, for instance, \citep{mrr:bbrri-05}). They can even be
pre-computed in the case of polynomials of degree 4
\citep{Emiris2004}. In order to compute the sign of a polynomial $g$
at a root $\alpha$ of $f ( \lambda ) = 0$, we use subresultant (or
Sturm-Habicht) sequences. We recall briefly the construction here
and refer to {\citep{Basu2003}} for more details.

Given two polynomials $f ( \lambda )$ and $g ( \lambda ) \in
\mathbb{A} [ \lambda ]$, where $\mathbb{A}$ is the ring of
coefficients, we compute the sub-resultant sequence in $\lambda$,
defined in terms of the minors of the Sylvester resultant matrix of
$f ( \lambda )$ and $f'(\lambda) g ( \lambda )$. This yields a
sequence of polynomials $\mathbf{R} ( \lambda ) = [ R_0 ( \lambda ),
R_1 ( \lambda ), \ldots, R_N ( \lambda ) ]$ with $R_i ( \lambda )
\in \mathbb{A}[ \lambda ]$, whose coefficients are in the same ring
$\mathbb{A}$.

In our case, we take $\mathbb{A} =\mathbb{Z}$. For any $a \in
\mathbb{R}$, we denote by $V_{f,g}( a )$ the number of sign
variation of $\mathbf{R} ( a )$. Then we have the following property
{\citep{Basu2003}}:
\begin{theorem}
\begin{eqnarray*}
V_{f,g}( a ) - V_{f,g} ( b ) &=&  \# \{ \alpha \in [a,b]\ {\rm root\ of}\ f( \lambda ) = 0\
     {\rm where} \ g( \alpha ) > 0 \} -\\
&& \# \{ \alpha\in [a,b]\ {\rm root\ of}\ f( \lambda ) = 0
     \  {\rm where} \ g( \alpha ) < 0 \} .
\end{eqnarray*}
\end{theorem}

In particular, if the interval $[ a, b ]$ is an isolating interval
for a root $\alpha$ of $c_0 ( \lambda ) = 0$, then $V_{f,g} ( a ) -
V_{f,g} ( b )$ gives the sign of $g ( \alpha )$. Taking $g(\lambda)$
to be the coefficients $c_i(\lambda)$ in
Theorem~\ref{thm:signature}, this method allows us to exactly
compute the signature of $\alpha A - B$, using only rational
arithmetic.

Efficient implementations of the algorithms presented here are
available in the library {\sc
synaps}\footnote{\texttt{http://www-sop.inria.fr/galaad/software/synaps/}}
and have been applied to classifying QSIC morphologies, based on the
signature sequences derived in this paper.

\subsection{List of QSIC morphologies}\label{sec:list of morphologies}

All 35 different morphologies of QSIC are listed in Tables 1 through
3. In the first column are the Segre characteristics with the
subscript indicating the number of real roots, not counting
multiplicities.
The index sequences and signature sequences are given in the second
column and the third column, respectively. Here, only one
representative is given for each equivalence class associated with
the corresponding QSIC morphology; in several cases, there are two
equivalence classes associated with one QSIC morphology.
The numeral label for each case, from 1 to 35, is given at the left
upper corner of each entry in the second column.
These labels are referred to in subsequent theorems establishing the
relation between the index sequence and the QSIC morphology.
Cases 4, 10 and 31 share the same index sequence $\langle 2
\rangle$, thus also the same signature sequence $( 2 )$.
Additional simple conditions based on minimal polynomials for
distinguishing these three cases are presented in
Section~\ref{sec:complete classification}.
Two different index sequences in cases 26 and 34 correspond to the
same signature sequences; the discrimination of these two cases is
also discussed in Section~\ref{sec:complete classification}.
In the illustration of each QSIC morphology in column four, a solid
line or curve stands for a real component and a dashed one depicts
an imaginary component. A solid dot indicates a real singular point,
which in many cases is a real intersection point of two or more
components of a QSIC.
An open or closed component is drawn as such in the illustration.
Note that, in addition to topological properties, we also take
algebraic properties into consideration in defining morphologically
different types. For example, a nonsingular QSIC may be vacuous in
$\mathbb{PR}^3$, so is a QSIC consisting two imaginary conics; these
two QSICs are defined to be morphologically different since the
former is irreducible algebraically but the latter is not.

\section{Classifications of nonsingular QSIC}
\label{sec:nonsingular}

\subsection {$[1111]_4$: f$(\lambda)=0$ has four distinct real roots}
\label{sec:four_real_roots}

\begin{theorem} \label{thm:four_real_roots} Given two quadrics ${\mathcal A}$: $X^TAX = 0$ and ${\mathcal
B}$: $X^TBX = 0$, if their characteristic equation $f (\lambda) = 0$
has four distinct real roots, then the only possible index sequences
are $\langle 1|2|1|2|3\rangle$ and $\langle 0|1|2|3|4\rangle$.
Furthermore,
\begin {enumerate}
\item (Case 1, Table 1) when the index sequence is $\langle 1|2|1|2|3\rangle$,
the QSIC has two closed components;
\item (Case 2, Table 1) when the index sequence is $\langle 0|1|2|3|4\rangle$, the QSIC is vacuous in
$\mathbb{PR}^3$.
\end {enumerate}
\end{theorem}

%\begin{proof}
{\bf Proof} Let $\lambda_i$, $i=1,2,3,4$, be the four distinct real
roots of $f(\lambda) =0$.
By Theorem 1, $A$ and $B$ are simultaneously congruent to
\[
\bar A = {\rm diag} (\varepsilon_1, \varepsilon_2, \varepsilon_3,
\varepsilon_4),\;\;\;{\rm and}\;\;\;
\bar B = {\rm diag} (\varepsilon_1\lambda_1, \varepsilon_2\lambda_2,
\varepsilon_3\lambda_3, \varepsilon_4\lambda_4),
\]
where $\varepsilon_i=\pm 1$, $i=1,2,3,4$.
Without loss of generality, we suppose that
$\lambda_1<\lambda_2<\lambda_3<\lambda_4$; this permutation of the
diagonal elements can be achieved by a further congruence
transformation to $\bar A$ and $\bar B$.

Clearly, the only possible index sequences are (up to the
equivalence rules of Section \ref{sec:index sequences}) $\langle
1|2|1|2|3\rangle$ and $\langle 0|1|2|3|4\rangle$. Since a pencil
with the second index sequence $\langle 0|1|2|3|4\rangle$ contains a
positive definite or negative definite quadric, i.e., with the index
being 4 or 0, we deduce that the intersection curve is empty in that
case.

\begin {figure}[htbp]
\begin{minipage}{0.16\textwidth}
\leftline{\epsfxsize=1.0\textwidth \epsfbox{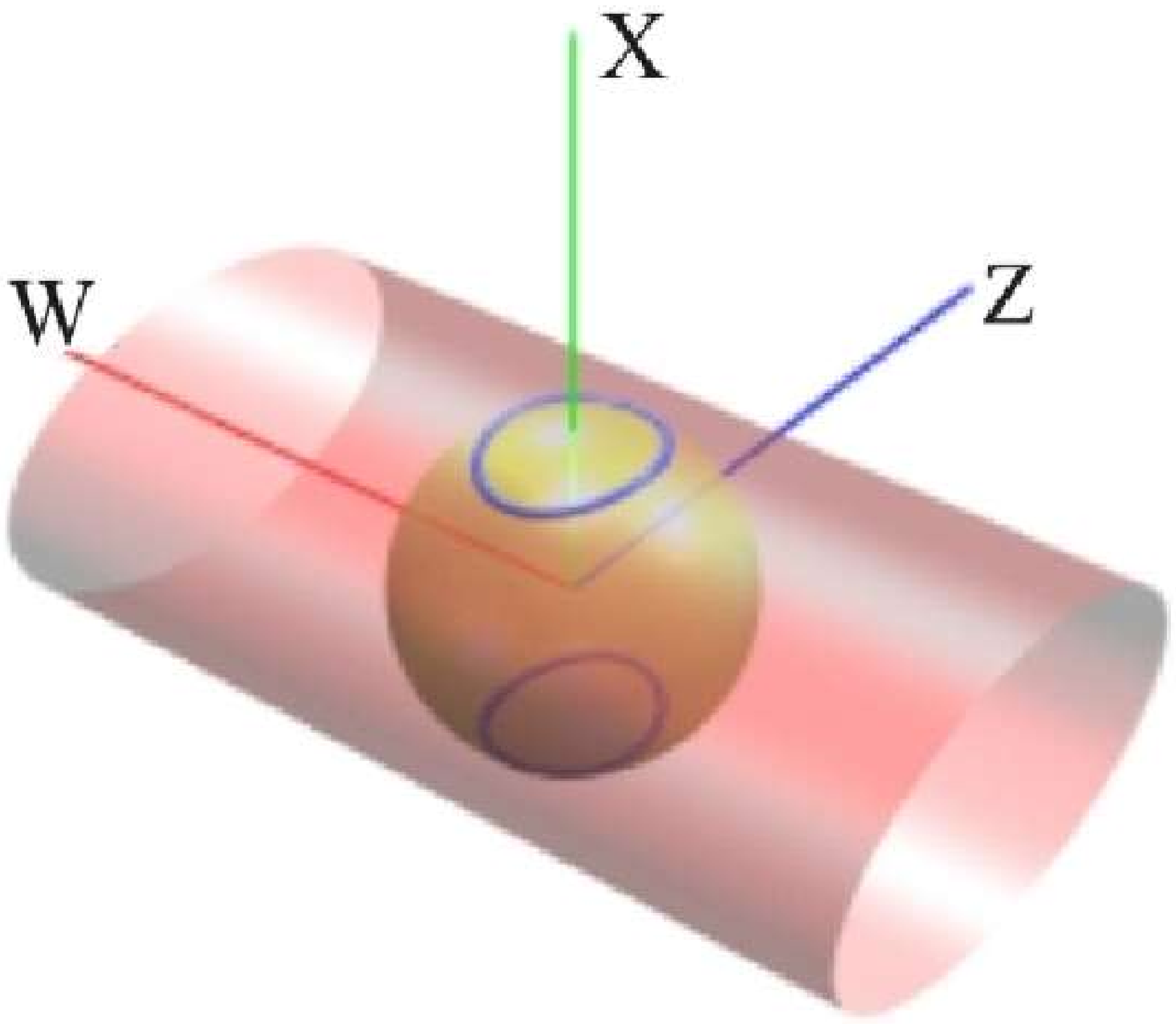}}
\end{minipage}
\begin{minipage}{0.14\textwidth}
\vspace {0.8cm} \rightline{\epsfxsize=1.0\textwidth
\epsfbox{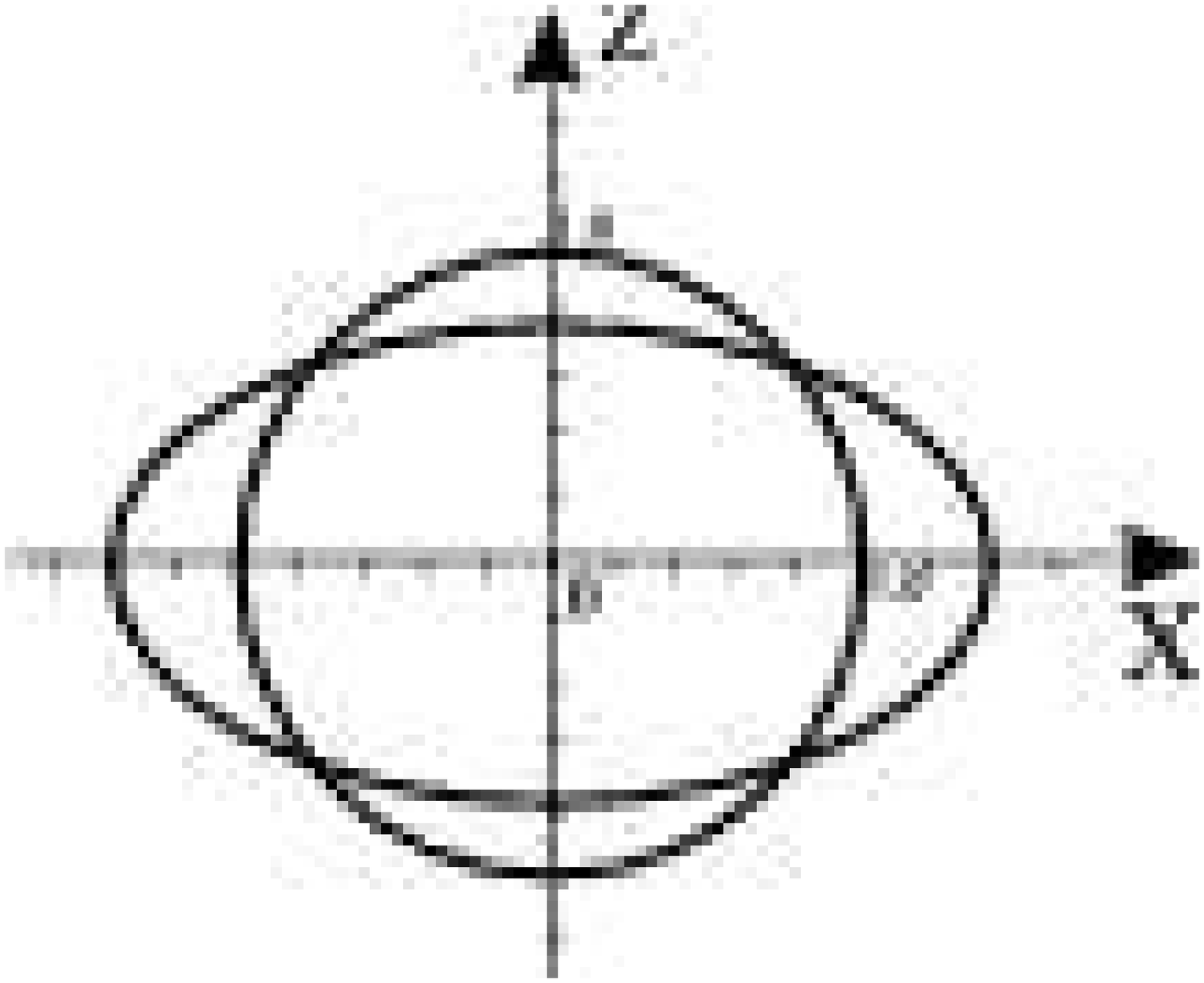}}
\end{minipage}
\begin{minipage}{0.16\textwidth}
\leftline{\epsfxsize=0.9\textwidth \epsfbox{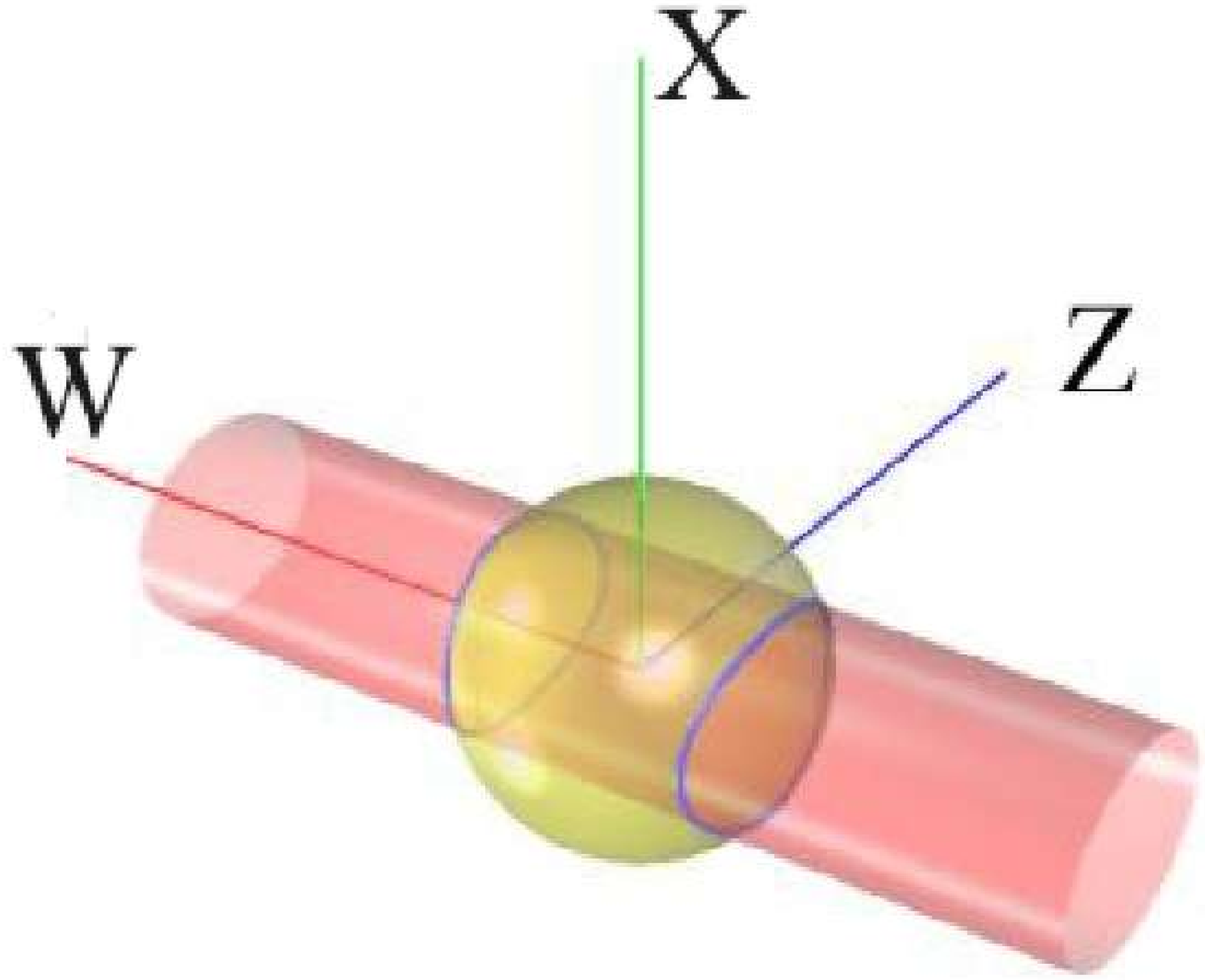}}
\end{minipage}
\begin{minipage}{0.14\textwidth}
\vspace {0.8cm} \rightline{\epsfxsize=1.0\textwidth
\epsfbox{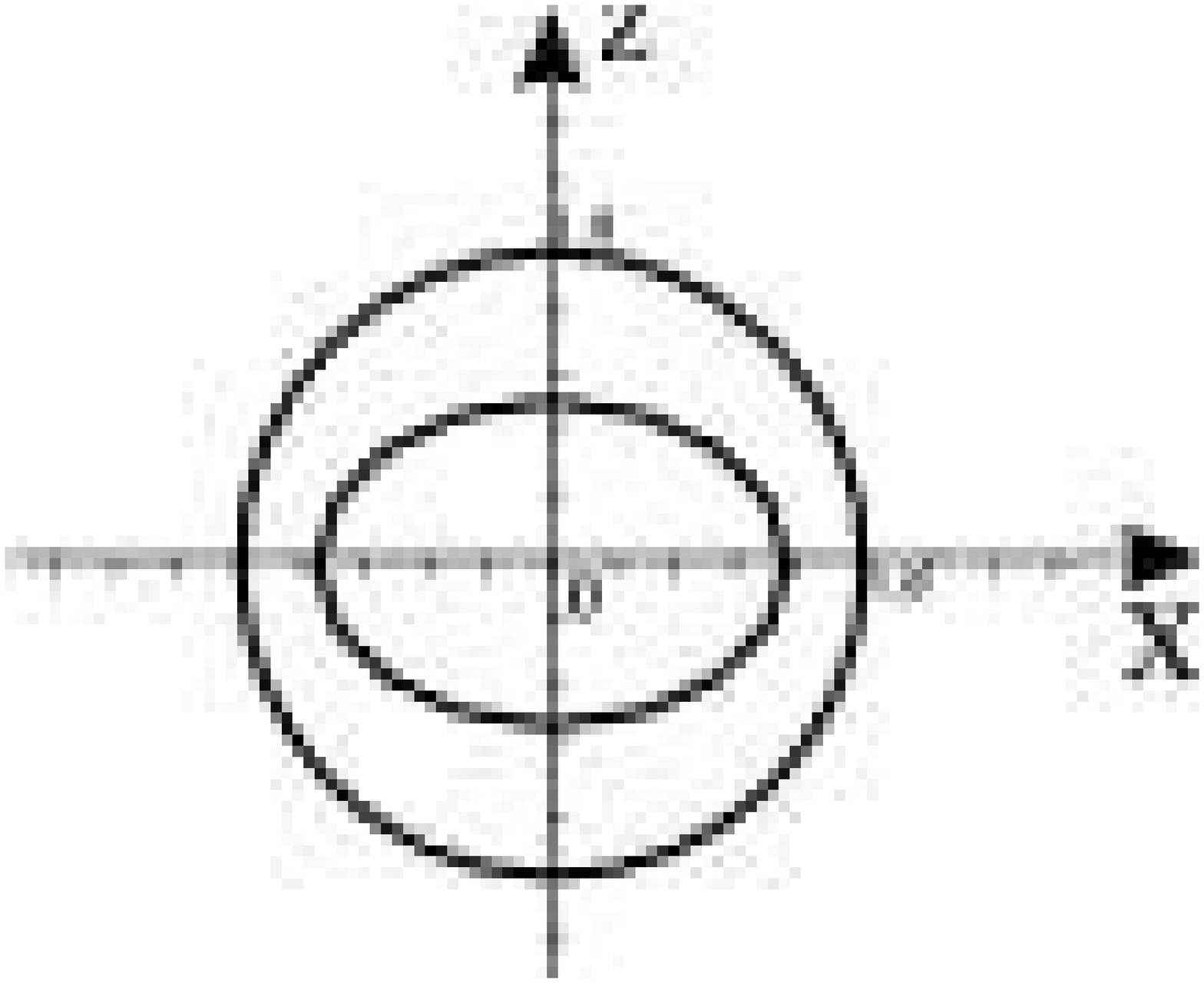}}
\end{minipage}
\begin{minipage}{0.22\textwidth}
\leftline{\epsfxsize=1.0\textwidth \epsfbox{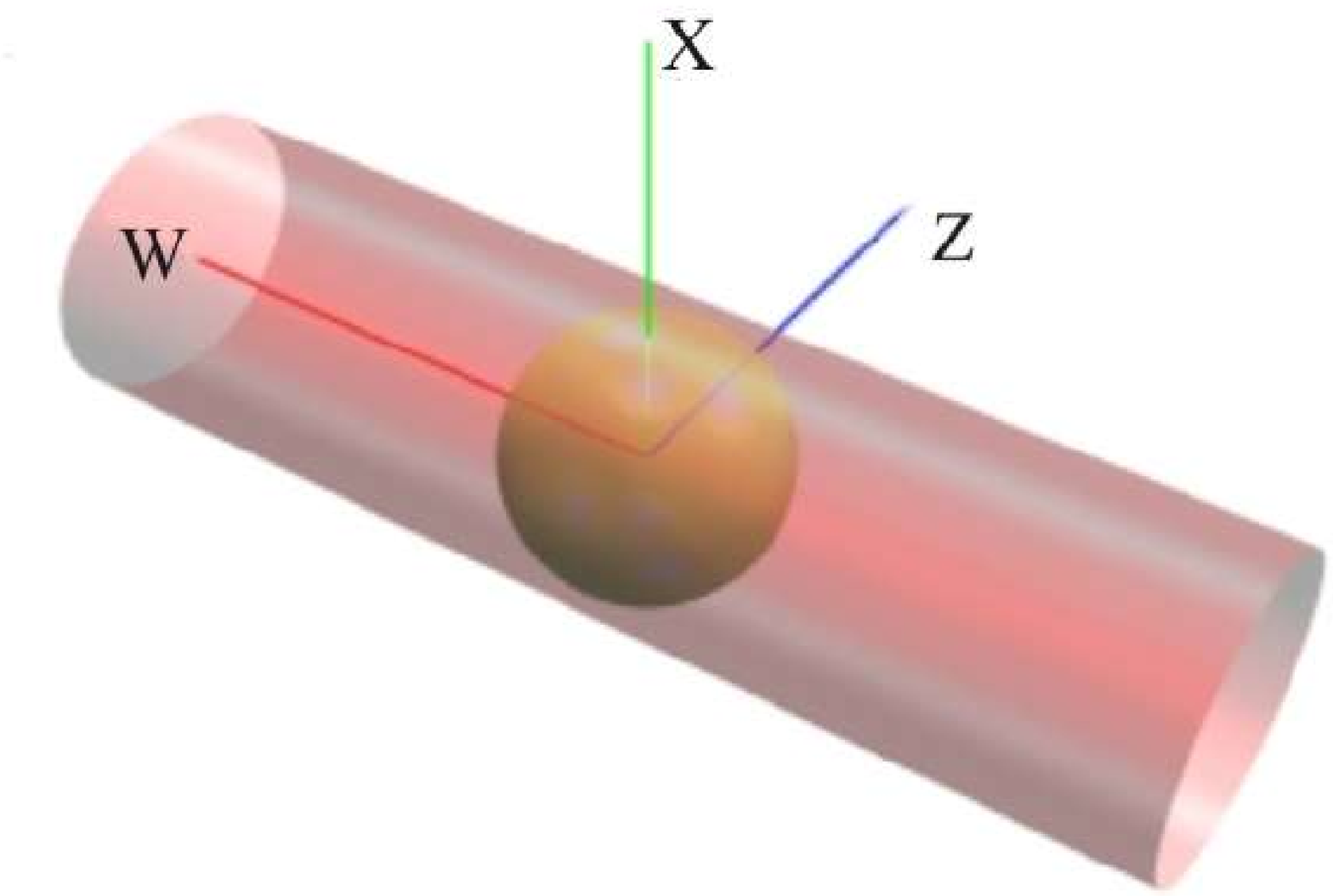}}
\end{minipage}
\begin{minipage}{0.14\textwidth}
\vspace {0.8cm} \rightline{\epsfxsize=1.0\textwidth
\epsfbox{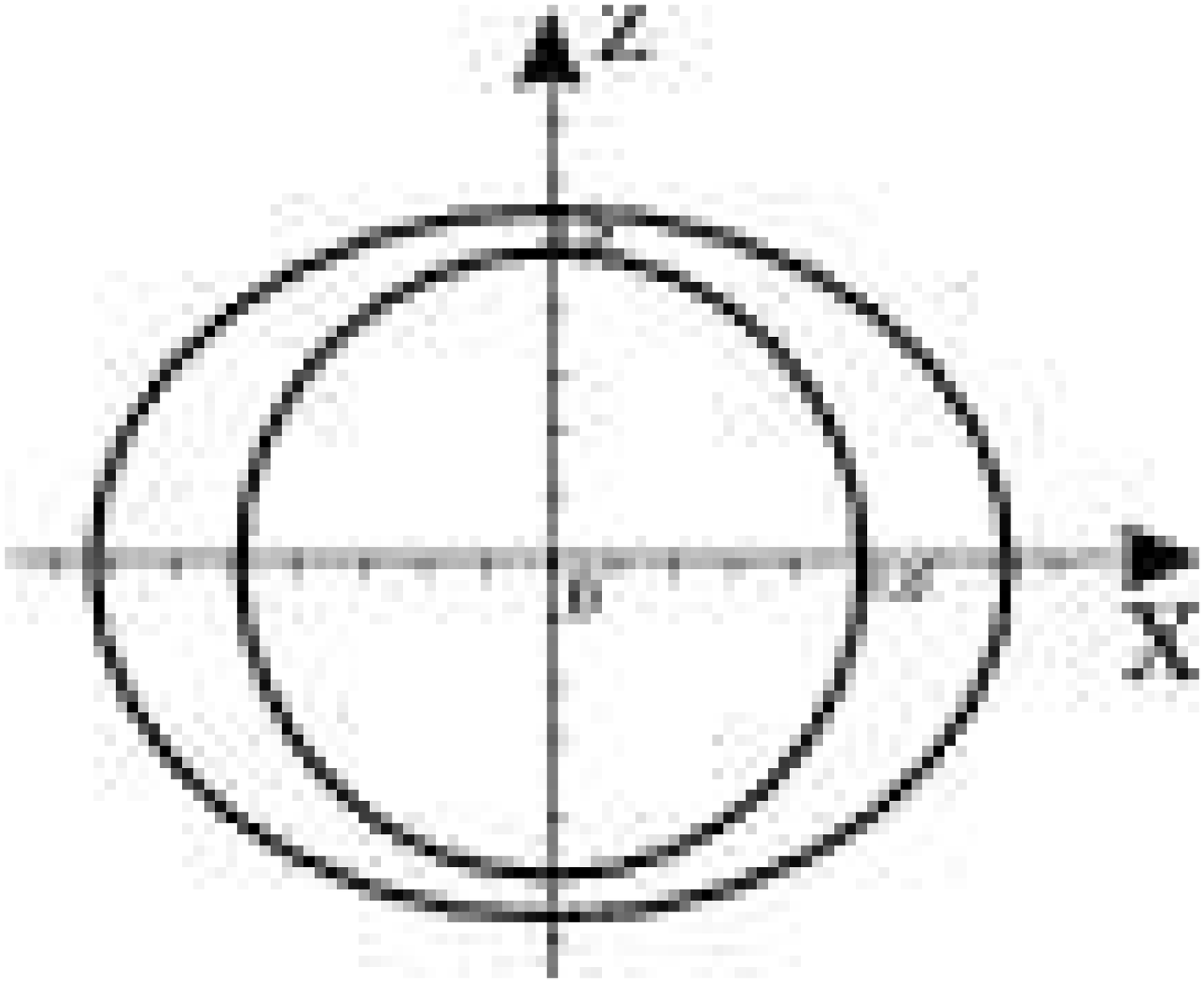}}
\end{minipage}\\
\caption{Three cases of an elliptic cylinder intersecting with a
unit sphere
 and their corresponding cross sections in the $x$-$z$ plane.}
\label{cylinder-sphere}
\end{figure}

For the first index sequence $\langle 1|2|1|2|3\rangle$, according
to Section \ref{signvariation}, the sign sequence in the
corresponding Quadric Pair Canonical Form is $(\varepsilon_{1}=1,
\varepsilon_{2}=-1, \varepsilon_{3}=1, \varepsilon_{4}=1)$.
Setting $\bar A$ to $A'$ and $\bar B-\lambda_4 \bar A$ to  $B'$, we
obtain
\[
A' = {\rm diag} (1, -1, 1, 1),
\]
\[
B' = {\rm diag} \left( (\lambda_1 - \lambda_4), -(\lambda_2 -
\lambda_4), (\lambda_3 - \lambda_4), 0\;\right).
\]
Consider the affine realization of $\mathbb{PR}^3$ by making $y=0$
the plane at infinity.
Then ${\mathcal A'}$ is a sphere, which intersects the $x$-$z$ plane
in a unit circle, while the quadric ${\mathcal B'}$ is an elliptic
cylinder with the $w$-axis being its central direction, which
intersects the $x$-$z$ plane in an ellipse, since
$\lambda_{i}<\lambda_{4}$, $i=1,2,3$.
Clearly, if one of the ellipse's semi-axes is smaller than 1 or both
are smaller than 1, the QSIC of ${\mathcal A'}$ and ${\mathcal B'}$
has two oval branches (see the left and middle configurations in
Figure~\ref{cylinder-sphere}).
If both of the ellipse's semi-axes are greater than 1, ${\mathcal
A'}$ and ${\mathcal B'}$ have no real intersection points (see the
right configuration in Figure~\ref{cylinder-sphere}). We recall the
following result from~\citep{Finsler1937,Uhlig1973b}: {\em Two
quadrics ${\mathcal A}:\;X^TAX=0$ and ${\mathcal B}:\;X^TBX=0$ in
${\mathbb {PR}}^3$ has no real points if and only if $\lambda_0 A-B$
is positive definite or negative definite for some real number
$\lambda_0$.} It implies that the index sequence of the pencil
cannot be $\langle 1|2|1|2|3\rangle$. This is a contradiction.
Hence, the QSIC has two ovals.

Note that none of the semi-axes can be of length $1$, since
$f(\lambda)=0$ is assumed to have no multiple roots.
We deduce that the QSIC has two closed components when the index
sequence is $\langle 1|2|1|2|3\rangle$ and is empty when the index
sequence is $\langle 0|1|2|3|4\rangle$. This completes the proof of
Theorem~\ref{thm:four_real_roots}.
%\end{proof}

%

\subsection{ $[1111]_2$: $f(\lambda)=0$ has two distinct real roots and a pair of complex conjugate roots}
\label{sec:two_roots}

\begin{theorem}\label{thm:two_roots} (Case 3, Table 1)  If $f (\lambda) = 0$ has two distinct real roots
and one pair of complex conjugate roots, then the index sequence of
the pencil $\lambda A - B$ is $\langle 1|2|3\rangle$, and the QSIC
comprises exactly one closed component in $\mathbb{PR}^3$.
\end{theorem}

%\begin{proof}
{\bf Proof} Wlog, we assume $A$ is nonsingular. Suppose that
$f(\lambda) =0$ has two real roots $\lambda_1 \neq \lambda_2$ and
two complex conjugate roots $\lambda_{3,4} = a\pm bi$.
First, it is easy to see that the only index sequence possible is
$\langle 1|2|3\rangle$.
We may suppose that $\lambda_{3,4} = \pm i$; this can be done by
setting $(B-aA)/b$ to $B$.
By Theorem~\ref{thm:canonical_form}, $A$ and $B$ are congruent to
\[
A' = (a'_{ij}) = {\rm diag}( E_1, \varepsilon_1, \varepsilon_2) =
{\rm diag} \left( \left(\begin{array}{cc} 0 & 1
\\ 1 & 0 \end{array} \right), \varepsilon_1, \varepsilon_2 \right),
\]
\[
B' = (b'_{ij}) =  {\rm diag} ( E_1J_1, \varepsilon_1 \lambda_1,
\varepsilon_2\lambda_2) =  {\rm diag} (-1, 1,
\varepsilon_1\lambda_1, \varepsilon_2\lambda_2).
\]
As the index sequence is $\langle 1|2|3\rangle$, we have
$\varepsilon_{1}=1, \varepsilon_{2}=1$.
Next we need consider two cases: (1) $\lambda_1\lambda_2 \neq 0$ and
(2) $\lambda_1\lambda_2 =0$.

{\bf Case 1 ($\lambda_1\lambda_2 \neq 0$)}: By a variable
transformation $\lambda' = - \lambda$ if necessary, we may assume
that at least one of $\lambda_1$ and $\lambda_2$ is positive.  Then
we denote $\lambda_1
>0$ and $\lambda_2 <0$ if only one of them is positive or denote
$\lambda_2
> \lambda_1 >0$ if both are positive.
It follows that $\frac{\lambda_1}{\lambda_2}<1$.
We then set $\lambda_1A'-B'$ to $A'$ and use a further simultaneous
congruence transformation to scale the diagonal elements of $B'$
into ${\pm 1}$.  For simplicity of notation, we use the same symbols
$A'$ and $B'$ for the resulting matrices and obtain
\[
A'= (a'_{ij}) =\left(\begin{array}{cccc} 1&\lambda_1&& \\
\lambda_1&-1&&\\
&&0&\\
&&&\beta_2 (\frac{\lambda_1}{\lambda_2}-1)\end{array} \right),\;
B' = (b'_{ij}) =\left(\begin{array}{cccc} -1&&&\\
&1&&\\
&& 1 &\\
&&&\beta_2 \end{array} \right).
\]
where $\beta_2 = \lambda_2/|\lambda_2| = \pm 1$.

If $\beta_2=1$, we swap $b'_{4,4}$ and $b'_{1,1}$, as well as
$a'_{4,4}$ and $a'_{1,1}$, to obtain
\[
A'=\left(\begin{array}{cccc} (\frac{\lambda_1}{\lambda_2}-1)&&& \\
&-1&&\lambda_1\\
&&0&\\
&\lambda_1&&1\end{array} \right),\;\;\;
B'=\left(\begin{array}{cccc} 1&&& \\
&1&&\\
&&1&\\
&&&-1\end{array} \right).
\]
Or, if $\beta_2=-1$, we swap $b'_{4,4}$ and $b'_{2,2}$, as well as
$a'_{4,4}$ and $a'_{2,2}$, to obtain
\[
A' =\left(\begin{array}{cccc} 1&&&\lambda_1\\
&(1-\frac{\lambda_1}{\lambda_2})&&\\
&&0&\\
\lambda_1&&&-1\end{array} \right),\;\;
B'=\left(\begin{array}{cccc} -1&&& \\
&-1&&\\
&&1&\\
&&&1\end{array} \right).
\]
Note that permuting diagonal elements can be achieved by a
congruence transformation.
Hence, whether $\beta_2=1$ or $\beta_2=-1$, after a proper
simultaneous congruence transformation, ${\mathcal B'}$ is the unit
sphere or a one-sheet hyperboloid with the $z$-axis as its central
axis.
Since $\frac{\lambda_1}{\lambda_2}<1$, $a'_{1,1}$ and $a'_{2,2}$
have the same sign. Therefore, ${\mathcal A'}$ is an elliptic
cylinder parallel to the $z$-axis.
Due to the symmetry of  ${\mathcal B'}$ and  ${\mathcal A'}$  about
the $x$-$y$ plane, we just need to analyze the relationship between
the two conic sections in which ${\mathcal A'}$ and ${\mathcal B'}$
intersect with the $x$-$y$ plane.

The quadric ${\mathcal B'}$ intersects the $x$-$y$ plane in the unit
circle $x^2+y^2=w^2$, and ${\mathcal A'}$ intersects the $x$-$y$
plane in the ellipse
\[
\frac{x^2}{a^2}+\frac{(y-cw)^2}{b^2}=w^2 \] when $\beta_2=1$, or in
the ellipse
\[
\frac{(x+cw)^2}{b^2}+\frac{y^2}{a^2}=w^2
\]
when $\beta_2=-1$. Here
$a=\sqrt{\frac{\lambda_2(1+\lambda_1^2)}{(\lambda_2-\lambda_1)}}$,
$b=\sqrt{1+\lambda_1^2}$, and $c=\lambda_1$.

\begin{figure}[htbp]

\begin{minipage}[t]{7.0cm}
\centering
\includegraphics[width=1.06in]{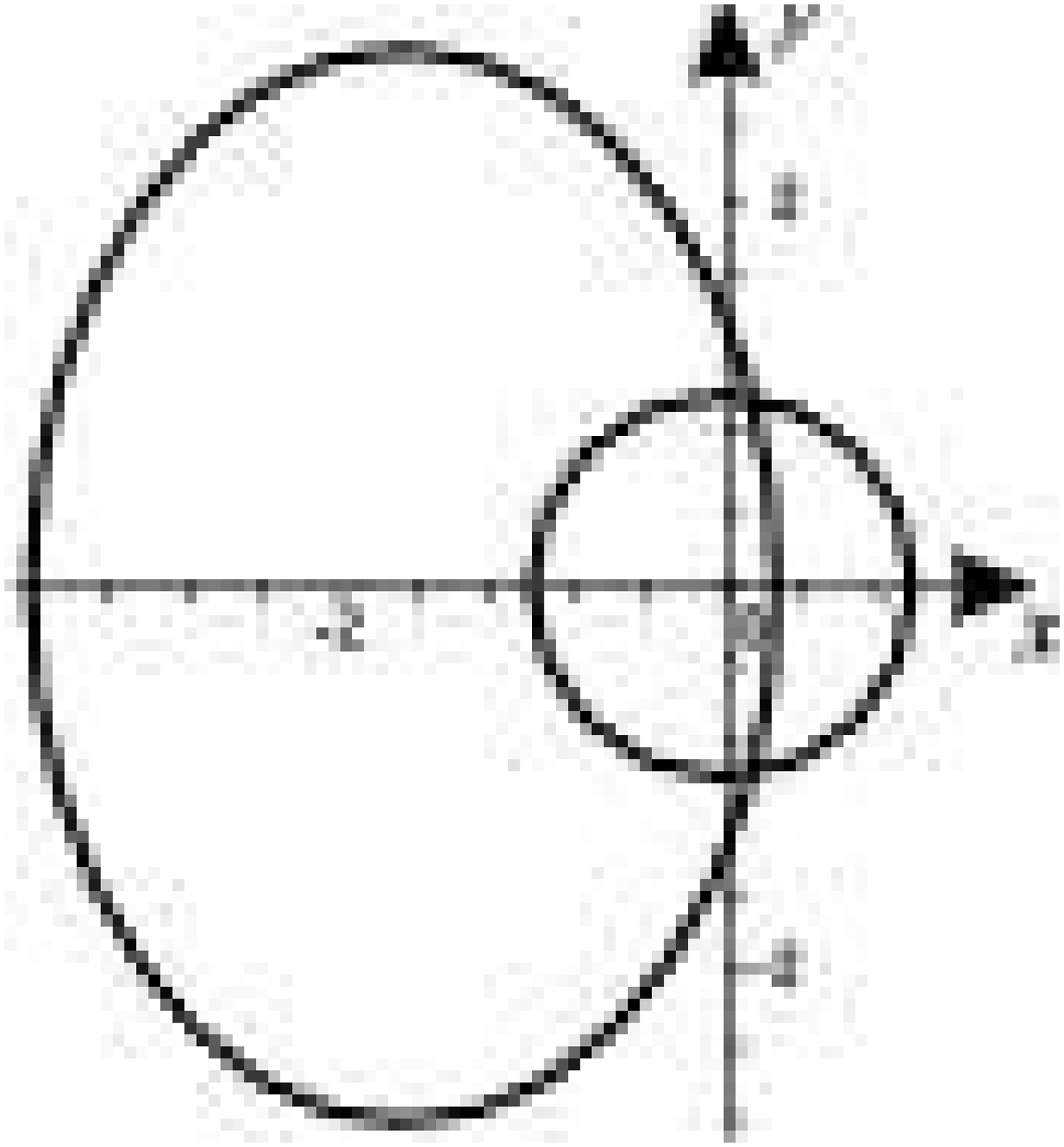}
\footnotesize \caption{The cross-sections of an elliptic cylinder
and a hyperboloid with one sheet in the $x$-$y$ plane.}
\label{ellipse}
\end{minipage}
\hspace{0.2cm}
\begin{minipage}[t]{7.0cm}
\centering
\includegraphics[width=1.46in]{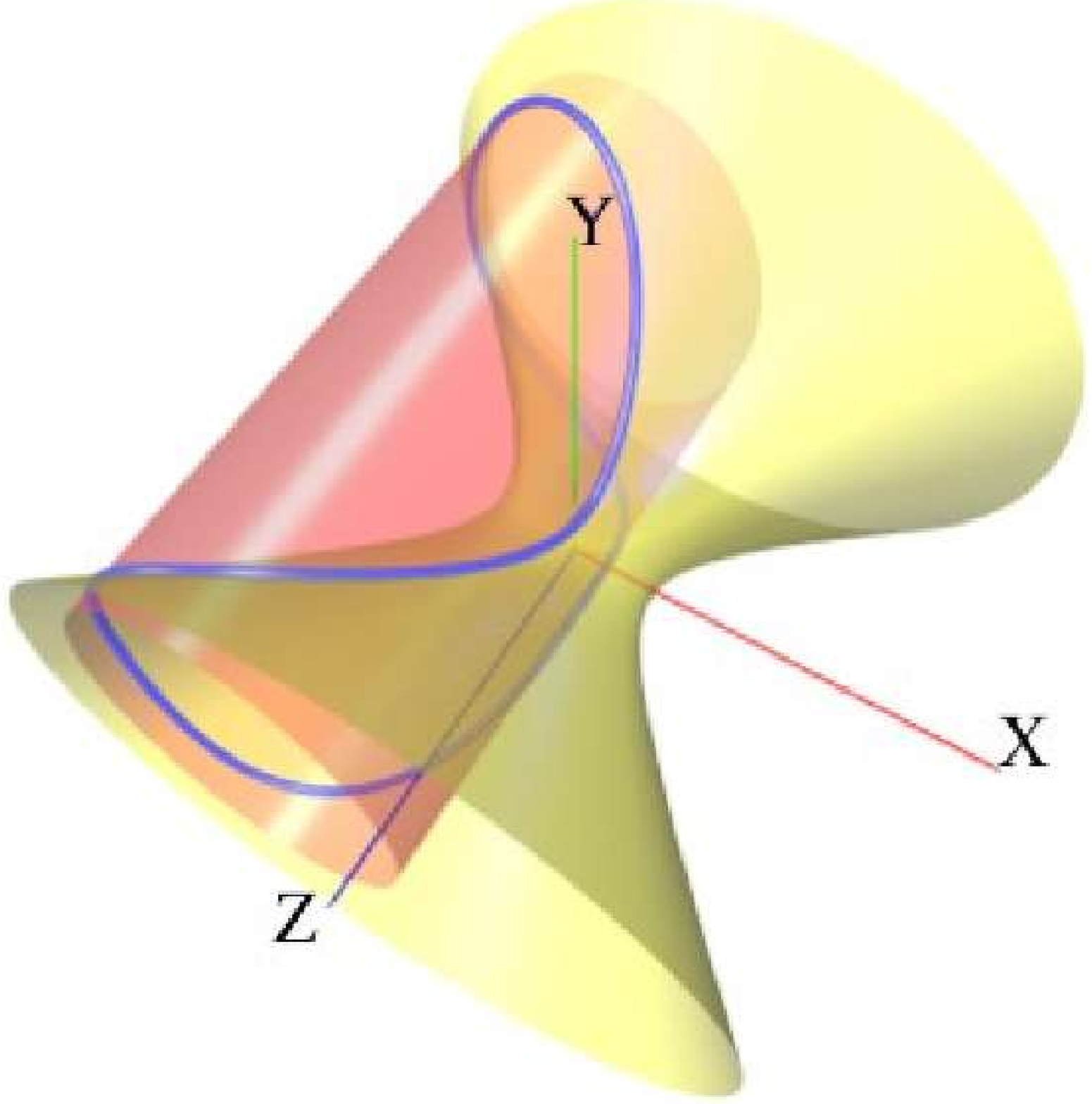}
\footnotesize \caption{The intersection curve referred to in
Figure~\ref{ellipse}}. \label{fig2}
\end{minipage}
\end{figure}

In both cases of $\beta_2=\pm 1$, the center of the ellipse shifts
from the origin (along the $x$ direction or $y$ direction) by the
distance $|\lambda_1|$, and the length of the ellipse's semi-axis in
the shift direction is $b=\sqrt{1+\lambda_1^2}$.
Then it is straightforward to verify that one of the ellipse's
extreme points of this axis is inside the unit circle, while the
other is outside the unit circle. (See Figure \ref{ellipse} for the
case of $\beta_2=-1$.)
In this case the QSIC of ${\mathcal A'}$ and ${\mathcal B'}$ has one
closed component in $\mathbb{PR}^3$ (see Figure \ref{fig2}).

{\bf Case 2 ($\lambda_1\lambda_2 =0$):} Wlog, we may suppose that
$\lambda_1 = 0$ and $\lambda_2 \neq 0$. Then, by Theorem 1, noting
that $\varepsilon_1 = \varepsilon_2 =1$, $A$ and $B$ are congruent
to
\[
A'=\left(\begin{array}{cccc} 0&1&&\\
1&0&& \\
&& 1&\\
&&&1 \end{array} \right),\;\;\;
B'=\left(\begin{array}{cccc} -1&0&& \\
0&1&&\\
&&0& \\
&&& \lambda_2\end{array} \right).
\]
First set  $A'-(1/\lambda_2)B'$ to be $A'$. Then we use a congruence
transformation to make the diagonal elements of $B'$ become ${\pm
1}$ and apply the same transformation to $A'$.
Denoting the resulting matrices again using $A'$ and $B'$, we
obtain
\[
A' = (a'_{ij}) =\left(\begin{array}{cccc} \frac{1}{\lambda_2}&1&&\\
1&-\frac{1}{\lambda_2}&&\\
&&1 &\\
&&&0\end{array} \right),\;\;\;\;
B' = (b'_{ij}) =\left(\begin{array}{cccc} -1&0&&\\
0&1&&\\
&&0& \\
&&& 1 \end{array} \right).
\]

We swap $b'_{4,4}$ and $b'_{1,1}$, as well as $a'_{4,4}$ and
$a'_{1,1}$, by a simultaneous congruence transformation to obtain
\[
A'=\left(\begin{array}{cccc} 0&&& \\
&-\frac{1}{\lambda_2}&&1\\
&& 1&\\
&1&&\frac{1}{\lambda_2}\end{array} \right),\;\;\;
B'=\left(\begin{array}{cccc} 1&&& \\
&1&&\\
&&0&\\
&&&-1\end{array} \right).
\]

\begin {figure}[htbp]
\begin{minipage}{0.24\textwidth}
\leftline{\epsfxsize=1.0\textwidth \epsfbox{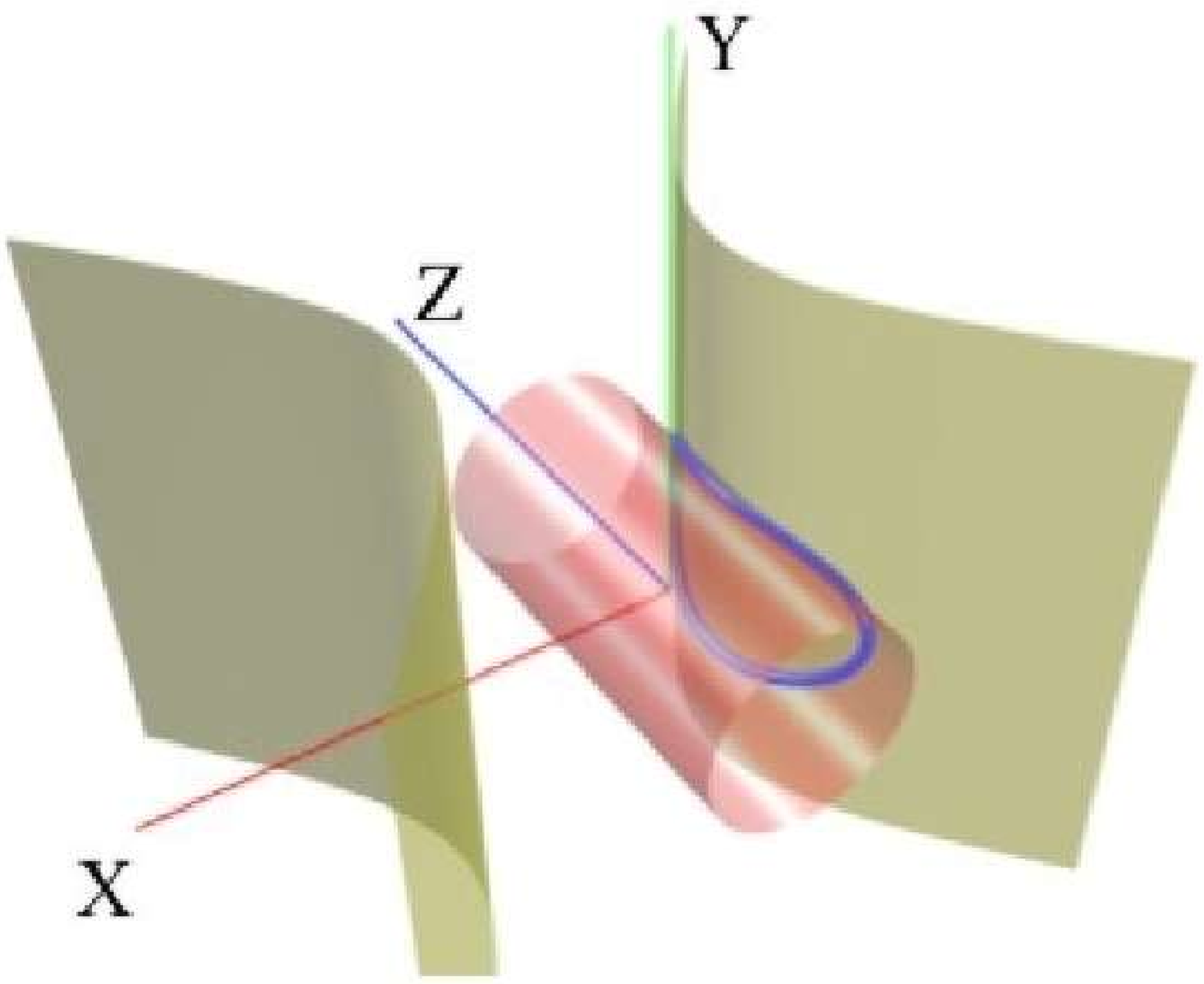}}
\end{minipage}
\begin{minipage}{0.24\textwidth}
\rightline{\epsfxsize=1.0\textwidth \epsfbox{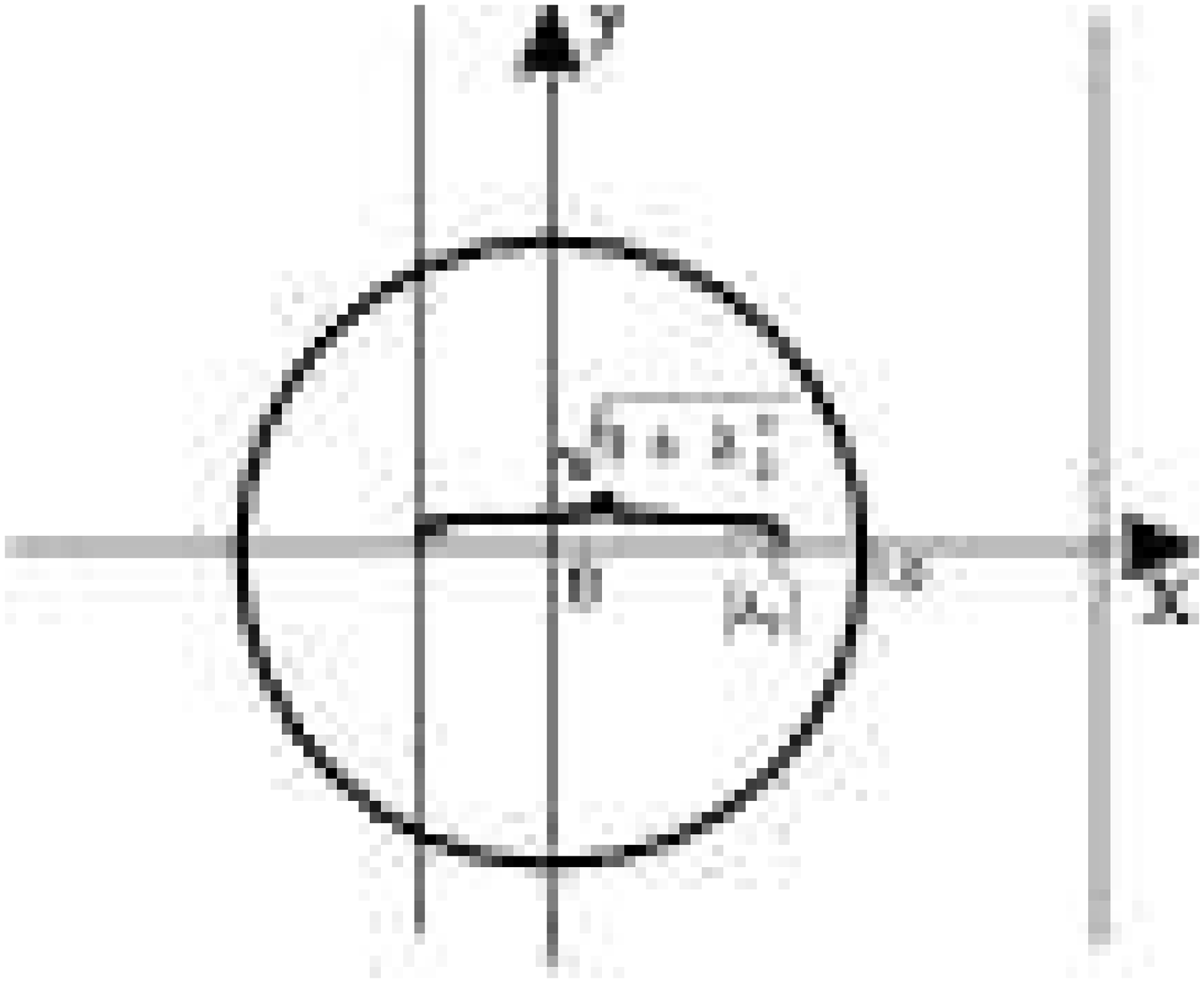}}
\end{minipage}
\begin{minipage}{0.24\textwidth}
\leftline{\epsfxsize=1.0\textwidth \epsfbox{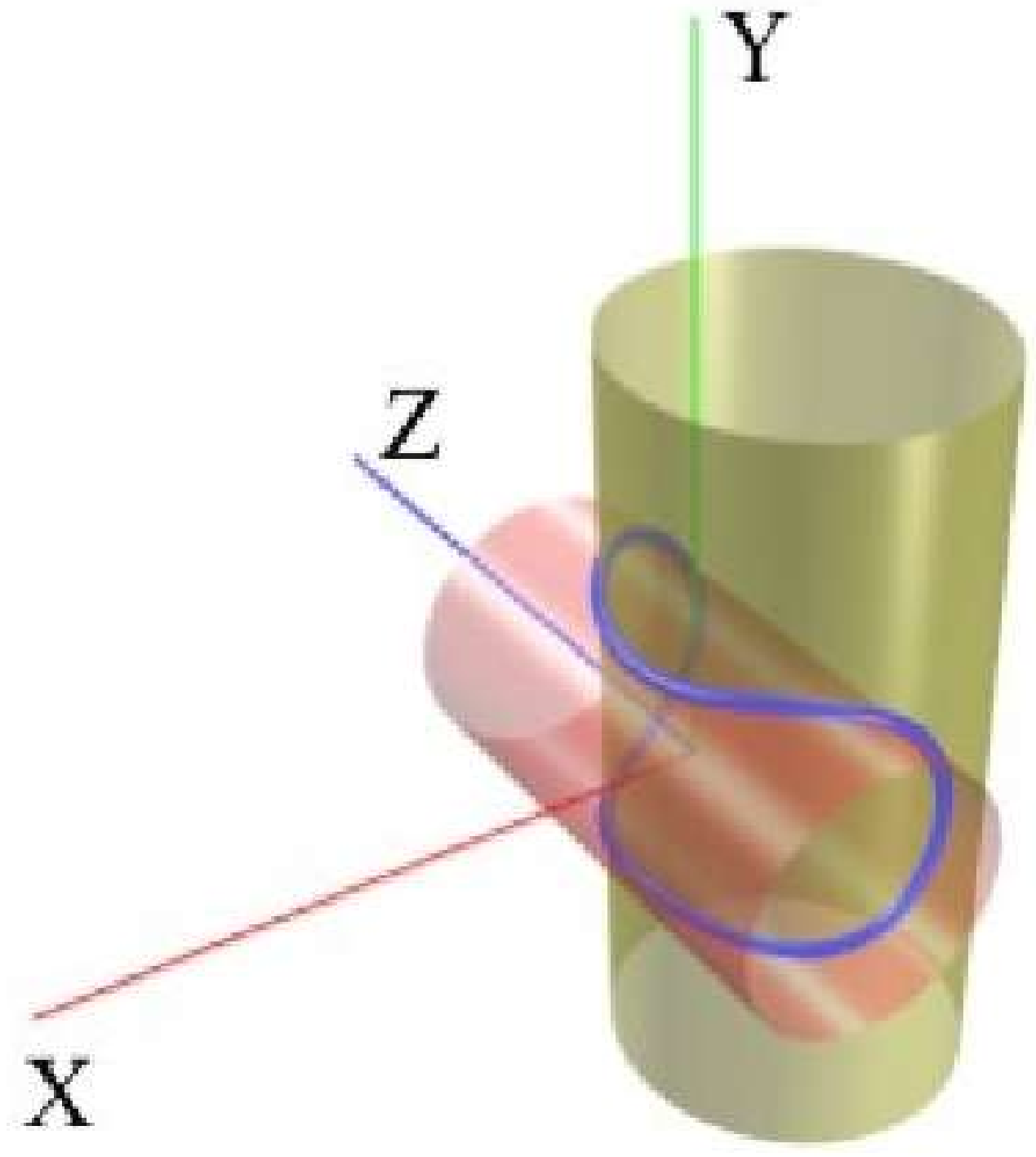}}
\end{minipage}
\begin{minipage}{0.24\textwidth}
\rightline{\epsfxsize=1.0\textwidth \epsfbox{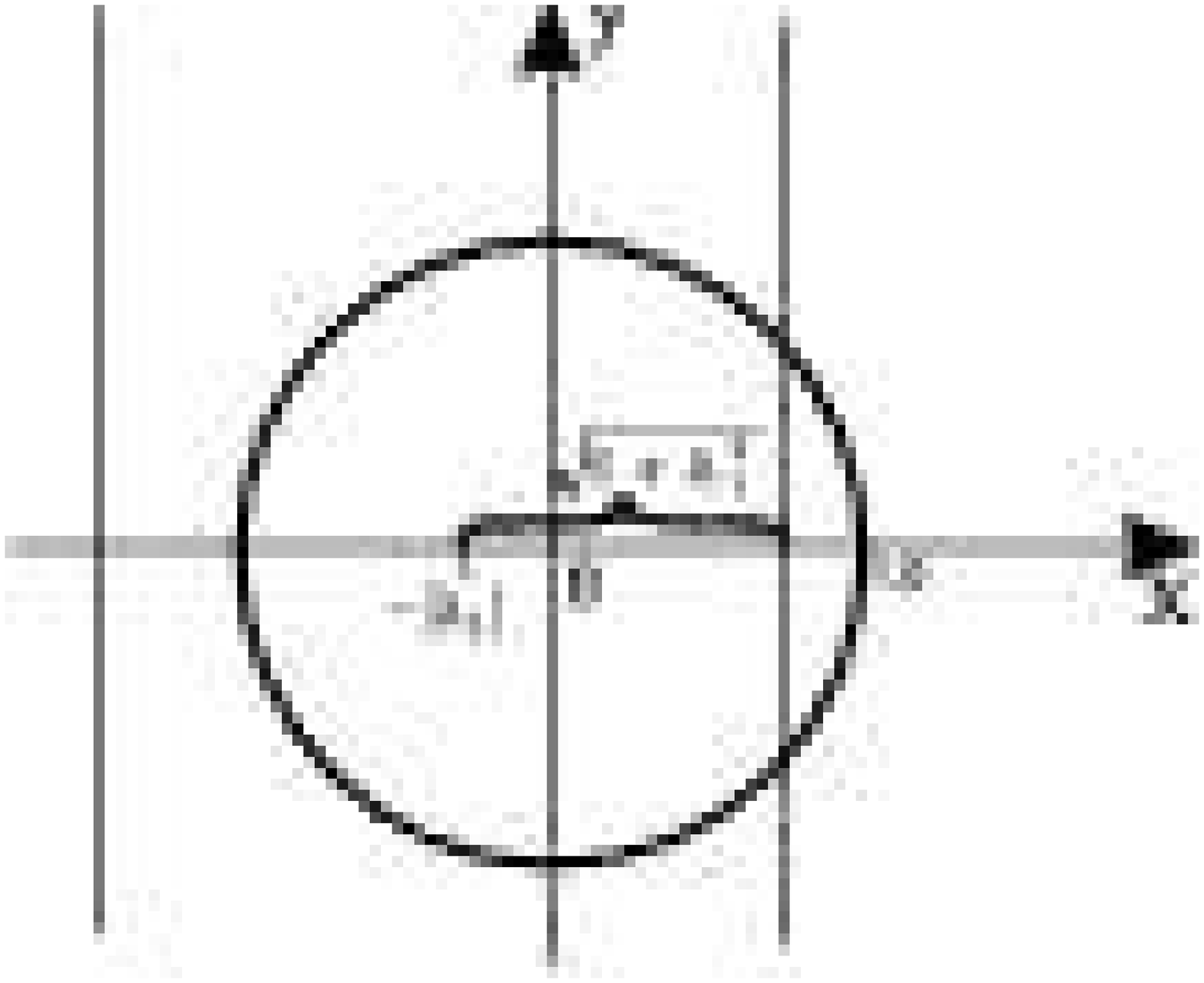}}
\end{minipage}\\
\caption{The intersection of a circular cylinder with a hyperbolic
cylinder or an elliptic cylinder.} \label{fig3}
\end{figure}

Thus, ${\mathcal B'}$ is a cylinder with the $z$-axis as its central
axis, and ${\mathcal A'}$ is either an elliptic cylinder or a
hyperbolic cylinder, depending on the sign of $\lambda_2$, and
${\mathcal A}'$ is parallel to the $y$-axis.
The equation of  ${\mathcal A'}$ is
\[
\frac{(y-cw)^2}{a^2} \pm \frac{z^2}{b^2}=w^2,
\]
% %
% when $\varepsilon_2=1$, or
% %
% \[
% \frac{(x+c)^2}{a^2} \pm \frac{z^2}{b^2}=1
% \]
% %
% when $\varepsilon_2=-1$,
where $a=\sqrt{1+\lambda_2^2}$,
$b=\sqrt{\frac{1+\lambda_2^2}{|\lambda_2|}}$, $c=\lambda_2$.
The cylinder ${\mathcal A'}$ shifts from the origin by the distance
$|\lambda_2|$ along the $x$-axis or the $y$-axis,
%(depending on the sign of $\varepsilon_2$),
and the length of its semi-axis in the
shift direction is $\sqrt{1+\lambda_2^2}$.
Clearly, in this case, the QSIC of the cylinders ${\mathcal A'}$ and
${\mathcal B'}$ has exactly one closed component in $\mathbb{PR}^3$.
(See Figure \ref{fig3}.) This completes the proof.

%\end{proof}

\subsection{$[1111]_0$: $f(\lambda)=0$ has two distinct pairs of complex conjugate roots}
\label{sec:two pairs}

\begin{theorem} \label{thm:two pairs} (Case 4, Table 1)
If ${\rm f} (\lambda) = 0$ has two distinct pairs of complex
conjugate roots, then the Segre characteristic is $[1111]$ and the
index sequence is $\langle 2\rangle$. In this case the QSIC
comprises two open components in $\mathbb{PR}^3$.
\end{theorem}

%\begin{proof}
{\bf Proof} Suppose that $f(\lambda)=0$ has the roots $a \pm bi$ and
$c \pm di$.
First, it is easy to see that the index sequence is $\langle
2\rangle$.
By setting $(B-cA)/d$ to be $B$, we transform conjugate roots $c \pm
di$ to $\pm i$.
Therefore, we suppose that $f(\lambda)=0$ has the roots $a \pm bi$
and $\pm i$.
Furthermore, we may suppose that $A$ and $B$ form a nonsingular pair
of real symmetric matrices.
Then, by Theorem 1, $A$ and $B$ have the following canonical forms
\[
A'= {\rm diag}\left(\left(\begin{array}{cc} 0&1\\
1&0 \end{array} \right),
\left(\begin{array}{cc} 0&1\\
1&0 \end{array} \right) \right)\;\; {\rm and} \;\;
B'={\rm diag}  \left(\left(\begin{array}{cc} -1& \\
&1 \end{array} \right), \left(\begin{array}{cc} \
-b&a \\
a&b\end{array} \right)\right).
\]
Here, $a \neq 0$ or $b\neq \pm 1$, since the roots $a\pm bi$ are
distinct from $\pm i$. Also, $b\neq 0$ since $a\pm bi$ are
imaginary. Wlog, we may assume $b >0$. In the following we will
derive a parameterization of the QSIC from which the topological
information about the QSIC can be deduced.
The quadric ${\mathcal A}': X^TA'X=0$ is a hyperbolic paraboloid and
can therefore be parameterized by ${\bf r}(u,v)={\bf g}(u)+{\bf
h}(u)v$ where
\[
{\bf g}(u)=( -u, 0, 0,  1)^T \;\; {\rm and}\;\;
{\bf h}(u)=( 0, 1, u, 0)^T.
\]
Substituting ${\bf r}(u,v)$ into $X^TB'X=0$ yields
%$${\bf g}(u)^TB'{\bf g}(u)+2{\bf g}(u)^TB'{\bf h}(u)v+{\bf h}(u)^TB'{\bf h}(u)v^2 = 0$$ or
%
\begin{equation}
v=\frac{-{\bf g}(u)^TB'{\bf h}(u)\pm \sqrt{s(u)}}{{\bf h}(u)^TB'{\bf h}(u)},
\label{solution_v}
\end{equation}
where
\[
\begin{split}
s(u)
&= [{\bf g}(u)^TB'{\bf h}(u)]^2-[({\bf g}(u)^TB'{\bf g}(u))({\bf h}(u)^TB'{\bf h}(u))]\\
&= -bu^4+(a^2+b^2+1)u^2-b.
\end{split}
\]
Substituting (\ref{solution_v}) into ${\bf r}(u,v)$ yields the
following parameterization of the QSIC,
\begin{equation} \label{qsic_param}
{\bf p}(u) = \left[bu^3-u,  -\left(au\pm \sqrt{s(u)}\right),
-u\left(au\pm \sqrt{s(u)}\right), 1-bu^2\right]^T.
\end{equation}

Since ${\bf p}(u)$ is a real point only when $s(u) \geq 0$, we are
going to identify the intervals in which $s(u) \geq 0$ holds.
We will first show that $s(u)=0$ always has four distinct real
roots.
The equation $$s(u) = -bu^4+(a^2+b^2+1)u^2-b = 0$$ is a quadratic
equation in $u^2$ with discriminant
\[
\Delta
=(a^2+b^2+1)^2-4b^2
=a^2(a^2+2b^2+2)+(b^2-1)^2 > 0,
\]
since $a \neq 0$ or $b\neq \pm 1$.
Therefore the two real solutions of $u^2$ are
\begin{equation}
u^2=\frac{(a^2+b^2+1)\pm\sqrt{\Delta}}{2b}.
\label{solution_u2}
\end{equation}
Since $\Delta=(a^2+b^2+1)^2-4b^2$ and $b\neq 0$, we have $(a^2+b^2+1)>\sqrt{\Delta}$.
It follows that the numerator and denominator in
(\ref{solution_u2}) are positive; recall that $b>0$ is assumed.
Then we get the four real solutions for $s(u)=0$ from
(\ref{solution_u2}), denoted by $\pm u_+$ and $\pm u_-$, with $u_+ >
u_- > 0$.

\begin{figure}[htbp]

\begin{minipage}[t]{7.0cm}
\centering
\includegraphics[width=2.0in]{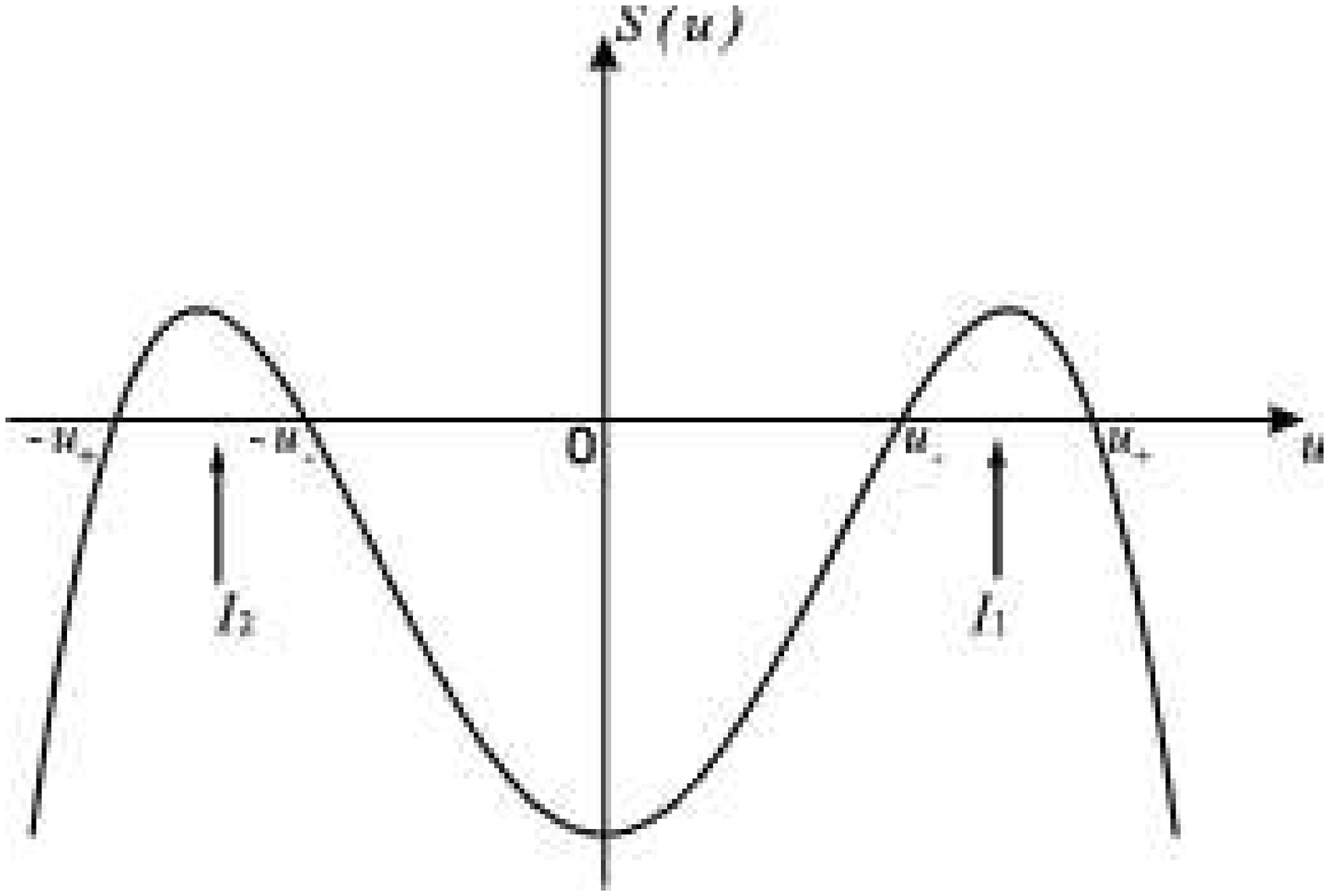}
\footnotesize \caption{The graph of $s(u)$.}\label{su}
\end{minipage}
\hspace{0.2cm}
\begin{minipage}[t]{7.0cm}
\centering
\includegraphics[width=1.8in]{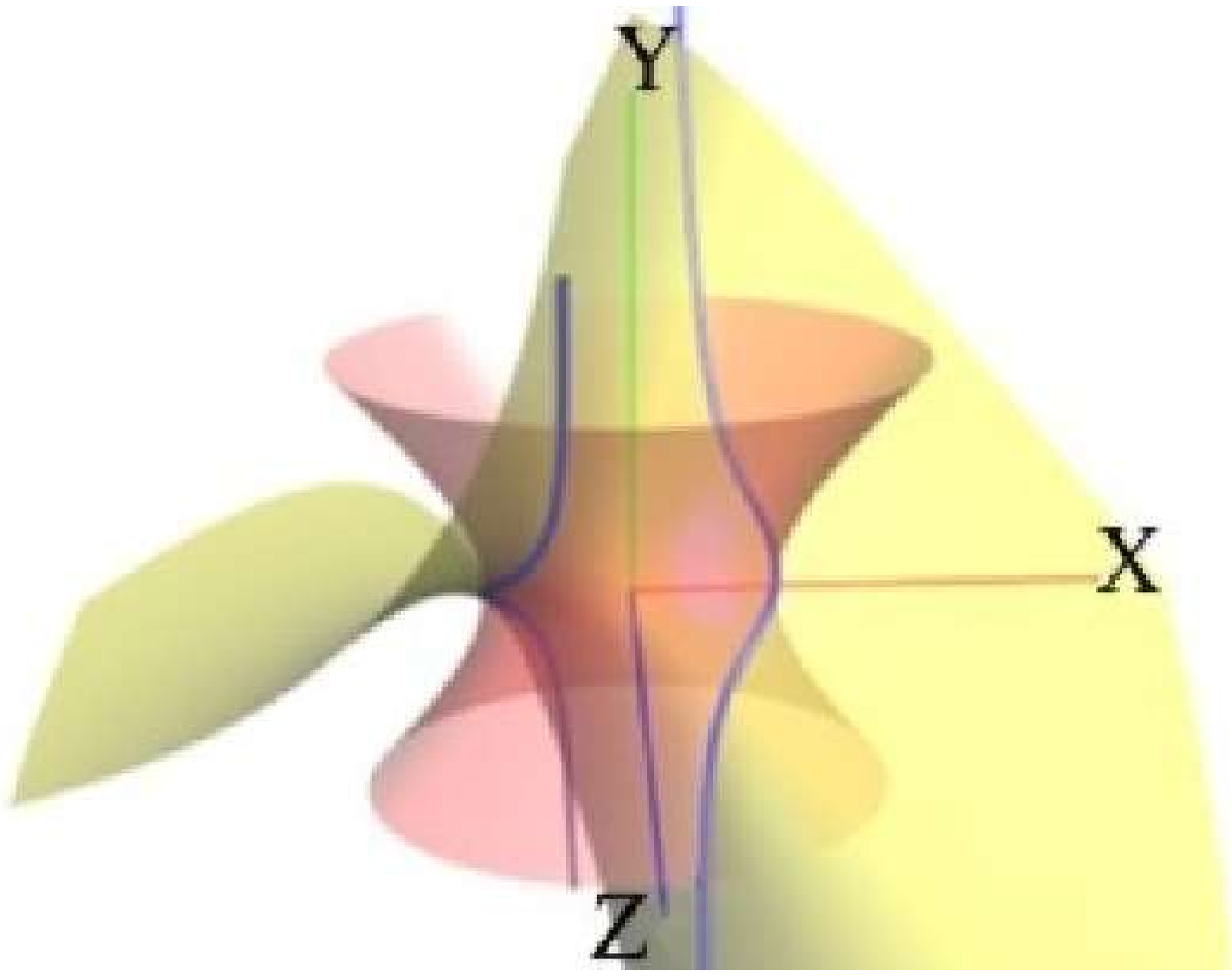}
\footnotesize \caption{The case of the QSIC has two affinely
infinite components.}\label{fig5}
\end{minipage}
\end{figure}

Define two intervals $I_1=[u_-, u_+]$, $I_2=[-u_+, -u_-]$. Since
$s(0) = -b < 0$, we have $s(u)\geq 0$ for $u \in I_1 \bigcup I_2$
and $s(u) < 0$ for the other values of $u$.
(See Figure {\ref{su}} for the graph of $s(u)$.)
This implies that the QSIC, given by ${\bf p}(u)$, has two connected
components, denoted by ${\mathcal V}_1$ and ${\mathcal V}_2$,
corresponding to the intervals $I_1$ and $I_2$: ${\mathcal P}_1$ is
defined by ${\bf p}(u)$ over the interval $I_1$, and ${\mathcal
V}_2$ is defined by ${\bf p}(u)$ over the interval $I_2$.

Next we are going to show that the two components ${\mathcal V}_1$
and ${\mathcal V}_2$ are open curves in $\mathbb{PR}^3$. Since
${\mathcal V}_1$ and ${\mathcal V}_2$ have the same parametric
expression ${\bf p}(u)$ but over different intervals, we will only
analyze the component ${\mathcal V}_1$; the  analysis for ${\mathcal
V}_2$ is similar. The key idea of the proof is to show that
${\mathcal V}_1$ has exactly one intersection point with the plane
$w=0$; it will then follows that ${\mathcal V}_1$ has an
intersection with every plane in $\mathbb{PR}^3$.

Consider the affine realization $\mathbb{AR}^3$ of $\mathbb{PR}^3$
by making the plane $w=0$ the plane at infinity.
The $w$-coordinate component of ${\bf p}(u)$ is $w(u) = 1-bu^2$,
which has two zeros $u_1 = 1/\sqrt{b}$ and $u_2 = -1/\sqrt{b}$,
and it is straightforward to verify that $u_1 = 1/\sqrt{b}\in I_1$
and $u_2 = -1/\sqrt{b}\in I_2$.
Therefore we will only consider the two points ${\bf p}(u_1)$ (i.e.,
with $\pm$ in Eqn.~(\ref{qsic_param})) on the component ${\mathcal
V}_1$.
Let ${\bf q}_0(u)$ and ${\bf q}_1(u)$ denote the two ``branches'' of
${\bf p}(u)$ corresponding to $\pm$ in front of $\sqrt{s(u)}$ in
Eqn.~(\ref{qsic_param}).
Then
\begin{eqnarray} \nonumber
{\bf q}_0(u) &=& \left(bu^3-u, -\left(au+\sqrt{s(u)}\right),
-u\left(au+\sqrt{s(u)}\right),
1-bu^2\right)^T\\
{\bf q}_1(u) &=& \left( bu^3-u, -\left(au-\sqrt{s(u)}\right),
-u\left(au-\sqrt{s(u)}\right), 1-bu^2 \right)^T. \label{q0q1}
\end{eqnarray}

There are now three cases to consider: ($i$) $a = 0$; ($ii$) $a
> 0$; and ($iii$) $a <0$.
First consider the case ($i$) $a = 0$. In this case,
\[
s(u) = -bu^4+(b^2+1)u^2-b = (u^2-b)(1-bu^2).
\]
It follows from Eqn. (\ref{q0q1}), after dropping a common factor
$\sqrt{|1-bu^2|}$, that
\begin{eqnarray*}
{\bf q}_0(u) &=& \left( -u\sqrt{|1-bu^2|}, -\sqrt{|u^2-b|},
-u\sqrt{|u^2-b|}, \sqrt{|1-bu^2|}\right)^T, \\
{\bf q}_1(u) &=& \left(-u\sqrt{|1-bu^2|}, \sqrt{|u^2-b|},
u\sqrt{|u^2-b|}, \sqrt{|1-bu^2|}\right)^T.
\end{eqnarray*}

Note that the two ends of $I_1$ are $1/\sqrt{b}$ and $\sqrt{b}$ when
$a=0$. It is easy to verify that, when $u = \sqrt{b}$, the two
branches $\q_0(u)$ and $\q_1(u)$ are joined together at the finite
point
\[
\q_0(\sqrt{b}) = \q_1(\sqrt{b}) = (-\sqrt{b|1-b^2|}, 0, 0,
\sqrt{|1-b^2|})^T.
\]
Next consider the behavior of ${\bf q}_0(u)$ and ${\bf q}_1(u)$ when
$u \rightarrow u_1 = 1/\sqrt{b}$, which is the other end of $I_1$.
Since
\[
\q_0(u_1) = (0, -\sqrt{|b^{-1} - b|}, -\sqrt{|b^{-1} - b|/b}, 0)^T
\]
and
\[
\q_1(u_1) = (0, \sqrt{|b^{-1} - b|}, \sqrt{|b^{-1} - b|/b}, 0)^T,
\]
$\q_0(u_1)$ and $\q_1(u_1)$ represent the same point at infinity.

Denote $\q_i(u) = (x_i,y_i, z_i, w_i)^T$, $i=0,1$.
To study the asymptotic behavior of the QSIC, let us consider the
limit of the affine coordinates of $\q_0(u)$ and $\q_1(u)$, i.e.,
$(x_0/w_0, y_0/w_0, z_0/w_0)^T$ and $(x_1/w_1, y_1/w_1, z_1/w_1)^T$,
as $u \rightarrow u_1 = 1/\sqrt{b}$.
Clearly,
\[
\lim_{u \rightarrow u_1} \frac{x_0(u)}{w_0(u)} = \lim_{u \rightarrow
u_1} \frac{x_1(u)}{w_1(u)} = - \frac{1}{\sqrt{b}},
\]
\[
\lim_{u \rightarrow u_1} \frac{y_0(u)}{w_0(u)} = - \lim_{u
\rightarrow u_1} \frac{y_1(u)}{w_1 (u)} = -\infty,
\]
and
\[
\lim_{u \rightarrow u_1} \frac{z_0(u)}{w_0(u)} = - \lim_{u
\rightarrow u_1} \frac{z_1(u)}{w_1 (u)} = -\infty.
\]
Therefore, the component curve ${\mathcal V}_1$ comprises one
connected component and its two ends extend to infinity in opposite
directions in ${\mathbb {AR}}^3$, with its asymptote line being the
intersection line of the two planes $x + u_1w =0$ and $-u_1y + z
=0$.
Clearly, ${\mathcal V}_1$ is intersected by every plane in
$\mathbb{PR}^3$. Hence, ${\mathcal V}_1$ is open in $\mathbb{PR}^3$.

Now we consider case ($ii)$: $a > 0$. In this case, $u_1 =
1/\sqrt{b} \in I_1 = (u_-, u_+)$.
Clearly, the two parts of ${\mathcal V}_1$ defined by $\q_0(u)$ and
$\q_1(u)$ are joined at the two finite points $q_0(u_-) = q_1(u_-)$
and $q_0(u_+) = q_1(u_+)$
We will show that, when $u = u_1 = 1/\sqrt{b}$, $\q_0(u_1)$ gives
the {\it only} infinite point on ${\mathcal V}_1$.
Since, from Eqn. (\ref{q0q1}),
\[
\q_0(u_1) = (0, -2a/\sqrt{b}, -2a/b, 0)^T,
\]
$\q_0(u_1)$ is a point at infinity.
To study how $\q_0(u)$ approach the infinite point $\q_0(u_1)$, let
us consider the $\lim_{u \rightarrow u_{1-}}\q_0(u)$ and $\lim_{u
\rightarrow u_{1+}}\q_0(u)$ when $u$ approaches $u_1$ from different
sides.
Using the affine coordinates of $\q_0(u)$, we have
\[
\lim_{u \rightarrow u_{1-}}\frac{x_0(u)}{w_0(u)}
 =  - \lim_{u \rightarrow u_{1+}}\frac{x_0(u)}{w_0(u)}  = - 1/\sqrt{b},
 \]
\[
\lim_{u \rightarrow u_{1-}}\frac{y_0(u)}{w_0(u)}  = - \lim_{u
\rightarrow u_{1+}}\frac{y_0(u)}{w_0(u)}  =  +\infty,
\]
and
\[
 \lim_{u \rightarrow u_{1-}}\frac{z_0(u)}{w_0(u)}
= - \lim_{u \rightarrow u_{1+}}\frac{z_0(u)}{w_0(u)}  = +\infty.
\]
Thus, the component ${\mathcal V}_1$ extends to infinity in opposite
directions, with its asymptote being the intersection line of the
two planes $x + u_1w =0$ and $-u_1y + z =0$.

Note that all other points of ${\mathcal V}_1$ are obviously finite,
except for the point $\q_1(u_1)$ whose $w$ component is zero.
But we will show that $\q_1(u_1)$ is, in fact, also a finite point.
Denote $g(u) = \sqrt{s(u)}$. Expanding $g(u)$ at $u = u_1$ by the
Taylor formula yields
\[
g(u) = \frac{a}{\sqrt{b}}+ g'(u_1)(u-u_1)+ o(u-u_1),
\]
where $o(u-u_1)$ is a term whose order is higher than $u-u_1$ when
$u \rightarrow u_1$.
Plugging the above $g(u)$ in $\q_1(u)$ yields
\begin{eqnarray*}
{\bf q}_1(u) &=& \left(bu^3-u, -(au-g(u)), -u(au-g(u)), 1-bu^2\right)^T \\
 &=& \left(\begin{array}{c} bu(u+1/\sqrt{b})(u-1/\sqrt{b})\\
-\left(au-a/\sqrt{b}-g'(u_1)(u-1/\sqrt{b}) \right)\\
-u\left(au-q/\sqrt{b}-g'(u_1)(u- 1/\sqrt{b})\right)\\
-b(u+ 1/\sqrt{b})(u- 1/\sqrt{b})\end{array} \right) + o(u-u_1) \\ \\
&= & \left(\begin{array}{c} bu(u+ 1/\sqrt{b})(u- 1/\sqrt{b})\\
(g'(u_1) - a)(u- 1/\sqrt{b})\\
u(g'(u_1) - a)(u- 1/\sqrt{b})\\
-b(u+ 1/\sqrt{b})(u- 1/\sqrt{b})\end{array} \right) + o(u-u_1).
\end{eqnarray*}
Dividing a common factor $u-u_1 = u - 1/\sqrt{b}$ to these
homogeneous coordinates, we have
\[
{\bf q}_1(u) = \left( bu(u+ 1/\sqrt{b}), g'(u_1) - a, u(g'(u_1)-a),
-b(u+ 1/\sqrt{b})\right)^T + o(1)
\]
Therefore,
\[
\lim_{u\rightarrow u_1}  {\bf q}_1(u) = \left(2, \;\;g'(u_1) -
a,\;\; (g'(u_1) -a)/\sqrt{b},\;\;-2\sqrt{b}\right)^T.
\]
It follows that ${\bf q}_1(u_1)$ is a finite point in
$\mathbb{AR}^3$.
Hence, ${\mathcal V}_1$ is an open curve in $\mathbb{PR}^3$, since
it is a continuous curve that extends to infinity in opposite
directions with an asymptote line.

In the third case of $a<0$, it can be proved similarly that ${\bf
q}_0(u_1)$ is a finite point in $\mathbb{AR}^3$ and ${\bf q}_1(u_1)$
is the only infinite point on ${\mathcal V}_1$.
Therefore, in this case ${\mathcal V}_1$ is also an open curve.
Finally, in all the three subcases (i.e., $a =0$, $a>0$ and $a <0$),
we can show similarly that the other component ${\mathcal V}_2$ of
the QSIC of ${\mathcal A}$ and ${\mathcal B}$ is also open in
$\mathbb{PR}^3$.
Hence, the QSIC of ${\mathcal A}$ and ${\mathcal B}$ has two open
components in $\mathbb{PR}^3$.
An example of such a QSIC is shown in Figure~\ref{fig5}.
This completes the proof of Theorem~\ref{thm:two pairs}.
%\end{proof}
%

\section{Classification of singular but non-planar QSIC}
\label{sec:singular}

\subsection{$[211]$: $f(\lambda)=0$ has one real double root and two other distinct roots}

\begin{theorem} ($[211]_3$) \label{thm:$[211]_3$} Given two quadrics ${\mathcal A}$:$X^TAX = 0$ and ${\mathcal
B}$:$X^TBX = 0$, if $f(\lambda) = 0$ has one double real root and
two distinct real roots with the Segre characteristic $[211]$, then
the only possible index sequences of the pencil $\lambda A - B$ are
$\langle 2 {\wr\wr}_{-}2 | 3 | 2 \rangle$, $\langle 2 {\wr\wr}_{+}2
| 3 | 2 \rangle$, $\langle 1 {\wr\wr}_{-} 1 | 2 |3 \rangle$ and
$\langle 1 {\wr\wr}_{+} 1 | 2 |3 \rangle$.
Furthermore,
\begin{enumerate}
\item (Case 5, Table 1) when the index sequence is $\langle 2 {\wr\wr}_{-} 2 | 3 | 2 \rangle$ or $\langle 2 {\wr\wr}_{+}2
| 3 | 2 \rangle$, the QSIC has one closed component with a crunode;
\item (Case 6, Table 1) when the index sequence is $\langle 1 {\wr\wr}_{-} 1 | 2 |3 \rangle$, the QSIC has a
closed component plus an acnode;
\item (Case 7, Table 1) when the index sequence is $\langle 1 {\wr\wr}_{+} 1 | 2 |3 \rangle$, the QSIC has only
one real point, which is an acnode.
\end{enumerate}
\end{theorem}

%\begin{proof}
{\bf Proof} Suppose that $f(\lambda) = 0$ has one double real root
$\lambda_0$ and two distinct real roots $\lambda_1$ and $\lambda_2$
with the Segre characteristic $[211]$.
First, it is easy to check that the only possible index sequences
are $\langle 2 {\wr\wr}_{-}2 | 3 | 2 \rangle$, $\langle 2
{\wr\wr}_{+}2 | 3 | 2 \rangle$, $\langle 1 {\wr\wr}_{-} 1 | 2 |3
\rangle$ and $\langle 1 {\wr\wr}_{+} 1 | 2 |3 \rangle$.

By setting $B-\lambda_0A$ to be $B$, we can transform the double
root $\lambda_0$ into $0$. With a further projective transform to
$\lambda$, we may assume that $0<\lambda_1<\lambda_2$.
According to Theorem~\ref{thm:canonical_form}, and wlog, assuming
$\varepsilon_0=1$, the two quadrics can be reduced simultaneously to
the following forms:
\[
A' =(a'_{ij}) =\left(\begin{array}{cccc} 0& 1 &0&0 \\
1 &0&0&0\\
0&0&\varepsilon_1&0\\
0&0&0&\varepsilon_2\end{array} \right),\;\;
B' = (b'_{ij}) =\left(\begin{array}{cccc} 0&0&0&0 \\
0& 1 &0&0\\
0&0&\varepsilon_1\lambda_1&0\\
0&0&0&\varepsilon_2\lambda_2\end{array}\right),
\]
where $\varepsilon_{1,2} = \pm 1$.
By swapping the position $a'_{1,1}$ and $a'_{4,4}$, as well as
$b'_{1,1}$ and $b'_{4,4}$, we obtain
\[
A' =\left(\begin{array}{cccc} \varepsilon_2&0&0&0 \\
0&0&0&1\\
0&0&\varepsilon_1&0\\
0&1&0&0\end{array} \right),\;\;
B' =\left(\begin{array}{cccc} \varepsilon_2\lambda_2&0&0&0 \\
0&1&0&0\\
0&0&\varepsilon_1\lambda_1&0\\
0&0&0&0\end{array}\right).
\]
There are now two cases to consider: ($i$) $\det(A')>0$ and ($ii$)
$\det(A')<0$. In case ($i$) (det$(A')>0$), we have
$\varepsilon_1\varepsilon_2=-1$.
Because Id($\infty$)= index($A'$)= 2 and the index jump of index
function at $\lambda_0=0$ is 0, the associated index sequence is
$\langle 2 {\wr\wr}_{-} 2 | 3 | 2 \rangle$ or $\langle 2
{\wr\wr}_{+}2 | 3 | 2 \rangle$.
Note that $\langle 2 {\wr\wr}_{-} 2 | 3 | 2 \rangle$ is equivalent
to $\langle 2 {\wr\wr}_{+} 2 | 1 | 2 \rangle$, and $\langle 2
{\wr\wr}_{+}2 | 3 | 2 \rangle$ is equivalent to $\langle 2
{\wr\wr}_{-}2 | 1 | 2 \rangle$.

By setting  $\frac{\lambda_1+\lambda_2}{2}A'-B'$ to be $A'$, we
obtain
\begin{equation}
A' =\left(\begin{array}{cccc} \varepsilon_2\frac{\lambda_1-\lambda_2}{2}&0&0&0 \\
0&-1&0&\varepsilon_1\frac{\lambda_1+\lambda_2}{2}\\
0&0&\frac{\lambda_2-\lambda_1}{2}&0\\
0&\varepsilon_1\frac{\lambda_1+\lambda_2}{2}&0&0\end{array}
\right),\;\;
B' =\left(\begin{array}{cccc} \varepsilon_2\lambda_2&0&0&0 \\
0&1&0&0\\
0&0&\varepsilon_1\lambda_1&0\\
0&0&0&0\end{array}\right)  \label{Eqn4-3}
\end{equation}

\begin{figure}[htbp]
\begin{minipage}[t]{7.0cm}
\centering
\includegraphics[width=2.2in]{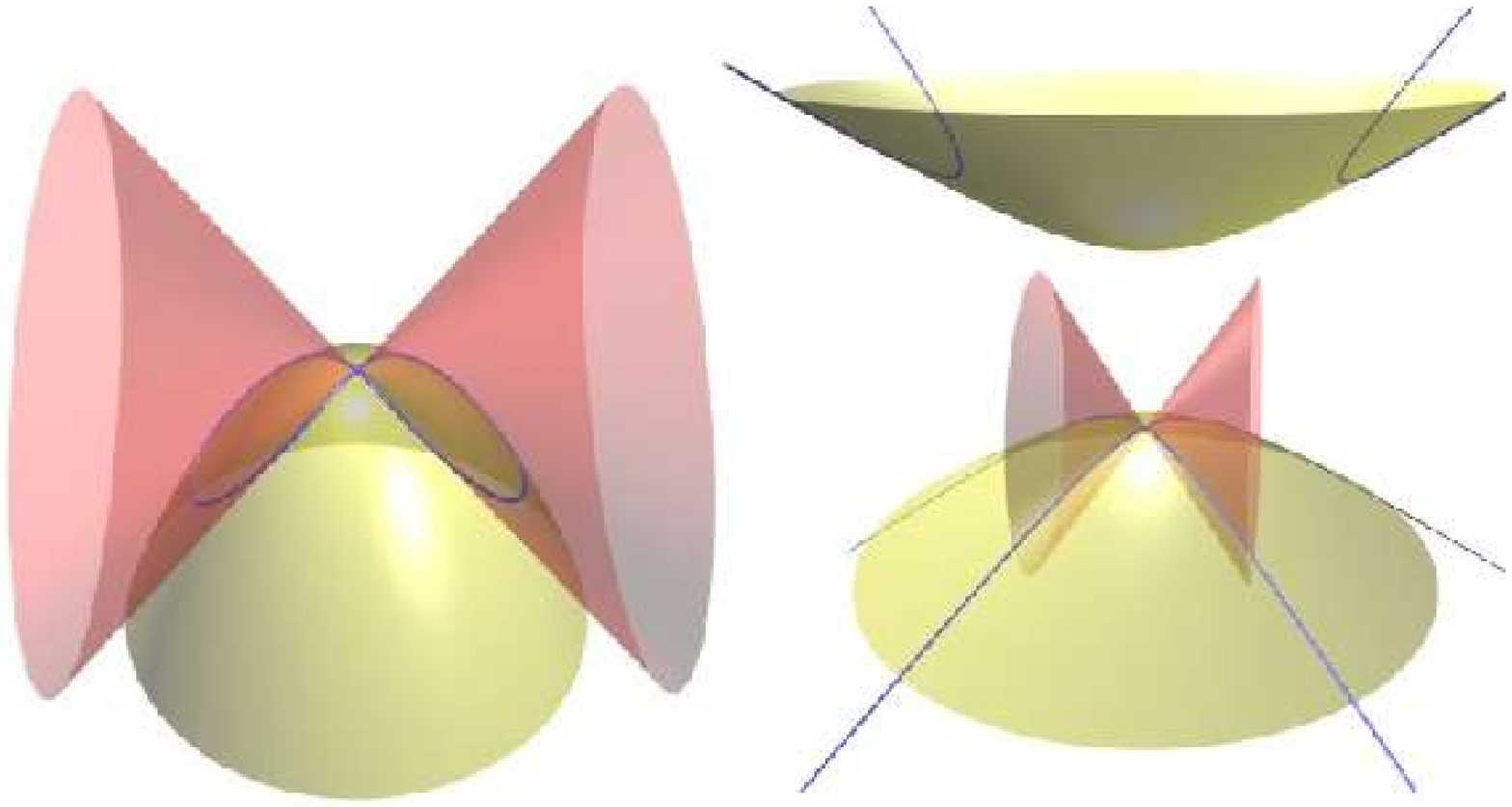}
\footnotesize \caption{Two cases of the QSIC having a
crunode.}\label{fig4-4}
\end{minipage}
\hspace{0.2in}
\begin{minipage}[t]{7.0cm}
\centering
\includegraphics[width=1.2 in]{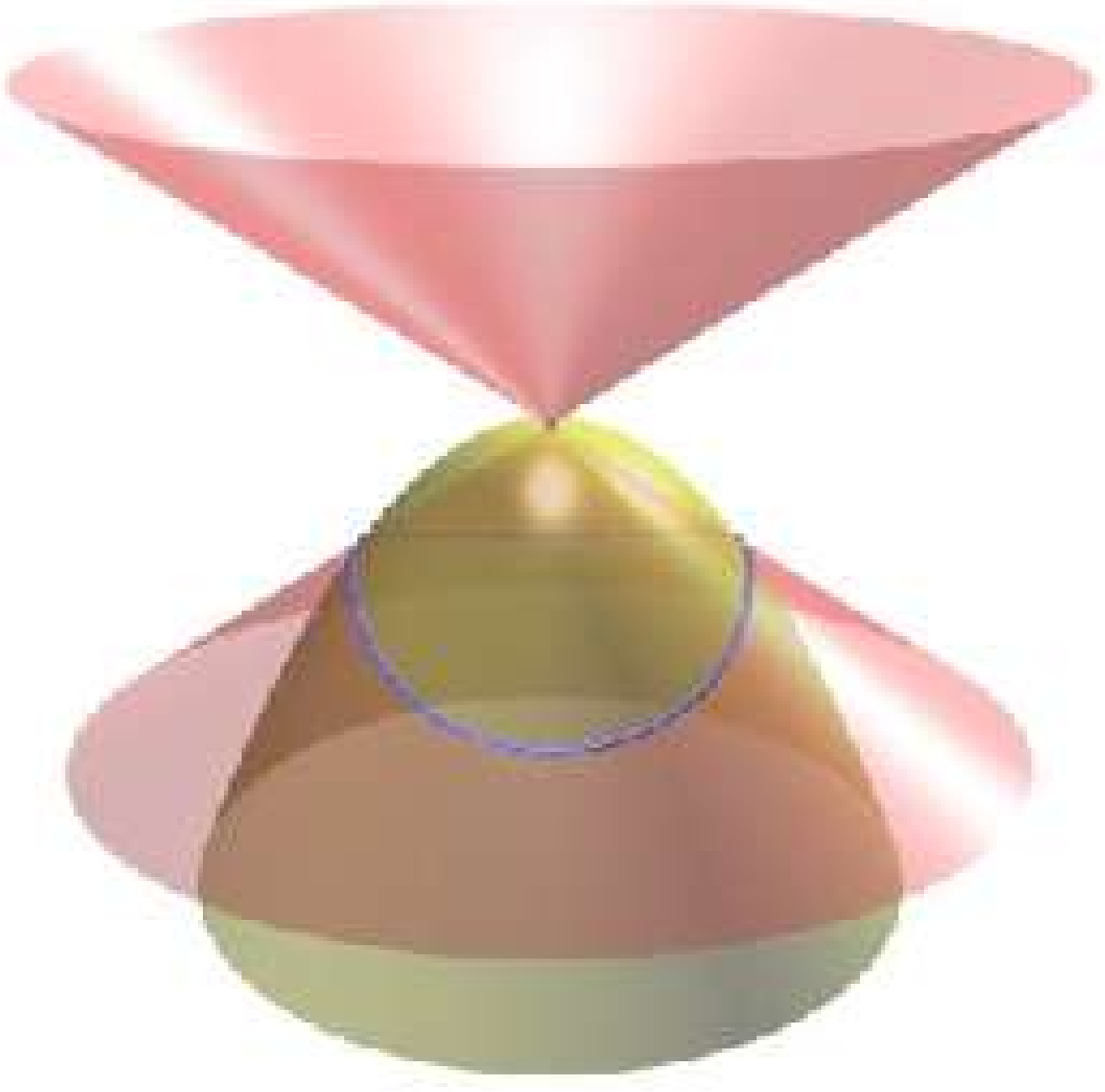}
\footnotesize \caption{The case of the QSIC having an
acnode.}\label{fig4-2}
\end{minipage}
\end{figure}

Clearly, the quadric ${\mathcal B'}$ is a cone passing through the
point $(0,0,0,1)^T$.
Since $\lambda_2-\lambda_1>0$, ${\mathcal A'}$ is an ellipsoid if
$\varepsilon_1=-1$ and $\varepsilon_2=1$, and is a two-sheet
hyperboloid if $\varepsilon_1=1$ and $\varepsilon_2=-1$.
In both cases ${\mathcal A'}$ passes through the point
$(0,0,0,1)^T$.
According to Eqn. (\ref{Eqn4-3}), ${\mathcal A'}$ and ${\mathcal
B'}$ are in one of the two cases shown in Figure~\ref{fig4-4}.
Thus, the QSIC is a singular quartic having one component with a
crunode.
Because the QSIC is contained in the ellipsoid or two-sheet
hyperboloid ${\mathcal A'}$, it is a closed curve in $\mathbb{
PR}^3$.
This proves the first item of Theorem~\ref{thm:$[211]_3$}.

In case ($ii$) (det$(A')<0$): $\varepsilon_1$ and $\varepsilon_2$
have the same sign and Id($\infty$) = Id($A'$)= 1 or 3. Also, the
index jump at $\lambda_0=0$ is $0$ because the size of its Jordan
block associated with $\lambda_0$ is 2. Therefore, the associated
index sequence is  $\langle 1{\wr\wr}_{-}1|2|3\rangle$ or $\langle
1{\wr\wr}_{+}1|2|3\rangle$.

If $\varepsilon_1=\varepsilon_2=-1$, the index of $\lambda_0A' - B'=
-B'$ is 2. Thus the index sequence is $\langle
3{\wr\wr}_{+}3|2|1\rangle$, which is equivalent to $\langle
1{\wr\wr}_{-}1|2|3\rangle$.
In this case the quadric ${\mathcal B'}$ is a cone with the $y$-axis
as its central axis, and the quadric ${\mathcal A'}$ is an elliptic
paraboloid also with $y$-axis as its central axis.
It is then easy to verify that the the QSIC has a closed component
plus an acnode, as shown in Figure \ref{fig4-2}.
This proves the second item of Theorem~\ref{thm:$[211]_3$}.

If $\varepsilon_1=\varepsilon_2=1$, the index of $\lambda_0A' - B'=
-B'$ is 0. Thus the index sequence $\langle 1{\wr\wr}1|2|3\rangle$
specializes to $\langle 1{\wr\wr}_{+}1|2|3\rangle$.
In this case, $B'$ is a positive semi-definite; thus, the QSIC of
${\mathcal A'}: X^TA'X = 0$ and ${\mathcal B'}: X^TB'X = 0$ has only
one real point $(0,0,0,1)$, which can be verified to be an acnode.
This proves the last item of Theorem~\ref{thm:$[211]_3$}.
%\end{proof}

\begin{theorem} ($[211]_1$: Case 8, Table 1) \label{thm:$[211]_1$}
If $f(\lambda) = 0$ has one double root and a pair of complex
conjugate roots, then the only possible index sequence is $\langle 2
{\wr\wr}_{-} 2 \rangle $ (or its equivalent $\langle 2 {\wr\wr}_{+}
2 \rangle $) and in this case the QSIC comprises one open component
with a crunode.
\end{theorem}

%\begin{proof}
{\bf Proof} Suppose that the pair of complex conjugate roots are $a
\pm bi$ and the double root is $\lambda_0$. Setting $B$ to be
$(B-aA)/b$, we transform the roots $a \pm bi$ to $\pm i$. By
Theorem~\ref{thm:canonical_form}, the two quadrics can be reduced to
the following forms:
\begin {equation}
A' =\left(\begin{array}{cccc} 0&\varepsilon&0&0\\
\varepsilon&0&0&0\\
0&0&0&1\\
0&0&1&0\end{array} \right),\;\;
B' =\left(\begin{array}{cccc} 0&\varepsilon\lambda_0&0&0 \\
\varepsilon\lambda_0&\varepsilon&0&0\\
0&0&-1&0\\
0&0&0&1\end{array}\right),  \label{eqn4-1}
\end {equation}
where $\varepsilon = \pm 1$.
Setting $B'-\lambda_0 A$ to be $B'$, and then swapping $A'_{1,1}$,
$A'_{4,4}$, as well as $B'_{1,1}$, $B'_{4,4}$, $A'$ and $B'$ are
transformed to
\[
A' =\left(\begin{array}{cccc} 0&0&1&0\\
0&0&0&\varepsilon\\
1&0&0&0\\
0&\varepsilon&0&0\end{array} \right),\;\;
B' =\left(\begin{array}{cccc} 1&0&-\lambda_0&0 \\
0&\varepsilon&0&0\\
-\lambda_0&0&-1&0\\
0&0&0&0\end{array}\right).
\]
Denote $k = 1/\sqrt{1+\lambda_0^2}$. Applying the congruence
transformation $C = P^TDP$ with
\[
P =\left(\begin{array}{cccc}1&0&k\lambda_0&0\\
0&1&0&0\\
0&0&k&0\\
0&0&0&1\end{array} \right).
\]
simultaneously to ${\mathcal A'}$ and ${\mathcal B'}$, we obtain the
transformed $A'$ and $B'$ as
\[
A' =\left(\begin{array}{cccc} 0&0&k&0\\
0&0&0&\varepsilon\\
k&0&2k^2\lambda_0&0\\
0&\varepsilon&0&0\end{array} \right),\;\;
B' =\left(\begin{array}{cccc} 1&0&0&0 \\
0&\varepsilon&0&0\\
0&0&-1&0\\
0&0&0&0\end{array}\right).
\]
The quadric ${\mathcal B}'$ is a cone and therefore can be
parameterized by
$${\bf r}(u,v)={\bf g}(u)+v{\bf h}(u),$$ where
${\bf g}(u)=(1-\varepsilon u^2,\; 2u,\; 1+\varepsilon u^2,\; 0)^T$
and ${\bf h}(u)=(0, 0, 0, 1)^T$.
Substituting ${\bf r}(u,v)$ into $X^TA'X=0$, we obtain a quadratic
equation whose two solutions are $v= \infty$, which is trivial, and
$v=-{\bf c_0}(u)/(2{\bf c_1}(u))$. Substituting the latter solution
of $v$ into ${\bf r}(u,v)$ yields the parameterization of the QSIC,
\begin{equation}\label{eqn:qsic12}
{\bf p}(u)=\left(2\varepsilon u(1-\varepsilon u^2),\; 4u^2,\;
2\varepsilon u(1+\varepsilon u^2),\; k(1-u^4)+k^2\lambda_0
(1+\varepsilon u^2)^2\;\right)^T
\end{equation}
From ${\bf p}(u)$, we see that the QSIC passes through the point
${\bf p}_0 =(0,0,0,1)^T$ twice, with $u=0$ or $u=\infty$. Hence,
${\bf p}_0$ is a singular point of the QSIC.
Furthermore, it is easy to verify that ${\bf p}_0$ is a crunode.

In the following we will show that the QSIC has two open branches
intersecting at the crunode.
Consider the intersection of the QSIC with the plane
$w=0$.
The last component $w(u)$ of ${\bf p}(u)$ in Eqn.~(\ref{eqn:qsic12})
is a quadratic polynomial in $u^2$, whose two zeros are
$$u^2=(-\varepsilon k\lambda_0+1)/(k\lambda_0-1)$$ and
$$u^2=(-\varepsilon k\lambda_0-1)/(k\lambda_0-1).$$
Recall that $k = 1/\sqrt{1+\lambda_0^2}$,
it is straightforward to verify that $w(u)$ has two real zeros in
$u$. These two real zeros are $u_{1,2} = \pm
(\lambda_0+\sqrt{1+\lambda_0^2})$ when $\varepsilon=1$ or $u_{1,2} =
\pm 1$ when $\varepsilon=-1$. We observe that $u_1$ and $u_2$ have
opposite signs and ${\bf p}(u_1)$ and ${\bf p}(u_2)$ are two
distinct points.

\begin{figure}[htbp]
\begin{minipage}[t]{7.0cm}
\centering
\includegraphics[width=2.0 in]{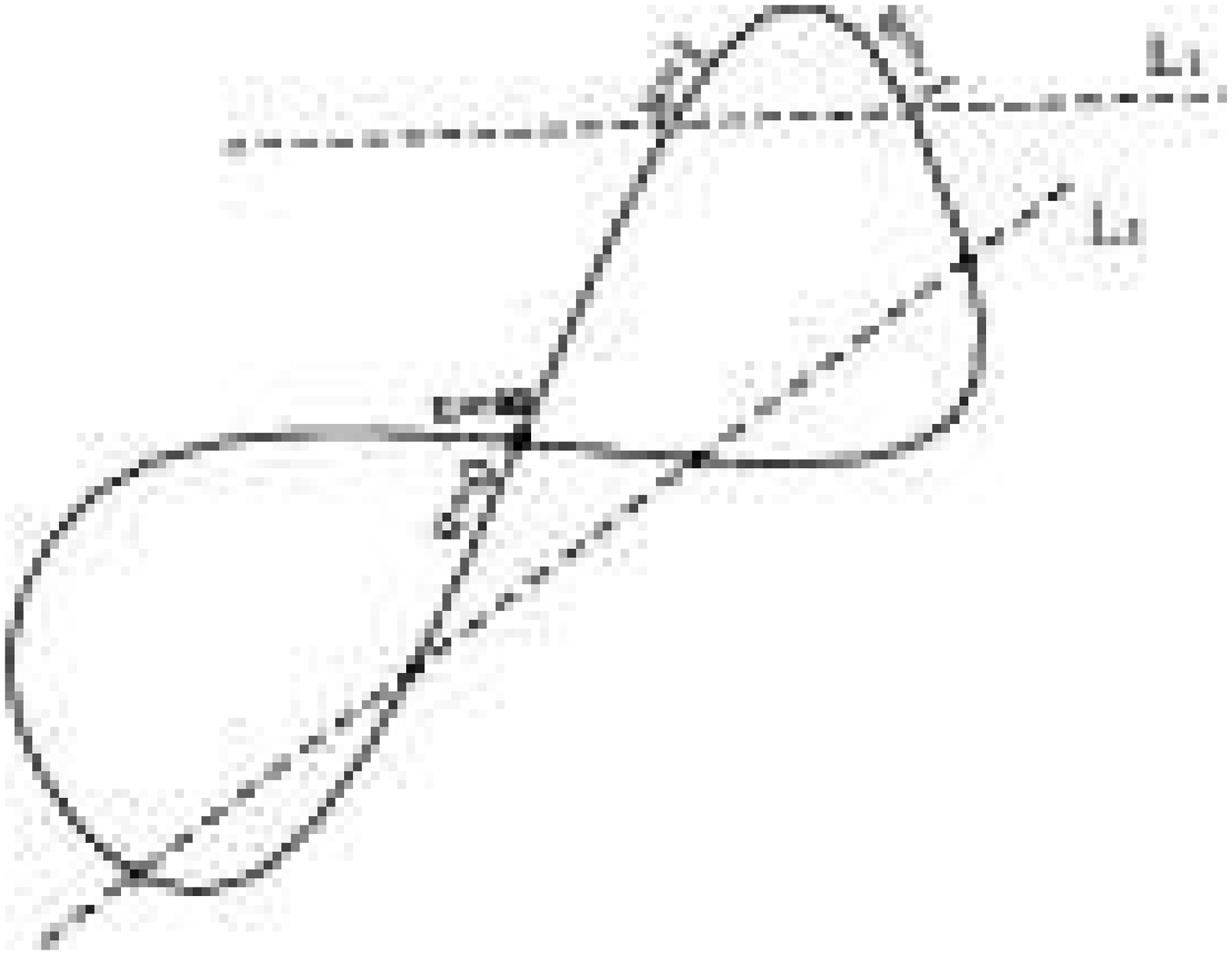}
\footnotesize \caption{The hypothetical topological shape of a
QSIC.} \label{fig4-5}
\end{minipage}
\hspace{0.2in}
\begin{minipage}[t]{7.0cm}
\centering
\includegraphics[width=1.7in]{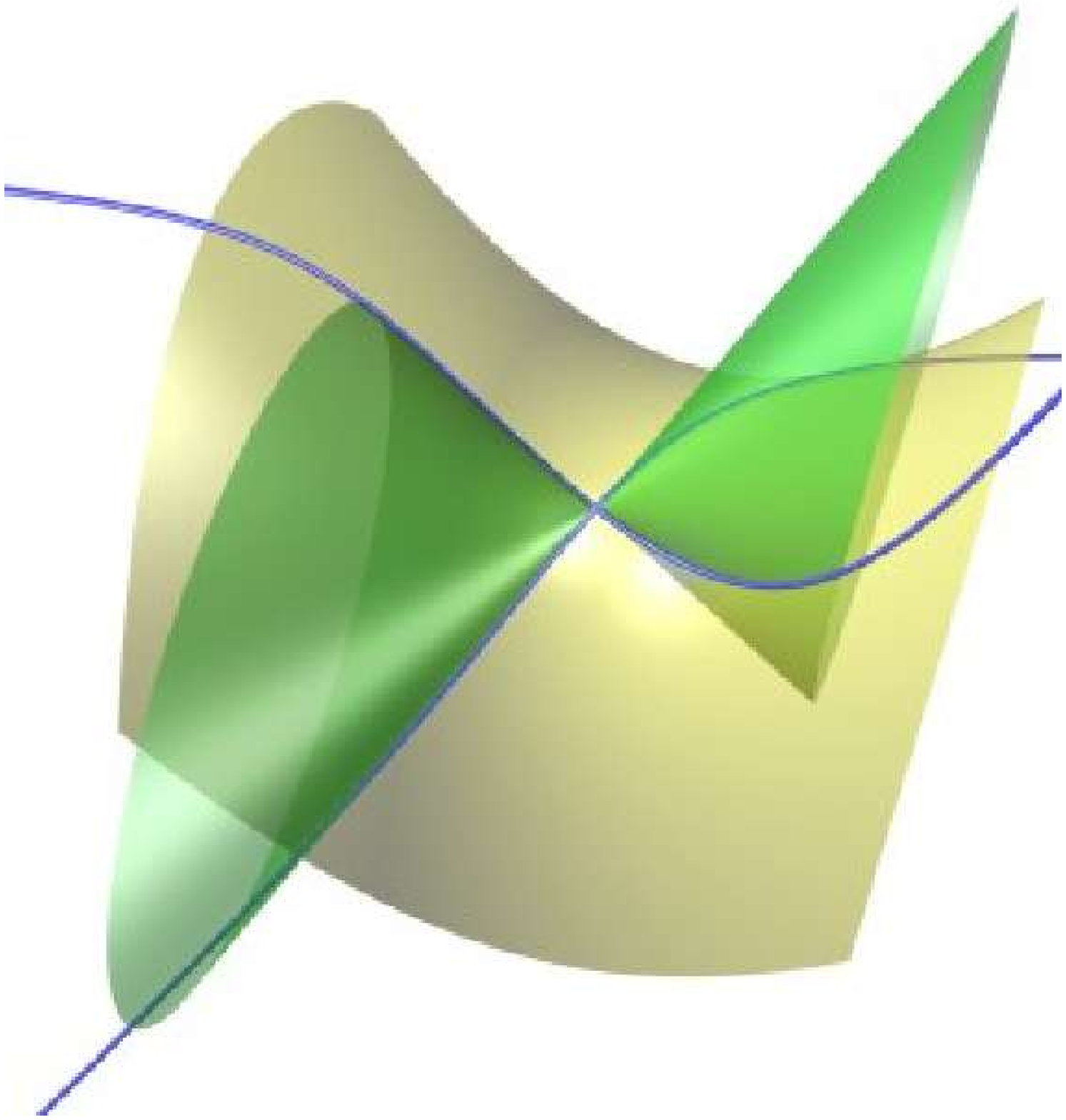}
\footnotesize \caption{A QSIC having one open component with a
crunode.} \label{fig4-6}
\end{minipage}
\end{figure}

Now we are going to show by contradiction that the QSIC cannot be
closed. Assume that the QSIC is closed, i.e., there is an affine
realization of ${\mathbb {PR}}^3$ in which the QSIC is compact.
Note that the plane $w=0$ is not necessarily the plane at infinity
in this affine realization.
Then the QSIC has a topology shown in Figure~\ref{fig4-5}, having
two closed loops joining at the crunode, i.e., like the figure of
``8''.
Since the crunode corresponds to two parameter values $0$ and
$\infty$ of $u$ under the parameterization ${\bf p}(u)$ in
Eqn.~(\ref{eqn:qsic12}), the two loops must be parameterized over
the positive interval $u\in (0, +\infty)$ and the negative interval
$u\in (-\infty, 0)$, respectively.
Now consider again the intersection of the QSIC with the plane
$w=0$.
If the plane $w=0$ intersects any loop of the QSIC, say the loop
defined over the positive interval, there must be at least two
intersection points, which should be given by two positive values of
$u$ through ${\bf p}(u)$.
However, from the preceding discussions we know that there are only
two intersections between the QSIC: ${\bf p}(u)$ and the plane
$w=0$, which correspond to one positive value and one negative value
of $u$.
This is a contradiction. Hence, there is no finite loop of the QSIC
in any affine realization of the projective space. That is, the QSIC
has one open component with a crunode. (An example of such a QSIC is
shown in Figure~\ref{fig4-6}.) This completes the proof of
Theorem~\ref{thm:$[211]_1$}.
%\end{proof}

\subsection{$[22]$: f$(\lambda) = 0$ has two double roots}

%[22]_2
\begin{theorem} ($[22]_2$: Case 9, Table 1) \label{thm:$[22]_2$} If $f(\lambda)=0$ has two real double roots with the Segre
characteristic $[22]$, then the only possible index sequences are
$\langle 2{\wr\wr}_{-} 2{\wr\wr}_{-} 2 \rangle $ and  $\langle
2{\wr\wr}_{-} 2{\wr\wr}_{+} 2 \rangle $, the QSIC comprises a real
line and a space cubic curve intersecting at two distinct real
points for both sequences.
\end{theorem}

\begin{theorem} ($[22]_0$: Case 10, Table 1)\label{thm:$[22]_0$} If $f(\lambda)=0$ has two pairs of identical complex conjugate roots
with the Segre characteristic $[22]$, then the index sequences are
$\langle 2 \rangle$ and the QSIC comprises a real line and a space
cubic curve that do not intersect at any real point.
\end{theorem}

%\begin{proof}
%{\bf Proof} Suppose that the two identical pairs of conjugate roots
%of $f(\lambda)=0$ are $a\pm bi$. In this case the only possible
%index sequence is $\langle 2 \rangle$.
%%
%Setting  $(B-aA)/b$ to be $B$, we transform the roots $a\pm bi$ to
%$\pm i$.
%%
%By Theorem~\ref{thm:canonical_form}, the two quadrics can be
%transformed to the following forms,
%%
%\[
%A' =\left(\begin{array}{cccc} 0&0&0&1\\
%0&0&1&0\\
%0&1&0&0\\
%1&0&0&0\end{array} \right),\;\;
%B' =\left(\begin{array}{cccc} 0&0&1&0 \\
%0&0&0&-1\\
%1&0&0&1\\
%0&-1&1&0\end{array}\right).
%\]
%Clearly, the QSIC contains the line $z=w=0$, and the remaining
%component is a cubic curve parameterized by
%\[
%{\bf q}(u) = \left(-u^2,\; u, \; -u(1+u^2), \; -(1+u^2)\;\right)^T.
%\]
%Since the last component function $-(1+u^2)$ does not vanish for any
%real value of $u$, the space cubic curve has no real intersection
%with the line $z=w=0$.
%%
%This completes the proof of Theorem~\ref{thm:$[22]_0$}.
%%\end{proof}

\subsection{$[31]$: $f(\lambda)=0$ has one real triple root and one real simple root}

\begin{theorem} (Case 11, Table 1) \label{thm:$[31]$} If ${\rm f}(\lambda) = 0$ has one
triple root and one simple real root with the Segre characteristic
$[31]$, then the index sequence is $\langle 1 {\wr\wr}{\wr}_{+} 2 |
3 \rangle$ and the QSIC comprises a closed component with a real
cusp.
\end{theorem}

%%\begin{proof}
%{\bf Proof} According to the discussion about the index sequences in
%Section~\ref{sec:index sequences}, the index jump is 1 across the
%real root with a $3\times 3$ Jordan block. Thus it is easy to see
%the only possible index sequence is $\langle 1 {\wr\wr}{\wr} 2 | 3
%\rangle$.
%
%That the QSIC comprises a closed component with a real cusp follows
%from the classification of QSIC in ${\mathbb {PC}}^3$ by the Segre
%characteristics. By complex conjugation it is easy to see that the
%cusp is real. Since the QSIC is contained in a projective ellipsoid
%in the quadric pencil (i.e., with the index being 1 or 3), it is
%closed in ${\mathbb {PR}}^3$.
%%
%This completes the proof of Theorem~\ref{thm:$[31]$}.
%%\end{proof}

\subsection{$[4]$: $f(\lambda)=0$ has one real quadruple root}

\begin{theorem} (Case 12, Table 1) \label{thm:$[4]$} If $f (\lambda) = 0$ has one quadruple root with
the Segre characteristic $[4]$, then the index sequence is $\langle
2 {\wr\wr}{\wr\wr}_{-} 2 \rangle$ or its equivalent form $\langle 2
{\wr\wr}{\wr\wr}_{+} 2 \rangle$, and the QSIC comprises a real line
and a real space cubic curve tangent to each other at a real point
in this case.
\end{theorem}

%%\begin{proof}
%{\bf Proof} According to the discussion about  index sequences in
%Section~\ref{sec:index sequences}, the index jump is 0 across the
%real root with a $4\times 4$ Jordan block. Thus, the only possible
%index sequence is $\langle 2 {\wr\wr}{\wr\wr}_{-} 2 \rangle$ or its
%equivalent form $\langle 2 {\wr\wr}{\wr\wr}_{+} 2 \rangle$.
%
%That the QSIC comprises a line and a space cubic curve tangent to
%each other at a point follows from the classification of QSIC in
%${\mathbb {PC}}^3$ by the Segre characteristics. By complex
%conjugation it is easy to see that the line and the cubic are both
%real and their tangent point is also real. This completes the proof
%of Theorem~\ref{thm:$[4]$}.
%%\end{proof}

\section{Classification of planar QSIC} \label{sec:planar}

\subsection{$[(11)11]$: $f(\lambda)=0$ has one real double root and two other distinct roots}

\begin{theorem} ($[(11)11]_3$) \label{thm:$[(11)11]_3} If f$(\lambda) = 0$ has one double real root and two
distinct real roots with the Segre characteristic $[(11)11]$, then
there are only five different possible index sequences and these
index sequences correspond to four different QSIC morphologies as
follows:
\begin {enumerate}
\item (Case 13, Table 2) $\langle 2 || 2 | 1 | 2 \rangle$ -
two real closed conics intersecting at two distinct real points;
\item (Case 14, Table 2) $\langle 1 || 3 | 2 | 3 \rangle$ -
two real conics not intersecting at any real points;
\item (Case 15, Table 2) $\langle 1 || 1 | 2 | 3 \rangle$ -
two imaginary conics intersecting at two distinct real points;
\item (Case 16, Table 2) $\langle 0 || 2  | 3 | 4\rangle$ or $\langle 1 || 3  | 4 | 3 \rangle$ -
two imaginary conics not intersecting at any real points.
\end {enumerate}
\end{theorem}

\begin{theorem} ($[(11)11]_1$) \label{thm:$[(11)11]_1$}
If $f(\lambda) = 0$ has a real double root $\lambda_0$ and a pair of
complex conjugate roots with the Segre characteristic $[(11)11]$,
then the possible index sequences of the pencil $\lambda A - B$ are
$\langle 1 || 3\rangle$ and $\langle 2 || 2\rangle$. Furthermore,
\begin {enumerate}
\item (Case 17, Table 2) when the index sequence is $\langle 1 || 3\rangle$, the QSIC comprises of two conics, one real and one imaginary;
\item (Case 18, Table 2) when the index sequence is $\langle 2 || 2\rangle$, the QSIC comprises of two real
conics which cannot both be ellipses simultaneously in any affine
realization of ${\mathbb {PR}}^3$.
\end{enumerate}
\end{theorem}

\subsection{$[(111)1]_2$: $f(\lambda)=0$ has one real triple root and a real simple root}

\begin{theorem} ($[(111)1]_2$) \label{thm:$[(111)1]_2$}
If ${\rm f}(\lambda)=0$ has one triple root and a simple root with
the Segre characteristic $[(111)1]$, then the only possible index
sequences are $\langle 1 ||| 2 |3\rangle$ and $\langle 0||| 3 |4
\rangle $.
Furthermore,
\begin{enumerate}
\item (Case 19, Table 2) when the index sequence is $\langle 1
||| 2 |3\rangle$, the QSIC is a real conic counted twice;
\item (Case 20, Table 2) when the index sequence is $\langle 0||| 3 |4 \rangle $,
the QSIC is an imaginary conic counted twice.
\end{enumerate}
\end{theorem}

%%\begin{proof}
%{\bf Proof} Clearly, all the roots of $f(\lambda)=0$ are necessarily
%real in this case. Wlog, we may assume the triple root $\lambda_0$
%to be zero. Then, by Theorem~\ref{thm:canonical_form}, the canonical
%matrix forms of the given quadrics are
%\[
%A' ={\rm diag}(\varepsilon_1,\; \varepsilon_2,\; \varepsilon_3,\;
%\varepsilon_4\;),\;\;\;\;
%%
%B' = {\rm diag}(\varepsilon_1\lambda_1,\;0,\; 0,\;0\;).
%\]
%Setting  $\lambda_1A'-B'$ to be $A'$ and then $\varepsilon_1 B'$ to
%be $B'$, we obtain
%\[
%A' = {\rm diag} (0,\; \varepsilon_2\lambda_1,\;
%\varepsilon_3\lambda_1,\; \varepsilon_4\lambda_1\;), \;\;
%%
%B' = {\rm diag} (\lambda_1,\; 0,\; 0,\; 0\;).
%\]
%The quadric ${\mathcal A'}$ is a cylinder (real or imaginary) with
%the $x$-axis being its central axis, while the quadric ${\mathcal
%B'}$ is a plane counted twice.
%%
%Thus the QSIC is a conic counted twice, real or imaginary.
%
%When $\varepsilon_2$, $\varepsilon_3$ and $\varepsilon_4$ have
%different signs, the only possible index sequence is $\langle
%1|||2|3 \rangle$ (or its equivalent form $\langle 3|||2|1 \rangle$).
%In this case ${\mathcal A'}$ is real, and the QSIC is a real conic
%counted twice.
%%
%When $\varepsilon_2$, $\varepsilon_3$ ad $\varepsilon_4$ have the
%same sign, the only possible index sequence is $\langle 0|||3|4
%\rangle$ (or its equivalent form $\langle 3|||0|1 \rangle$). In this
%case ${\mathcal A'}$ is imaginary, and the QSIC is an imaginary
%conic counted twice. This completes the proof of
%Theorem~\ref{thm:$[(111)1]_2$}.
%%\end{proof}

\subsection{$[(21)1]_2$: $f(\lambda)=0$ has one real triple root and a real simple root}

\begin{theorem} ($[(21)1]_2$) \label{thm:$[(21)1]_2$}
If ${\rm f}(\lambda)=0$ has one triple root and a simple root with
the Segre characteristic $[(21)1]$, then the only possible index
sequences are $\langle 1{\wr\wr}_{-}|2|3 \rangle$ and $\langle
1{\wr\wr}_{+}|2|3 \rangle$.
Furthermore,
\begin{enumerate}
\item (Case 21, Table 2) when the index sequence is $\langle 1{\wr\wr}_{-}|2|3 \rangle$, the
QSIC  comprises two real conics tangent to each other at one real
point;
\item (Case 22, Table 2) when the index sequence is $\langle
1{\wr\wr}_{+}|2|3 \rangle$, the QSIC comprises two imaginary conics
tangent to each other at one real point.
\end{enumerate}
\end{theorem}

\subsection {$[2(11)]_2$: $f(\lambda)=0$ has two real double roots}

\begin{theorem}\label{thm:$[2(11)]_2$}
If $f (\lambda)=0$ has two double roots with the Segre
characteristic $[2(11)]$, then the only possible index sequences are
$\langle 2{\wr\wr}_{-}2||2 \rangle$ (or its equivalent form $\langle
2{\wr\wr}_{+}2||2 \rangle$), $\langle 1{\wr\wr}_{+} 1 || 3 \rangle$
and $\langle 1{\wr\wr}_{-} 1 || 3\rangle$.
Furthermore,
\begin{enumerate}
\item (Case 23, Table 3) when the index sequence is $\langle 2{\wr\wr}_{-}2||2 \rangle$,
the QSIC consists of a real conic and two real lines which intersect
pairwise at three distinct real points;
\item (Case 24, Table 3) when the index sequence is $\langle 1{\wr\wr}_{-} 1 || 3
\rangle$, the QSIC consists of a real conic and a pair of complex
conjugate lines. The conic and the pair of lines do not intersect;
\item (Case 25, Table 3) when the index sequence is $\langle 1{\wr\wr}_{+} 1 || 3\rangle$,
the QSIC consists of an imaginary conic and a pair of complex
conjugate lines. The conic and the pair of lines do not intersect.
\end{enumerate}
\end{theorem}

\subsection{$[(31)]_1$: $f(\lambda)=0$ has one real quadruple root}

\begin{theorem}\label{thm:$[(31)]_2$}
If $f(\lambda)=0$ has one quadruple root with the Segre
characteristic $[(31)]$, then the only possible index sequences are
$\langle 2 {\wr\wr}{\wr}_{-}| 2 \rangle$ and $\langle 1
{\wr\wr}{\wr}_{+}| 3 \rangle $.
Furthermore,
\begin{enumerate}
\item (Case 26, Table 3) when the index sequence is $\langle 2 {\wr\wr}{\wr}_{-}| 2 \rangle$,
the QSIC consists of a real conic and two real lines, and these
three components intersect at a common real point;
\item (Case 27, Table 3) when the index sequence is $\langle 1 {\wr\wr}{\wr}_{+}| 3
\rangle $, the QSIC consists of a real conic and a pair of complex
conjugate lines, and these three components intersect at a common
real point.
\end{enumerate}
\end{theorem}

\subsection {$[(11)(11)]$: $f(\lambda)=0$ has two double roots}

\begin{theorem} ($[(11)(11)]_2$) \label{thm:$[(11)(11)]_2$}
If $f(\lambda)=0$ has two real double roots with the Segre
characteristic $[(11)(11)]$, then the only possible index sequences
are $\langle 2||2||2 \rangle $, $\langle 0||2||4 \rangle$ and $
\langle 1||1||3 \rangle $.
Furthermore,
\begin{enumerate}
\item (Case 28, Table 3) when the index sequence is $\langle 2||2||2 \rangle
$,
 the QSIC consists of four real lines, and these four lines
form a quadrangle in $\mathbb{PR}^3$;
\item (Case 29, Table 3) when the index sequence is $\langle 0||2||4 \rangle$, the QSIC consists
of four imaginary lines and has no real point;
\item (Case 30, Table 3) when the index sequence is $ \langle 1||1||3 \rangle $, the QSIC consists
of two pair of complex conjugate lines, with each pair intersecting
at a real point.
\end{enumerate}
\end{theorem}

\begin{theorem} ($[(11)(11)]_0$: Case 31, Table 3) \label{thm:$[(11)(11)]_0$}
If $f(\lambda)=0$ has two identical pairs of complex conjugate roots
with the Segre characteristic $[(11)(11)]$, the only possible index
sequences are $\langle 2 \rangle$, and in this case the QSIC
comprises two non-intersecting real lines and two non-intersecting
imaginary lines.
\end{theorem}

%%\begin{proof}
%By Theorem~\ref{thm:canonical_form},
%\[
%A' ={\rm diag} \left(\;\left(\begin{array}{cc} 0&1\\
%1&0 \end{array} \right),\;
%\left(\begin{array}{cc} 0&1\\
%1&0 \end{array} \right)\;\right),\;\;
%%
%B' = {\rm diag} (-1, \; 1,\; -1,\; 1).
%\]
%Then ${\mathcal A'}$ is a hyperbolic paraboloid $xy+zw=0$ and
%${\mathcal B'}$ is a hyperboloid $-x^2+y^2-z^2+w^2=0$.
%%
%It is easy to verify that the QSIC consists of the two
%non-intersecting real lines defined by ($x+w=0$, $y-z=0$) and
%($x-w=0$, $y+z=0$), and two non-intersecting imaginary lines defined
%($x-iz=0$, $y-iw=0$) and ($x+iz=0$, $y+iw=0$).
%%
%In this case, since ${f}(\lambda)=0$ has no real root, the index
%sequence is $\langle 2 \rangle$. This completes the proof of
%Theorem~\ref{thm:$[(11)(11)]_0$}.
%%\end{proof}

\subsection{$[(211)]$: $f(\lambda)=0$ has one real quadruple root}

\begin{theorem}\label{thm:$[(211)]$}
If $f(\lambda)=0$ has a quadruple root with the Segre characteristic
$[(211)]$, then the only possible index sequences are $\langle
2{\wr\wr}_{-}||2 \rangle$ and $\langle 1{\wr\wr}_{-} || 3 \rangle$.
Furthermore,
\begin{enumerate}
\item (Case 32, Table 3) when the index sequence is $\langle 2{\wr\wr}_{-}||2 \rangle$, the QSIC consists
of a pair of intersecting real lines, counted twice;
\item (Case 33, Table 3) when the index sequence is $\langle 1{\wr\wr}_{-} || 3 \rangle$,
 the QSIC consists of a pair of conjugate lines, counted twice.
\end{enumerate}
\end{theorem}

%%\begin{proof}
%{\bf Proof} The quadruple root, denoted by $\lambda_1$, of
%$f(\lambda)=0$ is necessarily real. Wlog, we may assume
%$\lambda_1=0$.
%%
%By Theorem~\ref{thm:canonical_form}, the canonical form of the two
%given quadrics are
%\[
%A' ={\rm diag} \left(\;\left(\begin{array}{cc}  & \varepsilon_1\\
%\varepsilon_1 & \end{array} \right),\; \varepsilon_2,\;
%\varepsilon_3 \right),\;\;
%%
%B' ={\rm diag} (0,\; \varepsilon_1,\; 0,\; 0\;).
%\]
%The quadric ${\mathcal B'}$ is a pair of the identical real plane
%$y=0$.
%%
%Substituting $y^2=0$ in the the quadric ${ \mathcal A'}$, we find
%that the QSIC is the intersection between $y^2=0$ and the pair of
%planes $z^2 = -\varepsilon_3/\varepsilon_2w^2$.
%%
%Clearly, the QSIC comprises two intersecting real lines, counted
%twice, if $\varepsilon_2\varepsilon_3 =-1$, or a pair of conjugate
%lines, counted twice, if  $\varepsilon_2\varepsilon_3 =1$.
%
%
%Clearly, the index jump of ${\rm Id}(\lambda A' - B')$ at
%$\lambda_1=0$ is 0 or $\pm 2$. When $\varepsilon_3/\varepsilon_2
%=-1$, $\det(A)=-\varepsilon_2\varepsilon_3>0$, therefore ${\rm
%index} (A')$ is 0, 2 or 4.  It follows that the index sequence is
%$\langle 2{\wr\wr}_{-}||2 \rangle$ or its equivalent form $\langle
%2{\wr\wr}_{+}||2 \rangle$.
%%
%When $\varepsilon_3/\varepsilon_2 =1$,
%$\det(A)=-\varepsilon_2\varepsilon_3<0$, therefore ${\rm index}
%(A')$ is 1 or 3.  It follows that the index sequence is $\langle
%1{\wr\wr}_{-} || 3 \rangle$ or its equivalent form $\langle
%1{\wr\wr}_{+} || 3 \rangle$. This completes the proof of
%Theorem~\ref{thm:$[(211)]$}.
%%\end{proof}

\subsection{$[(22)]$: $f(\lambda)=0$ has one real quadruple root}

\begin{theorem}\label{thm:$[(22)]$}
If $f(\lambda)=0$ has a quadruple root with the Segre characteristic
$[(22)]$, then the only possible index sequences are $\langle 2
{\wr\wr}_-{\wr\wr}_- 2 \rangle$ and $\langle 2 {\wr\wr}_-{\wr\wr}_+
2 \rangle$
Furthermore,
\begin{enumerate}
 \item (Case 34, Table 3) when the index sequence is $\langle 2 {\wr\wr}_{-}{\wr\wr}_{-} 2 \rangle$,
 the QSIC consists a real double line and two other non-intersecting imaginary
 lines. The two imaginary lines do not form a complex conjugate
 pair.
\item (Case 35, Table 3)  when the index sequence is $\langle 2 {\wr\wr}_{-} {\wr\wr}_{+} 2 \rangle$,
 the QSIC consists a real double line and two other non-intersecting real
 lines. Each of the latter two lines intersects the real double
 line.
 \end{enumerate}
\end{theorem}

\section{Classification by signature sequences} \label{sec:complete
classification}

Through the above analysis, we have put {\em different} QSIC
morphologies in correspondence to {\em different} characterizing
conditions, given by conditions in Theorem~\ref{thm:four_real_roots}
through Theorem~\ref{thm:$[(22)]$}. Hence, we conclude that all
these conditions are necessary and sufficient for the corresponding
QSIC morphologies. We may then check these conditions to classify
the QSIC of a given pair of quadrics in $\mathbb{PR}^3$.
Based on these conditions one could compute the index sequence of
two given quadrics for QSIC classification; however, this would
involve the difficult task of computing Jordan blocks.
To avoid computing Jordan blocks, we convert all index sequences to
their corresponding \textit{signature sequences}. The advantage of
using the signature sequence over using the index sequence is that
we just need to compute the multiplicity of a real root and
determine the signature of $\lambda A - B$ at the root; this is a
far simpler than computing Jordan blocks.

There are some cases which cannot be distinguished by using
signature sequences alone. We will show in this section that these
cases can easily be distinguished by the fact that their
corresponding minimal polynomials have different degrees.

Not all the signature sequences of the 35 different QSIC
morphologies are distinct: the three different QSIC morphologies
with the Segre characteristics $[ 1111 ]_0$, $[ 22 ]_0$ and $[ ( 11
) ( 11 ) ]_0$ (i.e., cases 4, 10 and 31) share the same index
sequence $\langle 2 \rangle$, thus leading to the same signature
sequence $( 2 )$. Furthermore, two different index sequences
$\langle 2 {\wr\wr}{\wr}_{-}|2 \rangle$ and $\langle 2
{\wr\wr}_{-}{\wr\wr}_{+} 2 \rangle$ (i.e., cases 26 and 35) are
mapped to the same signature sequence $\langle 2 ( ( ( ( 1, 1 ) ) )
)2 \rangle$. Thus, in total, there are only 32 distinct signature
sequences. In the following we explain how these cases can be
distinguished.

The signature sequence  $( 2 )$ can be given by the different Segre
characteristics $[ 1111 ]_0$, $[ 22 ]_0$ and $[ ( 11 ) ( 11 ) ]_0$.
Suppose that the signature sequence $( 2 )$ has been detected, i.e.,
$f(\lambda ) = 0$ has been found to have no real root. The case of
$[ 1111 ]_0$ is distinguished from the other two cases by the fact
that ${f} ( \lambda ) = 0$ has no multiple roots; this can be
detected by whether the discriminant of ${f} ( \lambda )$ vanishes,
i.e., whether ${\textrm{Disc}} ( f ) \equiv {\textrm{Res}}_{\lambda}
( f, f_{\lambda} ) = 0$.

Then the case of $[ 22 ]_0$ and the case of $[ ( 11 ) ( 11 ) ]_0$
can be distinguished by the fact that they have different minimal
polynomials. Suppose that the input quadrics are given in the real
symmetric matrices $A$ and $B$; and, wlog, assume that $A$ is
nonsingular. Since in the case of $[ 22 ]_0$ or $[ ( 11 ) ( 11 )
]_0$, $f( \lambda )$ is a squared polynomial, we suppose
\[ f ( \lambda ) = ( a \lambda^2 + b \lambda + c )^2, \]
whose square-free part is
\[ g ( \lambda ) = a \lambda^2 + b \lambda + c, \]
where $a, b, c \in \mathbb{R}$ and $b^2 - 4 ac < 0$. Then, by
Theorem~\ref{thm:canonical_form} and the Cauchy-Cayley Theorem, the
case of $[ ( 11 ) ( 11 ) ]_0$ occurs if $g ( \lambda )$ annihilates
$A^{- 1} B$, i.e., $g ( A^{- 1} B ) = 0$; otherwise, the case of $[
22 ]_0$ occurs.
Note that $g(\lambda)$ can be obtained as the GCD of $f(\lambda)$
and $f'(\lambda)$.

The remaining problem is that the two index sequences $\langle 2
{\wr\wr}{\wr}_{-}|2 \rangle$ and $\langle 2 {\wr\wr}_{-}
{\wr\wr}_{+} 2 \rangle$ are mapped to the same signature sequence
$\langle 2 ((((1,1))))2 \rangle$. For either of the two cases, $f (
\lambda ) = ( \lambda - a )^4$ for some $a \in \mathbb{R}$, but the
minimal polynomial for the case of $\langle 2 {\wr\wr}_{-}
{\wr\wr}_{+} 2 \rangle$ is $g ( \lambda ) = ( \lambda - a )^2$,
while the minimal polynomial for the case of $\langle
2{{\wr\wr}\wr}_{-}|2 \rangle$ is $h ( \lambda ) = ( \lambda - a
)^3$. Therefore, the case of $\langle 2 {\wr\wr}_{-} {\wr\wr}_{+} 2
\rangle$ occurs if $A^{- 1} B$ is annihilated by $g ( \lambda )$,
i.e., $g ( A^{- 1} B ) = 0$; otherwise, the case of $\langle 2
{\wr\wr}{\wr}_{-} |2 \rangle$ occurs. Note that $g ( \lambda ) = (
\lambda - a )^2$ can be obtained without solving for the root $a$.

Combining the preceding methods based on minimal polynomials with
the methods described in Section~\ref{EffectiveIssue} for exact
computation of the signature sequences, we have a complete algorithm
for exact classification of QSIC morphologies.

\noindent {\bf Example 1}: Now we use a running example to show the
procedure of using the signature sequence for QSIC morphology
classification. Consider two quadrics
\begin{eqnarray*}
\mathcal{A}: \;\; 20\,{x}^{2} &-& 12\,xy+48\,xz+76\,x+16\,{y}^{2}\\
& & -16\,yz-12\,y+42\,{z}^{2}+ 72\,z+58 =0, \\
\mathcal{B}: \;\; 28\,{x}^{2} &+& 16\,xy+80\,xz+56\,x+2\,{y}^{2}  \\
& & +24\,yz+20\,y+56\,{z}^{2}+72 \,z+14 =0.
\end{eqnarray*}
The equation of the eigenvalue curve $\mathcal{C}$ is
\begin{eqnarray*}
{u}^{4} &+& \left( -136\,\lambda+100 \right) {u}^{3}+ \left(
-1048-3612\,\lambda+2904 \,{\lambda}^{2} \right) {u}^{2} \\
  &+& \left( -10000\,{\lambda}^{3}+22616\,{\lambda}^{2}+28416 \,\lambda \right) u
\\
 &-& 170528\,{\lambda}^{2}+170528\,{\lambda}^{3}-85264\,{\lambda}^{4}=0.
\end{eqnarray*}
Substituting $u=0$ in this polynomial yields
$$
-85264\,{\lambda}^{4}+170528\,{\lambda}^{3}-170528\,{\lambda}^{2}
=0,
$$
whose only real root is the double root $\lambda=0$. Substituting
$\lambda=-1$ in the equation of $\mathcal{C}$ yields
$$
{u}^{4}+236\,{u}^{3}+5468\,{u}^{2}+4200\,u-426320,
$$
which has one sign change in its coefficients; therefore, by the
Descartes rule, it has only one positive root.
It follows that the signature sequence is $(1, ((1,1)), 3)$.
By Theorem~\ref{thm:$[(11)11]_1$}, the corresponding QSIC is the
union of a real conic and an imaginary one, which is case 17 in
Table 2.

\section{Conclusions}
\label{sec:conclusion}

To summarize, we have obtained the following result:
\begin{theorem}
There are in total 35 different QSIC morphologies with
non-degenerate pencils (see Tables 1, 2 and 3). The morphology of
the QSIC of a pencil $(A,B)$ is entirely classified by its signature
sequence and the degree of its minimal polynomial, using only
rational arithmetic computation.
\end{theorem}

Besides used for determining the QSIC morphology for enhancing
robust computation of QSIC in surface boundary evaluation, another
application of our results is to derive simple algebraic conditions
for interference analysis of quadrics.
For arrangement computation, it is an interesting problem to
classify all possible partitions of $\mathbb{R}^3$ that can be
formed by two ellipsoids.
It is also possible to apply the results here to derive efficient
algebraic conditions for collision detection between various types
of quadric surfaces, such as cones and cylinders, following the
framework in~\citep{Wenping2004}.

One could also use the idea developed here to study the
classification of a pencil of conics in $\mathbb{PR}^2$, which would
lead to a classification of QSIC with degenerate pencils in
$\mathbb{PR}^3$.
A more challenging problem is to use the signature sequence to
classify the intersection of two quadrics in higher dimensions,
$\mathbb{PR}^4$ say. Here the difficult issue is to deduce the
geometry of the QSIC associated with each possible Quadric Quadric Pair Canonical
Form, while it is should be straightforward to obtain the signature 
sequence of the normal form, based on the results presented in the
present paper.

Another direction of investigation would be the classification of
the net of three quadrics in $\mathbb{PR}^n$. In this case, given
three quadratic forms $A$, $B$ and $C$, the question is how to use
the invariants of the planar curve $f(\alpha, \beta, \gamma) \equiv
\det(\alpha A + \beta B + \gamma C)=0$ to characterize the geometric
properties of the net $X^T(\alpha A + \beta B + \gamma C)X=0$ or the
intersection of the quadrics $X^TAX=0$, $X^TBX=0$ and $X^TCX=0$.

%\bibliographystyle{acmtrans}
%\bibliography{paper}

\appendix{
{\bf \large Appendix: Proofs of Theorems 9 Through 22}
\label{sec:appendix}

%%[22]_2
%\begin{theorem} ($[22]_2$: Case 9, Table 1) \label{thm:$[22]_2$} If $f(\lambda)=0$ has two real double roots with the Segre
%characteristic $[22]$, then the only possible index sequences are
%$\langle 2{\wr\wr}_{-} 2{\wr\wr}_{-} 2 \rangle $ and  $\langle
%2{\wr\wr}_{-} 2{\wr\wr}_{+} 2 \rangle $, the QSIC comprises a real
%line and a space cubic curve intersecting at two distinct real
%points for both sequences.
%\end{theorem}

%\begin{proof}
{\bf Proof of Theorem \ref{thm:$[22]_2$}} Suppose that the two real
double roots are $\lambda_0$ and $\lambda_1$.
By setting  $(B-\lambda_1A)$ to be $B$, we transform the root
$\lambda_1$ to 0; the other root is still denoted by $\lambda_0$. By
Theorem~\ref{thm:canonical_form}, the two quadrics have the
canonical forms
\[
A' = (a'_{ij}) =\left(\begin{array}{cccc} 0&1&0&0 \\
1&0&0&0\\
0&0&0&\varepsilon\\
0&0&\varepsilon&0\end{array} \right),\;\;
B' = (b'_{ij}) =\left(\begin{array}{cccc} 0&0&0&0 \\
0&1&0&0\\
0&0&0&\varepsilon\lambda_0\\
0&0&\varepsilon\lambda_0&\varepsilon\end{array}\right),
\]
where $\varepsilon = \pm 1$.
Clearly, Id($A'$) = Id($\infty$) = 2 and the index jumps at both
roots $\lambda_0$ and $\lambda_1$ are 0. Therefore the only possible
index sequence takes the form $\langle 2{\wr\wr} 2{\wr\wr}2
\rangle$, covering the two nonequivalent index sequences $\langle
2{\wr\wr}_{-} 2{\wr\wr}_{-} 2 \rangle $ and  $\langle 2{\wr\wr}_{-}
2{\wr\wr}_{+} 2 \rangle $.
Note that $\langle 2{\wr\wr}_{-} 2{\wr\wr}_{-} 2 \rangle $ is
equivalent to $\langle 2{\wr\wr}_{+} 2{\wr\wr}_{+} 2 \rangle $, and
$\langle 2{\wr\wr}_{-} 2{\wr\wr}_{+} 2 \rangle $ is equivalent to
$\langle 2{\wr\wr}_{+} 2{\wr\wr}_{-} 2 \rangle $.

Swapping $a'_{4,4}$ and $a'_{1,1}$, as well as $b'_{4,4}$ and
$b'_{1,1}$, we obtain
\[
A' =\left(\begin{array}{cccc} 0&0&\varepsilon&0 \\
0&0&0&1\\
\varepsilon&0&0&0\\
0&1&0&0\end{array} \right),\;\;
B' =\left(\begin{array}{cccc} \varepsilon&0&\varepsilon\lambda_0&0 \\
0&1&0&0\\
\varepsilon\lambda_0&0&0&0\\
0&0&0&0\end{array}\right).
\]
Obviously, the QSIC contains the line $x=y=0$. From the
classification of QSIC by Segre characteristic in ${\mathbb
{PC}}^3$, we know that the remaining component is a cubic curve,
whose equation is found to be
\[
{\bf q}(u) = \left(2\varepsilon \lambda_0 u/(\varepsilon+u^2), \;
2\lambda_0 u^2/(\varepsilon+u^2), \; -u,\; 1\;\right)^T.
\]
It is easy to verify that the line and the cubic curve intersect at
two distinct real points $(0,0,0,1)^T$ and $(0,0,1,0)^T$. This
completes the proof of Theorem~\ref{thm:$[22]_2$}.
%\end{proof}

%%[22]_0
%
%\begin{theorem} ($[22]_0$: Case 10, Table 1)\label{thm:$[22]_0$} If $f(\lambda)=0$ has two pairs of identical complex conjugate roots
%with the Segre characteristic $[22]$, then the index sequences are
%$\langle 2 \rangle$ and the QSIC comprises a real line and a space
%cubic curve that do not intersect at any real point.
%\end{theorem}

%\begin{proof}
{\bf Proof of Theorem \ref{thm:$[22]_0$}} Suppose that the two
identical pairs of conjugate roots of $f(\lambda)=0$ are $a\pm bi$.
In this case the only possible index sequence is $\langle 2
\rangle$.
Setting  $(B-aA)/b$ to be $B$, we transform the roots $a\pm bi$ to
$\pm i$.
By Theorem~\ref{thm:canonical_form}, the two quadrics can be
transformed to the following forms,
\[
A' =\left(\begin{array}{cccc} 0&0&0&1\\
0&0&1&0\\
0&1&0&0\\
1&0&0&0\end{array} \right),\;\;
B' =\left(\begin{array}{cccc} 0&0&1&0 \\
0&0&0&-1\\
1&0&0&1\\
0&-1&1&0\end{array}\right).
\]
Clearly, the QSIC contains the line $z=w=0$, and the remaining
component is a cubic curve parameterized by
\[
{\bf q}(u) = \left(-u^2,\; u, \; -u(1+u^2), \; -(1+u^2)\;\right)^T.
\]
Since the last component function $-(1+u^2)$ does not vanish for any
real value of $u$, the space cubic curve has no real intersection
with the line $z=w=0$.
This completes the proof of Theorem~\ref{thm:$[22]_0$}.
%\end{proof}

%
%\subsection{$[31]$: $f(\lambda)=0$ has one real triple root and one real simple root}
%
%\begin{theorem} (Case 11, Table 1) \label{thm:$[31]$} If ${\rm f}(\lambda) = 0$ has one
%triple root and one simple real root with the Segre characteristic
%$[31]$, then the index sequence is $\langle 1 {\wr\wr}{\wr}_{+} 2 |
%3 \rangle$ and the QSIC comprises a closed component with a real
%cusp.
%\end{theorem}

%\begin{proof}
{\bf Proof of Theorem \ref{thm:$[31]$}} According to the discussion
about the index sequences in Section~\ref{sec:index sequences}, the
index jump is 1 across the real root with a $3\times 3$ Jordan
block. Thus it is easy to see the only possible index sequence is
$\langle 1 {\wr\wr}{\wr} 2 | 3 \rangle$.

That the QSIC comprises a closed component with a real cusp follows
from the classification of QSIC in ${\mathbb {PC}}^3$ by the Segre
characteristics. By complex conjugation it is easy to see that the
cusp is real. Since the QSIC is contained in a projective ellipsoid
in the quadric pencil (i.e., with the index being 1 or 3), it is
closed in ${\mathbb {PR}}^3$.
This completes the proof of Theorem~\ref{thm:$[31]$}.
%\end{proof}

%\subsection{$[4]$: $f(\lambda)=0$ has one real quadruple root}
%
%\begin{theorem} (Case 12, Table 1) \label{thm:$[4]$} If $f (\lambda) = 0$
%has one quadruple root with
%the Segre characteristic $[4]$, then the index sequence is $\langle
%2 {\wr\wr}{\wr\wr}_{-} 2 \rangle$ or its equivalent form $\langle 2
%{\wr\wr}{\wr\wr}_{+} 2 \rangle$, and the QSIC comprises a real line
%and a real space cubic curve tangent to each other at a real point
%in this case.
%\end{theorem}

%\begin{proof}
{\bf Proof of Theorem \ref{thm:$[4]$}} According to the discussion
about index sequences in Section~\ref{sec:index sequences}, the
index jump is 0 across the real root with a $4\times 4$ Jordan
block. Thus, the only possible index sequence is $\langle 2
{\wr\wr}{\wr\wr}_{-} 2 \rangle$ or its equivalent form $\langle 2
{\wr\wr}{\wr\wr}_{+} 2 \rangle$.

That the QSIC comprises a line and a space cubic curve tangent to
each other at a point follows from the classification of QSIC in
${\mathbb {PC}}^3$ by the Segre characteristics. By complex
conjugation it is easy to see that the line and the cubic are both
real and their tangent point is also real. This completes the proof
of Theorem~\ref{thm:$[4]$}.
%\end{proof}

%\section{Classification of planar QSIC} \label{sec:planar}
%
%\subsection{$[(11)11]$: $f(\lambda)=0$ has one real double root and two other distinct roots}
%
%
%\begin{theorem} ($[(11)11]_3$) \label{thm:$[(11)11]_3} If f$(\lambda) = 0$ has one double real root and two
%distinct real roots with the Segre characteristic $[(11)11]$, then
%there are only five different possible index sequences and these
%index sequences correspond to four different QSIC morphologies as
%follows:
%\begin {enumerate}
%\item (Case 13, Table 2) $\langle 2 || 2 | 1 | 2 \rangle$ -
%two real closed conics intersecting at two distinct real points;
%\item (Case 14, Table 2) $\langle 1 || 3 | 2 | 3 \rangle$ -
%two real conics not intersecting at any real points;
%\item (Case 15, Table 2) $\langle 1 || 1 | 2 | 3 \rangle$ -
%two imaginary conics intersecting at two distinct real points;
%\item (Case 16, Table 2) $\langle 0 || 2  | 3 | 4\rangle$ or $\langle 1 || 3  | 4 | 3 \rangle$ -
%two imaginary conics not intersecting at any real points.
%\end {enumerate}
%\end{theorem}

%\begin{proof}
{\bf Proof of Theorem \ref{thm:$[(11)11]_3}} Let the double zero be
$\lambda_0$. By setting $B-\lambda_0A$ to $B$, we transform the
double root $\lambda_0$ to 0.
Let $\lambda_1 \neq \lambda_2$ denote the other two roots. Wlog, we
may assume $0<\lambda_1<\lambda_2$.
By Theorem~\ref{thm:canonical_form}, the matrices $A$ and $B$ of two
given quadrics have the following canonical forms,
\[
A' ={\rm diag} (\varepsilon_1,\; \varepsilon_2, \; \varepsilon_3,
\varepsilon_4\;), \;\;\; B' = {\rm diag} (\varepsilon_1\lambda_1, \;
\varepsilon_2\lambda_2, \; 0, \;  0\;).
\]

Now we consider two cases: ($i$) $\det(A')>0$; and ($ii$)
$\det(A')<0$. In case ($i$), the following two subcases need to be
further distinguished: ($i$-$a$) $\varepsilon_1\varepsilon_2>0$ and
$\varepsilon_3\varepsilon_4>0$; and ($i$-$b$)
$\varepsilon_1\varepsilon_2<0$ and $\varepsilon_3\varepsilon_4<0$.

In subcase ($i$-$a$), since $\varepsilon_1\varepsilon_2>0$,
${\mathcal B}'$ consists of a pair of complex conjugate planes; thus
${\mathcal B}'$ intersects ${\mathcal A}'$ in two imaginary conics.
Since $\varepsilon_3\varepsilon_4>0$, the index jump of
Id($\lambda$) at $\lambda_0=0$ is $\pm 2$.
Since $\det(A')>0$,  ${\rm Id}(-\infty)$=${\rm index}(-A')$ = 0, 2,
or 4.
Therefore, all possible index sequences are $\langle
2||4|3|2\rangle$, $\langle 2||0|1|2\rangle$, $\langle
0||2|3|4\rangle$, or $\langle 4||2|1|0\rangle$.
Clearly, these sequences are equivalent; so we use  $\langle
0||2|3|4\rangle$ as the representative.
%\
Since there is a virtual quadric (i.e., one whose index is 0 or 4),
the QSIC has no real point.
This completes the first part of item 4 of the theorem.

In subcase ($i$-$b$), since $\varepsilon_3\varepsilon_4<0$, the
index jump of Id($\lambda$) at $\lambda_0=0$ is 0.
Since $\det(A')>0$, ${\rm Id}(-\infty)$= ${\rm index}(-A')$ = 0, 2,
or 4.
Thus, by a similar argument to case ($i$-$a$), the only possible
index sequence is $\langle 2||2|1|2\rangle$.
By swapping $\varepsilon_3$ and $\varepsilon_4$ in the matrix $A'$
if necessary, we may suppose that the quadric ${\mathcal A'}:
X^TA'X=0$ is a one-sheet hyperboloid with the $y$-axis being its
symmetric axis, i.e., $x^2 - y^2 + z^2 -w^2 =0$.
Recall that $0<\lambda_1<\lambda_2$. The two planes $y=\pm
(\lambda_1/\lambda_2)^{1/2}x$ given by the quadric ${\mathcal B'}:
X^TB'X=0$ intersect ${\mathcal A'}$ in two ellipses intersecting at
two real points, as shown in Figures~\ref{fig6-1} and~\ref{fig6-2}.
This completes the proof of item 1.

\begin{figure}[htbp]
\begin{minipage}[t]{7.0cm}
\centering
\includegraphics[width=1.6in]{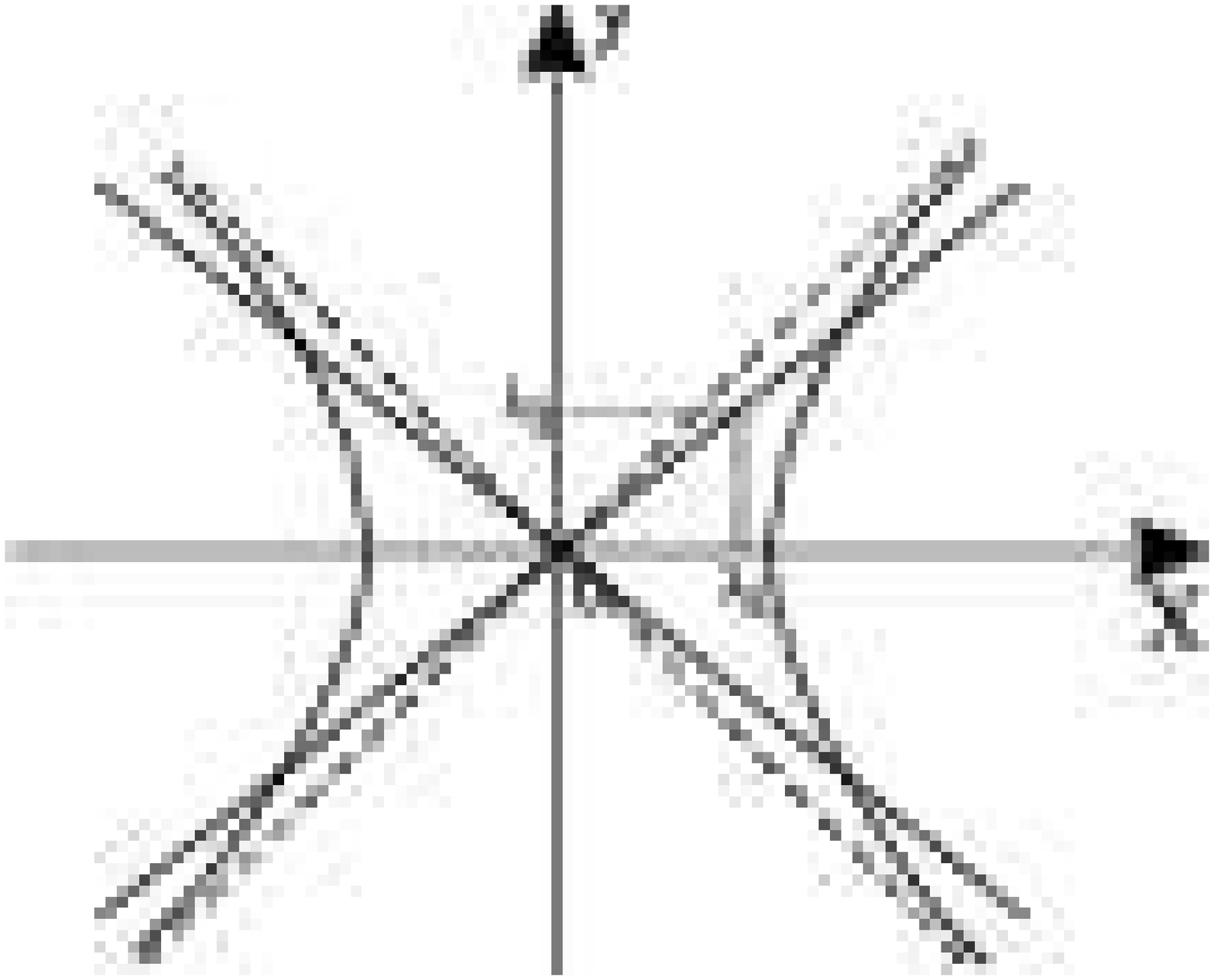}
\footnotesize \caption{Cross section of a one-sheet hyperboloid and
a pair of planes.} \label{fig6-1}
\end{minipage}
\hspace{0.2cm}
\begin{minipage}[t]{7.0cm}
\centering
\includegraphics[width=1.6in]{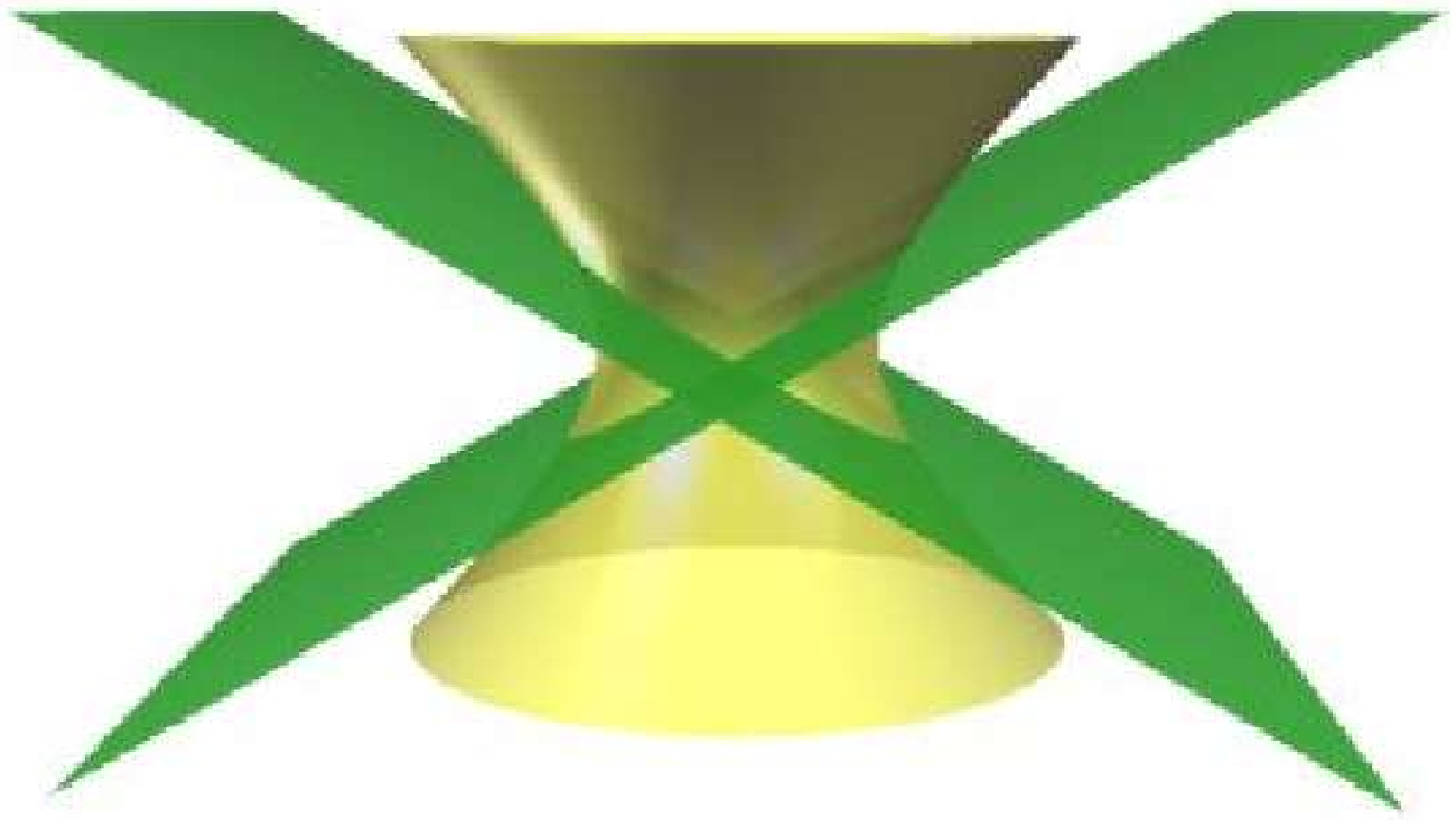}
\footnotesize \caption{A QSIC comprising two closed conics
 intersecting at two real points.} \label{fig6-2}
\end{minipage}
\end{figure}

Below we consider case ($ii$) of $\det(A')<0$.
Again we need to distinguish two subcases: ($ii$-$a$)
$\varepsilon_1\varepsilon_2<0$ and $\varepsilon_3\varepsilon_4>0$;
and ($ii$-$b$) $\varepsilon_1\varepsilon_2>0$ and
$\varepsilon_3\varepsilon_4<0$.

In subcase ($ii$-$a$), since $\varepsilon_3\varepsilon_4>0$, the
index jump of Id($\lambda$) at $\lambda_0=0$ is $\pm 2$.
Since $\det(A')<0$, ${\rm Id}(-\infty)$= ${\rm index}(-A')$ = 1 or
3.
Therefore, the only possible index sequences are $\langle
1||3|2|3\rangle$ and $\langle 1||3|4|3\rangle$.
Recall that $\varepsilon_1\varepsilon_2<0$.
It is then easy to verify that the index sequence is $\langle
1||3|2|3\rangle$ if $\varepsilon_1=-1$ and $\varepsilon_2=1$, and is
$\langle 1||3|4|3\rangle$ if $\varepsilon_1 =1$ and
$\varepsilon_2=-1$.

When $\varepsilon_1=-1$ and $\varepsilon_2=1$ (i.e., when the index
sequence is $\langle 1||3|2|3\rangle$),
by a projective transformation, we may transform ${\mathcal A'}$ to
the unit sphere $x^2 + y^2 + z^2 - w^2=0$ and ${\mathcal B'}$ to two
parallel planes $z=\pm (\lambda_1/\lambda_2)^{1/2}w$.
Since $0 < \lambda_1<\lambda_2$, the QSIC consists of two real
conics not intersecting each other in real points. This completes
the proof of item 2.

When $\varepsilon_1 =1$ and $\varepsilon_2=-1$ (i.e., when the index
sequence is $\langle 1||3|4|3\rangle$), there is a virtual quadric
in the pencil. Therefore the QSIC has no real points. Hence, the
QSIC consists of two imaginary conics not having common real points.
This proves the second part of item 4.

In subcase ($ii$-$b$), using a similar argument, we know that the
only possible index sequence is $\langle 1||1|2|3\rangle$.
The quadric ${\mathcal A'}$ is either two-sheet hyperboloid with the
$z$-axis being its centered axis (if $\varepsilon_1$,
$\varepsilon_2$, and $\varepsilon_4$ have the same sign ) or the
unit sphere centered at the origin (if $\varepsilon_1$,
$\varepsilon_2$ and $\varepsilon_3$ have the same sign).
The quadric ${\mathcal B'}$ comprises of a pair of imaginary
conjugate planes intersecting in a real line --- the $z$-axis, which
intersects ${\mathcal A'}$ at two real points.
Hence, the QSIC consists of two complex conjugate conics
intersecting at two real points. This completes the proof of item 3.
Hence, Theorem~\ref{thm:$[(11)11]_3} is proved.
%\end{proof}

%%[(11)11]_1
%\begin{theorem} ($[(11)11]_1$) \label{thm:$[(11)11]_1$}
%If $f(\lambda) = 0$ has a real double root $\lambda_0$ and a pair of
%complex conjugate roots with the Segre characteristic $[(11)11]$,
%then the possible index sequences of the pencil $\lambda A - B$ are
%$\langle 1 || 3\rangle$ and $\langle 2 || 2\rangle$. Furthermore,
%\begin {enumerate}
%\item (Case 17, Table 2) when the index sequence is $\langle 1 || 3\rangle$, the QSIC comprises of two conics, one real and one imaginary;
%\item (Case 18, Table 2) when the index sequence is $\langle 2 || 2\rangle$, the QSIC comprises of two real
%conics which cannot both be ellipses simultaneously in any affine
%realization of ${\mathbb {PR}}^3$.
%\end{enumerate}
%\end{theorem}

%\begin{proof}
{\bf Proof of Theorem \ref{thm:$[(11)11]_1$}} Wlog, by setting
$B-\lambda_0A$ to be $B$, we may assume the real double root
$\lambda_0$ to be 0. Let the other two roots be $a \pm bi$, $b\neq
0$. By Theorem~\ref{thm:canonical_form}, the two matrices $A$ and
$B$ have the following canonical forms
\[
A' ={\rm diag}\left(\varepsilon_1,\;\varepsilon_2,\;\left(\begin{array}{cc} 0&1\\
1&0\end{array}\right)\;\right),\;\;
B' ={\rm diag} \left(0,\;0,\;\left(\begin{array}{cc} -b&a\\
a&b\end{array}\right)\;\right).
\]
In the following we consider two subcases: ($i$) $a \neq 0$; and
($ii$) $a=0$.

In case ($i$) of $a \neq 0$, by setting  $aA'-B'$ to be $A'$, we
obtain
\[
A' ={\rm diag}\left(\; \varepsilon_1a,\; \varepsilon_2a,\;
\left(\begin{array}{cc} b&0\\0&-b\end{array} \right)\;\right),
\]
which is the quadric,
\[
\frac{\varepsilon_1a}{b}x^2 +\frac{\varepsilon_2a}{b}y^2+z^2-w^2=0.
\]
The quadric ${\mathcal B'}$ consists of the following two planes
\[
z=\left(\frac{a\pm\sqrt{a^2+b^2}}{b}\right) w.
\]
When $\varepsilon_1$ and $\varepsilon_2$ have the same sign, the
index sequence is $\langle 1||3 \rangle$ (or its equivalent form
$\langle 3||1 \rangle$). In this case, the quadric ${\mathcal A'}$
is either an ellipsoid or a two-sheet hyperboloid with two of its
tangent planes being $z \pm w =0$.
The quadric ${\mathcal B'}$ comprises two parallel planes
perpendicular to $z$-axis. Wlog, we assume that $b>0$. Then it is
easy to verify that
\[
\frac{a+\sqrt{a^2+b^2}}{b}>1\;\;{\rm
and}\;\;-1<\frac{a-\sqrt{a^2+b^2}}{b}<0\;\;{\rm if}\;\; a>0,
\]
or
\[
0<\frac{a+\sqrt{a^2+b^2}}{b}<1\;\; {\rm and}\;\;
\frac{a-\sqrt{a^2+b^2}}{b}<-1\;\;{\rm if}\;\;a<0.
\]
It follows that one of the planes of ${\mathcal B'}$ intersects
${\mathcal A'}$ in an ellipse and the other plane does not intersect
${\mathcal A'}$ at any real point, as shown by the two cases in
Figure~\ref{fig6-4}.

\begin{figure}[!htbp]
\begin{center}
\includegraphics[width = 1.5in]{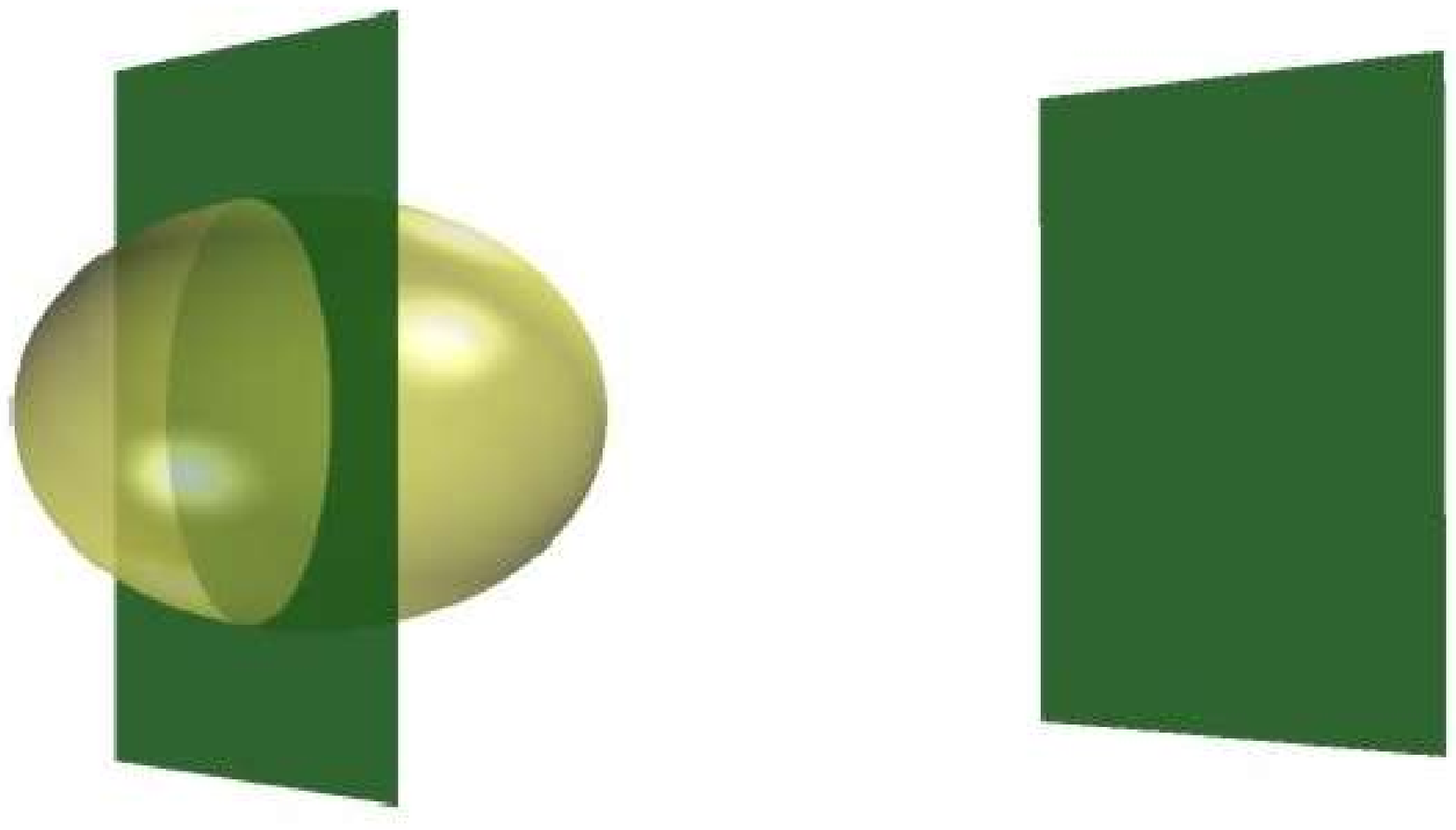}
\hspace{0.5in}
\includegraphics[width = 1.5in]{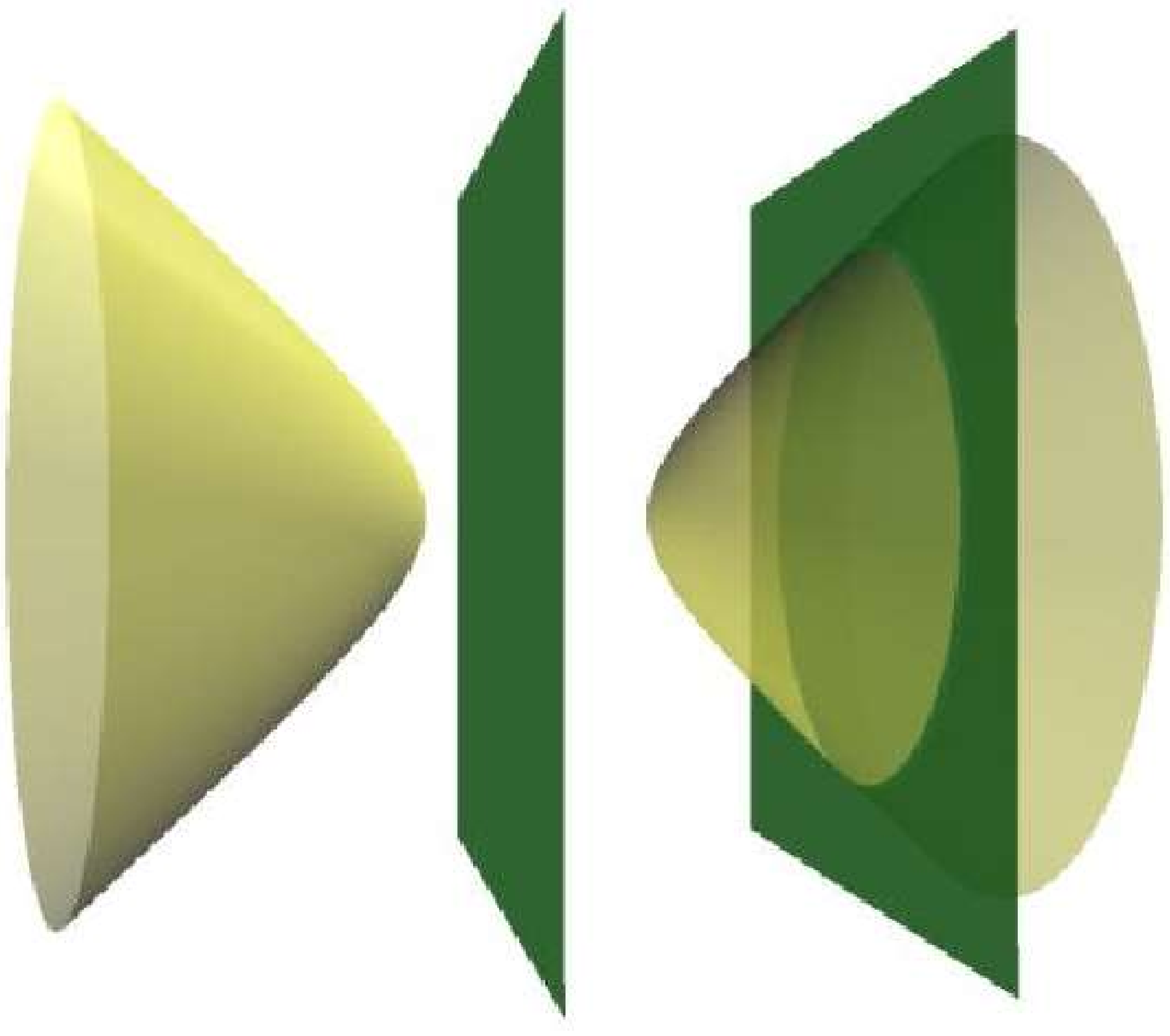}
\caption{The two cases in the proof of
Theorem~\ref{thm:$[(11)11]_1$} where the QSIC has one real conic and
 one imaginary conic.}\label{fig6-4}
\end{center}
\end{figure}

When $\varepsilon_1$ and $\varepsilon_2$ have opposite signs, the
index sequence is $\langle 2||2 \rangle$.
In this case the quadric ${\mathcal A'}$ is a one-sheet hyperboloid
and ${\mathcal B'}$ comprises two planes intersecting ${\mathcal
A'}$ in an ellipse and a hyperbola in the affine realization shown
in Figure~\ref{fig6-6}.
Since the ellipse intersects the hyperbola at its two branches, any
real plane in $\mathbb{PR}^3$ intersects at least one of the two
conics.

\begin{figure}[!htbp]
\begin{center}
\includegraphics[width = 1.5in]{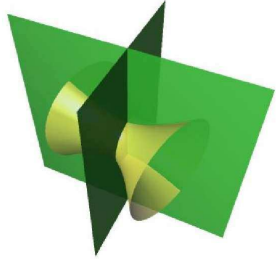}
\caption{The case of QSIC having two real conics intersecting in two
real points.} \label{fig6-6}
\end{center}
\end{figure}

Now consider case ($ii$) of $a = 0$.
We transform the matrices $A'$ and $B'$ into the following forms,
\[
A' ={\rm diag} \left(\;\left(\begin{array}{cc} 0&1\\
1&0 \end{array} \right), \; \varepsilon_1, \;
\varepsilon_2\;\right),\;\; {\rm and}\;\; B' = {\rm diag} (-b, \; b,
\; 0, \; 0\;).
\]
The quadric ${\mathcal B'}$ comprises of two planes.
When $\varepsilon_1\varepsilon_2>0$, the only possible index
sequence is $\langle 1||3 \rangle$ (or its equivalent form $\langle
3||1 \rangle$), the quadric ${\mathcal A'}$ is a two-sheet
hyperboloid, and the QSIC has one real and one imaginary conic, as
shown in Figure \ref{fig6-7}.

When $\varepsilon_1\varepsilon_2<0$, the only possible index
sequence is $\langle 2||2 \rangle$, the quadric ${\mathcal A'}$ is a
one-sheet hyperboloid, and the QSIC has two real conics intersecting
in two real points as shown in Figure \ref{fig6-8}.
Again, as in case ($i$) where the index sequence is $\langle 2||2
\rangle$, the two conic components of the QSIC cannot be both
ellipses in any affine realization of ${\mathbb {PR}}^3$.  This
completes the proof of Theorem~\ref{thm:$[(11)11]_1$}.
%\end{proof}

\begin{figure}[htbp]
\begin{minipage}[t]{7.0cm}
\centering
\includegraphics[width=1.6in]{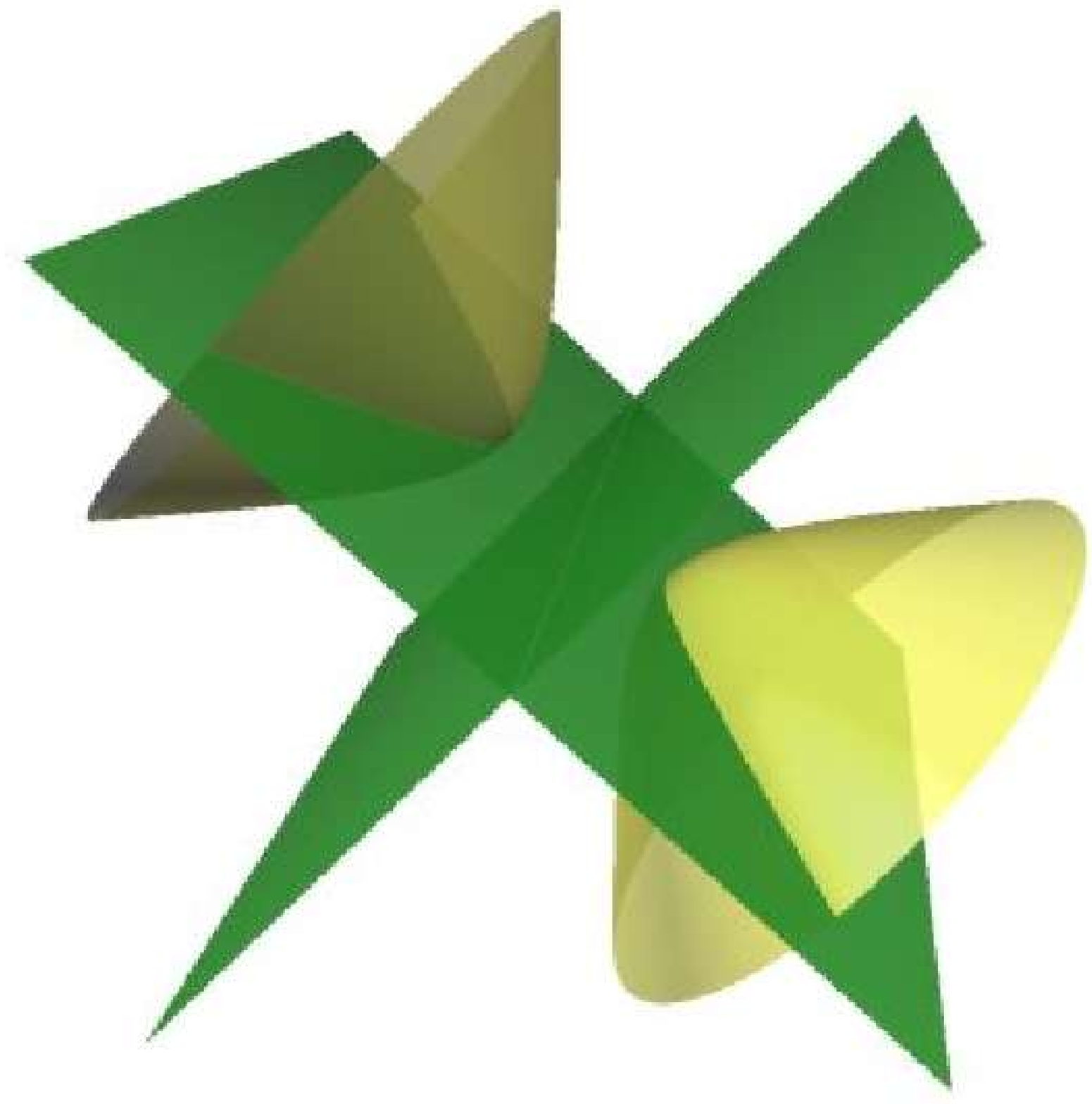}
\footnotesize \caption{The case of the QSIC having two conics, one
real and one imaginary.} \label{fig6-7}
\end{minipage}
\hspace{0.2cm}
\begin{minipage}[t]{7.0cm}
\centering
\includegraphics[width=1.6in]{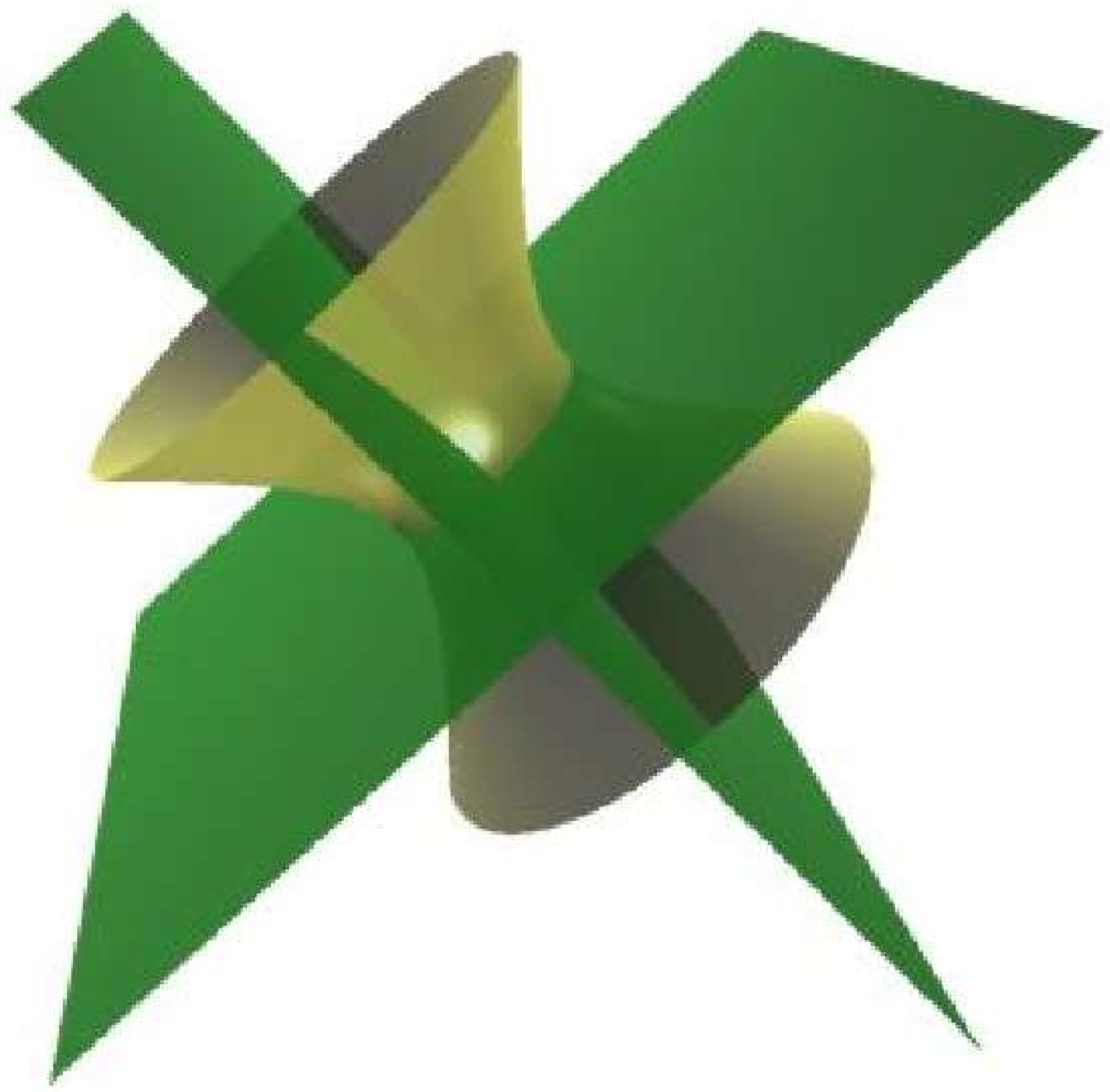}
\footnotesize \caption{The case of the QSIC having two real conics
intersecting in two real points.} \label{fig6-8}
\end{minipage}
\end{figure}

%\subsection{$[(111)1]_2$: $f(\lambda)=0$ has one real triple root and a real simple root}
%
%
%\begin{theorem} ($[(111)1]_2$) \label{thm:$[(111)1]_2$}
%If ${\rm f}(\lambda)=0$ has one triple root and a simple root with
%the Segre characteristic $[(111)1]$, then the only possible index
%sequences are $\langle 1 ||| 2 |3\rangle$ and $\langle 0||| 3 |4
%\rangle $.
%%
%Furthermore,
%\begin{enumerate}
%\item (Case 19, Table 2) when the index sequence is $\langle 1
%||| 2 |3\rangle$, the QSIC is a real conic counted twice;
%\item (Case 20, Table 2) when the index sequence is $\langle 0||| 3 |4 \rangle $,
%the QSIC is an imaginary conic counted twice.
%\end{enumerate}
%\end{theorem}

%\begin{proof}
{\bf Proof of Theorem \ref{thm:$[(111)1]_2$}} Clearly, all the roots
of $f(\lambda)=0$ are necessarily real in this case. Wlog, we may
assume the triple root $\lambda_0$ to be zero. Then, by
Theorem~\ref{thm:canonical_form}, the canonical matrix forms of the
given quadrics are
\[
A' ={\rm diag}(\varepsilon_1,\; \varepsilon_2,\; \varepsilon_3,\;
\varepsilon_4\;),\;\;\;\;
B' = {\rm diag}(\varepsilon_1\lambda_1,\;0,\; 0,\;0\;).
\]
Setting  $\lambda_1A'-B'$ to be $A'$ and then $\varepsilon_1 B'$ to
be $B'$, we obtain
\[
A' = {\rm diag} (0,\; \varepsilon_2\lambda_1,\;
\varepsilon_3\lambda_1,\; \varepsilon_4\lambda_1\;), \;\;
B' = {\rm diag} (\lambda_1,\; 0,\; 0,\; 0\;).
\]
The quadric ${\mathcal A'}$ is a cylinder (real or imaginary) with
the $x$-axis being its central axis, while the quadric ${\mathcal
B'}$ is a plane counted twice.
Thus the QSIC is a conic counted twice, real or imaginary.

When $\varepsilon_2$, $\varepsilon_3$ and $\varepsilon_4$ have
different signs, the only possible index sequence is $\langle
1|||2|3 \rangle$ (or its equivalent form $\langle 3|||2|1 \rangle$).
In this case ${\mathcal A'}$ is real, and the QSIC is a real conic
counted twice.
When $\varepsilon_2$, $\varepsilon_3$ ad $\varepsilon_4$ have the
same sign, the only possible index sequence is $\langle 0|||3|4
\rangle$ (or its equivalent form $\langle 3|||0|1 \rangle$). In this
case ${\mathcal A'}$ is imaginary, and the QSIC is an imaginary
conic counted twice. This completes the proof of
Theorem~\ref{thm:$[(111)1]_2$}.
%\end{proof}

%\subsection{$[(21)1]_2$: $f(\lambda)=0$ has one real triple root and a real simple root}
%
%\begin{theorem} ($[(21)1]_2$) \label{thm:$[(21)1]_2$}
%If ${\rm f}(\lambda)=0$ has one triple root and a simple root with
%the Segre characteristic $[(21)1]$, then the only possible index
%sequences are $\langle 1{\wr\wr}_{-}|2|3 \rangle$ and $\langle
%1{\wr\wr}_{+}|2|3 \rangle$.
%%
%Furthermore,
%\begin{enumerate}
%\item (Case 21, Table 2) when the index sequence is $\langle 1{\wr\wr}_{-}|2|3 \rangle$, the
%QSIC  comprises two real conics tangent to each other at one real
%point;
%\item (Case 22, Table 2) when the index sequence is $\langle
%1{\wr\wr}_{+}|2|3 \rangle$, the QSIC comprises two imaginary conics
%tangent to each other at one real point.
%\end{enumerate}
%\end{theorem}

%\begin{proof}
{\bf Proof of Theorem \ref{thm:$[(21)1]_2$}} The two roots of
$f(\lambda)=0$ are necessarily real. Let $\lambda_0$ denote the
triple root and $\lambda_1$ denote the simple root. Wlog, we assume
$ \lambda_0 =0$ and $\lambda_1>0$.
By Theorem~\ref{thm:canonical_form}, the canonical matrices of the
two given quadric are
\[
A' =\left(\begin{array}{cccc}
0&\varepsilon_1&\\
\varepsilon_1&0\\
& & \varepsilon_2&\\
&&&\varepsilon_3\end{array} \right),\;\;
B' =\left(\begin{array}{cccc} 0&&&\\
&\varepsilon_1&&\\
&&0&\\
&&& \varepsilon_3\lambda_1 \\
\end{array}\right).
\]
Since the index jump at a root of Jordan block of size $2\times 2$
is 0, the index jump at the triple root is $\pm 1$. It is then easy
to see that the only possible index sequence is $\langle
1{\wr\wr}|2|3 \rangle$ (or its equivalents). Thus, we have
$\varepsilon_{2}=1$,  $\varepsilon_{3}=1$.

The quadric ${\mathcal A'}$, as $2\,\varepsilon_{1}\, y + z^{2} +
w^{2}=0$, passes through the origin $(1,0,0,0)^{T}$ and its tangent
plane at this point is $y=0$.
The quadric ${\mathcal B'}$: $\varepsilon_1\,y^2 + \lambda_{1}\,
w^2=0$, comprises two planes intersecting at the line $(y=0, w=0)$,
which touches the quadric ${\mathcal A'}$ at the point
$(1,0,0,0)^T$.
Since the tangent plane $y=0$ of $\mathcal{A}$ at  $(1,0,0,0)^T$ is
different from either of these two planes, the QSIC comprises two
conics that are tangent to each other at the real point
$(1,0,0,0)^T$.

If the $\varepsilon_1 = 1$, ${\mathcal B'}$ consists of two
imaginary planes, so the QSIC consists of two imaginary conics. In
this case, the number of positive eigenvalues of $-B'$ is 0, and the
index sequence therefore becomes $\langle 1{\wr\wr}_{+}|2|3
\rangle$.
If $\varepsilon_1 = -1$, ${\mathcal B'}$ consists of two real
planes, so the QSIC consists of two real conics. In this case, the
number of positive eigenvalues of $-B'$ is 1, and the index sequence
therefore becomes $\langle 1{\wr\wr}_{-} |2|3 \rangle$. This
completes the proof of Theorem~\ref{thm:$[(21)1]_2$}.
%\end{proof}

%\subsection {$[2(11)]_2$: $f(\lambda)=0$ has two real double roots}
%
%\begin{theorem}\label{thm:$[2(11)]_2$}
%If $f (\lambda)=0$ has two double roots with the Segre
%characteristic $[2(11)]$, then the only possible index sequences are
%$\langle 2{\wr\wr}_{-}2||2 \rangle$ (or its equivalent form $\langle
%2{\wr\wr}_{+}2||2 \rangle$), $\langle 1{\wr\wr}_{+} 1 || 3 \rangle$
%and $\langle 1{\wr\wr}_{-} 1 || 3\rangle$.
%%
%Furthermore,
%\begin{enumerate}
%\item (Case 23, Table 3) when the index sequence is $\langle 2{\wr\wr}_{-}2||2 \rangle$,
%the QSIC consists of a real conic and two real lines which intersect
%pairwise at three distinct real points;
%%
%\item (Case 24, Table 3) when the index sequence is $\langle 1{\wr\wr}_{-} 1 || 3
%\rangle$, the QSIC consists of a real conic and a pair of complex
%conjugate lines. The conic and the pair of lines do not intersect;
%%
%\item (Case 25, Table 3) when the index sequence is $\langle 1{\wr\wr}_{+} 1 || 3\rangle$,
%the QSIC consists of an imaginary conic and a pair of complex
%conjugate lines. The conic and the pair of lines do not intersect.
%\end{enumerate}
%\end{theorem}

%\begin{proof}
{\bf Proof of Theorem \ref{thm:$[2(11)]_2$}} The two double roots of
$f(\lambda)=0$ are necessarily real. Up to equivalence relations,
the only possibilities for the index sequences are $\<2 {\wr\wr}_{-}
2 || 2\>$, $\<1 {\wr\wr}_{-} 1 || 3\>$ and $\<1 {\wr\wr}_{+} 1 ||
3\>$.
Note that $\<2 {\wr\wr}_{-} 2 || 2\>$ is equivalent to $\<2
{\wr\wr}_{+} 2 || 2\>$.
Let $\lambda_1$ denote the root associated with the $2\times 2$
Jordan block, and let $\lambda_2$ denote the other root. By setting
$B-\lambda_1\, A$ to $B$, we may assume $\lambda_1 =0$. Let
$\lambda_2$ denote the other double root; wlog, assume
$\lambda_{2}>0$.

 Then the canonical forms of the two quadrics are
\begin {equation}  \label {Eqn2(11)}
A' ={\rm diag}\left(\;\left(\begin{array}{cc} 0&\varepsilon_1\\
\varepsilon_1&0 \end{array} \right),\; \varepsilon_2,\;
\varepsilon_3\;\right),\;\;
B' ={\rm diag}\left(\left(\begin{array}{cc} 0&0 \\
0&\varepsilon_1\end{array}\right),\; \varepsilon_2\lambda_2,\;
\varepsilon_3\lambda_2\;\right).
\end {equation}
Due to the sign change rules (see Section \ref{signvariation}), for
the index sequence $\<2 {\wr\wr}_{-} 2 || 2\>$, we have
$\varepsilon_2\varepsilon_3<0$. For the index sequences $\<1
{\wr\wr}_{-} 1 || 3\>$ and $\<1 {\wr\wr}_{+} 1 || 3\>$, we have
$\varepsilon_2=\varepsilon_3=1$. Below we discuss these two cases:
($i$) $\varepsilon_2\varepsilon_3<0$; and ($ii$)
$\varepsilon_2=\varepsilon_3=1$.

In case ($i$) of $\varepsilon_2\varepsilon_3<0$, setting
$\varepsilon_1\lambda_2A'-\varepsilon_1B'$ to $\bar A$ and setting
$B'$ to $\bar B$, we obtain
\[
\bar A ={\rm diag} \left( \;\left( \begin{array}{cc} 0&\lambda_2\\
\lambda_2&-1\end{array} \right),\; 0,\; 0\;\right),\;\;
\bar B = {\rm diag} \left(\;\left(\begin{array}{cccc} 0&0 \\
0& \varepsilon_1 \end{array}\right),\; \varepsilon_2\lambda_2, \;
\varepsilon_3\lambda_2\; \right),
\]
Then the quadric $\bar {\mathcal A}$ consists of two planes $y=0$
and $y=2\lambda_2x$. The quadric $\bar {\mathcal B}$ is a real
cylinder with the $x$-axis as its central axis.
Clearly, $\bar {\mathcal B}$ intersects $y=0$ in two real lines,
denoted by $\ell_1$ and $\ell_2$, on the planes $z = \pm 1 $. These
two lines intersect at the point $(1,0,0,0)^T$.
The quadric $\bar {\mathcal B}$ also intersects the plane
$y=2\lambda_2x$ in a real conic, which intersects the two lines
$\ell_{1,2}$ in two distinct real points.
This completes the proof of item 1.

In case ($ii$) of $\varepsilon_2=\varepsilon_3=1$, ${\rm Id}(A')=3$,
and the index jump at the root $\lambda_2$ is $2$.  Below we
distinguish two subcases: ($ii$-$a$) $\varepsilon_1<0$; and
($ii$-$b$) $\varepsilon_1>0$.

In subcase ($ii$-$a$) of $\varepsilon_1<0$, the index sequence is
$\langle 1{\wr\wr}_{-} 1||3 \rangle$. The quadric ${\mathcal B'}$ is
a real hyperbolic cylinder with $x$-axis as central axis and
symmetric with the plane $y=0$. Therefore it intersects the plane
$y=0$ in two imaginary lines and intersects the plane
$y=2\lambda_2x$ in a real conic. The two lines intersect at the
point $(1,0,0,0)^T$. This completes the proof of item 2.

In subcase ($ii$-$b$) of $\varepsilon_1>0$, the index sequence is
$\langle 1{\wr\wr}_{+} 1||3 \rangle$.  The quadric ${\mathcal B'}$
is an imaginary cylinder; thus it intersects the quadric ${\mathcal
A'}$ in an imaginary conic and two complex conjugate lines; these
two lines intersect at the real point $(1,0,0,0)^T$. This completes
the proof of item 3, and hence, Theorem~\ref{thm:$[2(11)]_2$}.
%\end{proof}

%\subsection{$[(31)]_1$: $f(\lambda)=0$ has one real quadruple root}
%
%\begin{theorem}\label{thm:$[(31)]_2$}
%If $f(\lambda)=0$ has one quadruple root with the Segre
%characteristic $[(31)]$, then the only possible index sequences are
%$\langle 2 {\wr\wr}{\wr}_{-}| 2 \rangle$ and $\langle 1
%{\wr\wr}{\wr}_{+}| 3 \rangle $.
%%
%Furthermore,
%\begin{enumerate}
%\item (Case 26, Table 3) when the index sequence is $\langle 2 {\wr\wr}{\wr}_{-}| 2 \rangle$,
%the QSIC consists of a real conic and two real lines, and these
%three components intersect at a common real point;
%%
%\item (Case 27, Table 3) when the index sequence is $\langle 1 {\wr\wr}{\wr}_{+}| 3
%\rangle $, the QSIC consists of a real conic and a pair of complex
%conjugate lines, and these three components intersect at a common
%real point.
%\end{enumerate}
%\end{theorem}

%\begin{proof}
{\bf Proof of Theorem \ref{thm:$[(31)]_2$}} Wlog, we may assume the
quadruple root, denoted by $\lambda_1$, to be 0. Then, by
Theorem~\ref{thm:canonical_form}, the canonical form of the two
quadrics are
\begin {equation}
A' =\left(\begin{array}{cccc} 0&&\varepsilon_1&\\
&\varepsilon_1&&\\
\varepsilon_1&&0&\\
&&&\varepsilon_2\end{array} \right),\;\;
B' =\left(\begin{array}{cccc} 0&&& \\
&0&\varepsilon_1&\\
&\varepsilon_1&0&\\
&&&0\end{array}\right). \label {Eqn(31)}
\end {equation}
The quadric ${\mathcal B'}$ comprises two planes $y=0$ and $z=0$.
The plane $y=0$ intersects the quadric ${\mathcal A'}$ along a real
conic $xz=-{\varepsilon_2/\varepsilon_1w^2}$, denoted as ${\mathcal
C}$.
The plane $z=0$ intersects ${\mathcal A'}$ in two lines, defined by
the intersection of $z=0$ and $y^2 =
-{\varepsilon_2/\varepsilon_1}w^2$.
If ${\varepsilon_2/\varepsilon_1}=-1$, the two lines are real; if
${\varepsilon_2/\varepsilon_1}=1$, the two lines are imaginary.
In both cases, the two lines intersect the conic ${\mathcal C}$ at
the real point $(1,0,0,0)^T$.

On the other hand, when ${\varepsilon_2/\varepsilon_1}=-1$, we have
index($A'$)= 2 and the index jump of ${\rm Id}(\lambda A- B)$ at the
root $\lambda_1=0$ is 0. Therefore, the index sequence is $\langle
2{\wr\wr}{\wr}_{-}|2 \rangle$ or its equivalent form $\langle
2{\wr\wr}{\wr}_{+}|2 \rangle$. When
${\varepsilon_2/\varepsilon_1}=1$, we have $\mathrm{index}(A')= 1$
or $3$, and the index jump of ${\rm Id}(\lambda A- B)$ at
$\lambda_1=0$ is $\pm 2$. Therefore, the index sequence of the
pencil is $\langle 1 {\wr\wr}{\wr}| 3 \rangle$.
This completes the proof of Theorem~\ref{thm:$[(31)]_2$}.
%\end{proof}

%\subsection {$[(11)(11)]$: $f(\lambda)=0$ has two double roots}
%
%\begin{theorem} ($[(11)(11)]_2$) \label{thm:$[(11)(11)]_2$}
%If $f(\lambda)=0$ has two real double roots with the Segre
%characteristic $[(11)(11)]$, then the only possible index sequences
%are $\langle 2||2||2 \rangle $, $\langle 0||2||4 \rangle$ and $
%\langle 1||1||3 \rangle $.
%%
%Furthermore,
%\begin{enumerate}
%\item (Case 28, Table 3) when the index sequence is $\langle 2||2||2 \rangle
%$,
% the QSIC consists of four real lines, and these four lines
%form a quadrangle in $\mathbb{PR}^3$;
%%
%\item (Case 29, Table 3) when the index sequence is $\langle 0||2||4 \rangle$, the QSIC consists
%of four imaginary lines and has no real point;
%%
%\item (Case 30, Table 3) when the index sequence is $ \langle 1||1||3 \rangle $, the QSIC consists
%of two pair of complex conjugate lines, with each pair intersecting
%at a real point.
%\end{enumerate}
%\end{theorem}

%\begin{proof}
{\bf Proof of Theorem \ref{thm:$[(11)(11)]_2$}} Let the two roots be
$\lambda_1$ and $\lambda_2$. By setting $B'= B-\lambda_1A$, we may
assume $\lambda_1=0$. By Theorem~\ref{thm:canonical_form}, the
canonical form of the two quadrics are
\begin {equation} \label{Eqn(11)(11)F}
A' ={\rm diag} (\varepsilon_1,\; \varepsilon_2,\; \varepsilon_3,\;
\varepsilon_4 ),\;\;
B' = {\rm diag} (0,\; 0, \; \varepsilon_3\lambda_2,\;
\varepsilon_4\lambda_2\;).
\end {equation}
Setting $\lambda_2A'-B'$ to be $A'$, we obtain
\begin {equation} \label{Eqn(11)(11)S}
A' ={\rm diag} (\varepsilon_1\lambda_2,\; \varepsilon_2\lambda_2,\;
0\; 0),\;\;
B' = {\rm diag} (0,\; 0, \; \varepsilon_3\lambda_2,\;
\varepsilon_4\lambda_2\;).
\end {equation}

We consider the following three cases: ($i$)
$\varepsilon_1\varepsilon_2<0$ and $\varepsilon_3\varepsilon_4<0$;
($ii$) $\varepsilon_1\varepsilon_2>0$ and
$\varepsilon_3\varepsilon_4>0$; and ($iii$)
$(\varepsilon_1\varepsilon_2)(\varepsilon_3\varepsilon_4) < 0$.

In case ($i$), the index jumps at $\lambda_1$ and $\lambda_2$ are
both $0$. Hence, the only possible index sequence is $\langle
2||2||2 \rangle$. In this case, each of ${\mathcal A'}$ and
${\mathcal B'}$ consists of a pair of real planes intersecting at a
real line.
Since the two real lines on ${\mathcal A'}$ and ${\mathcal B'}$ do
not intersect, the QSIC consists of four real lines forming a
quadrangle. This quadrangle can be obtained from a tetrahedron
(defined by the four planes of ${\mathcal A'}$ and ${\mathcal B'}$)
by removing two of the six sides; the two removed sides are the
intersecting line of the plane pair ${\mathcal A'}$ and  the
intersecting line of the plane pair ${\mathcal B'}$. This completes
the proof of item 1.

In case ($ii$), the index jumps at $\lambda_1$ and $\lambda_2$ are
both $\pm 2$. Hence, the only possible index sequence is $\langle
0||2||4 \rangle$. In this case, each of ${\mathcal A'}$ and
${\mathcal B'}$ consists of a pair of complex conjugate planes
intersecting at a real line.
Since the two real lines on ${\mathcal A'}$ and ${\mathcal B'}$ do
not intersect, the QSIC consists of four imaginary lines and has no
real point. This completes the proof of item 2.

In case ($iii$), either $\varepsilon_1\varepsilon_2>0$ and
$\varepsilon_3\varepsilon_4 < 0$ or $\varepsilon_1\varepsilon_2<0$
and $\varepsilon_3\varepsilon_4 >0 $.
Thus the only possible index sequence is $\langle 1||1||3 \rangle$
or its equivalent forms.
At the same time, one of ${\mathcal A'}$ and ${\mathcal B'}$ is a
pair of real planes and the other is pair of conjugate planes.
Therefore, the QSIC consists two pairs of conjugate lines with each
pair intersecting at a real point. This completes item 3, and hence,
Theorem~\ref{thm:$[(11)(11)]_2$}.
%\end{proof}

%\begin{theorem} ($[(11)(11)]_0$: Case 31, Table 3) \label{thm:$[(11)(11)]_0$}
%If $f(\lambda)=0$ has two identical pairs of complex conjugate roots
%with the Segre characteristic $[(11)(11)]$, the only possible index
%sequences are $\langle 2 \rangle$, and in this case the QSIC
%comprises two non-intersecting real lines and two non-intersecting
%imaginary lines.
%\end{theorem}

%\begin{proof}
{\bf Proof of Theorem \ref{thm:$[(11)(11)]_0$}}
By Theorem~\ref{thm:canonical_form},
\[
A' ={\rm diag} \left(\;\left(\begin{array}{cc} 0&1\\
1&0 \end{array} \right),\;
\left(\begin{array}{cc} 0&1\\
1&0 \end{array} \right)\;\right),\;\;
B' = {\rm diag} (-1, \; 1,\; -1,\; 1).
\]
Then ${\mathcal A'}$ is a hyperbolic paraboloid $xy+zw=0$ and
${\mathcal B'}$ is a hyperboloid $-x^2+y^2-z^2+w^2=0$.
It is easy to verify that the QSIC consists of the two
non-intersecting real lines defined by ($x+w=0$, $y-z=0$) and
($x-w=0$, $y+z=0$), and two non-intersecting imaginary lines defined
($x-iz=0$, $y-iw=0$) and ($x+iz=0$, $y+iw=0$).
In this case, since ${f}(\lambda)=0$ has no real root, the index
sequence is $\langle 2 \rangle$. This completes the proof of
Theorem~\ref{thm:$[(11)(11)]_0$}.
%\end{proof}

%\subsection{$[(211)]$: $f(\lambda)=0$ has one real quadruple root}
%
%\begin{theorem}\label{thm:$[(211)]$}
%If $f(\lambda)=0$ has a quadruple root with the Segre characteristic
%$[(211)]$, then the only possible index sequences are $\langle
%2{\wr\wr}_{-}||2 \rangle$ and $\langle 1{\wr\wr}_{-} || 3 \rangle$.
%%
%Furthermore,
%\begin{enumerate}
%\item (Case 32, Table 3) when the index sequence is $\langle 2{\wr\wr}_{-}||2 \rangle$, the QSIC consists
%of a pair of intersecting real lines, counted twice;
%%
%\item (Case 33, Table 3) when the index sequence is $\langle 1{\wr\wr}_{-} || 3 \rangle$,
% the QSIC consists of a pair of conjugate lines, counted twice.
%\end{enumerate}
%\end{theorem}

%\begin{proof}
{\bf Proof of Theorem \ref{thm:$[(211)]$}} The quadruple root,
denoted by $\lambda_1$, of $f(\lambda)=0$ is necessarily real. Wlog,
we may assume $\lambda_1=0$.
By Theorem~\ref{thm:canonical_form}, the canonical form of the two
given quadrics are
\[
A' ={\rm diag} \left(\;\left(\begin{array}{cc}  & \varepsilon_1\\
\varepsilon_1 & \end{array} \right),\; \varepsilon_2,\;
\varepsilon_3 \right),\;\;
B' ={\rm diag} (0,\; \varepsilon_1,\; 0,\; 0\;).
\]
The quadric ${\mathcal B'}$ is a pair of the identical real plane
$y=0$.
Substituting $y^2=0$ in the the quadric ${ \mathcal A'}$, we find
that the QSIC is the intersection between $y^2=0$ and the pair of
planes $z^2 = -\varepsilon_3/\varepsilon_2w^2$.
Clearly, the QSIC comprises two intersecting real lines, counted
twice, if $\varepsilon_2\varepsilon_3 =-1$, or a pair of conjugate
lines, counted twice, if  $\varepsilon_2\varepsilon_3 =1$.

Clearly, the index jump of ${\rm Id}(\lambda A' - B')$ at
$\lambda_1=0$ is 0 or $\pm 2$. When $\varepsilon_3/\varepsilon_2
=-1$, $\det(A)=-\varepsilon_2\varepsilon_3>0$, therefore ${\rm
index} (A')$ is 0, 2 or 4.  It follows that the index sequence is
$\langle 2{\wr\wr}_{-}||2 \rangle$ or its equivalent form $\langle
2{\wr\wr}_{+}||2 \rangle$.
When $\varepsilon_3/\varepsilon_2 =1$,
$\det(A)=-\varepsilon_2\varepsilon_3<0$, therefore ${\rm index}
(A')$ is 1 or 3.  It follows that the index sequence is $\langle
1{\wr\wr}_{-} || 3 \rangle$ or its equivalent form $\langle
1{\wr\wr}_{+} || 3 \rangle$. This completes the proof of
Theorem~\ref{thm:$[(211)]$}.
%\end{proof}

%\subsection{$[(22)]$: $f(\lambda)=0$ has one real quadruple root}
%
%\begin{theorem}\label{thm:$[(22)]$}
%If $f(\lambda)=0$ has a quadruple root with the Segre characteristic
%$[(22)]$, then the only possible index sequences are $\langle 2
%{\wr\wr}_-{\wr\wr}_- 2 \rangle$ and $\langle 2 {\wr\wr}_-{\wr\wr}_+
%2 \rangle$
%%
%Furthermore,
%\begin{enumerate}
% \item (Case 34, Table 3) when the index sequence is $\langle 2 {\wr\wr}_{-}{\wr\wr}_{-} 2 \rangle$,
% the QSIC consists a real double line and two other non-intersecting imaginary
% lines. The two imaginary lines do not form a complex conjugate
% pair.
% %
%\item (Case 35, Table 3)  when the index sequence is $\langle 2 {\wr\wr}_{-} {\wr\wr}_{+} 2 \rangle$,
% the QSIC consists a real double line and two other non-intersecting real
% lines. Each of the latter two lines intersects the real double
% line.
% \end{enumerate}
%\end{theorem}

%\begin{proof}
{\bf Proof of Theorem \ref{thm:$[(22)]$}} The quadruple root of
$f(\lambda)=0$ is necessarily real, and may be assume to be $0$.
Since the index jump at a real root with a $2\times 2$ Jordan block
is 0, the index sequence is of the form $\langle 2
\widehat{\wr\wr}\widehat{\wr\wr} 2 \rangle$.
By Theorem~\ref{thm:canonical_form},  the canonical form of the two
quadrics are
\[
A' ={\rm diag} \left( \left(\begin{array}{cc} 0&\varepsilon_1\\
\varepsilon_1&0 \end{array} \right),\; \left(\begin{array}{cc} &\varepsilon_2\\
\varepsilon_2& \end{array} \right)\; \right),\;\;\;
B' ={\rm diag} (0,\; \varepsilon_1,\; 0,\; \varepsilon_2\;).
\]
The quadric ${\mathcal A'}$ is the hyperbolic paraboloid
$$
\varepsilon_1xy+\varepsilon_2zw=0,
$$
and ${\mathcal B'}$ is a pair of planes
$$\varepsilon_1y^2+\varepsilon_2w^2=0.$$

When $\varepsilon_1\varepsilon_2=1$, the index sequence is $\langle
2 {\wr\wr}_{-}{\wr\wr}_{-} 2 \rangle$ or its equivalent form
$\langle 2 {\wr\wr}_{+}{\wr\wr}_{+} 2 \rangle$.
In this case ${\mathcal B'}$ comprises two conjugate planes $y+iw=0$
and $y-iw=0$, which intersects ${\mathcal A'}$ in the real double
line ($y=0$, $w=0$) and two non-intersecting imaginary lines
($x-iz=0$, $y-iw=0$) and ($x+iz=0$, $y+iw=0$).

When $\varepsilon_1\varepsilon_2=-1$, the index sequence is $\langle
2 {\wr\wr}_{-}{\wr\wr}_{+} 2 \rangle$ or its equivalent form
$\langle 2 {\wr\wr}_{+}{\wr\wr}_{-} 2 \rangle$. In this case
${\mathcal B'}$ comprises two real planes $y-w=0$ and $y+w=0$, which
intersects ${\mathcal A'}$ in the real double line ($y=0$, $w=0$)
and two non-intersecting real lines ($x-z=0$, $y-w=0$) and ($x+z=0$,
$y+w=0$).

This completes the proof of Theorem~\ref{thm:$[(22)]$}.
%\end{proof}
%\bibliographystyle{plain}
%\bibliography{paper}

\end{document}